\documentclass[12pt, twoside]{article}
\usepackage{amsmath, amsfonts, amscd, amssymb}
\usepackage{amsthm}
\usepackage{fancybox, eufrak, euscript, oldgerm}
\makeatletter
 
\def\diagram{\m@th\leftwidth=\z@ \rightwidth=\z@ \topheight=\z@
\botheight=\z@ \setbox\@picbox\hbox\bgroup}
 
\def\enddiagram{\egroup\wd\@picbox\rightwidth\unitlength
\ht\@picbox\topheight\unitlength \dp\@picbox\botheight\unitlength
\hskip\leftwidth\unitlength\box\@picbox}
 
\def\bfig{\begin{diagram}}
\def\efig{\end{diagram}}
\newcount\wideness \newcount\leftwidth \newcount\rightwidth
\newcount\highness \newcount\topheight \newcount\botheight
 
\def\ratchet#1#2{\ifnum#1<#2 \global #1=#2 \fi}
 
\def\putbox(#1,#2)#3{%
\horsize{\wideness}{#3} \divide\wideness by 2
{\advance\wideness by #1 \ratchet{\rightwidth}{\wideness}}
{\advance\wideness by -#1 \ratchet{\leftwidth}{\wideness}}
\vertsize{\highness}{#3} \divide\highness by 2
{\advance\highness by #2 \ratchet{\topheight}{\highness}}
{\advance\highness by -#2 \ratchet{\botheight}{\highness}}
\put(#1,#2){\makebox(0,0){$#3$}}}
 
\def\putlbox(#1,#2)#3{%
\horsize{\wideness}{#3}
{\advance\wideness by #1 \ratchet{\rightwidth}{\wideness}}
{\ratchet{\leftwidth}{-#1}}
\vertsize{\highness}{#3} \divide\highness by 2
{\advance\highness by #2 \ratchet{\topheight}{\highness}}
{\advance\highness by -#2 \ratchet{\botheight}{\highness}}
\put(#1,#2){\makebox(0,0)[l]{$#3$}}}
 
\def\putrbox(#1,#2)#3{%
\horsize{\wideness}{#3}
{\ratchet{\rightwidth}{#1}}
{\advance\wideness by -#1 \ratchet{\leftwidth}{\wideness}}
\vertsize{\highness}{#3} \divide\highness by 2
{\advance\highness by #2 \ratchet{\topheight}{\highness}}
{\advance\highness by -#2 \ratchet{\botheight}{\highness}}
\put(#1,#2){\makebox(0,0)[r]{$#3$}}}

\def\adjust[#1]{} 
 
\newcount \coefa
\newcount \coefb
\newcount \coefc
\newcount\tempcounta
\newcount\tempcountb
\newcount\tempcountc
\newcount\tempcountd
\newcount\xext
\newcount\yext
\newcount\xoff
\newcount\yoff
\newcount\gap%
\newcount\arrowtypea
\newcount\arrowtypeb
\newcount\arrowtypec
\newcount\arrowtyped
\newcount\arrowtypee
\newcount\height
\newcount\width
\newcount\xpos
\newcount\ypos
\newcount\run
\newcount\rise
\newcount\arrowlength
\newcount\halflength
\newcount\arrowtype
\newdimen\tempdimen
\newdimen\xlen
\newdimen\ylen
\newsavebox{\tempboxa}%
\newsavebox{\tempboxb}%
\newsavebox{\tempboxc}%
 
\newdimen\w@dth
 
\def\setw@dth#1#2{\setbox\z@\hbox{\m@th$#1$}\w@dth=\wd\z@
\setbox\@ne\hbox{\m@th$#2$}\ifnum\w@dth<\wd\@ne \w@dth=\wd\@ne \fi
\advance\w@dth by 1.2em}
 
\def\t@^#1_#2{\allowbreak\def\n@one{#1}\def\n@two{#2}\mathrel
{\setw@dth{#1}{#2}
\mathop{\hbox to \w@dth{\rightarrowfill}}\limits
\ifx\n@one\empty\else ^{\box\z@}\fi
\ifx\n@two\empty\else _{\box\@ne}\fi}}
\def\t@@^#1{\@ifnextchar_{\t@^{#1}}{\t@^{#1}_{}}}
\def\to{\@ifnextchar^{\t@@}{\t@@^{}}}
 
\def\t@left^#1_#2{\def\n@one{#1}\def\n@two{#2}\mathrel{\setw@dth{#1}{#2}
\mathop{\hbox to \w@dth{\leftarrowfill}}\limits
\ifx\n@one\empty\else ^{\box\z@}\fi
\ifx\n@two\empty\else _{\box\@ne}\fi}}
\def\t@@left^#1{\@ifnextchar_{\t@left^{#1}}{\t@left^{#1}_{}}}
\def\toleft{\@ifnextchar^{\t@@left}{\t@@left^{}}}
 
\def\two@^#1_#2{\allowbreak
\def\n@one{#1}\def\n@two{#2}\mathrel{\setw@dth{#1}{#2}
\mathop{\vcenter{\lineskip\z@\baselineskip\z@
                 \hbox to \w@dth{\rightarrowfill}%
                 \hbox to \w@dth{\rightarrowfill}}%
       }\limits
\ifx\n@one\empty\else ^{\box\z@}\fi
\ifx\n@two\empty\else _{\box\@ne}\fi}}
\def\tw@@^#1{\@ifnextchar _{\two@^{#1}}{\two@^{#1}_{}}}
\def\two{\@ifnextchar ^{\tw@@}{\tw@@^{}}}
 
\def\tofr@^#1_#2{\def\n@one{#1}\def\n@two{#2}\mathrel{\setw@dth{#1}{#2}
\mathop{\vcenter{\hbox to \w@dth{\rightarrowfill}\kern-1.7ex
                 \hbox to \w@dth{\leftarrowfill}}%
       }\limits
\ifx\n@one\empty\else ^{\box\z@}\fi
\ifx\n@two\empty\else _{\box\@ne}\fi}}
\def\t@fr@^#1{\@ifnextchar_ {\tofr@^{#1}}{\tofr@^{#1}_{}}}
\def\tofro{\@ifnextchar^ {\t@fr@}{\t@fr@^{}}}

\def\mon{\mathop{\m@th\hbox to
      14.6\P@{\lasyb\char'51\hskip-2.1\P@$\arrext$\hss
$\mathord\rightarrow$}}\limits} 
\def\leftmono{\mathrel{\m@th\hbox to
14.6\P@{$\mathord\leftarrow$\hss$\arrext$\hskip-2.1\P@\lasyb\char'50%
}}\limits} 
\mathchardef\arrext="0200       

\def\settypes(#1,#2,#3){\arrowtypea#1 \arrowtypeb#2 \arrowtypec#3}
\def\settoheight#1#2{\setbox\@tempboxa\hbox{#2}#1\ht\@tempboxa\relax}%
\def\settodepth#1#2{\setbox\@tempboxa\hbox{#2}#1\dp\@tempboxa\relax}%
\def\settokens`#1`#2`#3`#4`{%
     \def\tokena{#1}\def\tokenb{#2}\def\tokenc{#3}\def\tokend{#4}}
\def\setsqparms[#1`#2`#3`#4;#5`#6]{%
\arrowtypea #1
\arrowtypeb #2
\arrowtypec #3
\arrowtyped #4
\width #5
\height #6
}
\def\setpos(#1,#2){\xpos=#1 \ypos#2}

\def\settriparms[#1`#2`#3;#4]{\settripairparms[#1`#2`#3`1`1;#4]}%
 
\def\settripairparms[#1`#2`#3`#4`#5;#6]{%
\arrowtypea #1
\arrowtypeb #2
\arrowtypec #3
\arrowtyped #4
\arrowtypee #5
\width #6
\height #6
}
 
\def\resetparms{\settripairparms[1`1`1`1`1;500]\width 500}
 
\resetparms
 
\def\mvector(#1,#2)#3{
\put(0,0){\vector(#1,#2){#3}}%
\put(0,0){\vector(#1,#2){26}}%
}
\def\evector(#1,#2)#3{{
\arrowlength #3
\put(0,0){\vector(#1,#2){\arrowlength}}%
\advance \arrowlength by-30
\put(0,0){\vector(#1,#2){\arrowlength}}%
}}
 
\def\horsize#1#2{%
\settowidth{\tempdimen}{$#2$}%
#1=\tempdimen
\divide #1 by\unitlength
}
 
\def\vertsize#1#2{%
\settoheight{\tempdimen}{$#2$}%
#1=\tempdimen
\settodepth{\tempdimen}{$#2$}%
\advance #1 by\tempdimen
\divide #1 by\unitlength
}
 
\def\putvector(#1,#2)(#3,#4)#5#6{{%
\ifnum3<\arrowtype
\putdashvector(#1,#2)(#3,#4)#5\arrowtype
\else
\ifnum\arrowtype<-3
\putdashvector(#1,#2)(#3,#4)#5\arrowtype
\else
\xpos=#1
\ypos=#2
\run=#3
\rise=#4
\arrowlength=#5
\ifnum \arrowtype<0
    \ifnum \run=0
        \advance \ypos by-\arrowlength
    \else
        \tempcounta \arrowlength
        \multiply \tempcounta by\rise
        \divide \tempcounta by\run
        \ifnum\run>0
            \advance \xpos by\arrowlength
            \advance \ypos by\tempcounta
        \else
            \advance \xpos by-\arrowlength
            \advance \ypos by-\tempcounta
        \fi
    \fi
    \multiply \arrowtype by-1
    \multiply \rise by-1
    \multiply \run by-1
\fi
\ifcase \arrowtype
\or \put(\xpos,\ypos){\vector(\run,\rise){\arrowlength}}%
\or \put(\xpos,\ypos){\mvector(\run,\rise)\arrowlength}%
\or \put(\xpos,\ypos){\evector(\run,\rise){\arrowlength}}%
\fi\fi\fi
}}
 
\def\putsplitvector(#1,#2)#3#4{
\xpos #1
\ypos #2
\arrowtype #4
\halflength #3
\arrowlength #3
\gap 140
\advance \halflength by-\gap
\divide \halflength by2
\ifnum\arrowtype>0
   \ifcase \arrowtype
   \or \put(\xpos,\ypos){\line(0,-1){\halflength}}%
       \advance\ypos by-\halflength
       \advance\ypos by-\gap
       \put(\xpos,\ypos){\vector(0,-1){\halflength}}%
   \or \put(\xpos,\ypos){\line(0,-1)\halflength}%
       \put(\xpos,\ypos){\vector(0,-1)3}%
       \advance\ypos by-\halflength
       \advance\ypos by-\gap
       \put(\xpos,\ypos){\vector(0,-1){\halflength}}%
   \or \put(\xpos,\ypos){\line(0,-1)\halflength}%
       \advance\ypos by-\halflength
       \advance\ypos by-\gap
       \put(\xpos,\ypos){\evector(0,-1){\halflength}}%
   \fi
\else \arrowtype=-\arrowtype
   \ifcase\arrowtype
   \or \advance \ypos by-\arrowlength
       \put(\xpos,\ypos){\line(0,1){\halflength}}%
       \advance\ypos by\halflength
       \advance\ypos by\gap
       \put(\xpos,\ypos){\vector(0,1){\halflength}}%
   \or \advance \ypos by-\arrowlength
       \put(\xpos,\ypos){\line(0,1)\halflength}%
       \put(\xpos,\ypos){\vector(0,1)3}%
       \advance\ypos by\halflength
       \advance\ypos by\gap
       \put(\xpos,\ypos){\vector(0,1){\halflength}}%
   \or \advance \ypos by-\arrowlength
       \put(\xpos,\ypos){\line(0,1)\halflength}%
       \advance\ypos by\halflength
       \advance\ypos by\gap
       \put(\xpos,\ypos){\evector(0,1){\halflength}}%
   \fi
\fi
}
 
\def\putmorphism(#1)(#2,#3)[#4`#5`#6]#7#8#9{{%
\run #2
\rise #3
\ifnum\rise=0
  \puthmorphism(#1)[#4`#5`#6]{#7}{#8}#9%
\else\ifnum\run=0
  \putvmorphism(#1)[#4`#5`#6]{#7}{#8}#9%
\else
\setpos(#1)%
\arrowlength #7
\arrowtype #8
\ifnum\run=0
\else\ifnum\rise=0
\else
\ifnum\run>0
    \coefa=1
\else
   \coefa=-1
\fi
\ifnum\arrowtype>0
   \coefb=0
   \coefc=-1
\else
   \coefb=\coefa
   \coefc=1
   \arrowtype=-\arrowtype
\fi
\width=2
\multiply \width by\run
\divide \width by\rise
\ifnum \width<0  \width=-\width\fi
\advance\width by60
\if l#9 \width=-\width\fi
\putbox(\xpos,\ypos){#4}
{\multiply \coefa by\arrowlength
\advance\xpos by\coefa
\multiply \coefa by\rise
\divide \coefa by\run
\advance \ypos by\coefa
\putbox(\xpos,\ypos){#5} }%
{\multiply \coefa by\arrowlength
\divide \coefa by2
\advance \xpos by\coefa
\advance \xpos by\width
\multiply \coefa by\rise
\divide \coefa by\run
\advance \ypos by\coefa
\if l#9%
   \putrbox(\xpos,\ypos){#6}%
\else\if r#9%
   \putlbox(\xpos,\ypos){#6}%
\fi\fi }%
{\multiply \rise by-\coefc
\multiply \run by-\coefc
\multiply \coefb by\arrowlength
\advance \xpos by\coefb
\multiply \coefb by\rise
\divide \coefb by\run
\advance \ypos by\coefb
\multiply \coefc by70
\advance \ypos by\coefc
\multiply \coefc by\run
\divide \coefc by\rise
\advance \xpos by\coefc
\multiply \coefa by140
\multiply \coefa by\run
\divide \coefa by\rise
\advance \arrowlength by\coefa
\ifcase\arrowtype
\or \put(\xpos,\ypos){\vector(\run,\rise){\arrowlength}}%
\or \put(\xpos,\ypos){\mvector(\run,\rise){\arrowlength}}%
\or \put(\xpos,\ypos){\evector(\run,\rise){\arrowlength}}%
\fi}\fi\fi\fi\fi}}

\newcount\numbdashes \newcount\lengthdash \newcount\increment
 
\def\howmanydashes{
\numbdashes=\arrowlength \lengthdash=40
\divide\numbdashes by \lengthdash
\lengthdash=\arrowlength
\divide\lengthdash by \numbdashes
\increment=\lengthdash
\multiply\lengthdash by 3
\divide\lengthdash by 5
}
 
\def\putdashvector(#1)(#2,#3)#4#5{%
\ifnum#3=0 \putdashhvector(#1){#4}#5
\else
\ifnum#2=0
\putdashvvector(#1){#4}#5\fi\fi}
 
\def\putdashhvector(#1,#2)#3#4{{%
\arrowlength=#3 \howmanydashes
\multiput(#1,#2)(\increment,0){\numbdashes}%
{\vrule height .4pt width \lengthdash\unitlength}
\arrowtype=#4 \xpos=#1
\ifnum\arrowtype<0 \advance\arrowtype by 7 \fi
\ifcase\arrowtype
\or \advance\xpos by 10
    \put(\xpos,#2){\vector(-1,0){\lengthdash}}
    \advance\xpos by 40
    \put(\xpos,#2){\vector(-1,0){\lengthdash}}
\or \advance \xpos by 10
    \put(\xpos,#2){\vector(-1,0){\lengthdash}}
    \advance\xpos by  \arrowlength
    \advance\xpos by  -50
    \put(\xpos,#2){\vector(-1,0){\lengthdash}}
\or \advance\xpos by 10
    \put(\xpos,#2){\vector(-1,0){\lengthdash}}
\or \advance\xpos by \arrowlength
    \advance\xpos by -\lengthdash
    \put(\xpos,#2){\vector(1,0){\lengthdash}}
\or {\advance\xpos by 10
    \put(\xpos,#2){\vector(1,0){\lengthdash}}}
    \advance\xpos by \arrowlength
    \advance\xpos by -\lengthdash
    \put(\xpos,#2){\vector(1,0){\lengthdash}}
\or \advance\xpos by \arrowlength
    \advance\xpos by -\lengthdash
    \put(\xpos,#2){\vector(1,0){\lengthdash}}
    \advance\xpos by -40
    \put(\xpos,#2){\vector(1,0){\lengthdash}}
   \fi
}}
 
\def\putdashvvector(#1,#2)#3#4{{%
\arrowlength=#3 \howmanydashes
\ypos=#2 \advance\ypos by -\arrowlength
\multiput(#1,#2)(0,\increment){\numbdashes}%
    {\vrule width .4pt height \lengthdash\unitlength}
\arrowtype=#4 \ypos=#2
\ifnum\arrowtype<0 \advance\arrowtype by 7 \fi
\ifcase\arrowtype
\or \advance\ypos by \arrowlength \advance\ypos by -40
    \put(#1,\ypos){\vector(0,1){\lengthdash}}
    \advance\ypos by -40
    \put(#1,\ypos){\vector(0,1){\lengthdash}}
\or \advance\ypos by 10
    \put(#1,\ypos){\vector(0,1){\lengthdash}}
    \advance\ypos by \arrowlength \advance\ypos by -40
    \put(#1,\ypos){\vector(0,1){\lengthdash}}
\or \advance\ypos by \arrowlength \advance\ypos by -40
    \put(#1,\ypos){\vector(0,1){\lengthdash}}
\or \advance\ypos by 10
    \put(#1,\ypos){\vector(0,-1){\lengthdash}}
\or \advance\ypos by 10
    \put(#1,\ypos){\vector(0,-1){\lengthdash}}
    \advance\ypos by \arrowlength \advance\ypos by -40
    \put(#1,\ypos){\vector(0,-1){\lengthdash}}
\or \advance\ypos by 10
    \put(#1,\ypos){\vector(0,-1){\lengthdash}}
    \advance\ypos by 40
    \put(#1,\ypos){\vector(0,-1){\lengthdash}}
\fi
}}
 
\def\puthmorphism(#1,#2)[#3`#4`#5]#6#7#8{{%
\xpos #1
\ypos #2
\width #6
\arrowlength #6
\arrowtype=#7
\putbox(\xpos,\ypos){#3\vphantom{#4}}%
{\advance \xpos by\arrowlength
\putbox(\xpos,\ypos){\vphantom{#3}#4}}%
\horsize{\tempcounta}{#3}%
\horsize{\tempcountb}{#4}%
\divide \tempcounta by2
\divide \tempcountb by2
\advance \tempcounta by30
\advance \tempcountb by30
\advance \xpos by\tempcounta
\advance \arrowlength by-\tempcounta
\advance \arrowlength by-\tempcountb
\putvector(\xpos,\ypos)(1,0)\arrowlength\arrowtype
\divide \arrowlength by2
\advance \xpos by\arrowlength
\vertsize{\tempcounta}{#5}%
\divide\tempcounta by2
\advance \tempcounta by20
\if a#8 %
   \advance \ypos by\tempcounta
   \putbox(\xpos,\ypos){#5}%
\else
   \advance \ypos by-\tempcounta
   \putbox(\xpos,\ypos){#5}%
\fi}}
 
\def\putvmorphism(#1,#2)[#3`#4`#5]#6#7#8{{%
\xpos #1
\ypos #2
\arrowlength #6
\arrowtype #7
\settowidth{\xlen}{$#5$}%
\putbox(\xpos,\ypos){#3}%
{\advance \ypos by-\arrowlength
\putbox(\xpos,\ypos){#4}}%
{\advance\arrowlength by-140
\advance \ypos by-70
\ifdim\xlen>0pt
   \if m#8%
      \putsplitvector(\xpos,\ypos)\arrowlength\arrowtype
   \else
   \putvector(\xpos,\ypos)(0,-1)\arrowlength\arrowtype
   \fi
\else
   \putvector(\xpos,\ypos)(0,-1)\arrowlength\arrowtype
\fi}%
\ifdim\xlen>0pt
   \divide \arrowlength by2
   \advance\ypos by-\arrowlength
   \if l#8%
      \advance \xpos by-40
      \putrbox(\xpos,\ypos){#5}%
   \else\if r#8%
      \advance \xpos by40
      \putlbox(\xpos,\ypos){#5}%
   \else
      \putbox(\xpos,\ypos){#5}%
   \fi\fi
\fi
}}
 
\def\putsquarep<#1>(#2)[#3;#4`#5`#6`#7]{{%
\setsqparms[#1]%
\setpos(#2)%
\settokens`#3`%
\puthmorphism(\xpos,\ypos)[\tokenc`\tokend`{#7}]{\width}{\arrowtyped}b%
\advance\ypos by \height
\puthmorphism(\xpos,\ypos)[\tokena`\tokenb`{#4}]{\width}{\arrowtypea}a%
\putvmorphism(\xpos,\ypos)[``{#5}]{\height}{\arrowtypeb}l%
\advance\xpos by \width
\putvmorphism(\xpos,\ypos)[``{#6}]{\height}{\arrowtypec}r%
}}
 
\def\putsquare{\@ifnextchar <{\putsquarep}{\putsquarep%
   <\arrowtypea`\arrowtypeb`\arrowtypec`\arrowtyped;\width`\height>}}
\def\square{\@ifnextchar< {\squarep}{\squarep
   <\arrowtypea`\arrowtypeb`\arrowtypec`\arrowtyped;\width`\height>}}
\def\squarep<#1>[#2`#3`#4`#5;#6`#7`#8`#9]{{
\setsqparms[#1]
\diagram
\putsquarep<\arrowtypea`\arrowtypeb`\arrowtypec`
\arrowtyped;\width`\height>
(0,0)[#2`#3`#4`{#5};#6`#7`#8`{#9}]
\enddiagram
}}                                                 
\def\putptrianglep<#1>(#2,#3)[#4`#5`#6;#7`#8`#9]{{%
\settriparms[#1]%
\xpos=#2 \ypos=#3
\advance\ypos by \height
\puthmorphism(\xpos,\ypos)[#4`#5`{#7}]{\height}{\arrowtypea}a%
\putvmorphism(\xpos,\ypos)[`#6`{#8}]{\height}{\arrowtypeb}l%
\advance\xpos by\height
\putmorphism(\xpos,\ypos)(-1,-1)[``{#9}]{\height}{\arrowtypec}r%
}}
 
\def\putptriangle{\@ifnextchar <{\putptrianglep}{\putptrianglep
   <\arrowtypea`\arrowtypeb`\arrowtypec;\height>}}
\def\ptriangle{\@ifnextchar <{\ptrianglep}{\ptrianglep
   <\arrowtypea`\arrowtypeb`\arrowtypec;\height>}}
\def\ptrianglep<#1>[#2`#3`#4;#5`#6`#7]{{
\settriparms[#1]
\diagram
\putptrianglep<\arrowtypea`\arrowtypeb`
\arrowtypec;\height>
(0,0)[#2`#3`#4;#5`#6`{#7}]
\enddiagram
}}                                            
 
\def\putqtrianglep<#1>(#2,#3)[#4`#5`#6;#7`#8`#9]{{%
\settriparms[#1]%
\xpos=#2 \ypos=#3
\advance\ypos by\height
\puthmorphism(\xpos,\ypos)[#4`#5`{#7}]{\height}{\arrowtypea}a%
\putmorphism(\xpos,\ypos)(1,-1)[``{#8}]{\height}{\arrowtypeb}l%
\advance\xpos by\height
\putvmorphism(\xpos,\ypos)[`#6`{#9}]{\height}{\arrowtypec}r%
}}
 
\def\putqtriangle{\@ifnextchar <{\putqtrianglep}{\putqtrianglep
   <\arrowtypea`\arrowtypeb`\arrowtypec;\height>}}
\def\qtriangle{\@ifnextchar <{\qtrianglep}{\qtrianglep
   <\arrowtypea`\arrowtypeb`\arrowtypec;\height>}}
\def\qtrianglep<#1>[#2`#3`#4;#5`#6`#7]{{
\settriparms[#1]
\width=\height                                
\diagram
\putqtrianglep<\arrowtypea`\arrowtypeb`
\arrowtypec;\height>
(0,0)[#2`#3`#4;#5`#6`{#7}]
\enddiagram
}}
 
\def\putdtrianglep<#1>(#2,#3)[#4`#5`#6;#7`#8`#9]{{%
\settriparms[#1]%
\xpos=#2 \ypos=#3
\puthmorphism(\xpos,\ypos)[#5`#6`{#9}]{\height}{\arrowtypec}b%
\advance\xpos by \height \advance\ypos by\height
\putmorphism(\xpos,\ypos)(-1,-1)[``{#7}]{\height}{\arrowtypea}l%
\putvmorphism(\xpos,\ypos)[#4``{#8}]{\height}{\arrowtypeb}r%
}}
 
\def\putdtriangle{\@ifnextchar <{\putdtrianglep}{\putdtrianglep
   <\arrowtypea`\arrowtypeb`\arrowtypec;\height>}}
\def\dtriangle{\@ifnextchar <{\dtrianglep}{\dtrianglep
   <\arrowtypea`\arrowtypeb`\arrowtypec;\height>}}
\def\dtrianglep<#1>[#2`#3`#4;#5`#6`#7]{{
\settriparms[#1]
\width=\height                                
\diagram
\putdtrianglep<\arrowtypea`\arrowtypeb`
\arrowtypec;\height>
(0,0)[#2`#3`#4;#5`#6`{#7}]
\enddiagram
}}
 
\def\putbtrianglep<#1>(#2,#3)[#4`#5`#6;#7`#8`#9]{{%
\settriparms[#1]%
\xpos=#2 \ypos=#3
\puthmorphism(\xpos,\ypos)[#5`#6`{#9}]{\height}{\arrowtypec}b%
\advance\ypos by\height
\putmorphism(\xpos,\ypos)(1,-1)[``{#8}]{\height}{\arrowtypeb}r%
\putvmorphism(\xpos,\ypos)[#4``{#7}]{\height}{\arrowtypea}l%
}}
 
\def\putbtriangle{\@ifnextchar <{\putbtrianglep}{\putbtrianglep
   <\arrowtypea`\arrowtypeb`\arrowtypec;\height>}}
\def\btriangle{\@ifnextchar <{\btrianglep}{\btrianglep
   <\arrowtypea`\arrowtypeb`\arrowtypec;\height>}}
\def\btrianglep<#1>[#2`#3`#4;#5`#6`#7]{{
\settriparms[#1]
\width=\height                               
\diagram
\putbtrianglep<\arrowtypea`\arrowtypeb`
\arrowtypec;\height>
(0,0)[#2`#3`#4;#5`#6`{#7}]
\enddiagram
}}
 
\def\putAtrianglep<#1>(#2,#3)[#4`#5`#6;#7`#8`#9]{{%
\settriparms[#1]%
\xpos=#2 \ypos=#3
{\multiply \height by2
\puthmorphism(\xpos,\ypos)[#5`#6`{#9}]{\height}{\arrowtypec}b}%
\advance\xpos by\height \advance\ypos by\height
\putmorphism(\xpos,\ypos)(-1,-1)[#4``{#7}]{\height}{\arrowtypea}l%
\putmorphism(\xpos,\ypos)(1,-1)[``{#8}]{\height}{\arrowtypeb}r%
}}
 
\def\putAtriangle{\@ifnextchar <{\putAtrianglep}{\putAtrianglep
   <\arrowtypea`\arrowtypeb`\arrowtypec;\height>}}
\def\Atriangle{\@ifnextchar <{\Atrianglep}{\Atrianglep
   <\arrowtypea`\arrowtypeb`\arrowtypec;\height>}}
\def\Atrianglep<#1>[#2`#3`#4;#5`#6`#7]{{
\settriparms[#1]
\width=\height                                     
\diagram
\putAtrianglep<\arrowtypea`\arrowtypeb`
\arrowtypec;\height>
(0,0)[#2`#3`#4;#5`#6`{#7}]
\enddiagram
}}
 
\def\putAtrianglepairp<#1>(#2)[#3;#4`#5`#6`#7`#8]{{%
\settripairparms[#1]%
\setpos(#2)%
\settokens`#3`%
\puthmorphism(\xpos,\ypos)[\tokenb`\tokenc`{#7}]{\height}{\arrowtyped}b%
\advance\xpos by\height
\puthmorphism(\xpos,\ypos)[\phantom{\tokenc}`\tokend`{#8}]%
{\height}{\arrowtypee}b%
\advance\ypos by\height
\putmorphism(\xpos,\ypos)(-1,-1)[\tokena``{#4}]{\height}{\arrowtypea}l%
\putvmorphism(\xpos,\ypos)[``{#5}]{\height}{\arrowtypeb}m%
\putmorphism(\xpos,\ypos)(1,-1)[``{#6}]{\height}{\arrowtypec}r%
}}
 
\def\putAtrianglepair{\@ifnextchar <{\putAtrianglepairp}{\putAtrianglepairp%
   <\arrowtypea`\arrowtypeb`\arrowtypec`\arrowtyped`\arrowtypee;\height>}}
\def\Atrianglepair{\@ifnextchar <{\Atrianglepairp}{\Atrianglepairp%
   <\arrowtypea`\arrowtypeb`\arrowtypec`\arrowtyped`\arrowtypee;\height>}}
 
\def\Atrianglepairp<#1>[#2;#3`#4`#5`#6`#7]{{
\settripairparms[#1]
\settokens`#2`
\width=\height                                
\diagram
\putAtrianglepairp                            
<\arrowtypea`\arrowtypeb`\arrowtypec`
\arrowtyped`\arrowtypee;\height>
(0,0)[{#2};#3`#4`#5`#6`{#7}]
\enddiagram
}}
 
\def\putVtrianglep<#1>(#2,#3)[#4`#5`#6;#7`#8`#9]{{%
\settriparms[#1]%
\xpos=#2 \ypos=#3
\advance\ypos by\height
{\multiply\height by2
\puthmorphism(\xpos,\ypos)[#4`#5`{#7}]{\height}{\arrowtypea}a}%
\putmorphism(\xpos,\ypos)(1,-1)[`#6`{#8}]{\height}{\arrowtypeb}l%
\advance\xpos by\height
\advance\xpos by\height
\putmorphism(\xpos,\ypos)(-1,-1)[``{#9}]{\height}{\arrowtypec}r%
}}
 
\def\putVtriangle{\@ifnextchar <{\putVtrianglep}{\putVtrianglep
   <\arrowtypea`\arrowtypeb`\arrowtypec;\height>}}
\def\Vtriangle{\@ifnextchar <{\Vtrianglep}{\Vtrianglep
   <\arrowtypea`\arrowtypeb`\arrowtypec;\height>}}
\def\Vtrianglep<#1>[#2`#3`#4;#5`#6`#7]{{
\settriparms[#1]
\width=\height                                 
\diagram
\putVtrianglep<\arrowtypea`\arrowtypeb`
\arrowtypec;\height>
(0,0)[#2`#3`#4;#5`#6`{#7}]
\enddiagram
}}
 
\def\putVtrianglepairp<#1>(#2)[#3;#4`#5`#6`#7`#8]{{
\settripairparms[#1]%
\setpos(#2)%
\settokens`#3`%
\advance\ypos by\height
\putmorphism(\xpos,\ypos)(1,-1)[`\tokend`{#6}]{\height}{\arrowtypec}l%
\puthmorphism(\xpos,\ypos)[\tokena`\tokenb`{#4}]{\height}{\arrowtypea}a%
\advance\xpos by\height
\puthmorphism(\xpos,\ypos)[\phantom{\tokenb}`\tokenc`{#5}]%
{\height}{\arrowtypeb}a%
\putvmorphism(\xpos,\ypos)[``{#7}]{\height}{\arrowtyped}m%
\advance\xpos by\height
\putmorphism(\xpos,\ypos)(-1,-1)[``{#8}]{\height}{\arrowtypee}r%
}}
 
\def\putVtrianglepair{\@ifnextchar <{\putVtrianglepairp}{\putVtrianglepairp%
    <\arrowtypea`\arrowtypeb`\arrowtypec`\arrowtyped`\arrowtypee;\height>}}
\def\Vtrianglepair{\@ifnextchar <{\Vtrianglepairp}{\Vtrianglepairp%
    <\arrowtypea`\arrowtypeb`\arrowtypec`\arrowtyped`\arrowtypee;\height>}}
\def\Vtrianglepairp<#1>[#2;#3`#4`#5`#6`#7]{{
\settripairparms[#1]
\settokens`#2`
\diagram
\putVtrianglepairp                             
<\arrowtypea`\arrowtypeb`\arrowtypec`
\arrowtyped`\arrowtypee;\height>
(0,0)[{#2};#3`#4`#5`#6`{#7}]
\enddiagram
}}

\def\putCtrianglep<#1>(#2,#3)[#4`#5`#6;#7`#8`#9]{{%
\settriparms[#1]%
\xpos=#2 \ypos=#3
\advance\ypos by\height
\putmorphism(\xpos,\ypos)(1,-1)[``{#9}]{\height}{\arrowtypec}l%
\advance\xpos by\height
\advance\ypos by\height
\putmorphism(\xpos,\ypos)(-1,-1)[#4`#5`{#7}]{\height}{\arrowtypea}l%
{\multiply\height by 2
\putvmorphism(\xpos,\ypos)[`#6`{#8}]{\height}{\arrowtypeb}r}%
}}
 
\def\putCtriangle{\@ifnextchar <{\putCtrianglep}{\putCtrianglep
    <\arrowtypea`\arrowtypeb`\arrowtypec;\height>}}
\def\Ctriangle{\@ifnextchar <{\Ctrianglep}{\Ctrianglep
    <\arrowtypea`\arrowtypeb`\arrowtypec;\height>}}
\def\Ctrianglep<#1>[#2`#3`#4;#5`#6`#7]{{
\settriparms[#1]
\width=\height                               
\diagram
\putCtrianglep<\arrowtypea`\arrowtypeb`
\arrowtypec;\height>
(0,0)[#2`#3`#4;#5`#6`{#7}]
\enddiagram
}}                                           
\def\putDtrianglep<#1>(#2,#3)[#4`#5`#6;#7`#8`#9]{{%
\settriparms[#1]%
\xpos=#2 \ypos=#3
\advance\xpos by\height \advance\ypos by\height
\putmorphism(\xpos,\ypos)(-1,-1)[``{#9}]{\height}{\arrowtypec}r%
\advance\xpos by-\height \advance\ypos by\height
\putmorphism(\xpos,\ypos)(1,-1)[`#5`{#8}]{\height}{\arrowtypeb}r%
{\multiply\height by 2
\putvmorphism(\xpos,\ypos)[#4`#6`{#7}]{\height}{\arrowtypea}l}%
}}
 
\def\putDtriangle{\@ifnextchar <{\putDtrianglep}{\putDtrianglep
    <\arrowtypea`\arrowtypeb`\arrowtypec;\height>}}
\def\Dtriangle{\@ifnextchar <{\Dtrianglep}{\Dtrianglep
   <\arrowtypea`\arrowtypeb`\arrowtypec;\height>}}
\def\Dtrianglep<#1>[#2`#3`#4;#5`#6`#7]{{
\settriparms[#1]
\width=\height                              
\diagram
\putDtrianglep<\arrowtypea`\arrowtypeb`
\arrowtypec;\height>
(0,0)[#2`#3`#4;#5`#6`{#7}]
\enddiagram
}}                                          
\def\setrecparms[#1`#2]{\width=#1 \height=#2}%
 
\def\recursep<#1`#2>[#3;#4`#5`#6`#7`#8]{{\m@th
\width=#1 \height=#2
\settokens`#3`
\settowidth{\tempdimen}{$\tokena$}
\ifdim\tempdimen=0pt
  \savebox{\tempboxa}{\hbox{$\tokenb$}}%
  \savebox{\tempboxb}{\hbox{$\tokend$}}%
  \savebox{\tempboxc}{\hbox{$#6$}}%
\else
  \savebox{\tempboxa}{\hbox{$\hbox{$\tokena$}\times\hbox{$\tokenb$}$}}%
  \savebox{\tempboxb}{\hbox{$\hbox{$\tokena$}\times\hbox{$\tokend$}$}}%
  \savebox{\tempboxc}{\hbox{$\hbox{$\tokena$}\times\hbox{$#6$}$}}%
\fi
\ypos=\height
\divide\ypos by 2
\xpos=\ypos
\advance\xpos by \width
\bfig
\putCtrianglep<-1`1`1;\ypos>(0,0)[`\tokenc`;#5`#6`{#7}]%
\puthmorphism(\ypos,0)[\tokend`\usebox{\tempboxb}`{#8}]{\width}{-1}b%
\puthmorphism(\ypos,\height)[\tokenb`\usebox{\tempboxa}`{#4}]{\width}{-1}a%
\advance\ypos by \width
\putvmorphism(\ypos,\height)[``\usebox{\tempboxc}]{\height}1r%
\efig
}}
 
\def\recurse{\@ifnextchar <{\recursep}{\recursep<\width`\height>}}
 
\def\puttwohmorphisms(#1,#2)[#3`#4;#5`#6]#7#8#9{{%
\puthmorphism(#1,#2)[#3`#4`]{#7}0a
\ypos=#2
\advance\ypos by 20
\puthmorphism(#1,\ypos)[\phantom{#3}`\phantom{#4}`#5]{#7}{#8}a
\advance\ypos by -40
\puthmorphism(#1,\ypos)[\phantom{#3}`\phantom{#4}`#6]{#7}{#9}b
}}
 
\def\puttwovmorphisms(#1,#2)[#3`#4;#5`#6]#7#8#9{{%
\putvmorphism(#1,#2)[#3`#4`]{#7}0a
\xpos=#1
\advance\xpos by -20
\putvmorphism(\xpos,#2)[\phantom{#3}`\phantom{#4}`#5]{#7}{#8}l
\advance\xpos by 40
\putvmorphism(\xpos,#2)[\phantom{#3}`\phantom{#4}`#6]{#7}{#9}r
}}
 
\def\puthcoequalizer(#1)[#2`#3`#4;#5`#6`#7]#8#9{{%
\setpos(#1)%
\puttwohmorphisms(\xpos,\ypos)[#2`#3;#5`#6]{#8}11%
\advance\xpos by #8
\puthmorphism(\xpos,\ypos)[\phantom{#3}`#4`#7]{#8}1{#9}
}}
 
\def\putvcoequalizer(#1)[#2`#3`#4;#5`#6`#7]#8#9{{%
\setpos(#1)%
\puttwovmorphisms(\xpos,\ypos)[#2`#3;#5`#6]{#8}11%
\advance\ypos by -#8
\putvmorphism(\xpos,\ypos)[\phantom{#3}`#4`#7]{#8}1{#9}
}}
 
\def\putthreehmorphisms(#1)[#2`#3;#4`#5`#6]#7(#8)#9{{%
\setpos(#1) \settypes(#8)
\if a#9 %
     \vertsize{\tempcounta}{#5}%
     \vertsize{\tempcountb}{#6}%
     \ifnum \tempcounta<\tempcountb \tempcounta=\tempcountb \fi
\else
     \vertsize{\tempcounta}{#4}%
     \vertsize{\tempcountb}{#5}%
     \ifnum \tempcounta<\tempcountb \tempcounta=\tempcountb \fi
\fi
\advance \tempcounta by 60
\puthmorphism(\xpos,\ypos)[#2`#3`#5]{#7}{\arrowtypeb}{#9}
\advance\ypos by \tempcounta
\puthmorphism(\xpos,\ypos)[\phantom{#2}`\phantom{#3}`#4]{#7}{\arrowtypea}{#9}
\advance\ypos by -\tempcounta \advance\ypos by -\tempcounta
\puthmorphism(\xpos,\ypos)[\phantom{#2}`\phantom{#3}`#6]{#7}{\arrowtypec}{#9}
}}
 
\def\setarrowtoks[#1`#2`#3`#4`#5`#6]{%
\def\toka{#1}
\def\tokb{#2}
\def\tokc{#3}
\def\tokd{#4}
\def\toke{#5}
\def\tokf{#6}
}
\def\hex{\@ifnextchar <{\hexp}{\hexp<1000`400>}}
\def\hexp<#1`#2>[#3`#4`#5`#6`#7`#8;#9]{%
\setarrowtoks[#9]
\yext=#2 \advance \yext by #2
\xext=#1 \advance\xext by \yext
\bfig
\putCtriangle<-1`0`1;#2>(0,0)[`#5`;\tokb``\tokd]
\xext=#1 \yext=#2 \advance \yext by #2
\putsquare<1`0`0`1;\xext`\yext>(#2,0)[#3`#4`#7`#8;\toka```\tokf]
\advance \xext by #2
\putDtriangle<0`1`-1;#2>(\xext,0)[`#6`;`\tokc`\toke]
\efig
}
\makeatother

\newcommand{\aconn}{\mathcal{A}}
\newcommand{\qaconn}{\vec{\aconn}}
\newcommand{\as}{\mathfrak{P}}
\newcommand{\ashlew}{\overleftarrow{{\mathcal{M}}}}
\newcommand{\rz}{\mathfrak{Z}}

\newcommand{\aut}{{\mathcal{A}}ut}
\newcommand{\qaut}{\overrightarrow{\aut}}
\newcommand{\con}{\EuScript{D}}
\newcommand{\conn}{\mathcal{D}}
\newcommand{\conne}{\mathfrak{D}}
\newcommand{\conf}{\vec{\con}}

\newcommand{\caus}{\vec{P}}
\newcommand{\qaus}{\vec{\omg}}
\newcommand{\cons}{\mathbf{C}}

\newcommand{\curv}{R}
\newcommand{\ric}{{\mathcal{R}}}
\newcommand{\fric}{\vec{\ric}}
\newcommand{\ricci}{\EuScript{R}}
\newcommand{\fricci}{\vec{\ricci}}
\newcommand{\gelsp}{\mathfrak{M}}

\newcommand{\eh}{\mathfrak{E}\mathfrak{H}}
\newcommand{\qeh}{\overrightarrow{\eh}}
\newcommand{\fcurv}{\vec{\curv}}
\newcommand{\gauge}{\mathcal{U}}
\newcommand{\qgauge}{\vec{\gauge}}
\newcommand{\gl}{\mathcal{G}\mathcal{L}}
\newcommand{\gel}{\mathfrak{C}}
\newcommand{\gelt}{\mathfrak{T}}
\newcommand{\hil}{{\mathcal{H}}}
\newcommand{\Hom}{{\mathcal{H}}om}
\newcommand{\kd}{\text{\texttt{d}}}

\newcommand{\mink}{{\mathcal{M}}}
\newcommand{\inv}{\overleftarrow{\EuScript{P}}}
\newcommand{\inveinst}{\overleftarrow{\EuScript{E}}}
\newcommand{\invmod}{\overleftarrow{\EuScript{M}}}
\newcommand{\invs}{\overleftarrow{\EuScript{S}}}
\newcommand{\invsconn}{\overleftarrow{\mathfrak{A}}}
\newcommand{\invel}{\overleftarrow{\mathfrak{E}}}
\newcommand{\inveh}{\overleftarrow{\EuScript{EH}}}
\newcommand{\invcq}{\overleftarrow{\EuScript{Z}}}
\newcommand{\invtriad}{\stackrel{\rightleftarrows}{\EuScript{T}}}
\newcommand{\finv}{\overleftarrow{{\EuScript{G}}}}
\newcommand{\diromg}{\overrightarrow{\mathfrak{R}}}
\newcommand{\lie}{\EuScript{L}}
\newcommand{\lsh}{{\mathcal{L}}}
\newcommand{\grouv}{{\mathcal{G}}}

\newcommand{\modl}{\mathbf{\mathcal{E}}}
\newcommand{\modll}{\mathbf{\mathcal{E}}^{\uparrow}}
\newcommand{\qmod}{\vec{\mathcal{M}}}
\newcommand{\qmodl}{\vec{\modl}}
\newcommand{\qmodll}{\qmodl^{\uparrow}}
\newcommand{\qmoddll}{\qmodl^{\uparrow *}}
\newcommand{\qrho}{\vec{\rho}}

\newcommand{\omg}{\Omega}
\newcommand{\Omg}{\mathbf{\Omega}}
\newcommand{\Qaus}{{\vec{\Omg}}}
\newcommand{\orthl}{{\mathfrak{L}}^{+}}
\newcommand{\princ}{{\mathcal{P}}}
\newcommand{\pee}{\vec{\princ}}
\newcommand{\peel}{\vec{\princ}^{\uparrow}}

\newcommand{\cont}{\mathcal{C}^{0}}
\newcommand{\smooth}{\mathcal{C}^{\infty}}
\newcommand{\sconn}{\textsf{A}}
\newcommand{\fconn}{\vec{\EuScript{A}}}
\newcommand{\striad}{\large\texttt{T}_{\infty}}
\newcommand{\infconn}{\EuScript{A}_{\infty}}
\newcommand{\ssmooth}{\EuScript{C}^{\infty}}
\newcommand{\stre}{{\mathcal{F}}}
\newcommand{\struc}{\mathbf{A}}
\newcommand{\struct}{\mathcal{G}}
\newcommand{\triad}{{\mathfrak{T}}}
\newcommand{\ctriad}{{\mathfrak{D}}\triad}
\newcommand{\sstriad}{\triad_{\infty}}
\newcommand{\qstruct}{\vec{\struct}}
\newcommand{\vol}{\varpi}
\newcommand{\wee}{\,{\scriptstyle\wedge}\,}
\newcommand{\bull}{{\scriptstyle\bullet}}
\newcommand{\ym}{\mathfrak{Y}\mathfrak{M}}
\newcommand{\com}{\mathbb{C}}
\newcommand{\mapto}{\longrightarrow}

\newcommand{\PI}{{\mathcal{Z}}}
\newcommand{\QPI}{\vec{\PI}_{i}}
\newcommand{\alg}{\mathbb{A}}
\newcommand{\qalg}{\vec{\alg}}
\newcommand{\ring}{\mathbb{D}}
\newcommand{\qring}{\vec{\ring}}
\newcommand{\Qalg}{\vec{\mathbf{A}}}
\newcommand{\Qring}{\vec{\mathbf{D}}}
\newcommand{\N}{\mathbb{N}}
\newcommand{\Z}{\mathbb{Z}}
\newcommand{\R}{\mathbb{R}}
\newcommand{\K}{\mathbb{K}}

\title{\bf Finitary, Causal and Quantal\\ Vacuum Einstein Gravity}

\author{Anastasios Mallios\thanks{Algebra and Geometry Section,
Department of Mathematics, University of Athens,
Panepistimioupolis Zografou 157 84, Athens, Greece; e-mail:
amallios@cc.uoa.gr} and Ioannis Raptis\thanks{Theoretical Physics
Group, Blackett Laboratory, Imperial College of Science,
Technology and Medicine, Prince Consort Road, South Kensington,
London SW7 2BZ, UK; e-mail: i.raptis@ic.ac.uk}}

\date{}

\begin{document}

{\catcode`\ =13\global\let =\ \catcode`\^^M=13
\gdef^^M{\par\noindent}}
\def\verbatim{\tt
\catcode`\^^M=13 \catcode`\ =13 \catcode`\\=12 \catcode`\{=12
\catcode`\}=12 \catcode`\_=12 \catcode`\^=12 \catcode`\&=12
\catcode`\~=12 \catcode`\#=12 \catcode`\%=12 \catcode`\$=12
\catcode`|=0 }

\maketitle

\pagestyle{myheadings}\markboth{\centerline {\small {\sc
{Anastasios Mallios and Ioannis Raptis}}}}{\centerline
{\footnotesize {\sc {Finitary, Causal and Quantal Vacuum Einstein
Gravity}}}}

\pagenumbering{arabic}

\begin{abstract}

\noindent{\small We continue recent work \cite{malrap1,malrap2}
and formulate the gravitational vacuum Einstein equations over a
locally finite spacetime by using the basic axiomatics,
techniques, ideas and working philosophy of Abstract Differential
Geometry. The main kinematical structure involved, originally
introduced and explored in \cite{malrap1}, is a curved principal
finitary spacetime sheaf of incidence algebras, which have been
interpreted as quantum causal sets, together with a non-trivial
locally finite spin-Loretzian connection on it which lays the
structural foundation for the formulation of a covariant dynamics
of quantum causality in terms of sheaf morphisms. Our scheme is
innately algebraic and it supports a categorical version of the
principle of general covariance that is manifestly independent of
a background $\smooth$-smooth spacetime manifold $M$. Thus, we
entertain the possibility of developing a `fully covariant' path
integral-type of quantum dynamical scenario for these connections
that avoids {\it ab initio} various problems that such a dynamics
encounters in other current quantization schemes for
gravity---either canonical (Hamiltonian), or covariant
(Lagrangian)---involving an external, base differential spacetime
manifold, namely, the choice of a diffeomorphism-invariant measure
on the moduli space of gauge-equivalent (self-dual) gravitational
spin-Lorentzian connections and the (Hilbert space) inner product
that could in principle be constructed relative to that measure in
the quantum theory---the so-called `inner product problem', as
well as the `problem of time' that also involves the
$\mathrm{Diff}(M)$ `structure group' of the classical
$\smooth$-smooth spacetime continuum of general relativity. Hence,
by using the inherently algebraico-sheaf-theoretic and
calculus-free ideas of Abstract Differential Geometry, we are able
to draw preliminary, albeit suggestive, connections between
certain non-perturbative (canonical or covariant) approaches to
quantum general relativity ({\it eg}, Ashtekar's new variables and
the loop formalism that has been developed along with them) and
Sorkin {\it et al.}'s causal set program---as it were, we
`noncommutatively algebraize', `differential geometrize' and, as a
result, dynamically vary causal sets. At the end, we anticipate
various consequences that such a scenario for a locally finite,
causal and quantal vacuum Einstein gravity might have for the
obstinate from the viewpoint of the smooth continuum problem of
$\smooth$-smooth spacetime singularities, thus we prepare the
ground for a forthcoming paper \cite{malrap3}.}

\vskip 0.1in



\end{abstract}

\newpage

\setlength{\textwidth}{15.9cm} 
\setlength{\oddsidemargin}{0cm}  
\setlength{\evensidemargin}{0cm} 
\setlength{\topskip}{0pt}  
\setlength{\textheight}{21.6cm} 

\setlength{\footskip}{-2cm}

\setlength{\topmargin}{0pt}

\begin{quotation}
\noindent ``{\small{\em ...the theory that space is continuous is
wrong, because we get...infinities} [viz. `singularities'] {\em
and other similar difficulties} ...[while] {\em the simple ideas
of geometry, extended down to infinitely small, are wrong...}}''
\cite{feyn1}
\end{quotation}

\vskip 0.05in

\centerline{..................................}

\vskip 0.05in

\begin{quotation}
\noindent ``{\small{\em ...at the Planck-length scale, classical
differential geometry is simply incompatible with quantum
theory}...[so that] {\em one will not be able to use differential
geometry in the true quantum-gravity theory...}}'' \cite{ish}
\end{quotation}

\vskip 0.1in

\section{Prologue cum Physical Motivation}

In the last century, the path that we have followed to unite
quantum mechanics with general relativity into a coherent, both
technically and conceptually, quantum theory of gravity has been a
long and arduous one, full of unexpected twists and turns,
surprising detours, branchings and loops---even disheartening
setbacks and impasses, as well as hopes, disappointments or even
disillusionments at times. Certainly though, the whole enterprize
has been supported and nurtured by impressive technical ingenuity,
and creative imagination coming from physicists and mathematicians
alike. All in all, it has been a trip of adventure, discovery and
intellectual reward for all who have been privileged to be
involved in this formidable quest. Arguably then, the attempt to
arrive at a conceptually sound and `calculationally' finite
quantum gravity must be regarded and hailed as one of the most
challenging and inspired endeavors in theoretical physics research
that must be carried over and be zestfully continued in the new
millenium.

Admittedly, however, a cogent theoretical scenario for quantum
gravity has proved to be stubbornly elusive not least because
there is no unanimous agreement about what ought to qualify as the
`proper' approach to a quantum theoresis of spacetime and gravity.
Generally speaking, most of the approaches fall into the following
three categories:\footnote{These categories should by no means be
regarded as being mutually exclusive or exhaustive, and they
certainly reflect only these authors' subjective criteria and
personal perspective on the general characteristics of various
approaches to quantum gravity. This coarse classification will be
useful for the informal description of our finitary and causal
approach to Lorentzian vacuum quantum gravity to be discussed
shortly.}

\begin{enumerate}

\item {\it `General relativity conservative'}: The general aim of
the approaches falling into this category is to quantize classical
gravity somehow. Thus, the mathematical theory on which general
relativity---in fact, any field theory whether classical or
quantum---is based, namely, the differential geometry of
$\smooth$-manifolds ({\it ie}, the usual differential calculus on
manifolds), is essentially retained\footnote{That is, in general
relativity spacetime is modelled after a $\smooth$-smooth
manifold. Purely mathematically speaking, approaches in this
category could also be called `$\smooth$-smoothness or
differential manifold conservative'.} and it is used to treat the
gravitational field quantum field-theoretically. Both the
non-perturbative canonical and covariant ({\it ie}, path integral
or `action-weighed sum-over-histories') approaches to `quantum
general relativity', topological quantum field theories, as well
as, to a large extent, higher dimensional (or extended objects')
theories like (super)string and membrane schemes arguably belong
to this category.

\item {\it `Quantum mechanics conservative'}: The general spirit
here is to start from general quantum principles such as algebraic
operationality, noncommutativity and finitism (`discreteness')
about the structure of spacetime and its dynamics, and then try to
derive somehow general relativistic attributes, as it were, from
within the quantum framework. Such approaches assume up-front that
quantum theory is primary and fundamental, while the classical
geometrical smooth spacetime continuum and its dynamics secondary
and derivative (emergent) from the deeper quantum dynamical realm.
For instance, Connes' noncommutative geometry
\cite{connes,kastler} and, perhaps more notably, Finkelstein's
quantum relativity \cite{df1,sel4}\footnote{In fact, Finkelstein
maintains that ``{\em all is quantum. Anything that appears to be
classical has not yet been resolved into its quantum elements}''
(David Finkelstein in private communication).} may be classified
here.

\item {\it `Independent'}: Approaches in this category
assume neither quantum mechanics nor general relativity as a
fundamental, `fixed' background theory relative to which the other
must be modified to suit. Rather, they start independently from
principles that are neither quantum mechanical nor general
relativistic {\it per se}, and proceed to construct a theory and a
suitable mathematical formalism to accompany it that later may be
interpreted as a coherent amalgamation (or perhaps even extension)
of both. It is inevitable with such `iconoclastic' schemes that in
the end both general relativity and quantum mechanics may appear
to be modified to some extent. One could assign to this category
Penrose's combinatorial spin-networks \cite{pen,rovelli1} and its
current relativistic spin-foam descendants
\cite{baez1,barrett,perez}, Regge's homological spacetime
triangulations and simplicial gravity \cite{regge}, as well as
Sorkin {\it et al.}'s causal sets
\cite{bomb87,sork1,sork2,ridesork,sork3}.

\end{enumerate}

\noindent It goes without saying that this is no place for us to
review in any detail the approaches mentioned above.\footnote{For
reviews of and different perspectives on the main approaches to
quantum gravity, the reader is referred to
\cite{ish1,ash1,rovelli}. In the last, most recent reference, one
notices a similar partition of the various approaches to quantum
gravity into three classes called {\em covariant}, {\em canonical}
and {\em sum-over-histories}. Then one realizes that presently we
assigned all these three classes to category 1, since our general
classification criterion is which approaches, like general
relativity, more-or-less preserve a $\smooth$-smooth base
spacetime manifold hence use the methods of the usual differential
geometry on it, and which do not. Also, by `covariant' we do not
mean what Rovelli does. `Covariant' for us is synonymous to
`action-weighed sum-over-histories' or `path integral'.
Undoubtedly, there is arbitrariness and subjectivity in such
denominations, so that the boundaries of those distinctions are
rather fuzzy.} Rather, we wish to continue a finitary, causal and
quantal sheaf-theoretic approach to spacetime and vacuum
Lorentzian gravity that we have already started to develop in
\cite{malrap1,malrap2}. This approach, as we will argue
subsequently, combines characteristics from all three categories
above and, in particular, the mathematical backbone which supports
it, {\em Abstract Differential Geometry} (ADG)
\cite{mall1,mall2,mall3,mall4,mall7}, was originally conceived in
order to evade the $\smooth$-smooth spacetime manifold $M$ (and
consequently its diffeomorphism group $\mathrm{Diff}(M)$)
underlying (and creating numerous problems for) the various
approaches in 1. For, it must be emphasized up-front, {\em ADG is
an axiomatic formulation of differential geometry which does not
use any $\smooth$-notion from the usual differential
calculus---the classical differential geometry of smooth
manifolds}.

To summarize briefly what we have already accomplished in this
direction,\footnote{For a recent, concise review of our work so
far on this sheaf-theoretic approach to discrete Lorentzian
quantum gravity, as well as on its possible topos-theoretic
extension, the reader is referred to \cite{rap6}. In particular,
the topos-theoretic viewpoint is currently being elaborated in
\cite{rap7}.} in \cite{malrap1} we combined ideas from the second
author's work on {\em finitary spacetime
sheaves}\footnote{Throughout this paper, the epithets `finitary'
and `locally finite' will be used interchangeably.} (finsheaves)
\cite{rap2} and on an algebraic quantization scenario for Sorkin's
causal sets (causets) \cite{rap1} with the first author's ADG
\cite{mall1,mall2}, and we arrived at a locally finite, causal and
quantal version of the kinematical structure of Lorentzian
gravity. The latter pertains to the definition of a curved
principal finsheaf $\peel_{i}$ of incidence Rota algebras
modelling {\em quantum causal sets} (qausets) \cite{rap1}, having
for structure group a locally finite version of the continuous
orthochronous Lorentz group $SO(1,3)^{\uparrow}$ of local
symmetries (isometries) of general relativity, together with a
non-trivial ({\it ie}, non-flat) locally finite
$so(1,3)^{\uparrow}_{i}\simeq sl(2,\com)_{i}$-valued
spin-Lorentzian connection $\conf_{i}$\footnote{From
\cite{malrap1} we note that only the gauge potential
$\vec{\aconn}_{i}$ part of the reticular
$\conf_{i}=\vec{\partial}_{i} +\vec{\aconn}_{i}$ is
spin-Lorentzian proper ({\it ie}, discrete
$so(1,3)^{\uparrow}_{i}\simeq sl(2,\com)_{i}$-valued), but here
too we will abuse terminology and refer to either $\conf_{i}$ or
its part $\vec{\aconn}_{i}$ as `the spin-Lorentzian connection'.
(The reader should also note that the arrows over the various
symbols will be justified in the sequel in view of the causal
interpretation that our incidence algebra finsheaves have; while,
the subscript `$i$' is the so-called `finitarity', `resolution',
or `localization index' \cite{rap2,malrap1,malrap2}, which we will
also explain in the sequel.)} which represents the localization or
gauging and concomitant dynamical variability of the qausets in
the sheaf due to a finitary, causal and quantal version of
Lorentzian gravity in the absence of matter ({\it ie}, vacuum
Einstein gravity). We also gave the following quantum particle
interpretation to this reticular scheme: a so-called {\em
causon}---the elementary particle of the field of dynamical
quantum causality represented by $\vec{\aconn}_{i}$---was
envisioned to dynamically propagate in the reticular curved
spacetime vacuum represented by the finsheaf of qausets under the
influence of finitary Lorentzian (vacuum) quantum gravity.

In the sequel \cite{malrap2}, by using the universal constructions
and the powerful sheaf-cohomological tools of ADG together with
the rich differential structure with which the incidence algebras
modelling qausets are equipped
\cite{rapzap1,rap1,rapzap2,zap1,zap3}, we showed how basic
differential geometric ideas and results usually thought of as
being vitally dependent on $\smooth$-smooth manifolds for their
realization, as for example the standard \v{C}ech-de Rham
cohomology, carry through virtually unaltered to the finitary
regime of the curved finsheaves of qausets. For instance, we gave
finitary versions of important $\smooth$-theorems such as de
Rham's, Weil's integrality and the Chern-Weil theorem and, based
on certain robust results from the application of ADG to the
theory of geometric (pre)quantization \cite{mall2,mall5,mall6}, we
carried out a sheaf-cohomological classification of the associated
line sheaves bearing the finitary spin-Lorentzian $\qaconn_{i}$s
whose quanta were referred to as causons above---the elementary
(bosonic) particles carrying the dynamical field of quantum
causality whose (local) states correspond precisely to (local)
sections of those line sheaves. By this virtually complete
transcription of the basic $\smooth$-constructions, concepts and
results to the locally finite and quantal realm of the curved
finsheaves of qausets, we highlighted that for their formulation
the classical smooth background spacetime continuum is essentially
of no contributing value. Moreover, we argued that since the
$\smooth$-smooth spacetime manifold can be regarded as the main
culprit for the singularities that plague general relativity as
well as for the weaker but still troublesome infinities that
assail the flat quantum field theories of matter, its
evasion---especially by the finitistic-algebraic means that we
employed---should be most welcome for the formulation of a
`calculationally' and, in a sense to be explained later,
`inherently finite' and `fully covariant' quantum theory of
gravity.

With respect to the aforementioned three categories of approaches
to quantum gravity, our scheme certainly has attributes of 2 as it
employs finite dimensional non-abelian incidence algebras to model
(dynamically variable) qausets in the stalks of the relevant
finsheaves, which qausets have a rather natural quantum-theoretic
(because algebraico-operational) physical interpretation
\cite{rapzap1,rap1,malrap1,rapzap2}. It also has traits of
category 3 since the incidence algebras are, by definition, of
combinatorial and `directed simplicial' homological character and,
in particular, Sorkin's causet theory was in effect its principal
physical motivation \cite{rap1,malrap2}. Finally, regarding
category 1, the purely mathematical, ADG-based aspect of our
approach was originally motivated by a need to show that {\em all
the `intrinsic' differential mechanism of the usual calculus on
manifolds is independent of $\smooth$-smoothness}, in fact, {\em
of any notion of `space' supporting the usual differential
geometric concepts and constructions},\footnote{Thus, as we will
time and again stress in the sequel, with the development of ADG
we have come to realize that the main operative role of the
$\smooth$-smooth manifold is to provide us with {\em a} convenient
(and quite successful in various applications to both classical
and quantum physics!), {\em but by no means unique}, differential
mechanism, namely, that accommodated by the algebra $\smooth(M)$
of infinitely differentiable functions `coordinatizing' the
(points of the) differential manifold $M$. However, the latter
algebra's pathologies in the form of singularities made us ponder
on the question whether the differential mechanism itself is
`innate' to $\smooth(M)$ and the manifold supporting these
`generalized arithmetics' (this term is borrowed straight from
ADG). As alluded to above, ADG's answer to the latter is an
emphatic `{\em No}!' \cite{mall1,mall2,mall7}. For example, one
can do differential geometry over very (in fact, {\em the most}!)
singular from the point of view of the $\smooth$-smooth $M$ spaces
and their `arithmetic algebras', such as Rosinger's non-linear
distributions---the so-called differential algebras of generalized
functions) \cite{malros1,malros2,ros}. As a matter of fact, the
last two papers, together with the duet \cite{malrap1,malrap2},
are examples of two successful applications of ADG proving its
main point, that: ``{\em differentiability is independent of
$\smooth$-smoothness}'' (see slogan 2 at the end of
\cite{malrap2}).} thus entirely avoid, or better, manage to
integrate or `absorb' into the (now generalized) abstract
differential geometry, the `anomalies' ({\it ie}, the
singularities and other `infinity-related pathologies') that
plague the classical $\smooth$-smooth continuum case
\cite{mall1,mall2,mall7}. Arguably then, our approach is an
amalgamation of elements from 1--3.

Let us now move on to specifics. In the present paper we continue
our work in \cite{malrap1,malrap2} and formulate the dynamical
vacuum Einstein equations in $\peel_{i}$. On the one hand, this
extends our work on the kinematics of a finitary and causal scheme
for Lorentzian quantum gravity developed in \cite{malrap1} as it
provides a suitable dynamics for it, and on the other, it may be
regarded as another concrete physical application of ADG to the
locally finite, causal and quantum regime, and all this {\em in
spite of the $\smooth$-smooth spacetime manifold}, in accord with
the spirit of \cite{malrap2}. Our work here is the second physical
application of ADG to vacuum Einstein gravity, the first having
already involved the successful formulation of the vacuum Einstein
equations over spaces with singularities concentrated on arbitrary
closed nowhere dense sets---arguably, {\em the} most singular
spaces when viewed from the featureless $\smooth$-smooth spacetime
manifold perspective \cite{mall3,malros1,malros2,mall7,ros}.

The paper is organized as follows: in the following section we
recall the basic ideas about connections in ADG focusing our
attention mainly on Yang-Mills (Y-M) and Lorentzian connections on
finite dimensional vector sheaves, on principal sheaves (whose
associated sheaves are the aforementioned vector sheaves), their
curvatures, symmetries and (Bianchi) identities, as well as the
affine spaces that they constitute. In section 3 we discuss the
connection-based picture of gravity---the way in which general
relativity may be thought of as a Y-M-type of gauge theory in the
manner of ADG \cite{mall4}. Based mainly on \cite{mall3}, we
present vacuum Einstein gravity {\it \`{a} la} ADG and explore the
relevant gravitational moduli spaces of spin-Lorentzian
connections. In section 4 we remind the reader of some basic
kinematical features of our curved principal finsheaves of qausets
from \cite{malrap1,malrap2} and, in particular, based on recent
results of Papatriantafillou and Vassiliou
\cite{pap1,pap2,vas1,vas2,vas3}, we describe in a categorical way
inverse (projective) and direct (inductive) limits of such
principal finsheaves and their reticular connections. We also
comment on the use of the real ($\R$) and complex ($\com$) number
fields in our manifold-free, combinatory-algebraic theory, and
compare it with some recent critical remarks of Isham \cite{ish3}
about the {\it a priori} assumption---one that is essentially
based on the classical manifold model of spacetime---of the $\R$
and $\com$ continua in conventional quantum theory {\it
vis-\`a-vis} its application to quantum gravity. Section 5 is the
focal area of this paper as it presents a locally finite, causal
and quantal version of the vacuum Einstein equations for
Lorentzian gravity. The idea is also entertained of developing a
possible covariant quantization scheme for finitary Lorentzian
gravity involving a path integral-type of functional over the
moduli space $\fconn_{i}/\grouv_{i}$ of all reticular
gauge-equivalent spin-Lorentzian connections $\qaconn_{i}$. Based
on the `innate' finiteness of our model, we discuss how such a
scenario may on the one hand avoid {\it ab initio} the choice of
measure for $\fconn_{i}$ that troubles the continuum functional
integrals over the infinite dimensional, non-linear and with a
`complicated' topology moduli space $\infconn^{(+)}/\grouv$ of
smooth, (self-dual) Lorentzian connections in the standard
covariant approach to the quantization of (self-dual) Lorentzian
gravity, and on the other, how our up-front avoiding of
$\mathrm{Diff}(M)$ may cut the `Gordian knot' that the problems of
time and of the inner product in the Hilbert space of physical
states present to the non-perturbative canonical approach to
quantum gravity based on Ashtekar's new variables and the holonomy
(Wilson loop) formalism associated with them. Ultimately, all this
points to the fact that {\em our theory is genuinely
$\smooth$-smooth spacetime background independent} and, perhaps
more importantly, {\em regardless of the perennial debate whether
classical (vacuum) gravity should be quantized covariantly or
canonically}. This makes us ask---in fact, altogether
doubt---whether quantizing classical spacetime and gravity by
using the constructions and techniques of the usual differential
geometry of smooth manifolds is the `right' approach to quantum
spacetime and gravity, thus align ourselves more with the
categories 2 and 3 above, and less with 1. As a matter of fact,
and in contradistinction to the `iconoclastic' approaches in
category 3 (most notably, in contrast to the theory of causal
sets), in developing our entirely algebraico-sheaf-theoretic
approach to finitary Lorentzian quantum gravity based on ADG, we
have come to question altogether whether the notion of (an inert
geometrical background) `spacetime'---whether it is modelled after
a continuous or a discrete base space---makes any physical sense
in the ever dynamically fluctuating quantum deep where the vacuum
is `filled' solely by (the dynamics of) causons and where there is
no `ambient' or surrounding spacetime that actively participates
into or influences in any way that dynamics.\footnote{Of course,
we will see that there is a base topological `localization
space'---a stage on which we solder our algebraic structures, but
this space is of an ether-like character, a surrogate scaffolding
of no physical significance whatsoever as it does not actively
engage into the quantum dynamics of the causons---the quanta of
the field $\qaconn_{i}$ of quantum causality that is localized
(gauged) and dynamically propagates on `it'.} We thus infer that
both our finitary vacuum Einstein equations for the causon and the
path integral-like quantum dynamics of our reticular (self-dual)
spin-Lorentzian connections $\qaconn_{i}^{(+)}$ is `genuinely', or
better, `fully' covariant since they both concern directly and
solely the objects (the quanta of causality, {\it ie}, the
dynamical connections $\qaconn_{i}$) that live on that base
`space(time)', and not at all that external, passive and
dynamically inert `space(time) arena' itself. We also make
comments on geometric (pre)quantization \cite{mall2,mall5,mall6}
in the light of our application here of ADG to finitary and causal
Lorentzian gravity \cite{malrap2} and we stress that our scheme
may be perceived as being, in a strong sense, `already' or
`inherently' quantum, meaning that it is in no need of the
(formal) process of quantization of the corresponding classical
theory (here, general relativity on a $\smooth$-smooth spacetime
manifold). This seems to support further our doubts about the
quantization of classical spacetime and gravity mentioned above.
Furthermore, motivated by the `full covariance' and `inherent
quantumness' of our theory, we draw numerous close parallels
between our scenario and certain ideas of Einstein about the
so-called (post general relativity) `new ether' concept, the
unitary field theory that goes hand in hand with the latter, but
more importantly, about the possible abandonement altogether, in
the light of singularities and quantum discontinuities, of this
continuous field theory and the $\smooth$-spacetime continuum
supporting it for ``{\em a purely algebraic description of
reality}'' \cite{einst3}. {\it In toto}, we argue that ADG,
especially in its finitary and causal application to Lorentzian
quantum gravity in the present paper, may provide the basis for
the ``{\em organic}'' \cite{einst2}, ``{\em algebraic}''
\cite{einst3} theory that Einstein was searching for in order to
replace the multiply assailed by unmanageable singularities,
unphysical infinities and other anomaliles geometric spacetime
continuum of macroscopic physics. At the same time, we will
maintain that this abandonement of the spacetime manifold for a
more finitistic-algebraic theory can be captured to a great extent
by the mathematical notion of Gel'fand duality---a notion that
permeates the general sheaf-theoretic methods of ADG effectively
ever since its inception \cite{mall0,mall,mall1} as well its
particular finitary, causal and quantal applications thereafter
\cite{rapzap1,rap1,rap2,malrap1,rapzap2,malrap2,rap3,rap4,rap6}.
The paper concludes with some remarks on $\smooth$-smooth
singularities---some of which having already been presented in a
slightly different, purely ADG-theoretic, guise in
\cite{mall7}---that anticipate a paper currently in preparation
\cite{malrap3}.

\section{Connections in Abstract Differential Geometry}

Connections, {\it alias} `generalized differentials', are {\em
the} central objects in ADG which purports to abstract from, thus
axiomatize and effectively generalize, the usual differential
calculus on $\smooth$-manifolds. In this section we give a brief
{\it r\'{e}sum\'{e}} of both the local and global ADG-theoretic
perspective on linear (Koszul), pseudo-Riemannian (Lorentzian)
connections and their associated curvatures. For more details and
completeness of exposition, the reader is referred to
\cite{mall1,mall2,mall4}.

\subsection{Basic Definitions about Linear Connections}

The main notion here is that of {\em differential triad}
$\triad=(\struc_{X} ,\partial ,\Omg(X))$, which consists of a
sheaf $\struc_{X}$ of (complex) abelian algebras $A$ over an in
general {\em arbitrary topological space $X$} called the {\em
structure sheaf} or the {\em sheaf of coefficients} of the
triad,\footnote{The pair $(X,\struc_{X})$ is called a {\em
$\cons$-algebraized space}, where $\cons$ corresponds to the
constant sheaf of the complex numbers $\com$ over $X$, which is
naturally injected into $\struc_{X}$ ({\it ie},
$\cons\stackrel{\subset}{\rightarrow}{\struc}_{X}$ and, plainly,
$\com=\Gamma(X,\cons)\equiv\cons(X)$). It is tacitly assumed that
for every open set $U$ in $X$, the algebra $\struc(U)$ of
continuous local sections of $\struc_{X}$ is a unital, commutative
and associative algebra over $\com$. It must be noted here however
that one could start with a $\mathbf{K}$-algebraized space
($\mathbf{K}=\mathbf{R},\cons$) in which the structure sheaf
$\struc_{X}$ would consist of unital, abelian and associative
algebras over the fields $\K=\R ,\com$ respectively. Here we have
just fixed $\K$ to the complete field of complex numbers, but in
the future we are going to discuss also the real case. Also, in
either case $\struc_{X}$ is assumed to be {\em fine}. In the
sequel, when it is rather clear what the base topological space
$X$ is, we will omit it from $\struc_{X}$ and simply write
$\struc$.} a sheaf $\Omg$ of (differential) $A$-modules $\omg$
over $X$, and a $\cons$-derivation $\partial$ defined as the {\em
sheaf morphism}

\begin{equation}\label{eq1}
\partial :~\struc\mapto\Omg
\end{equation}

\noindent which is $\cons$-linear and satisfies Leibniz's rule

\begin{equation}\label{eq2}
\partial(s\cdot t)=s\cdot\partial(t)+t\cdot\partial(s)
\end{equation}

\noindent for any local sections $s$ and $t$ of $\struc$ ({\it
ie}, $s,t\in\Gamma(U,\struc)\equiv\struc(U)$, with $U\subseteq X$
open). It can be shown that the usual differential operator
$\partial$ in (\ref{eq1}) above is {\em the} prototype of a {\em
flat $\struc$-connection} \cite{mall1,mall2}.

The aforementioned generalization of the usual differential
operator $\partial$ to an (abstract) $\struc$-connection $\conn$
involves two steps emulating the definition of $\partial$ above.
First, one identifies $\conn$ with a suitable ($\cons$-linear)
sheaf morphism as in (\ref{eq1}), and second, one secures that the
Leibniz condition is satisfied by $\conn$, as in (\ref{eq2})
above. So, given a differential triad $\triad=(\struc,\partial
,\Omg)$, let $\modl$ be an $\struc$-module sheaf on $X$. Then, the
first step corresponds to defining $\conn$ as a map

\begin{equation}\label{eq3}
\conn:~\modl\mapto\modl\otimes_{\struc}\Omg\cong
\Omg\otimes_{\struc}\modl\equiv\Omg(\modl)
\end{equation}

\noindent which is a $\cons$-linear morphism of the complex vector
sheaves involved, while the second, that this map satisfies the
following condition

\begin{equation}\label{eq4}
\conn(\alpha \cdot s)=\alpha\cdot\conn(s)+s\otimes\partial(\alpha)
\end{equation}

\noindent for $\alpha\in\struc(U)$,
$s\in\modl(U)\equiv\Gamma(U,\modl)$, and $U$ open in $X$.

The connection $\conn$ as defined above may be coined a `{\em
Koszul linear connection}' and its existence on the vector sheaf
$\modl$ is crucially dependent on both the base space $X$ and the
structure sheaf $\struc$. For $X$ a {\em paracompact} and {\em
Hausdorff} topological space, and for $\struc_{X}$ a {\em fine}
sheaf on it, the existence of $\conn$ is well secured, as for
instance in the case of $\smooth$-smooth manifolds
\cite{mall1,mall2}.

\subsubsection{The local form of $\conn$}

Given a local gauge $e^{U}\equiv\{ U;~(e_{i})_{0\leq i\leq n-1}\}$
of the vector sheaf $\modl$ of rank $n$,\footnote{We recall from
\cite{mall1,mall2,malrap2} that in ADG, $\gauge=\{
U_{\alpha}\}_{\alpha\in I}$ is called {\em a local frame} or {\em
a coordinatizing open cover of}, or even {\em a local choice of
basis (or gauge!) for} $\modl$. The $e_{i}$s in $e^{U}$ are local
sections of $\modl$ ({\it ie}, elements of $\Gamma(U,\modl)$)
constituting a basis of $\modl(U)$. We also mention that for the
$\struc$-module sheaf $\modl$, regarded as a vector sheaf of rank
$n$, one has by definition the following
$\struc|_{U}$-isomorphisms:
$\modl|_{U}=\struc^{n}|_{U}=(\struc|_{U})^{n}$ and, concomitantly,
the following equalities section-wise:
$\modl(U)=\struc^{n}(U)=\struc(U)^{n}$ (with $\struc^{n}$ the
$n$-fold Whitney sum of $\struc$ with itself). Thus, $\modl$ is
{\em a locally free $\struc$-module of finite rank $n$}---an
appellation synonymous to {\em vector sheaf} in ADG
\cite{mall1,mall2}. For $n=1$, the vector sheaf $\modl$ is called
a {\em line sheaf} and it is symbolized by $\lsh$.} every
continuous local section $s\in\modl(U)~(U\in\gauge)$ can be
expressed as a unique superposition $\sum_{i=1}^{n}s_{i}e_{i}$
with coefficients $s_{i}$ in $\struc(U)$. The action of $\conn$ on
these sections reads

\begin{equation}\label{eq5}
\conn(s)=\sum_{i=1}^{n}(s_{i}\conn(e_{i})+e_{i}\otimes\partial(s_{i}))
\end{equation}

\noindent with

\begin{equation}\label{eq6}
\conn(e_{i})=\sum_{i=1}^{n}e_{i} \otimes\omega_{ij},~1 \leq
i,j\leq{n}
\end{equation}

\noindent for some unique $\omega_{ij}\in\Omg(U)~(1\leq i,j\leq
n)$, which means that $\omega\equiv(\omega_{ij})\in
M_{n}(\Omg(U))=M_{n}(\Omg)(U)$ is an $n\times n$ matrix of
sections of local $1$-forms. Thus, (\ref{eq5}) reads via
(\ref{eq6})

\begin{equation}\label{eq7}
\conn(s)=\sum_{i=1}^{n}e_{i}\otimes(\stackrel{\partial}{\overbrace{\partial(s_{i})}}+
\sum_{i=1}^{n}\stackrel{\omega}{\overbrace{s_{j}\omega_{ij}}})\equiv(\partial+\omega)(s)
\end{equation}

\noindent So that, {\it in toto}, every connection $\conn$ can be
written locally as

\begin{equation}\label{eq8}
\conn=\partial+\omega
\end{equation}

\noindent with (\ref{eq8}) effectively expressing the procedure
commonly known in physics as {\em localizing} or {\em gauging} the
usual (flat) differential $\partial$ to the (curved) {\em
covariant derivative} $\conn$. Thus, the (non-flat) $\omega$ part
of $\conn$, called {\em the gauge potential} in physics, measures
the deviation from differentiating flatly ({\it ie}, by
$\partial$), when one differentiates `covariantly' by
$\conn$.\footnote{In the sequel we will symbolize the gauge
potential part of $\conn$ in (\ref{eq8}) by $\aconn$ instead of
$\omega$ in order to be in agreement with our notation in the
previous papers \cite{malrap1,malrap2}, as well as with the
standard notation for the spin-Lorentzian connection in current
Lorentzian quantum gravity research
\cite{ashish,ashlew1,ashlew2,baez}.}

\subsubsection{Local gauge transformations of $\conn$}

We investigate here, in the context of ADG, the behavior of the
gauge potential part $\aconn$ of $\conn$ under local gauge
transformations---the so-called `{\em transformation law of
potentials}' in \cite{mall1,mall2}.

Thus, let $\modl$ be an $\struc$-module or a vector sheaf of rank
$n$. Let $e^{U}\equiv\{ U;~ e_{i=1\cdots n}\}$ and $f^{V}\equiv\{
V;~ f_{i=1\cdots n}\}$ be local gauges of $\modl$ over the open
sets $U$ and $V$ of $X$ which, in turn, we assume have non-empty
intersection $U\cap V$. Let us denote by $g\equiv(g_{ij})$ the
following {\em change of local gauge matrix}

\begin{equation}\label{eq9}
f_{j}=\sum_{i=1}^{n}g_{ij}e_{i}
\end{equation}

\noindent which, plainly, is a local ({\it ie}, relative to $U\cap
V$) section of the `natural' structure group sheaf $\gl(n,\struc)$
of $\modl$\footnote{We will present some rudiments of structure
group (or principal or $\struct$-) sheaves of associated vector
sheaves $\modl$ in the next subsection. One may recognize
$\gl(n,\struc)$ above as the local version of the automorphism
group sheaf $\aut\modl$ of $\modl$. The adjective `local' here
pertains to the fact mentioned earlier that ADG assumes that
$\modl$ is locally isomorphic to $\struc^{n}$.}---that is,
$g_{ij}\in\mathrm{GL}(n,\struc(U\cap V))=\gl(n,\struc)(U\cap V)$.

Without going into the details of the derivation, which can be
found in \cite{mall1,mall2}, we note that under such a local gauge
transformation $g$, the gauge potential part $\omega\equiv\aconn$
of $\conn$ in (\ref{eq8}) transforms as follows

\begin{equation}\label{eq10}
\aconn^{'}=g^{-1}\aconn g+g^{-1}\partial g
\end{equation}

\noindent a way we are familiar with from the usual differential
geometry of the smooth fiber bundles of gauge theories.  For
completeness, it must be noted here that, in (\ref{eq10}),
$\aconn\equiv(\aconn_{ij})\in
M_{n}(\Omg^{1}(U))=M_{n}(\Omg^{1})(U)$ and
$\aconn^{'}\equiv(\aconn^{'}_{ij})\in
M_{n}(\Omg^{1}(V))=M_{n}(\Omg^{1})(V)$. The transformation of
$\aconn$ under local gauge changes is called {\em affine} or {\em
inhomogeneous} in the usual gauge-theoretic parlance precisely
because of the term $g^{-1}\partial g$. We will return to this
affine term in subsection 2.3 and subsequently in section 5 where
we will comment on the essentially non-geometrical ({\it ie},
non-tensorial) character of connection. Also, anticipating our
discussion of moduli spaces of gauge-equivalent connections in the
next section, we note that (\ref{eq10}) expresses an equivalence
relation `$\stackrel{g}{\sim}$' between the gauge potentials
$\aconn$ and $\aconn^{'}$.

\subsection{Pseudo-Riemannian (Lorentzian) Metric Connections}

In this subsection we are interested in endowing a vector sheaf
$\modl$ of finite rank $n\in\N$ with an indefinite $\struc$-valued
symmetric inner product $\rho$, and, concomitantly, study
$\struc$-connections $\conn$ that are compatible with the
(indefinite) metric $g$ associated with $\rho$---the so-called
{\em metric connections}. With an eye towards the applications to
Lorentzian (quantum) gravity in the sequel, we are particularly
interested in metric $\conn$s relative to Lorentzian metrics of
signature $\mathrm{diag}(g)=(-,+,+,\cdots)$. Also, continuing our
work \cite{malrap1} which dealt with {\em principal Lorentzian
finsheaves of qausets}, we are interested in the {\em group
sheaves $\aut_{\struc}(\modl)$ of $\struc$-automorphisms of
$\modl$}---the {\em principal sheaves of structure symmetries of
$\modl$}.\footnote{Commonly known as $\struct$-sheaves in the
mathematical literature \cite{mall1}.} In the case of a real ({\it
ie}, $\K=\R$ and $\mathbf{R}$-algebraized space) Lorentzian vector
sheaf $(\modl ,\rho)$ of rank $4$,\footnote{We would like to
declare up-front that in this paper we provide no argument
whatsoever for assuming that the dimensionality (rank) $n$ of our
vector sheaves is the `empirical' (or better, `conventional') $4$
of the spacetime manifold of `macroscopic experience' (or better,
of the classical theory). In the course of this work the reader
will realize that all our constructions are manifestly independent
of the classical $4$-dimensional, locally Euclidean,
$\smooth$-smooth, Lorentzian spacetime manifold of general
relativity so that we will time and again doubt whether the
latter, and the host of (mathematical) structures that classically
it is thought of as carrying ({\it eg}, its uncountably infinite
cardinality of events, its dimensionality, its topological,
differential and metric structures), is a physically meaningful
concept. For example, we will maintain that dimensionality and the
metric are free mathematical choices of ({\it ie}, fixed by) the
theorist and not Nature's own, while that the topology and
differential structure are inherent in the dynamical objects
(fields) that may be thought of as living and propagating on
`spacetime', not by that inert background `spacetime' itself,
which is devoid of any physical meaning. Moreover, all this will
be expressed in an algebraic, locally finite setting quite remote
from the uncountable continuous infinity of events of the
manifold.} the stalks of the corresponding $\struct$-sheaves will
`naturally' be assumed to host the group
$SO(1,3)^{\uparrow}$---the orthochronous Lorentz group of (local)
isometries of $(\modl ,\rho)$ which, in turn, is locally
isomorphic to the spin-group $SL(2,\com)$.\footnote{In the sense
that their corresponding Lie algebras are isomorphic:
$so(1,3)^{\uparrow}\simeq sl(2,\com)$ \cite{malrap1}.} We thus
catch a first glimpse of the spin-Lorentzian connections
considered in the context of curved finsheaves of qausets in
\cite{malrap1}, which will be dealt with in more detail in section
4.

Thus, let $\modl$ be a vector sheaf. By an {\em $\struc$-valued
pseudo-Riemannian inner product $\rho$ on $\modl$} (over $X$) we
mean a {\em sheaf} morphism

\begin{equation}\label{eq11}
\rho :~\modl\oplus\modl\mapto\struc
\end{equation}

\noindent which is i) {\em $\struc$-bilinear} between the
$\struc$-modules concerned, ii) symmetric ({\it ie},
$\rho(s,t)=\rho(t,s),~s,t\in\modl(U)$) and of indefinite
signature, as well as iii) {\em strongly non-degenerate}. That is,
we assume that $\rho(s,t)$, for any two local sections $s$ and $t$
in $\modl(U)$,\footnote{It is important to notice here that the
$\struc$-metric $\rho$ is not a (bilinear) map assigned to the
points of the base space $X$ {\it per se} (which is only assumed
to be a topological, not a differential, let alone a metric,
space), but to the fibers (stalks) of the relevant module or
vector sheaves which are inhabited by the geometrical objects that
live on $X$. As noted in a previous footnote, in our scheme,
metric and, as we shall see later, topological and differential
properties concern the objects that live on `space(time)', not the
supporting space(time) itself. This recalls Gauss' and Riemann's
original labors with endowing the linear fiber spaces tangent to a
sphere with a bilinear quadratic form---a metric. They ascribed a
metric to the linear fibers, not to the supporting sphere itself
which, anyway, is manifestly `non-linear' \cite{mall7}. What we
wish to highlight by these remarks is that {\em space(time)
carries no metric}. Equally important is to note that the
$\struc$-valued metric $\rho$ is imposed on these objects {\em by
us} and it is intimately tied to ({\it ie}, takes values in) our
own measurements (arithmetics) in $\struc$ (see comparison between
the notions of connection and curvature in 2.3.5). $\rho$ is not a
property of space(time), which does not exist (in a physical
sense) anyway; rather, it is an attribute related to our own
measurements of `it all'. These remarks are important for our
subsequent physical interpretation of ADG in its application to
finitary Lorentzian quantum gravity in the next four sections. It
is a preliminary indication that in our theory the base
space(time) is an ether-like `substance' without any physical
significance. See remarks about `gravity as a gauge theory' in the
next section, about the `physical insignificance' or
`non-physicality' of spacetime in 5.1.1 and about `the relativity
of differentiability' in 6.2, as well as some similar
anticipations in \cite{malrap1,malrap2}.} is given via the
canonical isomorphism

\begin{equation}\label{eq12}
\modl\stackrel{\tilde{\rho}}{\cong}\modl^{*}
\end{equation}

\noindent between $\modl$ and its dual $\modl^{*}$, as

\begin{equation}\label{eq13}
\tilde{\rho}(s)(t):=\rho(s,t)
\end{equation}

\noindent with (\ref{eq12}) being true up to an
$\struc$-isomorphism\footnote{The epithet `strongly' to
`non-degenerate' above indicates that $\tilde{\rho}$ in
(\ref{eq12}) is also {\em onto}.}.

We further assume that for the vector sheaf $\modl$ (of finite
rank $n\in\N$) endowed with the $\struc$-connection $\conn$, the
vector sheaf $\Omg$ in the given differential triad
$\triad=(\struc ,\partial, \Omg)$ is the dual of $\modl$ appearing
in (\ref{eq12}) ({\it ie},
$\Omg=\modl^{*}\equiv{\Hom}_{\struc}(\modl ,\struc)$). Thus, in
line with the usual Christoffel theory \cite{mall1,mall2}, we can
define a {\em linear connection} $\nabla$, as follows

\begin{equation}\label{eq14}
\nabla :~\modl\times\modl\mapto\modl
\end{equation}

\noindent acting section-wise on $\modl(U)$ as

\begin{equation}\label{eq15}
\nabla(s,t)\equiv\nabla_{s}(t):=\conn(t)(s)
\end{equation}

Now, one says that {\em $\conn$ is a pseudo-Riemannian
$\struc$-connection} or that it is compatible with the indefinite
metric $g$ of the inner product $\rho$ in (\ref{eq11}), whenever
it fulfills the following two conditions:

\begin{itemize}

\item {\em Riemannian symmetry}: $\nabla(s,t)-\nabla(t,s)=[s,t]$;
for $s,t\in\modl(U)$ and $[\, .\, ,\, .\,  ]$ the usual Lie
bracket (product).

\item {\em Ricci identity}:
$\partial(\rho(s,t))(u)=\rho(\nabla(u,s),t)+\rho(s,\nabla(u,t))$;
for $s,t,u\in\modl(U)$, as usual.

\end{itemize}

In particular, for a Lorentzian $\rho$ and its associated
$g$,\footnote{With respect to a {\em local (coordinate) gauge}
$e^{U}\equiv\{ U;~(e_{i})_{0\leq i\leq n-1}\}$ of the vector sheaf
$\modl$ of rank $n$,
$\rho(e_{i},e_{j})=g_{ij}=\mathrm{diag}(-1,+1,\cdots)$
\cite{mall1,mall2}.} an $\struc$-connection $\conn$ is said to be
compatible with the Lorentz $\struc$-inner product $\rho$ on
$\modl$\footnote{Such a metric connection is commonly known as
{\em Levi-Civita connection}.} when its associated Christoffel
$\nabla$ in (\ref{eq14}) satisfies

\begin{equation}\label{eq16}
\nabla\rho=0
\end{equation}

\noindent which, in turn, is equivalent to the following `{\em
horizontality}' condition for the canonical isomorphism
$\tilde{\rho}$ in (\ref{eq12}) relative to the {\em connection}
$\conn_{\modl\otimes_{\struc}\modl^{*}}$ in the tensor product
vector sheaf $\Hom_{\struc}(\modl
,\modl^{*})=(\modl\otimes_{\struc}\modl)^{*}=\modl^{*}\otimes_{\struc}\modl^{*}$
induced by the $\struc$-connection $\conn$ on $\modl$

\begin{equation}\label{eq17}
\conn_{\Hom_{\struc}(\modl ,\modl^{*})}(\tilde{\rho})=0
\end{equation}

\noindent It is worth reminding the reader who is familiar with
the usual theory that (\ref{eq17}) above implies that the
Levi-Civita $\struc$-connection $\conn$ induced by the Lorentz
$\struc$-metric $\rho$ is {\em torsion-free} \cite{mall3}.

\subsubsection{Connections on (Lorentzian) principal sheaves}

As mentioned in the beginning of this subsection, of special
interest in our study is the case of a (real) Lorentzian vector
sheaf $(\modl ,\rho)$ of rank $4$ whose $\struc$-automorphism
sheaf $\aut_{\struc}\modll$ bears
$G=L^{\uparrow}:=SO(1,3)^{\uparrow}$---the orthochronous
$\rho$-preserving $\struc$-automorphisms of $\modl$ in its
stalks.\footnote{One may wish to symbolize the pair $(\modl,\rho)$
by $\modll$, thus $\aut_{\struc}\modll$ by $\orthl$. In the
sequel, when it is clear from the context that we are talking
about a Lorentzian vector sheaf $\modll=(\modl ,\rho)$, we may use
the symbols $\modl$ and $\modll$ for it interchangeably hopefully
without confusion. For a general vector sheaf $\modl$,
$\aut_{\struc}\modl$ is a subsheaf of ${\modl}nd\modl$, in fact,
for a given open $U\subseteq X$,
$\aut_{\struc}(\modl)(U)\simeq{\mathrm{End}}_{\struc}(\modl|_{U})^{^\bull}$---the
upper dot denoting {\em invertible} endomorphisms. We thus write
in general: $\aut_{\struc}(\modl)\equiv\aut\modl:=(\modl
nd\modl)^{^\bull}$.} $\orthl$ is the {\em principal sheaf of
structure symmetries of $\modll$}. In turn, $\modll$ is called the
{\em $\orthl$-associated vector sheaf}.\footnote{Henceforth we
will assume that every principal sheaf acts on the typical stalk
of its associated sheaf on the left (see below).}

But let us first give a brief discussion of connections on
principal sheaves {\it \`a la} ADG and then focus on
spin-Lorentzian (metric) connections. The reader will have to wait
until section 4 where we recall in more detail from \cite{malrap1}
the curved principal finsheaves $\peel_{i}$ of qausets and their
non-trivial connections $\conf_{i}$. For the material that is
presented below, we draw information mainly from
\cite{vas1,vas2,vas3}.

Let $\grouv$ be a sheaf of groups\footnote{By abuse of notation,
and hopefully without confusing the reader, in the sequel we will
also symbolize the groups that dwell in the stalks of $\grouv$ by
`$\grouv$'.} over $X$. Let $\modl$ be an $\struc$-module and
$\sigma$ a representation of $\struct$ in $\modl$, that is to say,
a {\em a continuous group sheaf morphism}

\begin{equation}\label{eq18}
\sigma :~\grouv\mapto\aut\modl
\end{equation}

\noindent effecting local ({\it ie}, $U$-wise in $X$) continuous
left-actions of $\struct$ on $\modl$ as follows

\begin{equation}\label{eq19}
\grouv(U)\times\modl(U)\mapto\modl\, :\,
(g,v)\rightarrowtail[\sigma(g)](v),~v\in\modl(U),\, g\in\struct(U)
\end{equation}

\noindent Also, by letting $\Omg^{1}$ be a sheaf of (first order)
differential $\struc$-modules over $\modl$,
$\Omg^{1}(\modl):=\Omg^{1}\otimes_{\struc}\modl$ as in
(\ref{eq3}), we define a {\em Lie sheaf of groups}
$\grouv$\footnote{The reader should note that in the present paper
we symbolize the gauge (structure) group of both Y-M theory and
gravity also by $\grouv$, hopefully without causing any confusion
between it and the abstract Lie sheaf of groups above.} to be the
quadruple $(\lie ,\modl ,\sigma ,\dot{\partial})$, where $\lie$ is
an $\struc$-module of Lie algebras,\footnote{By assuming that the
group sheaf $\struct$ in (\ref{eq18}) is a sheaf of Lie groups, we
may take $\lie$ to be the corresponding sheaf of Lie algebras.}
$\sigma$ a representation of $\lie$ in $\modl$, and
$\dot{\partial}$ the following $\struc$-module sheaf morphism

\begin{equation}\label{eq20}
\dot{\partial}:~\lie\mapto\Omg^{1}(\modl)
\end{equation}

\noindent which reminds one of the flat connection $\partial$ in
(\ref{eq1}). $\dot{\partial}$, called the {\em Maurer-Cartan
differential of $\grouv$ relative to
$\sigma$},\footnote{$\dot{\partial}$ is also known as the {\em
logarithmic differential} of $\grouv$.} satisfies

\begin{equation}\label{eq21}
\dot{\partial}:~(s\cdot
t)=\sigma(t^{-1})\cdot\dot{\partial}s+\dot{\partial}t
\end{equation}

\noindent It must be noted here that in the same way that
ADG---the differential geometry of vector sheaves---represents an
abstraction and generalization of the usual calculus on vector
bundles over $\smooth$-smooth manifolds to the effect that {\em no
calculus, in the usual sense, is employed at all}
\cite{mall1,mall2}, Lie sheaves of groups are the abstract
analogues of the usual Lie groups that play a central role in the
classical differential geometry of principal fiber bundles over
differential manifolds \cite{vas1,vas2,vas3}.

Thus, let $\grouv$ be a Lie sheaf of groups as above. Formally
speaking, by a {\em principal sheaf $\princ$ with structure group}
$\struct$ {\em relative to} $\grouv=(\lie ,\modl ,\sigma
,\dot{\partial})$\footnote{Where $\lie$ is the sheaf of Lie
algebras of the Lie group sheaf $\struct$. $\lie$ is supposed to
represent the {\em local structural type} of $\princ$
\cite{vas2}.} we mean a quadruple $(\princ ,\lie ,X,\pi)$
consisting of a sheaf of sets $\princ$\footnote{$\princ$ may be
thought of as `coordinatizing' the principal sheaf, thus we use
the same symbol `$\princ$' for the principal sheaf and its
coordinatizing sheaf of sets. $\pi$ is the usual projection map
from $\princ$ to the base space $X$. For more details, refer to
\cite{vas1,vas2,vas3}.} such that:

\begin{enumerate}

\item There is a continuous right-action of $\lie$ on
$\princ$.

\item There is an open gauge $\gauge=\{ U_{\alpha}\}_{\alpha\in I}$ of $X$
and isomorphisms of sheaves of sets ({\it ie}, coordinate
mappings)

\begin{equation}\label{eq22}
\phi_{\alpha}:~{\princ}|_{U_{\alpha}}\stackrel{\cong}{\mapto}\lie|_{U_{\alpha}}
\end{equation}

\end{enumerate}

\noindent satisfying

\begin{equation}\label{eq23}
\phi_{\alpha}(s\cdot g)=\phi_{\alpha}(s)\cdot
g;~s\in{\princ}(U_{\alpha}), g\in\lie(U_{\alpha})
\end{equation}

\noindent Given $\princ$, a vector sheaf $\modl$ and the
representation $\sigma :~\lie\mapto\aut\modl$, one obtains the
so-called {\em associated sheaf of
$\sigma({\princ})$},\footnote{Otherwise called {\em the
$\princ$-}, or even, {\em the} $\lie${\em -associated vector
sheaf}.} which is a sheaf of vector spaces locally of type $\modl$
in the sense that, relative to a coordinate gauge $\gauge$ for
$X$, there are coordinate maps

\begin{equation}\label{eq24}
\Phi_{\alpha}:~\sigma({\princ})|_{U_{\alpha}}\stackrel{\cong}{\mapto}\modl|_{U_{\alpha}}
\end{equation}

We assume that the associated vector sheaves $\modl$ of the
$\struct$-sheaves $\princ$ presented above are of the type
mentioned before in the context of ADG, namely, {\em locally free
$\struc$-modules of finite rank} ({\it ie}, locally isomorphic to
$\struc^{n}$) \cite{mall1,mall2}. We thus come to the main
definition of a connection $\dot{\conn}$ on a principal sheaf
$\princ$ generalizing the Maurer-Cartan differential
$\dot{\partial}$ in (\ref{eq20}) in a way analogous to how $\conn$
on a vector sheaf $\modl$ in (\ref{eq3}) generalized the flat
differential $\partial$ in (\ref{eq1}). Thus,

\begin{equation}\label{eq25}
\dot{\conn}:~{\princ}\mapto\Omg^{1}(\modl)\footnote{This morphism
can be equivalently written as
$\dot{\conn}:~{\princ}\mapto\Omg^{1}\otimes_{\struc}\lie(\equiv\Omg^{1}(\lie))$,
to manifest the usual statement that a connection on a principal
sheaf is a Lie algebra-valued $1$-form. Time and again we will
encounter this definition below.}
\end{equation}

\noindent is a morphism of sheaves of sets satisfying

\begin{equation}\label{eq26}
\dot{\conn}(s\cdot
g)=\sigma(g^{-1})\cdot\dot{\conn}s+\dot{\partial}g;~s\in{\princ}(U)~\mathrm{and}~g\in\lie(U)
\end{equation}

\noindent Locally ({\it ie}, $U$-wise in $X$), one can show, in
complete analogy to the local decomposition $\partial+\aconn$ of
the $\struc$-connection $\conn$ on $\modl$ in(\ref{eq8}), that
$\dot{\conn}$ too can be written as

\begin{equation}\label{eq27}
\dot{\conn}=\dot{\partial}+\dot{\aconn}
\end{equation}

\noindent and that, for a given coordinate gauge $\gauge=\{
U_{\alpha}\}_{\alpha\in I}$ for $X$ with {\em natural local
coordinate sections of $\princ$}
$s_{\alpha}:=\phi_{\alpha}^{-1}\circ
\mathbf{1}|_{U_{\alpha}}\in\princ(U_{\alpha})$,

\begin{equation}\label{eq28}
(\dot{\aconn})_{\alpha}=\dot{\conn}(s_{\alpha})\in\Omg^{1}(\modl)(U_{\alpha})
\end{equation}

\noindent in complete analogy to the local gauge potential
$1$-forms $\aconn$ of connections $\conn$ on vector sheaves
presented in (\ref{eq5})--(\ref{eq8}).\footnote{Furthermore, one
can show that for a local change of gauge $g$ as in (\ref{eq9}),
the $\dot{\aconn}$s obey a transformation law of potentials
completely analogous to the one obeyed by the $\aconn$s in
(\ref{eq10}). Without going into any details, it reads:
$\dot{\aconn}^{'}=\sigma(g)^{-1}\dot{\aconn}\sigma(g)+\sigma(g)^{-1}
\dot{\partial}g,~(\sigma(g^{-1})\equiv\sigma(g)^{-1})$
\cite{vas1,vas2,vas3}.}

Now, the essential point in this presentation of connections
$\dot{\conn}$ on principal sheaves $\princ$ in relation to our
presentation of $\struc$-connections $\conn$ on vector sheaves
$\modl$ earlier, is that when the latter are the
$\princ$-associated sheaves relative to corresponding
representations $\sigma :~\lie\mapto\aut\modl$, the following
`commutative diagram' may be used to picture formally the {\em
`$\sigma$-induced projection $\hat{\sigma}$'} of $\dot{\conn}$ on
$\princ$ to $\conn$ on $\modl$

\vskip 0.04in

\begin{equation}\label{eq29}
\square[\princ`\struc`\Omg^{1}(\lie)`\Omg^{1}(\modl);\hat{\sigma}`\dot{\conn}`\conn`\mathrm{id}]
\end{equation}

\vskip 0.04in

\noindent where $\hat{\sigma}$ may be regarded a morphism between
$\princ$ and $\struc$ regarded simply as sheaves of structureless
sets.\footnote{That is to say, by forgetting both the group
structure of the $\struct$-sheaf $\princ$ and the algebra
structure of the structure sheaf $\struc$. The inverse procedure
of building the principal sheaf $\princ$ and the connection
$\dot{\conn}$ on it from its associated vector sheaf $\modl$ and
the connection $\conn$ on it, may be loosely called {\em
`$\sigma$-induced lifting $\hat{\sigma}^{-1}$'} of $(\modl
,\conn)$ to $(\princ ,\dot{\conn})$. The $\sigma^{-1}$-lifting is
a forgetful correspondence since, in going from a vector sheaf to
its structure group sheaf, the linear structure of the former is
lost---something which is in fact reflected on that, while $\conn$
is $\cons$-linear, $\dot{\conn}$ is not. However, for more details
about commutative diagrams like (\ref{eq29}) between principal
sheaves $(\princ_{1},\dot{\conn}_{1})$ and
$(\princ_{2},\dot{\conn}_{2})$, their corresponding associated
sheaves $(\modl_{1},\conn_{1})$ and $(\modl_{1},\conn_{1})$, as
well as the respective projections $\hat{\sigma}$ of the former to
the latter, the reader is referred to \cite{vas3}.}

To make an initial contact with \cite{malrap1}, we can now
particularize the general ADG-based presentation of principal
sheaves $\princ$ above to (real) {\em Lorentzian} $\struct${\em
-sheaves}. As briefly noted earlier, the structure group $G$
dwelling in the stalks of the latter is taken to be
$L^{\uparrow}:=SO(1,3)^{\uparrow}$---the Lie group of
orthochronous Lorentz $\struc$-isometries, so that $\princ$ in
this case is denoted by $\orthl$. The $\orthl$-associated sheaf
$\modll=(\modl ,\rho)$ is a (real) vector sheaf of rank $4$,
equipped with an $\struc$-metric $\rho$ of absolute trace equal to
$2$. Thus, there is a local homomorphism (representation) $\sigma$
of the Lie algebra $so(1,3)^{\uparrow}\simeq sl(2,\com)$ of the
structure group $L^{\uparrow}$ in $\orthl$ into the `Lie algebra'
sheaf $aut_{\struc}(\modll)$ of the group sheaf
$\aut_{\struc}(\modll)$ of invertible $\struc$-endomorphisms of
$\modl$ preserving the Lorentzian $\struc$-metric $\rho$---that
is, the $\struc$-metric $\rho$ symmetries (isometries) of
$\modll$.

Collecting information from our presentation of connections on
$\struct$-sheaves and their associated vector sheaves, we are in a
position now to recall from \cite{malrap1} that, in the particular
case of the $\orthl$-associated vector sheaf $\modll$,

\begin{quotation}
\noindent {\em the gauge potential part $\aconn$ of an
$\struc$-connection $\conn$ on $\modll$ is an
$so(1,3)^{\uparrow}\simeq sl(2,\com)$-valued $1$-form on
$\orthl$.}
\end{quotation}

\noindent the so-called {\em spin-Lorentzian connection $1$-form}.

After we discuss the affine space $\sconn$ of Y-M and Lorentzian
gravitational $\struct$-connections from an ADG-theoretic
perspective in 2.4, as well as present the connection-based vacuum
Einstein equations ADG-theoretically in the next section, we are
going to return to the kinematical spin-Lorentzian connections on
principal finsheaves of qausets and their associated vector
sheaves studied in \cite{malrap1} in section 4, then we will
formulate their dynamical vacuum Einstein equations in 5, and
finally, in the same section, we will discuss a possible covariant
({\it ie}, action-based, path integral-type of) quantum dynamics
for them.

\subsection{Curvatures of $\struc$-Connections}

In ADG, the curvature $\curv$ of an $\struc$-connection $\conn$,
like $\conn$ itself, is defined as an {\em $\struc$-module sheaf
morphism}. More analytically, let $\triad=(\struc ,\partial ,
\Omg)$ be a differential triad as before. Define `inductively' the
following hierarchy of sheaves of $\Z_{+}$-graded $\struc$-modules
$\Omg^{i}~(i\in\Z_{+}\equiv\N\cup\{0\})$ of exterior ({\it ie},
Cartan differential) forms over $X$

\begin{equation}\label{eq30}
\Omg^{0}:=\struc,~\Omg\equiv\Omg^{1}:=\struc\wedge_{\struc}\Omg
,~\Omg^{2}=\struc\wedge_{\struc}\Omg^{1}\wedge_{\struc}\Omg^{1},\cdots\Omg^{i}\equiv(\Omg^{1})^{i}:=
\wedge^{i}_{\struc}\Omg^{1}
\end{equation}

\noindent and, in the same way that $\partial$($\equiv d^{0}$) is
a $\cons$-linear morphism between $\struc\equiv\Omg^{0}$ and
$\Omg\equiv\Omg^{1}$ as depicted in (\ref{eq1}), define a second
differential operator $\kd$($\equiv d^{1}$) again as the following
$\cons$-linear $\struc$-module sheaf morphism

\begin{equation}\label{eq31}
\kd:~\Omg^{1}\mapto\Omg^{2}
\end{equation}

\noindent obeying relative to $\partial$

\begin{equation}\label{eq32}
\kd\circ\partial=0~{\mathrm{and}}~\kd(\alpha\cdot
s)=\alpha\cdot\kd s-s\wee\partial\alpha
,~(\alpha\in\struc(U),s\in\Omg(U),U~\mathrm{open~in}~X)
\end{equation}

\noindent and called {\em the 1st exterior
derivation}.\footnote{In (\ref{eq30}), `$\wee_{\struc}$' is the
completely antisymmetric $\struc$-respecting tensor product
`$\otimes_{\struc}$'.}

Then, in complete analogy to the `extension' of the flat
connection $\partial$ to $\kd$ above, given a $\struc$-module
$\modl$ endowed with an $\struc$-connection $\conn$, one can
define the {\em 1st prolongation of $\conn$} to be the following
$\cons$-linear vector sheaf morphism

\begin{equation}\label{eq33}
\conn^{1}:~\Omg^{1}(\modl)\mapto\Omg^{2}(\modl)
\end{equation}

\noindent satisfying section-wise relative to $\conn$

\begin{equation}\label{eq34}
\conn^{1}(s\otimes t):=s\otimes\kd t-t\wee\conn
s,~(s\in\modl(U),t\in\Omg^{1}(U),U~\mathrm{open~in}~X)
\end{equation}

We are now in a position to define the curvature $\curv$ of an
$\struc$-connection $\conn$ by the following commutative diagram

\begin{equation}\label{eq35}
\Vtriangle[\modl`\Omg^{1}(\modl)\equiv\modl\otimes_{\struc}\Omg^{1}`\Omg^{2}(\modl)\equiv\modl\otimes_{\struc}
\Omg^{2};\conn`\curv\equiv\conn^{1}\circ\conn`\conn^{1}]
\end{equation}

\noindent from which we read that

\begin{equation}\label{eq36}
\curv\equiv \curv(\conn):=\conn^{1}\circ\conn
\end{equation}

\noindent Therefore, any time we have the $\cons$-linear morphism
$\conn$ and its prolongation $\conn^{1}$ at our disposal, we can
define the curvature $\curv(\conn)$ of the connection
$\conn$.\footnote{In connection with (\ref{eq36}), one can justify
our earlier remark that the standard differential operator
$\partial$, regarded as an $\struc$-connection as in (\ref{eq1})
({\it ie}, as the sheaf morphism $\partial
:~\struc\mapto\Omg^{1}=\struc\otimes_{\struc}\Omg^{1}\equiv\Omg^{1}(\struc)$),
is {\em flat}, since: $\curv(\partial)=\kd\circ\partial=d^{1}\circ
d^{0}\equiv d^{2}=0$ (which is secured by the nilpotency of the
usual Cartan-K\"{a}hler (exterior) differential operator $d$
\cite{malrap2}). In the latter paper, and in a sheaf-cohomological
fashion, it was shown that it is exactly $\conn$'s deviation from
nilpotency ({\it ie}, from flatness), which in turn {\em defines}
a non-vanishing curvature $\curv(\conn)=\conn^{2}\not= 0$, that
prevents a sequence
$\cdots\stackrel{\conn^{i-1}}{\mapto}\Omg^{i}\stackrel{\conn^{i}}{\mapto}\Omg^{i+1}
\stackrel{\conn^{i+1}}{\mapto}\cdots$ of differential
$\struc$-module sheaves $\Omg^{i}$ and $\cons$-linear sheaf
morphisms $\conn^{i}$ between them from being a {\em complex}.
($\conn^{i}$, $i\geq2,$ stand for high-order prolongations of the
$\conn^{0}\equiv\conn$ and $\conn^{1}$ connections above
\cite{mall1,mall2}.)} By defining a {\em curvature space} as the
finite sequence $(\struc ,\partial, \Omg^{1},\kd ,\Omg^{2})$ of
$\struc$-modules and $\cons$-linear morphisms between them, we can
distill the last statement to the following:

\begin{quotation}
\noindent {\em we can always define the curvature $\curv$ of a
given $\struc$-connection $\conn$, provided we have a curvature
space}.
\end{quotation}

As a matter of fact, it is rather straightforward to see that, for
$\modl$ a vector sheaf, $\curv(\conn)$ is an $\struc$-morphism of
$\struc$-modules, in the following sense

\begin{equation}\label{eq37}
\begin{array}{c}
\curv\in{\mathrm{Hom}}_{\struc}(\modl
,\Omg^{2}(\modl))=\Hom_{\struc}(\modl,\Omg^{2}(\modl))(X)\cr
\Omg^{2}({\modl}nd\modl)(X)=Z^{0}(\gauge,\Omg^{2}({\modl}nd\modl))
\end{array}
\end{equation}

\noindent where $\gauge=\{ U_{\alpha}\}_{\alpha\in I}$ is an open
cover of $X$ and $Z^{0}(\gauge,\Omg^{2}(\mathcal{E}nd\modl))$ the
$\struc(U)$-module of $0$-{\em cocycles} of
$\Omg^{2}(\mathcal{E}nd\modl)$ relative to the
$\gauge$-coordinatization of $X$.\footnote{One may wish to recall
that, for a vector sheaf $\modl$ like the one involved in
(\ref{eq37}), ${\modl}nd\modl\equiv{\mathcal{H}}om_{\struc}(\modl
,\modl)\cong\modl\otimes_{\struc}\modl^{*}=\modl^{*}\otimes_{\struc}\modl$.}

\subsubsection{The local form of $\curv$}

Motivated by (\ref{eq37}) and the last remarks, we are in a
position to give the local form for the curvature $\curv$ of a
given $\struc$-connection $\conn$. Thus, let $\modl$ be a vector
sheaf of rank $n$, $\conn$ an $\struc$-connection on it and
$\gauge=\{ U_{\alpha}\}_{\alpha\in I}$ a local coordinatization
frame of it. By virtue of (\ref{eq37}) we have

\begin{equation}\label{eq38}
\begin{array}{c}
\curv(\conn)=\curv=(\curv^{(\alpha)}_{ij})\equiv((\curv^{\alpha}_{ij}))\in
Z^{0}(\gauge ,\Omg^{2}({\modl}nd\modl))\cr
\subseteq\prod_{\alpha}\Omg^{2}({\modl}nd\modl)(U_{\alpha})=\prod_{\alpha}M_{n}(\Omg^{2}(U_{\alpha}))
\end{array}
\end{equation}

\noindent so that we are led to remark that:

\begin{quotation}
\noindent {\em the curvature $\curv$ of an $\struc$-connection
$\conn$ on a vector sheaf $\modl$ of rank $n$ is a $0$-cocycle of
local $n\times n$ matrices having for entries local sections of
$\Omg^{2}$---{\it ie}, local $2$-forms on $X$}.
\end{quotation}

\subsubsection{Local gauge transformations of $\curv$}

We wish to investigate here the behavior of the curvature
$\curv(\conn)$ of an $\struc$-connection $\conn$ under local gauge
transformations---the so-called `{\em transformation law of field
strengths}' in the usual gauge-theoretic parlance and in ADG
\cite{mall1,mall2}.

Thus, let $g\equiv g_{ij}\in\gl(n,\struc)(U\cap V)$ be the
change-of-gauge matrix we considered in (\ref{eq9}) in connection
with the transformation law of gauge potentials. Again, without
going into the details of the derivation, we bring forth from
\cite{mall1,mall2} the following local transformation law of gauge
field strengths

\begin{equation}\label{eq39}
\begin{array}{c}
\mathrm{for~a~local~frame~change :~}
e^{U}\stackrel{g}{\mapto}e^{V} (U,V~\mathrm{open~gauges~in}~X),\cr
\mathrm{the~curvature~transforms~as:}~\curv\stackrel{g}\mapto\curv^{'}=g^{-1}\curv
g
\end{array}
\end{equation}

\noindent which we are familiar with from the usual differential
geometric ({\it ie}, smooth fiber bundle-theoretic) treatment of
gauge theories. For completeness, we remind ourselves here that,
in (\ref{eq39}), $\curv^{U\cap V}\equiv(\curv^{U\cap V}_{ij})\in
M_{n}(\Omg^{2}(U\cap V))$---an $n\times n$ matrix of sections of
local $2$-forms. The transformation of $\curv$ under local gauge
changes is called {\em homogeneous} or {\em covariant} in the
usual gauge-theoretic parlance. We will return to this term in
2.3.5 and subsequently in section 5 where we will comment on the
geometrical ({\it ie}, tensorial) character of curvature.

\subsubsection{Cartan's structural equation---Bianchi identities}

We wish to express in ADG-theoretic terms certain well known, but
important, (local) identities about curvature. We borrow material
mainly from \cite{mall2}.

So, let $\modl$ be a vector sheaf and assume that $\gauge=\{
U_{\alpha}\}_{\alpha\in I}$ provides a coordinatization for it, as
above. The usual Cartan's structural equation reads in our case

\begin{equation}\label{eq40}
\curv^{(\alpha)}\equiv(\curv^{(\alpha)}_{ij})=\kd\aconn^{(\alpha)}+\aconn^{(\alpha)}\wee\aconn^{(\alpha)}\in
M_{n}(\Omg^{2}(U_{\alpha}))
\end{equation}

\noindent and similarly in the case of a sheaf $\modl$ of
$\struc$-modules and $U$ open in $X$

\begin{equation}\label{eq41}
\curv=\kd\aconn+\aconn\wee\aconn ;~(\aconn_{ij})\in
M_{n}(\Omg^{1}(U))
\end{equation}

\noindent (\ref{eq41}) can be also written in the {\em
Maurer-Cartan form}

\begin{equation}\label{eq42}
\curv=\kd\aconn+\frac{1}{2}[\aconn ,\aconn]
\end{equation}

\noindent by setting $[\aconn ,\aconn]\equiv
\aconn\wee\aconn-\aconn\wee\aconn$. For a one-dimensional vector
sheaf $\modl$ ({\it ie}, a line sheaf $\lsh$) equipped with an
$\struc$-connection $\conn$, the commutator in (\ref{eq41})
vanishes and we obtain the curvature as the following $0$-cocycle

\begin{equation}\label{eq43}
\curv=(\kd\aconn_{a})\in Z^{0}(\gauge
,\kd\Omg^{1})=(\kd\Omg^{1})(X)\subseteq\Omg^{2}(X)\subseteq\prod_{\alpha}\Omg^{2}(U_{\alpha})
\end{equation}

\noindent with $(\aconn_{\alpha})\in C^{0}(\gauge
,\Omg^{1})=\prod_{\alpha}\omg^{1}(U_{\alpha})$ the corresponding
(local) $\struc$-connection $0$-cochain of $\conn$.

To express the familiar Bianchi identities obeyed by the curvature
$\curv( \conn)$, and similarly to the extension of $\partial\equiv
d^{0}$ to the nilpotent Cartan-K\"{a}hler differential $\kd\equiv
d^{1}$ in 2.3, we need the extension of $d^{1}$ to a {\em second
exterior derivation} $\mathbf{d}\equiv d^{2}$ which again is a
$\cons$-linear sheaf morphism of the respective exterior
$\struc$-modules\footnote{In the sequel, following the
cohomological custom in \cite{malrap2}, we identify $\partial$,
$\kd$ and $\mathbf{d}$ (and all higher order exterior derivations)
with the generic Cartan differential $d$, specifying its order
only when necessary and by writing generically $d^{i}$
($i\geq0$).}

\begin{equation}\label{eq44}
\mathbf{d}:~\Omg^{2}\mapto\Omg^{3}
\end{equation}

\noindent acting (local) section-wise as follows

\begin{equation}\label{eq45}
\mathbf{d}(s\wee t):=\kd s\wee t-s\wee\kd t,~\forall
s,t\in\Omg^{1}(U);~U\subseteq X\, \mathrm{open}
\end{equation}

\noindent and being nilpotent

\begin{equation}\label{eq46}
d^{2}\circ d^{1}\equiv d\circ d\equiv d^{2}=0
\end{equation}

\noindent As a result of the extension of $\kd$ to $\mathbf{d}$,
the aforementioned {\em curvature space} $(\struc ,\partial,
\Omg^{1},\kd ,\Omg^{2})$, when enriched with the $\struc$-module
sheaf $\Omg^{3}$ as well as with the nilpotent $\cons$-linear
morphism $\mathbf{d}$ in (\ref{eq44}), becomes a so-called {\em
Bianchi space}.

In a Bianchi space, the usual {\em second Bianchi identity} holds

\begin{equation}\label{eq47}
\mathbf{d}\curv\equiv d\curv=[\curv ,\aconn]\equiv
\curv\wee\aconn-\aconn\wee\curv
\end{equation}

\noindent where $\mathbf{d}$ is understood to effect
coordinate-wise: $\mathbf{d}:~ M_{n}(\Omg^{2})\mapto
M_{n}(\Omg^{3})$.

In the case of a line sheaf $\lsh$, one can easily show by using
(\ref{eq30}) and the nilpotency of $d$ that

\begin{equation}\label{eq48}
d\curv=0
\end{equation}

\noindent which is usually referred to as the {\em homogeneous
field equation}. The latter, in turn, translates to the following
cohomological statement:

\begin{quotation}
\noindent {\em the curvature $\curv$ of an $\struc$-connection
$\conn$ on a line sheaf $\lsh$ over $X$ provides a closed $2$-form
on $X$}.
\end{quotation}

\noindent which came very handy in the sheaf-cohomological
classification of the curved associated line sheaves of qausets
and their quanta---the so-called `{\em causons}'---performed in
\cite{malrap2}.

Finally, one can also show that the second prolongation
$\conn^{2}_{{{\modl}}nd\modl}$ of the induced $\struc$-connection
$\conn_{{\modl}nd\modl}$ on ${\modl}nd\modl
\cong\modl\otimes_{\struc}\modl^{*}$ satisfies the following
`covariant version' of the second Bianchi identity (\ref{eq47})
above

\begin{equation}\label{eq49}
\conn^{2}_{{{\modl}}nd\modl}(\curv)=0
\end{equation}

\noindent where
$\conn^{2}_{{\modl}nd\modl}:~\Omg^{2}({\modl}nd\modl)\mapto\Omg^{3}({\modl}nd\modl)$.
Thus, similarly to (\ref{eq47}), one also shows that

\begin{equation}\label{eq50}
\conn_{{\modl}nd\modl}\curv=d\curv+[\aconn ,\curv]
\end{equation}

\noindent which proves the {\em equivalence} of the second
(exterior differential) Bianchi identity on $\modl$ and its
induced (covariant differential) version on ${\modl}nd\modl$.

\subsubsection{The Ricci tensor, scalar and the
Einstein-Lorentz (curvature) space}

Given a (real) Lorentzian vector sheaf $(\modl ,\rho)$ of rank $n$
equipped with a non-flat $\struc$-connection $\con$,\footnote{The
reader should note that below, and only in the vacuum Einstein
case, we will symbolize the connections involved by $\con$ instead
of the calligraphic $\conn$ we have used so far to denote the
general $\struc$-connections in ADG.}, one can define, in view of
(\ref{eq37}) the following {\em Ricci curvature operator} $\ric$
relative to a local gauge $U$ of $\modl$

\begin{equation}\label{eq51}
\ric(\, .\, ,\,\! s)t\in (\modl nd\modl)(U)=M_{n}(\struc(U))
\end{equation}

\noindent for local sections $s$ and $t$ of $\modl$ in
$\modl(U)=\struc^{n}(U)=\struc(U)^{n}$. $\ric$ is an $\modl
nd\modl$-valued operator.\footnote{Due to this, $\ric$ has been
called a {\em curvature endomorphism} in \cite{mall3}.}

Since $\ric$ is matrix-valued, as (\ref{eq51}) depicts, one can
take its trace thus define the following {\em Ricci scalar
curvature operator} $\ricci$

\begin{equation}\label{eq52}
\ricci(s,t):=tr(\ric(\, .\, ,\,\! s)t)
\end{equation}

\noindent which, plainly, is $\struc(U)$-valued.

We have built a suitable conceptual background to arrive now at a
central notion in this paper. A (real) Lorentzian vector sheaf
$\modll=(\modl ,\rho)$ over a $\mathbf{R}$-algebraized space
$(X,\struc)$ such that:

\begin{enumerate}

\item it is supported by a differential triad $\triad=(\struc ,\partial, \Omg^{1})$
relative to which (\ref{eq12}) holds, that is,
$\modl^{*}\equiv\Omg^{1}$,

\item there is a $\mathbf{R}$-linear Lorentzian connection $\con$ on it satisfying
(\ref{eq17}) ({\it ie}, a metric connection) and, furthermore,

\item it is a curvature space $(\struc ,\partial, \Omg^{1},\kd ,\Omg^{2})$ supporting
a {\em null $\ricci$}, that is to say, {\em a Ricci scalar
operator satisfying the vacuum Einstein equations}

\begin{equation}\label{eq53}
\ricci(\modl)=0
\end{equation}

\end{enumerate}

\noindent is called an {\em Einstein-Lorentz (E-L) space}, while
the corresponding base space $X$, an {\em Einstein space}
\cite{mall3}.\footnote{In the next section, where we will cast
Lorentzian gravity as a Y-M-type of gauge theory {\it \`a la} ADG,
we will also define a {\em Yang-Mills space} analogous to the
Einstein space above.} Of course, it has been implicitly assumed
that, for an appropriate choice of structure sheaf $\struc$,
equation (\ref{eq53}) {\em can be actually derived from the
variation of the corresponding Lagrangian density} ({\it alias},
Einstein-Hilbert action functional $\eh$). We will return to this
assumption in the next section.

In connection with the definition of an Einstein space $X$, it is
worth noting that

\begin{quotation}
\noindent{\em the only structural requirement that ADG places on
the Einstein base space $X$ is that it is, merely, a topological
space---in fact, an arbitrary topological space, without any
assumptions whatsoever about its differential, let alone its
metric, structure.}
\end{quotation}

\noindent This prompts us to emphasize, once again
\cite{mall1,mall2,mall3,malros1,malros2,malrap1,malrap2,mall7},
the essential `working philosophy' of ADG:

\begin{quotation}
\noindent{\em to actually do differential geometry one need not
assume any `background differentiable space' $X$, for
differentiability derives from the algebraic structure of the
objects (structure algebras) that live on that `space'. The only
role of the latter is a secondary, auxiliary and, arguably, a
`physically atrophic' one in comparison to the active role played
by those objects (in particular, the algebra $\struc(U)$ of local
sections of $\struc$) themselves: $X$ merely provides an inert,
ether-like scaffolding for the localization and the dynamical
interactions (`algebraically and sheaf-theoretically modelled
interrelations') of those physically significant objects---a
passive substrate of no physical significance whatsoever, since it
does not actively participate into the algebraico-dynamical
relations between the objects themselves.\footnote{Its arbitrary
character---again, $X$ is assumed to be simply an {\em arbitrary
topological space}---reflects precisely its physical
insignificance. This non-physicality, the `algebraic inactivity'
and `dynamically non-participatory character' so to speak, of the
background space will become transparent subsequently when we
formulate the dynamical equations for vacuum gravity entirely in
terms of {\em sheaf morphisms between the objects---{\it ie},
virtually the sections---that live on $X$} (the main sheaf
morphism being the connection $\conn$---arguably the central
operator with which one actually does differential geometry!). At
this point we would like to further note, according to
\cite{mall1}, that {\em a sheaf morphism is} actually reduced to
{\em a family of (local) morphisms between} (the complete
presheaves of) {\em local sections} $Mor(\modl ,{\mathcal{F}}) \ni
\phi\longleftrightarrow(\phi_{U})\in
Mor(\Gamma(\modl),\Gamma({\mathcal{F}})$---a {\em category
equivalence} through (the {\em section functor}) $\Gamma$. In the
last section we will return to the inert, passive, ether-like
character of the base space in the particular case that $X$ is (a
region of) a $\smooth$-smooth spacetime manifold. There we will
argue how {\em ADG `relativizes' the `differential properties' of
space(time)}.} All in all, the basic objects that ADG works with
is the sections of the sheaves in focus---that is, the entities
that live in the stalks of the relevant sheaves, and not with the
underlying base space $X$, so that any notion of
`differentiability' according to ADG derives its sense from the
algebraic relations between ({\it ie}, the algebraic structure of)
those (local) sections, with the apparently `intervening between'
or `permeating through these objects' background space $X$ playing
absolutely no role in it.}
\end{quotation}

\subsubsection{A fundamental difference between $\conn$ and
$\curv(\conn)$ and its physical interpretation}

At this point it is worth stressing a characteristic difference
between an $\struc$-connection $\conn$ and its curvature
$\curv(\conn)$---a difference that is emphasized by ADG, it has a
significant bearing on the physical interpretation of our theory,
and it has been already highlighted in both \cite{malrap1} and
\cite{malrap2}; namely that,

\vskip 0.1in

\centerline{\em while $\curv$ is an $\struc$-morphism, $\conn$ is
only a $\mathbf{K}$-morphism $(\mathbf{K}=\mathbf{R},\cons)$.}

\vskip 0.1in

\noindent This means that, since the structure sheaf $\struc$
corresponds to `geometry' in our algebraic scheme, in the sense
that $\struc(U)$---the algebra of local sections of
$\struc$---represents {\em the algebra of local operations of
measurement} (of the quantum system `space-time') {\em relative to
the local laboratory} (frame, or gauge, or even `observation
device') $U$ \cite{rap2,malrap1,malrap2}, it effectively encodes
{\em our} geometrical information about the physical system in
focus.\footnote{As mentioned before, $\struc_{X}$ is the abelian
algebra sheaf of `{\em generalized arithmetics}' in ADG
generalizing the usual commutative coordinate sheaf
$^{\R}\smooth_{M}$ of the smooth manifold---the sheaf of abelian
rings ${}^{\R}\smooth(M)$ of infinitely differentiable,
real-valued functions on the differential manifold $M$. We tacitly
assume in our theory that {\em `geometry' is synonymous to
`measurement'}; hence, in the quantum context, it is intimately
related to `observation' (being, in fact, the result of it).
Furthermore, since the results of observation arguably lie on the
classical side of the quantum divide (the so-called {\em
Heisenberg Schnitt}), $\struc$ must be a sheaf of {\em abelian}
algebras. This is supposed to be a concise ADG-theoretic
encodement of Bohr's correspondence principle, namely, that {\em
the numbers that we obtain upon measuring the properties of a
quantum mechanical system (the so-called $q$-numbers) must be
commutative (the so-called $c$-numbers)}. In other words, {\em the
acts of measurement yield $c$-numbers from $q$-numbers}, so that
`geometry'---the structural analysis of (the algebras of our local
measurements of) `space'---deals, by definition, with commutative
numbers and the (sheaves of) abelian algebras into which the
latter are effectively encoded. See also closing remarks in
\cite{mall2} for a similar discussion of `geometry {\it \`a la}
ADG' in the sense above, as well as our remarks about {\em
Gel'fand duality} in 5.5.1} Consequently,

\begin{quotation}
\noindent {\em $\curv$, which, being an $\struc$-morphism,
respects our local measurements---the `geometry-encoding
(measuring) apparatus' $\struc$ of ADG so to speak---is a
geometrical object ({\it ie}, a tensor) in our theory and lies on
the classical side of the quantum divide. On the other hand,
$\conn$, which respects only the constant sheaf
$\mathbf{K}(=\mathbf{R},\cons)$ but not our (local) measurements
in $\struc$, is not a geometrical object\footnote{Another way to
say this is that {\em the notion of connection is algebraic ({\it
ie}, analytic), not geometrical}. In short, $\conn$ {\em is not a
tensor}. That $\curv$ is a tensor while $\conn$ is not is
reflected in their (local) gauge transformation laws that we saw
earlier: $\aconn$ transforms affinely or inhomogeneously
(non-tensorially), while $\curv$ covariantly or homogeneously
(tensorially) under a (local) change of gauges.} and it lies on
the quantum ({\it ie}, the purely algebraic, {\it \`a la} Leibniz}
\cite{mall7}{\em), side of Heisenberg's cut}.\footnote{Although it
must be also stressed that $\conn$, like the usual notion of
derivative $\partial$ that it generalizes, has a {\em geometrical
interpretation}. As the derivative of a function (of a single
variable) is usually interpreted in a Newtonian fashion as the
slope (gradient) of the tangent to the curve (graph) of the
function, so $\conn$ can be interpreted geometrically as a
parallel transporter of objects (here, $\struc$-tensors) along
geometrical curves (paths) in space(time). However, it is rather
inappropriate to think of $\conn$ as a geometrical object proper
and at the same maintain a geometrical interpretation for it, for
{\em does it not sound redundant to ask for the geometrical
interpretation of an `inherently geometrical' object, like the
triangle or the circle, for instance?} In other words, {\em if the
notion of connection was `inherently geometrical', it would
certainly be superfluous to also have a geometrical interpretation
for it}.}
\end{quotation}

\subsection{The Affine Space {\rm $\sconn$} of $\struc$-Connections}

We fix the $\mathbf{K}$-algebraized space $(X,\struc)$ and the
differential triad $\triad=(\struc ,\partial ,\Omg)$ on it with
which we are working, and we let $\modl$ be an $\struc$-module on
$X$. We denote by

\begin{equation}\label{eq54}
\sconn_{\struc}(\modl)
\end{equation}

\noindent {\em the set of $\struc$-connections on $\modl$}. By
definition (\ref{eq3}), $\sconn_{\struc}(\modl)$ is a subset of
${\mathrm{Hom}}_{\mathbf{K}}(\modl ,\Omg(\modl))$
($\Omg\equiv\Omg^{1}$) whose zero element may be regarded as the
zero $\struc$-connection in $\sconn_{\struc}(\modl)$. However, by
(\ref{eq4}), one infers that $\partial$ is also zero in this case,
thus we will exclude altogether the zero $\struc$-connection from
$\sconn_{\struc}(\modl)$. Since any connection may be taken to
serve as an `origin' for the space of $\struc$-connections, we
conclude that

\begin{quotation}
\noindent $\sconn_{\struc}(\modl)$ {\em is an affine space
modelled after the $\struc(X)$-module
${\mathrm{Hom}}_{\mathbf{K}}(\modl ,\Omg(\modl))$. For a vector
sheaf $\modl$, ${\mathrm{Hom}}_{\mathbf{K}}(\modl ,\Omg(\modl))$
becomes $\Omg({\modl}nd\modl)(X)$.}
\end{quotation}

\noindent Now, in connection with the statement above, let $\conn$
be an $\struc$-connection in
$\sconn_{\struc}(\modl)\equiv{\mathrm{Hom}}_{\mathbf{K}}(\modl
,\Omg(\modl))$. Then, it can be shown \cite{mall1,mall2} that any
other connection $\conn^{'}$ in $\sconn_{\struc}(\modl)$ is of the
form

\begin{equation}\label{eq55}
\conn^{'}=\conn+u
\end{equation}

\noindent for a uniquely defined
$u\in{\mathrm{Hom}}_{\struc}(\modl ,\Omg^{1}(\modl))$. For $\modl$
a vector sheaf, $u$ belongs to $\Omg^{1}({\modl}nd\modl)(X)$.
Thus, for a given $\conn\in\sconn_{\struc}(\modl)$ we can formally
write (\ref{eq55}) as:
$\sconn_{\struc}(\modl)=\conn+{\mathrm{Hom}}_{\struc}(\modl
,\Omg^{1}(\modl))$, within a bijection. Interestingly enough,
(\ref{eq55}) tells us that the difference of two connections,
which are $\mathbf{K}$-linear sheaf morphisms, is an
$\struc$-morphism like the curvature; hence, in view of the
comparison between $\conn$ and $\curv(\conn)$ above, we can say
that $\conn^{'}-\conn$ is a geometrical object since it respects
our measurements in $\struc$ by transforming homogeneously
(tensorially) under (local) gauge transformations.\footnote{The
reader could verify that $u$ transforms covariantly under (local)
changes of gauge.}

In the particular case of a line sheaf $\lsh$,

\vskip 0.1in

\centerline{$\sconn_{\struc}(\lsh)$ {\em can be identified with
$\Omg^{1}(X)$---the $\struc(X)$-module of `$1$-forms' on $X$.}}

\vskip 0.1in

\noindent Thus, given any connection $\conn$ in
$\sconn_{\struc}(\lsh)$, any other connection $\conn^{'}$ on
$\lsh$ can be written as $\conn^{'}=\conn+\omega$ for some unique
$\omega$ in $\Omg^{1}(X)$. This result was used in \cite{malrap2}
for the sheaf-cohomological classification of the line sheaves
associated with the curved principal finsheaves of qausets and the
non-trivial connections on them in \cite{malrap1}.

We will return to $\sconn_{\struc}(\modl)$ in the next section
where we will factor it by the structure (gauge) group
$\struct={\mathrm {Aut}}(\modl)$ of $\modl$ to obtain the orbifold
or moduli space $\sconn_{\struc}(\modl)/\struct$ of
gauge-equivalent connections on $\modl$ of a Y-M or gravitational
type depending on $\struct$.

\section{Vacuum Einstein Gravity as a Y-M-type of Gauge Theory \`a la ADG}

In this section we present the usual vacuum Einstein gravity in
the language of ADG, that is, as a Y-M-type of gauge theory
describing the dynamics of a Lorentzian connection on a suitable
principal Lorentzian sheaf and its associated vector sheaf, in
short, on an E-L space as defined above. We present only the
material that we feel is relevant to our subsequent presentation
of finitary vacuum Lorentzian gravity encouraging the reader to
refer to \cite{mall1,mall2,mall3,mall4} for more analytical
treatment of Y-M theories and gravity {\it \`{a} la} ADG. But let
us first motivate in a rather general way this conception of {\em
gravity as a gauge theory}.

\subsection{Physical Motivation}

It is well known that the original formulation of general
relativity was in terms of a pseudo-Riemannian metric $g_{\mu\nu}$
on a $\smooth$-smooth spacetime manifold $M$. For Einstein, the
ten components of the metric represented the gravitational
potentials---the pure gravitational dynamical degrees of freedom
so to speak. However, very early on it was realized that there was
an equivalent formulation of general relativity involving the
dynamics of the so-called {\em spin-connection} $\omega$. This
approach came to be known as {\em Einstein-Cartan theory}
\cite{gosch} and arguably it was the first indication, long before
the advent of the Y-M gauge theories of matter, that gravity
concealed some sort of gauge invariance which was simply masked by
the metric formulation.\footnote{Recently, after reading
\cite{kostro}, the present authors have become aware of a very
early attempt by Eddington at formulating general relativity (also
entertaining the possibility of unifying gravity with
electromagnetism) based solely on the affine connection and not on
the metric, which is treated as a secondary structure,
`derivative' in some sense from the connection. Indicatively,
Kostro writes: ``{\em ...{\rm [Eddington's]} approach relied on
affine geometry. In this geometry, connection, and not metric, is
considered to be the basic mathematical entity. The metric
$g_{\mu\nu}(x)$ needed for the description of gravitational
interactions, appears here as something secondary, which is
derived from connection...}'' (bottom of page 99 and references
therein).} In fact, Feynman, in an attempt to view gravity purely
field-theoretically and, {\it in extenso}, quantum gravity as a
quantum field theory ({\it ie}, in an attempt to quantize gravity
using a language and techniques more familiar to a particle
physicist than a general relativist,\footnote{Such an approach was
championed a decade later by Weinberg in a celebrated book
\cite{wein}.}) he essentially `downplayed', or at least
undermined, the differential geometric picture of general
relativity and instead he concentrated on its gauge-theoretic
attributes. Brian Hatfield nicely reconstructed Feynman's attitude
towards (quantum) gravity in \cite{feyn2},\footnote{See Hatfield's
preamble titled `{\itshape Quantum Gravity}'.} as follows

\begin{quotation}
\noindent\small{``...Thus it is no surprise that Feynman would
recreate general relativity from a non-geometrical viewpoint. The
practical side of this approach is that one does not have to learn
some `fancy-schmanzy' (as he liked to call it) differential
geometry in order to study gravitational physics. (Instead, one
would just have to learn some quantum field theory.) However, when
the ultimate goal is to quantize gravity, Feynman felt that the
geometrical interpretation just stood in the way. From the field
theoretic viewpoint, one could avoid actually defining---up
front---the physical meaning of quantum geometry, fluctuating
topology, space-time foam, {\it etc.}, and instead look for the
geometrical meaning after quantization...Feynman certainly felt
that the geometrical interpretation is marvellous, but `{\em the
fact that a massless spin-$2$ field can be interpreted as a metric
was simply a coincidence that might be understood as representing
some kind of gauge invariance'.}''\footnote{Our emphasis of
Feynman's words as quoted by Hatfield.}}
\end{quotation}

\noindent Feynman's `negative' attitude towards the standard
differential geometry and the smooth spacetime continuum that
supports it,\footnote{The reader must have realized by now that by
the epithets `standard', or `usual', or more importantly,
`classical', to `differential geometry' we mean the differential
geometry of $\smooth$-smooth manifolds---the so-called `calculus
on differential manifolds'.} especially if we consider the
unrenormalizable infinities that plague quantum gravity when
treated as another quantum field theory, is quite understandable
if we recall from the beginning of the present paper his earlier
position---repeated once again, that ``{\em the theory that space
is continuous is wrong, because we get...infinities...the simple
ideas of geometry, extended down to infinitely small, are
wrong!}''.\footnote{In the closing section we will return to
comment thoroughly, in the light of ADG, on this remark by Feynman
and the similar one of Isham also quoted in the beginning of the
paper.}

However, it must be noted that Feynman's `unconventional' attempt
in the early 60s to tackle the problem of quantum gravity gauge
quantum field-theoretically was preceded by Bergmann's ingenious
recasting of the Einstein-Cartan theory in terms of $2$-component
spinors, thus effectively showing that the main dynamical field
involved in that theory---the spin connection $\omega$---is an
$sl(2,\com)$-valued $1$-form \cite{berg}.\footnote{More precisely,
in Bergmann's theoretical scenario for classical Lorentzian
gravity, $g_{\mu\nu}$ is replaced by a field of four $2\times 2$
Pauli spin-matrices which is locally invariant when conjugated by
a member of $SL(2,\com)$---the double cover of the Lorentz group.}
All in all, it is remarkable indeed that such a connection-based
approach to general relativity, classical or quantum, has been
revived in the last fifteen years or so in the context of {\em
non-perturbative canonical quantum gravity}. We refer of course to
Ashtekar's modification of the Palatini {\it vierbein} or comoving
$4$-frame-based formalism by using new canonical variables to
describe the phase space of general relativity and in which
variables the gravitational constraints are significantly
simplified \cite{ash}. Interestingly enough, and in relation to
Bergmann's work mentioned briefly above, in Ashtekar's scheme the
principal dynamical variable is an $sl(2,\com)$-valued {\em
self-dual spin-Lorentzian connection $1$-form
$\aconn^{+}$}\footnote{Later in the present section we will
discuss briefly self-dual connections from ADG's point of view.}
\cite{ash}.

But after this lengthy preamble let us get on with our main aim in
this section to present the classical vacuum Lorentzian gravity as
a Y-M-type of gauge theory in the manner of ADG.

\subsection{Y-M Theory \`{a} la ADG---Y-M Curvature Space}

Let $(\modl ,\rho)$ be a (real) Lorentzian vector sheaf of finite
rank $n$ associated with a differential triad $\triad=(\struc
,\partial ,\Omg^{1})$, which in turn is associated with the
$\mathbf{R}$-algebraized space $(X,\struc)$,\footnote{With $X$ a
paracompact Hausdorff topological space and $\struc$ a {\em fine}
unital commutative algebra sheaf (over $\R$) on it, as usual.} and
$\conn$ a non-trivial Lorentzian $\struc$-connection on it ({\it
ie}, $\curv(\conn)\not= 0$). In ADG, the pair $(\modl ,\conn)$ is
generically referred to as a {\em Y-M field}, the triplet $(\modl
,\rho ,\conn)$ as a {\em Lorentz-Yang-Mills (L-Y-M) field}, and it
has been shown \cite{mall1,mall2,mall3} that\footnote{In the
sequel, and similarly to how we used different symbols for the
(vacuum) gravitational connection $\con$ and its Y-M counterpart
$\conn$, we will use $\stre$ for the curvature of the latter
instead of $\curv$ ($\ric$ and $\ricci$) that we used for the
former. In the Y-M context the curvature of a connection is
usually referred to as the (gauge) field strength.}

\begin{quotation}
\noindent{\em every Lorentzian vector sheaf yields a (non-trivial)
L-Y-M field $(\modl ,\rho ,\conn)$ on $X$ the (non-vanishing)
field strength of which is $\stre(\conn)$.}
\end{quotation}

\noindent As in the definition of the E-L space earlier, in case
the curvature $\stre$ of the connection $\conn$ of a L-Y-M field
satisfies the free Y-M equations, which we write as
follows\footnote{In (\ref{eq56}), `$\delta$' is the {\em
coderivative} \cite{gosch} and $\Delta$ the {\em Laplacian
operator}, which we will define in an ADG-theoretic manner
shortly. These are two equivalent expressions of the free Y-M
equations. Their equivalence, which is a consequence of the
covariant differential Bianchi identity (\ref{eq50}), has been
shown in \cite{mall1}.}

\begin{equation}\label{eq56}
\delta^{2}_{\modl nd\modl}(\stre)=0~\mathrm{or}~\Delta_{\modl
nd\modl}^{2}(\stre)=0
\end{equation}

\noindent and which, in turn, we assume that can be obtained from
the variation of a corresponding Y-M action functional
$\ym$,\footnote{We will discuss this derivation in more detail
shortly.} the curvature space $(\struc ,\partial, \Omg^{1},\kd
,\Omg^{2})$ associated with the L-Y-M field is called a {\em L-Y-M
curvature space},\footnote{A particular kind of Bianchi space
defined earlier.} while the supporting $X$, a {\em L-Y-M
space}.\footnote{In order for the reader not to be misled by our
terminology, it must be noted here that, in contrast to the usual
term `(free) Yang-Mills field' by which one understands the field
strength of a gauge potential which is a solution to the (free)
Y-M equations (\ref{eq56}), in ADG, admittedly with a certain
abuse of language, a Y-M field is just the pair $(\modl ,\conn)$,
without necessarily implying that $\stre(\conn)$ satisfies
(\ref{eq56}). On the other hand, the Y-M space $X$ supporting the
Y-M curvature space $(\struc ,\partial, \Omg^{1},\kd ,\Omg^{2})$
associated with a Y-M field $(\modl ,\conn)$, is supposed to refer
directly to solutions $\stre(\conn)$ of (\ref{eq56})---as it were,
it represents the `solution space' of (\ref{eq56}). This is in
complete analogy to the Einstein-Lorentz space and Einstein space
$X$ defined in connection with the vacuum Einstein equations for
Lorentzian gravity in (\ref{eq53}). We will return to comment
further on this conception of a curvature space as a geometrical
`solution space' in section 5 when we express (\ref{eq53}) in
finitary terms.} In connection with the said derivation of the Y-M
equations from $\ym$, we note that\footnote{In fact, the statement
that follows is a {\em theorem} in ADG \cite{mall2,mall3,mall4}.
We will return to it in 3.3.}

\begin{quotation}
\noindent{\em the solutions of the Y-M equations that correspond
to a given Y-M field $(\modl ,\conn)$ are precisely the critical
or stationary points (or extrema) of $\ym$ that can be associated
with $\modl$.}
\end{quotation}

In order to make sense of (\ref{eq56}) ADG-theoretically, we need
to define the coderivative and the Laplacian of a given L-Y-M
field $(\modl ,\rho ,\conn)$. We do this below.

\subsubsection{The adjoint $\delta$ and the Laplacian $\Delta$
of an $\struc$-connection in ADG}

Let $\triad=(\struc ,\partial ,\Omg^{1})$ be the differential
triad we are working with and $\rho$ a Lorentzian $\struc$-metric
on it, as usual. Let also $\modl$ be a Lorentzian vector sheaf of
finite rank $n$ and $\conn$ a Lorentzian Y-M connection on it. By
emulating the classical situation sheaf-theoretically, as it is
customary in ADG, one can define the {\em adjoint derivation}
$\delta$ of $\conn$ relative to $\rho$ as the following
$\struc$-morphism of the vector sheaves involved

\begin{equation}\label{eq57}
\delta^{1}\equiv\delta
:~\Omg^{1}(\modl)\mapto\modl(\equiv\Omg^{0}(\modl))
\end{equation}

\noindent satisfying

\begin{equation}\label{eq58}
\rho(\conn(s),t)=\rho(s,\delta(t))
\end{equation}

\noindent with the obvious identifications: $\forall
s\in\modl(U)$, $t\in\Omg^{1}(\modl)(U)$, and $U$ a common open
gauge of $\modl$ and $\Omg^{1}(\modl)$. $\delta$ is uniquely
defined through the $\struc$-metric isomorphism
$\modl\simeq\modl^{*}$ we saw in (\ref{eq12}).

To define the Laplacian $\Delta$ associated with $\conn$, apart
from the connection $\conn\equiv\conn^{0}$ and the coderivative
$\delta$, we also need $\conn^{1}$ (the first prolongation of
$\conn$, as in (\ref{eq33})) and
$\delta^{2}:~\Omg^{2}(\modl)\mapto\Omg^{1}$ (the second
contraction relative to $\conn$,\footnote{Which can be defined in
complete analogy to (\ref{eq58}).}) as follows

\begin{equation}\label{eq59}
\Delta\equiv\Delta^{1}:=\delta^{2}\circ\conn^{1}+\conn^{0}\circ\delta^{1}
\equiv\delta\conn+\conn\delta:~\Omg^{1}(\modl)\mapto\Omg^{1}(\modl)
\end{equation}

\noindent Higher order Laplacians $\Delta^{i}$, generically
referred to as $\Delta$, can be similarly defined as
$\mathbf{K}$-linear vector sheaf morphisms

\begin{equation}\label{eq60}
\Delta^{i}:~\Omg^{i}(\modl)\mapto\Omg^{i}(\modl),~i\in\N
\end{equation}

\noindent and they read via the corresponding higher order
connections $\conn^{i}$ and coderivatives $\delta^{i}$

\begin{equation}\label{eq61}
\Delta^{i}:=\delta^{i+1}\circ\conn^{i}+\conn^{i-1}\circ\delta^{i},~i\in\N
\end{equation}

\noindent with the higher order analogues of (\ref{eq58}) being

\begin{equation}\label{eq62}
\rho(\conn^{p}(s),t)=\rho(s,\delta^{p+1}(t)),~p\in\Z_{+}
\end{equation}

\noindent where $\rho$ is the $\struc$-metric on the vector sheaf
$\Omg^{p}(\modl)$ and the `exterior' analogue of (\ref{eq12})
reading

\begin{equation}\label{eq63}
\begin{CD}
\Omg^{p}(\modl)@>\subset>\tilde{\rho}>(\Omg^{p}(\modl))^{*}
\end{CD}
\end{equation}

\noindent Having defined $\Delta$ and $\delta$, the reader can now
return to (\ref{eq56}) understanding $\delta^{2}_{\modl nd\modl}$
and $\Delta_{\modl nd\modl}^{2}$ as the maps $\delta^{2}_{\modl
nd\modl}:~\Omg^{2}(\modl nd\modl)\mapto\Omg^{2}(\modl nd\modl)$
and $\Delta_{\modl nd\modl}^{2}=~\delta^{3}_{\modl
nd\modl}\circ\conn^{2}_{\modl
nd\modl}+\conn^{1}\circ\delta^{2}_{\modl nd\modl}:~\Omg^{2}(\modl
nd\modl)\mapto\Omg^{2}(\modl nd\modl)$,
respectively.\footnote{Always remembering that the field strength
$\stre$ of the L-Y-M connection $\conn$ is an $\struc$-morphism
between the $\struc$-modules $\modl$ and $\Omg^{2}(\modl)$ ({\it
ie}, a member of $\Hom_{\struc}(\modl ,\Omg^{2}(\modl))(X)$), as
(\ref{eq37}) depicts.} By abusing notation, we may rewrite the
free Y-M equations (\ref{eq56}) as

\begin{equation}\label{eq64}
\delta(\stre)=0~~\mathrm{or}~~\Delta(\stre)=0
\end{equation}

\noindent hopefully without sacrificing understanding.

Our ADG-theoretic exposition of the Y-M equations so far, together
with a quick formal comparison that one may wish to make between
the aforedefined (vacuum) E-L and the (free) L-Y-M curvature
spaces, reveals our central contention in this section, namely
that

\begin{quotation}
\noindent{\em in ADG, vacuum Einstein Lorentzian gravity is a
Yang-Mills-type of gauge theory involving the dynamics of a
Lorentzian connection $\con$ on an Einstein space $X$. In complete
analogy to the L-Y-M case above, the corresponding triplet $(\modl
,\rho ,\con)$ (whose Ricci scalar curvature $\ricci$ is)
satisfying} (\ref{eq53}), {\em is called a (vacuum)
Einstein-Lorentz field. For rank $n=4$, structure group
$\mathrm{Aut}(\modll)=L^{\uparrow}$ and principal sheaf $\orthl$,
the associated vacuum Einstein-Lorentz field is written as
$(\modll ,\con)$ $(\modll=(\modl ,\rho))$. Locally in the Einstein
space $X$, $\con=\partial+\aconn$, with $\aconn$ an
$sl(2,\com)\simeq so(1,3)^{\uparrow}$-valued $1$-form representing
the vacuum gravitational gauge potential.}
\end{quotation}

\subsection{The Einstein-Hilbert Action Functional $\eh$}

Now that we have established with the help of ADG the close
structural similarity between vacuum Einstein Lorentzian gravity
and free Y-M theory, we will elaborate for a while on our remark
earlier that both (\ref{eq53}) and (\ref{eq56}) or (\ref{eq64})
derive from the extremization of an action functional---the E-H
$\eh$ in the first case, and the Y-M $\ym$ in the second. Since
only vacuum Einstein gravity interests us here, we will discuss
only the variation of $\eh$, leaving the variation of $\ym$ for
the reader to read from \cite{mall1,mall2,mall4}.

As it has been transparent in the foregoing presentation, from the
ADG-theoretic point of view, the main dynamical variable in vacuum
Einstein Lorentzian gravity is the spin-Lorentzian
$\struc$-connection $\con$, or equivalently, its gauge potential
part $\aconn$ on the vector sheaf $\modll=(\modl ,\rho)$. Thus,
one naturally anticipates that

\begin{quotation}
\noindent {\em the E-H action $\eh$ is a functional on the affine
space} $\sconn_{\struc}(\modll)$ {\em of Lorentzian metric ({\it
ie}, $\rho$-compatible) $\struc$-connections on $\modll$.}
\end{quotation}

\noindent Indeed, we define $\eh$ as the following map

\begin{equation}\label{eq65}
\eh:~\sconn_{\struc}(\modll)\mapto\struc(X)
\end{equation}

\noindent reading `point-wise'

\begin{equation}\label{eq66}
\con\mapsto\eh(\con)\, :=\ricci(\con)=:\, tr\ric(\con)
\end{equation}

\noindent where. plainly, $\ricci$ is a global section of the
structure sheaf of coefficients $\struc$ ({\it ie},
$\ricci\in\struc(X)$).

Our main contention (in fact, a theorem in ADG
\cite{mall1,mall2,mall3}) in 2.3, as well as in 3.2 in connection
with Y-M theory, was that

\begin{quotation}
\noindent{\em the solutions of the vacuum Einstein field
equations} (\ref{eq53}) {\em that correspond to a given E-L field
$(\modll ,\con)$ are obtained from extremizing $\eh$---that is,
they are the critical or stationary points of the functional $\eh$
associated with $\modll$ in} (\ref{eq65}) {\em and} (\ref{eq66})
{\em above.}
\end{quotation}

\noindent In what follows we will recall briefly how ADG deals
with this statement.

The critical points of $\eh$ can be obtained by first restricting
it on a curve $\gamma(t)$ in connection space ({\it ie}, $\gamma
:~t\in\R\mapto \gamma(t)\in\sconn_{\struc}(\modll)$) and then by
infinitesimally varying it around its `initial' value
$\eh[\con_{0}]\equiv\eh[\gamma(0)]$. Alternatively, and following
the rationale in \cite{mall3}, in order to find the stationary
points of $\eh$, one has to find the `tangent vector' at time
$t=0$ to a path $\gamma(t)$ in the affine space
$\sconn_{\struc}(\modll)$ of $\struc$-connections of $\modll$, on
which path $\eh$ is constrained to take values in $\struc(X)$ as
(\ref{eq65}) dictates. All in all, one must evaluate

\begin{equation}\label{eq67}
\stackrel{\centerdot}{\overbrace{\eh(\gamma(t))}}(0)\equiv
\stackrel{\centerdot}{\overbrace{\eh(\gamma)}}(0)
\end{equation}

\noindent where $\dot{x}$ is Newton's notation for
$\frac{dx}{dt}$.

For a given Lorentzian metric connection $\con$, one can take the
path $\gamma$ in connection space to be

\begin{equation}\label{eq68}
\gamma(t)\equiv\con_{t}=\con+t\conne\in\sconn_{\struc}(\modll),\,
t\in\R
\end{equation}

\noindent where $\conne\in\Omg^{1}(\modl nd\modll)(X)$ as
mentioned earlier in (\ref{eq55}). $\con_{t}$ may be regarded as
the $\struc$-connection on $\modll$ compatible with the Lorentzian
metric $\rho_{t}=\rho+t\rho^{'}$, with $\rho^{'}$ an arbitrary
symmetric $\struc$-metric on $\modll$.

So, given the usual E-H action (without a cosmological constant)

\begin{equation}\label{eq69}
\eh(\con)=\int\ricci(\con)\vol
\end{equation}

\noindent with $\vol$ the volume element associated with
$\rho$,\footnote{We will return to define $\vol$ shortly.}
(\ref{eq67}) reads

\begin{equation}\label{eq70}
\frac{d}{dt}(\eh(\con_{t}))|_{t=0}\equiv
\stackrel{\centerdot}{\overbrace{\eh(\con_{t})}}(0)=\int\frac{d}{dt}(\ricci\vol)|_{t=0}
\end{equation}

\noindent By setting
$\stackrel{\centerdot}{\overbrace{\eh(\con_{t})}}(0)$ in
(\ref{eq70}) equal to zero, one arrives at the vacuum Einstein
equations (\ref{eq53}) for Lorentzian gravity.

\subsubsection{A brief note on the topology of $\sconn_{\struc}(\modll)$}

In the introduction we alluded to the general fact that the space
of connections is non-linear ({\it ie}, it is not a vector space)
with a `complicated' topology. Below we would like to comment
briefly on the issue of the topology of the space
$\sconn_{\struc}(\modll)$ of spin-Lorentzian connections on
$\modll$. This issue is of relevance here since one would like to
make sense of the $\frac{d}{dt}$-differentiation of $\eh$ in
(\ref{eq67}). Thus, in connection with (\ref{eq67}), the crucial
question appears to be:

\begin{quotation}
\noindent{\em with respect to what topology} (on
$\sconn_{\struc}(\modll)$) {\em does one take the limit so as to
define the (`variational') derivative of} $\eh$ {\em with respect
to $t$ ({\it ie}, with respect to $\con$)} {\em in}
(\ref{eq67})?\footnote{This question would also be of relevance if
for instance one asked whether the map (path) $\gamma$ in
(\ref{eq68}) is continuous.}
\end{quotation}

\noindent ADG answers this question by first translating it to an
equivalent question about convergence in the structure sheaf
$\struc$. That is to say,

\vskip 0.07in

\centerline{\em can one define limits and convergence in the sheaf
$\struc$ of coefficients?}

\noindent To see that this translation is effective, one should
realize that {\em in order to define the derivative of $\eh$ one
need only be able to take limits and study convergence in the
space where the latter takes values, which, according to}
(\ref{eq65}), {\em is $\struc(X)$}! Thus, ADG has given so far the
following two answers to the question when
$\stackrel{\centerdot}{\overbrace{\eh(\gamma)}}$ is well defined:

\begin{enumerate}

\item When $\struc$ is a topological algebra sheaf
\cite{mall1,mall2,mall3,mall4}.

\item When $\struc$ is Rosinger's algebra of generalized functions
\cite{mall3,mall4}.

\end{enumerate}

\medskip For in both cases $\struc$ has a well defined topology and the related notion
of convergence.

In section 5, where we give a finitary, causal and quantal version
of the vacuum Einstein equations for Lorentzian gravity
(\ref{eq53})---them too derived from a variation of a reticular
E-H action functional $\qeh_{i}$, we will give a third example of
algebra sheaves---the finsheaves of incidence algebras---in which
the notions of convergence, limits and topology (the so-called
Rota topology) are well defined so as to `justify' the
corresponding differentiation (variation)
$\stackrel{\centerdot}{\overbrace{\qeh}}_{i}$.

The discussion above prompts us to make the following
clarification:

\begin{quotation}
\noindent {\em to `justify' the derivation of Einstein's equations
from varying $\eh$ with respect to $\con$, one need not study the
topology of} $\sconn_{\struc}(\modll)$ {\em per se. Rather, all
that one has to secure is that there is a well defined notion of
(local) convergence in $\struc$}.\footnote{This is another example
of the general working philosophy of ADG according to which the
underlying space or `domain' so to speak (here
$\sconn_{\struc}(\modll)$) is of secondary importance for studying
`differentiability'. For the latter, what is of primary importance
is the algebraic structure of the objects that live on that
domain. For the notion of derivative, and differentiability in
general, one should care more about the structure of the `target
space' or `range' (here the structure sheaf space $\struc$) than
that of the `source space' or `domain' (here the base space
$X$)---after all, the generic base `localization' space $X$
employed by ADG is assumed to be just a topological space without
having been assigned {\it a priori} any sort of differential
structure whatsoever. Of course, {\em in the classical case, $X$
is completely characterized, as a differential manifold, by the
corresponding structure sheaf $\struc_{X}\equiv\smooth_{X}$ of
infinitely differentiable (smooth) functions} (in particular, see
our comments on Gel'fand duality in 5.5.1). In other words, the
classical differential geometric notions `differential ({\it ie},
$\smooth$-smooth) manifold' and `the topological algebra
$\smooth(X)$' are tautosemous ({\it ie}, semantically equivalent)
notions. Alas, other more general kinds of differentiability, may
come from algebraic structures $A$ other than $\smooth(X)$ that
one may localize sheaf-theoretically (as structure sheaves
$\struc_{X}$) on an arbitrary topological space $X$. This is the
very essence of ADG and will recur time and again in the sequel.}
\end{quotation}

\noindent This is how ADG essentially evades the problem of
dealing directly with the `complicated' topology of
$\sconn_{\struc}(\modll)$.

\medskip

We conclude this discussion of the E-H action functional $\eh$ and
its variation yielding the vacuum gravitational equations, by
giving a concise ADG-theoretic statement about the {\em (gauge)
invariance of the first} which in turn amounts to the (gauge)
covariance of the second. Let $\modll=(\modl ,\rho)$ be our usual
(real) E-L vector sheaf (of rank $4$) and $\con$ a spin-Lorentzian
gravitational metric connection on it whose curvature $\ricci$ is
involved in $\eh(\con)$ above. Then,

\begin{quotation}
\noindent {\em the Einstein-Hilbert functional $\eh$ is invariant
under the action of a (local) $\rho$-preserving gauge
transformation, by which we mean a (local) element ({\it ie},
local section) of the structure group sheaf
$\aut_{\struc}\modll\equiv\orthl:=\aut_{\rho}\modl$ of
$\modll=(\modl ,\rho)$, which, in turn, is a subsheaf of
$\aut_{\struc}\modl$}, where locally,
$\aut_{\struc}\modl(U)=\mathrm{GL}(4,\struc(U))=\mathcal{G}\mathcal{L}(4,\struc)(U)$.
\end{quotation}

\subsubsection{A brief note on $\vol$, the Hodge-$\star$
operator, and on self-duality in ADG}

Below, we discuss briefly {\it \`a la} ADG the volume element or
measure $\vol$ appearing in the E-H action integral (\ref{eq69}),
as well as the Hodge-$\star$ operator and the self-dual Lorentzian
connections $\aconn^{+}$ associated with it, thus prepare the
ground for a brief comparison we are going to make subsequently
between our locally finite, causal and quantal vacuum Einstein
gravity and an approach to non-perturbative canonical quantum
gravity based on Ashtekar's new variables \cite{ash}.

\begin{enumerate}

\item {\sl Volume element.} Let $(X,\struc)$ be our usual $\mathbf{K}$-algebraized space
and $\modl$ a free $\struc$-module of finite rank $n$ over $X$,
which is locally isomorphic to the `standard' one $\struc^{n}$.
Let also $\rho$ be a strongly non-degenerate (and indefinite, in
our case of interest) metric on $\modl$, which makes it a {\em
pseudo-Riemannian free $\struc$-module of finite rank $n$ over
$X$}. Then, one considers the sequence
$\epsilon\equiv(\epsilon_{i})_{1\leq i\leq n}$ of global sections
of $\modl\simeq\struc^{n}$ ({\it ie},
$\epsilon_{i}\in\struc^{n}(X)=\struc(X)^{n}$)---the so-called {\em
Kronecker gauge of} $\struc^{n}$.\footnote{In ADG, this
appellation for $\epsilon$ is reserved for positive definite
(Riemannian) metrics $\rho$ \cite{mall1}, but here we extend the
nomenclature to include indefinite metrics as well.} Then, the
volume element $\vol$ associated with the given $\struc$-metric
$\rho$ is defined to be

\begin{equation}\label{eq71}
\vol:=\sqrt{|\rho|}\epsilon_{1}\wee\cdots\wee\epsilon_{n}
\in(\wedge^{n}\struc^{n})(X)\equiv(\det\struc^{n})(X)=\struc(X)
\end{equation}

\noindent That is to say,

\begin{quotation}
\noindent {\em the volume element $\vol$ is a nowhere vanishing
(because $\rho$ is non-degenerate) global section of the structure
sheaf $\struc$. Moreover, since
$(\wedge^{n}\struc^{n})^{*}(X)=(\wedge^{n}(\struc^{n})^{*})(X)=(\det\struc^{n})^{*}(X)=\struc(X)$,
$\vol$ can be viewed as an $\struc(X)$-linear morphism on}
$\det(\struc^{n})$ and, as such, as {\em a map of $\struc$ into
itself}: $\vol\in(\modl
nd\struc)(X)=\mathrm{End}\struc=\struc(X)$.
\end{quotation}

The crux of the argument here is that the definition (\ref{eq71})
of $\vol$ readily applies to the case where $X$ is an Einstein
space and $(\modll ,\rho)$ our usual (real) Lorentzian vector
sheaf on it. This is so because, as mentioned earlier, $\modll$ is
a locally free $\struc$-module of rank $4$, that is, locally ({\it
ie}, $U$-wise) in $X$: $\modll\simeq\struc^{4}$. Hence, the volume
element $\vol$ appearing in (\ref{eq69}) is now an element of
$\struc(U)$. Of course, since, by definition, $\struc$ is a {\em
fine sheaf}, here too $\vol$ can be promoted to a global section
of $\struc$ ($\vol\in\struc(X)$).

\item {\sl Hodge-$\star$.} As with the volume element $\vol$, let
$(\modl ,\rho)$ be a pseudo-Riemannian (Lorentzian) free
$\struc$-module of rank $n$ and recall from (\ref{eq12}) the
canonical $\struc$-isomorphism $\tilde{\rho}$ between the
$\struc$-modules $\modl$ and its dual $\modl^{*}$ induced by
$\rho$. That is to say,
$\modl\stackrel{\tilde{\rho}}{\cong}\modl^{*}\equiv\Hom_{\struc}(\modl
,\struc)$.

We define the following $\struc$-isomorphism $\star$ of
$\struc$-modules

\begin{equation}\label{eq72}
\star :~\wedge^{p}\modl^{*}\mapto\wedge^{n-p}\modl^{*}
\end{equation}

\noindent To give $\star$'s section-wise action, we need to define
first, for any $v\in\wedge^{n-p}\modl(X)$,

\begin{equation}\label{eq73}
v^{\divideontimes}:=(\wedge^{n-p}\tilde{\rho})(v)\in\wedge^{n-p}\modl^{*}(X)=(\wedge^{n-p}\modl(X))^{*}
\end{equation}

\noindent so that then we can define

\begin{equation}\label{eq74}
(\star u)(v):=\vol(u\wee v^{\divideontimes})\equiv(u\wee
v^{\divideontimes})\cdot\vol\in\struc(X)
\end{equation}

\noindent for $u\in\wedge^{p}\modl^{*}(X)=\wedge^{p}\modl(X)^{*}$.

Two things can be mentioned at this point: first, that for the
{\em identity or unit global section $\mathbf{1}$ of $\struc$},
$\star\mathbf{1}=\vol$, and second, that $\star$ entails an
$\struc$-isomorphism of the $\struc$-module defined by the
exterior algebra of $\modl^{*}$, $\wedge\modl^{*}$, into itself.
The latter means, in turn, that $\star$ is an element of
$\aut_{\struc}(\wedge\modl^{*})$.

\begin{quotation}
\noindent{\em The map $\star$ of} (\ref{eq72}) {\em and}
(\ref{eq74}) {\em is the ADG-theoretic version of the usual
Hodge-$\star$ operator induced by the $\struc$-metric $\rho$.}
\end{quotation}

\item {\sl Self-dual Lorentzian connections $\aconn^{+}$.} Now
that we have $\star$ at our disposal, we can define a particular
class of Y-M $\struc$-connections $\conn^{+}$ on vector sheaves,
the so-called {\em self-dual connections}, whose gauge potential
parts $\aconn^{+}$ are coined {\em self-dual gauge fields}. So, we
let $(\modl ,\rho ,\conn)$ be a L-Y-M field on a L-Y-M space $X$.
The definition of $\conn^{+}$s pertains to the property that their
curvatures, $\stre^{+}:=\stre(\conn^{+})$, satisfy relative to the
Hodge-$\star$ duality operator

\begin{equation}\label{eq75}
\star\stre^{+}=\stre^{+}
\end{equation}

\noindent hence their name {\em self-dual}.

In view of (\ref{eq75}) and the second Bianchi identity
(\ref{eq49}), we have

\begin{equation}\label{eq76}
\begin{array}{c}
\delta^{2}_{\modl nd\modl}(\stre^{+})=((-1)^{n\cdot
3+1}\star\conn^{n-2}\star)(\stre^{+})=(-1)^{1+3n}\star\conn^{n-2}(\stre^{+})=\cr
=(-1)^{1+3n}\star\conn^{2}_{\modl nd\modl}(\stre^{+})=0
\end{array}
\end{equation}

\noindent the point being that the (field strengths $\stre^{+}$ of
the) self-dual connections $\conn^{+}$ also satisfy the Y-M
equations. We will return to self-dual connections in section 5
where we will discuss the close affinity between our finitary,
causal and quantal version of vacuum Einstein Lorentzian gravity
and a recent approach to non-perturbative quantum gravity which
uses Ashtekar's new (canonical) variables \cite{ash}.

\end{enumerate}

\subsection{Y-M and Gravitational Moduli Space: $\struct$-Equivalent Connections}

In the present subsection we will give a short account of the
ADG-theoretic perspective on moduli spaces of L-Y-M connections,
focusing our attention on the corresponding moduli spaces of
spin-Lorentzian (vacuum) gravitational connections that are of
special interest to our investigations in this paper.

To initiate our presentation, we consider a (real) Lorentzian
vector sheaf $\modll=(\modl ,\rho)$ and we recall from 2.4 the
affine space $\sconn_{\struc}(\modl)$ of metric
$\struc$-connections on it (\ref{eq54}). From our discussion of
$\struct$-sheaves in 2.2, we further suppose that $\modll$ is the
associated sheaf of the principal sheaf $\orthl :=
\aut_{\struc}\modll\equiv\aut_{\rho}\modl$---the group sheaf of
$\rho$-preserving $\struc$-automorphisms of $\modl$ (the structure
group sheaf of $\modll$, which is also the (local) invariance
group of the free Y-M action functional $\ym(\conn)$
\cite{mall1,mall2}).\footnote{In the case of the functional
$\eh(\con)$ on $(\modll ,\con)$, we saw in the previous subsection
that its (local) invariance (structure) group is precisely
$(\mathrm{Aut}_{\struc}\modll)(U):=\Gamma(U,\aut_{\struc}\modll)\equiv(\mathrm{Aut}_{\rho}\modl)(U)=:\orthl(U)\simeq
L^{\uparrow}$.} Our main contention in this section is that

\begin{quotation}
\noindent{\em the (global) gauge group
$\aut_{\struc}\modll(X)\equiv\mathrm{Aut}_{\struc}\modll\equiv\orthl(X):=\mathrm{Aut}_{\rho}\modl$
acts on the affine space} $\sconn_{\struc}(\modll)$ {\em of metric
$\struc$-connections on the Lorentzian vector sheaf $\modll=(\modl
,\rho)$.}
\end{quotation}

\noindent Let us elaborate a bit on the statement above, which
will subsequently lead us to define moduli spaces of
gauge-equivalent connections.

We have already alluded to the fact, in connection with the
(local) transformation law of gauge potentials $\aconn$ of
$\struc$-connections $\conn$ on general vector sheaves $\modl$ at
the end of 2.1, that one may be able to establish an equivalence
relation $\aconn\stackrel{g}{\sim}\aconn^{'}$ between them, for
$g$ a local gauge transformation ({\it ie}, a local section of the
structure $\struct$-sheaf $\aut_{\struc}(\modl)$ of $\modl$;
$g\in\aut_{\struc}(\modl)(U)=\mathcal{G}\mathcal{L}(n,\struc)(U)$).
We can extend this equivalence relation from the gauge potentials
$\aconn$ to their full connections $\conn$, as follows.

Schematically, and in general, for an $\struc$-module $\modl$ we
say that two connections $\conn$ and $\conn^{'}$ on it are
gauge-equivalent if there exists an element
$g\in\mathrm{Aut}(\modl)$ making the following diagram commutative

\begin{equation}\label{eq77}
\square[\modl`\Omg(\modl)`\modl`\Omg(\modl);\conn`g`g
\otimes\mathbf{1}_{\Omg}\equiv g\otimes\mathbf{1}`\conn^{'}]
\end{equation}

\noindent which is read as

\begin{equation}\label{eq78}
\conn^{'}\circ g=(g\otimes\mathbf{1})\circ\conn\Leftrightarrow
\conn^{'}=(g\otimes\mathbf{1})\circ\conn\circ g^{-1}
\end{equation}

\noindent or in terms of the adjoint representation
$\mathrm{Ad}(\struct)$ of the structure group $\struct\ni g$

\begin{equation}\label{eq79}
\conn^{'}=g\circ\conn\circ g^{-1}\equiv g\conn
g^{-1}=:\mathrm{Ad}(g)\conn
\end{equation}

\noindent It is now clear that

\begin{quotation}
\noindent (\ref{eq78}) and (\ref{eq79}) {\em define an equivalence
relation $\stackrel{g}{\sim}$ on} $\sconn_{\struc}(\modl)$:
$\conn\stackrel{g}{\sim}\conn^{'},~g\in\mathrm{Aut}\modl$.
$\stackrel{g}{\sim}$ {\em is precisely the equivalence relation
defined by the action of the structure group $\mathrm{Aut}\modl$
of $\modl$ on} $\sconn_{\struc}(\modl)$, {\em as alluded to
above.}
\end{quotation}

\noindent Thus, it is natural to consider the following
$\struct$-action $\alpha$ on $\sconn_{\struc}(\modl)$

\begin{equation}\label{eq80}
\alpha
:~\mathrm{Aut}\modl\times\sconn_{\struc}(\modl)\mapto\sconn_{\struc}(\modl)
\end{equation}

\noindent defined point-wise by

\begin{equation}\label{eq81}
(g,\conn)\mapsto\alpha(g,\conn)\equiv g\cdot\conn\equiv
g(\conn):=g\conn g^{-1}\equiv\mathrm{Ad}(g)\conn
\end{equation}

\noindent with the straightforward identification from
(\ref{eq78})

\begin{equation}\label{eq82}
g(\conn)\equiv g\conn g^{-1}\equiv(g\otimes1)\circ\conn\circ
g^{-1}\in\mathrm{Hom}_{\cons}(\modl ,\Omg(\modl))
\end{equation}

\noindent In turn, for a given $\conn\in\sconn_{\struc}(\modl)$,
$\alpha$ delimits the following set in $\sconn_{\struc}(\modl)$

\begin{equation}\label{eq83}
\begin{array}{c}
\mathcal{O}_{\conn}:=\{
g\cdot\conn\in\sconn_{\struc}(\modl):~g\in\mathrm{Aut}\modl\}=\cr
=\{\conn^{'}\in
\sconn_{\struc}(\modl):~\conn^{'}\stackrel{g}{\sim}\conn,
\mathrm{for~some}~g\in\mathrm{Aut}\modl\}
\end{array}
\end{equation}

\noindent called {\em the orbit of an $\struc$-connection $\conn$
on $\modl$ under the action $\alpha$ of the gauge group}
$\struct=\mathrm{Aut}\modl$ {\em on} $\sconn_{\struc}(\modl)$.
$\mathcal{O}_{\conn}$ consists of all connections $\conn^{'}$ in
$\sconn_{\struc}(\modl)$ that are gauge-equivalent to $\conn$.

Following \cite{mall1,mall2}, we would also like to note that it
can be shown that the gauge-orbit ${\mathcal{O}}_{\conn}$ in
(\ref{eq83}) can be equivalently written in terms of the induced
connection $\conn_{\modl nd\modl}$ as follows

\begin{equation}\label{eq84}
{\mathcal{O}}_{\conn}=\{\conn-\conn_{\modl
nd\modl}(g)g^{-1}:~g\in\mathrm{Aut}\modl\}
\end{equation}

\noindent At the same time, the {\em stability group}
${\mathcal{O}}(\conn)$ of $\conn\in\sconn_{\struc}(\modl)$ under
the action of $\mathrm{Aut}(\modl)$ is, by definition, the set of
all $g\in\mathrm{Aut}\modl$ such that $g\cdot\conn=\conn$, so that

\begin{equation}\label{eq85}
\begin{array}{c}
{\mathcal{O}}(\conn)=\mathrm{ker}(\conn_{\modl nd\modl |
\mathrm{Aut}\modl})\equiv\{ g\in\mathrm{Aut}\modl:~\conn_{\modl
nd\modl}(g)=0\}=\cr \{ g\in\mathrm{Aut}\modl :~[\conn ,g]:=\conn
g-g\conn=0\}
\end{array}
\end{equation}

\noindent which means that the stability group of the connection
$\conn\in\sconn_{\struc}(\modl)$ consists of all those (gauge)
transformations of $\modl$ ($g\in\mathrm{Aut}\modl$) that commute
with $\conn$.

At this point, and before we define moduli spaces of
gauge-equivalent connections ADG-theoretically, we would like to
digress a bit and make a few comments on the possibility of
developing differential geometric ideas (albeit, not of a
classical, geometrical $\smooth$-smooth sort, but of an algebraic
ADG kind) on the affine space $\sconn_{\struc}(\modl)$. The
remarks below are expressed in order to prepare the reader for
comments on the possibility of developing differential geometry on
the gauge moduli space of gravitational connections that we are
going to make in 5.3 in connection with some problems ({\it eg},
Gribov's ambiguity) people have encountered in trying to quantize
general relativity (regarded as a gauge theory) both canonically
({\it ie}, in a Hamiltonian fashion) and covariantly ({\it ie}, in
a Lagrangian fashion). It is exactly due to these problems that
others have also similarly felt the need of developing
differential geometric concepts and constructions (albeit, of the
classical, $\smooth$-sort) on moduli spaces of Y-M and
gravitational connections \cite{ashlew1,ashlew2}.

As a first differential geometric idea on
$\sconn_{\struc}(\modl)$, we would first like to define a set of
objects (to be regarded as abstract `tangent vectors') that would
qualify as the {\em `tangent space' of} $\sconn_{\struc}(\modl)$
{\em at any of its points} $\conn$, and then, after we define
moduli spaces of gauge-equivalent connections below, we would also
like to define an analogous {\em `tangent space' to the moduli
space at a gauge-orbit} ${\mathcal{O}}_{\conn}$ {\em of a
connection} $\conn\in\sconn_{\struc}(\modl)$.

We saw earlier (2.4) that for $\modl$ a vector sheaf of rank $n$,
the affine space $\sconn_{\struc}(\modl)$ can be modelled after
$\Omg^{1}(\modl nd\modl)(X)$. {\em We actually define the latter
space to be the sought after `tangent space' of}
$\sconn_{\struc}(\modl)$ {\em at any of its `points'} $\conn$.
That is to say,

\begin{equation}\label{eq86}
T(\sconn_{\struc}(\modl),\conn):=\Omg^{1}(\modl nd\modl)(X)
\end{equation}

\noindent and we recall from the foregoing that $\Omg^{1}(\modl
nd\modl)(X)$ is itself an $\struc(X)$-module which locally,
relative to a gauge $U$, becomes the $n\times n$-matrix of
$1$-forms $\struc(U)$-module
$M_{n}(\Omg^{1}(U))=M_{n}(\Omg^{1})(U)$.\footnote{As a matter of
fact, one can actually prove (\ref{eq86}) along classical
lines---for example, by fixing a point $\conn$ in the affine space
$\sconn_{\struc}(\modl)$, regard it as `origin' ({\it ie}, the
zero vector $0$), let a curve $\gamma(t)$ in
$\sconn_{\struc}(\modl)$ pass through it ({\it ie},
$\conn\equiv\conn_{0}=\gamma(0)$), and then find the vector
$\dot{\gamma}(t)$ tangent to $\gamma$. This proof has been shown
to work in the particular case the structure sheaf $\struc$ is a
topological vector space sheaf \cite{mall2,mall4} (and in section
5 we will see that it also works in the case of our finsheaves of
incidence algebras for deriving the locally finite, causal and
quantal vacuum Einstein equations for Lorentzian gravity); in
fact, we used it in (\ref{eq67}) and (\ref{eq68}) to derive the
vacuum Einstein equations from a variational principle on the
space of Lorentzian connections.}

We are now in a position to define the {\em global moduli space or
gauge orbit space of the $\struc$-connections on $\modl$}, as
follows

\begin{equation}\label{eq87}
M(\modl)\equiv\sconn_{\struc}(\modl)/
\mathrm{Aut}\modl:=\!\!\!\!\bigcup_{\conn\in\sconn_{\struc}(\modl)}\!\!\!\!\!\mathcal{O}_{\conn}
=\sum_{\conn}\mathcal{O}_{\conn}
\end{equation}

\noindent The epithet `global' above indicates that the quotient
in (\ref{eq87}) can be actuallly localized---something that comes
in handy when one, as we do, works with a vector sheaf $\modl$ on
$X$ and the latter is gauged relative to a local frame $\gauge=\{
U\}$. The localization of $M(\modl)$ means essentially that one
uses the {\em sheaf of germs of moduli spaces of the
$\struc$-connections of the module or vector sheaf $\modl$ in
focus}. To see this, the reader must realize that, as $U$ ranges
over the open subsets of $X$, one deals with a (complete) presheaf
of orbit spaces equipped with the obvious restriction maps. To
follow this line of thought, one first observes the inclusion

\begin{equation}\label{eq88}
\sconn_{\struc}(\modl)|_{U}\subseteq\sconn_{\struc|_{U}}(\modl_{U})
\end{equation}

\noindent and a similar restriction of the structure group sheaf
$\struct\equiv\aut\modl$. Then, section-wise over $U$ one has

\begin{equation}\label{eq89}
\begin{array}{c}
(\aut\modl)|_{U}=(\aut\modl)(U)=\mathrm{Isom}_{\struc|_{U}}(\modl|_{U},\modl|_{U})=\cr
=\mathcal{I}som_{\struc|_{U}}(\modl|_{U},\modl|_{U})(U)\equiv\aut(\modl|_{U})(U)=\mathrm{Aut}(\modl|_{U})
\end{array}
\end{equation}

\noindent thus, {\it in toto}, the following local equality

\begin{equation}\label{eq90}
\aut\modl(U)=\mathrm{Aut}(\modl|_{U})
\end{equation}

\noindent for every open $U$ in $X$.

So, in complete analogy to (\ref{eq81}), one has the action of
$\mathrm{Aut}(\modl|_{U})$ on the local sets
$\sconn_{\struc}(\modl)|_{U}$ of $\struc$-connections in
(\ref{eq88})

\begin{equation}\label{eq91}
\mathrm{Aut}(\modl|_{U})\times\sconn_{\struc}(\modl)|_{U}\mapto\sconn_{\struc}(\modl)|_{U}
\end{equation}

\noindent entailing the following `orbifold sheaf' of
gauge-equivalent $\struc$-connections on $\modl$

\begin{equation}\label{eq92}
{\mathcal{M}}(\modl)=\sconn_{\struc}(\modl)/\aut\modl
\end{equation}

\noindent $\mathcal{M}(\modl)$ is the aforesaid sheaf of germs of
moduli spaces of $\struc$-connections on $\modl$.

Finally, it must also be mentioned here, in connection with the
local isomorphism $\modl\simeq\struc^{n}$ of a vector sheaf
$\modl$ mentioned earlier, that $(\aut_{\struc}\modl)(U)$ above
reduces locally to
$\mathcal{G}\mathcal{L}(n,\struc)(U)=\mathrm{GL}(n,\struc(U))$, as
follows\footnote{In the case of $\modll$, the local reduction
below has already been anticipated earlier.}

\begin{equation}\label{eq93}
\begin{array}{c}
(\aut_{\struc}\modl)(U)=\mathrm{Aut}(\modl|_{U})=\aut(\struc^{n}|_{U})=(\aut\struc^{n})(U)=\cr
=M_{n}(\struc)^{^{\bull}}(U)\equiv\mathcal{G}\mathcal{L}(n,\struc)(U)\equiv\mathrm{GL}(n,\struc(U))
\end{array}
\end{equation}

\noindent We can distill this to the following remark

\begin{quotation}
\noindent {\em any local automorphism of a given vector sheaf
$\modl$ of rank $n$ over one of its local gauges $U$ is
effectively given by a local automorphism of $\struc^{n}$---that
is to say, by an element of}
$\mathrm{GL}(n,\struc(U))=\mathcal{G}\mathcal{L}(n,\struc)(U)\equiv\mathrm{GL}(n,\struc|_{U})$.
\end{quotation}

\noindent so that the gauge (structure) group $\aut_{\struc}\modl$
of $\modl$ is locally ({\it ie}, $U$-wise) reduced to the group
sheaf $\mathcal{GL}(n,\struc)$,\footnote{Or equivalently, to its
complete presheaf of sections $\Gamma(\mathcal{GL}(n,\struc))$.}
as it has been already anticipated, for example, in 2.1.2 in
connection with the transformation law of gauge
potentials,\footnote{See remarks after (\ref{eq9}).} and earlier
in connection with vacuum Einstein Lorentzian gravity on $\modll$.

As noted before, now that we have defined moduli spaces of
gauge-equivalent connections, and similarly to the `tangent space'
$T(\sconn_{\struc}(\modl),\conn)$ in (\ref{eq86}), we would like
to define $T({\mathcal{O}}_{\conn},\conn)$---{\em the `tangent
space' to a gauge-orbit of an element}
$\conn\in\sconn_{\struc}(\modl)$ and, {\it in extenso},
$T(M(\modl),{\mathcal{O}}_{\conn})$---{\em the `tangent space' to
the moduli space of $\modl$ at an orbit of}
$\conn\in\sconn_{\struc}(\modl)$. We have seen how the induced
$\struc$-connection of the vector sheaf $\modl nd\modl$

\begin{equation}\label{eq94}
\conn_{\modl nd\modl}:~\modl nd\modl\mapto\Omg^{1}(\modl nd\modl)
\end{equation}

\noindent can be viewed as the `covariant differential' of the
connection $\conn$ in $\sconn_{\struc}(\modl)$. By defining the
induced coderivative $\delta^{1}_{\modl nd\modl}$ adjoint to
$\conn_{\modl nd\modl}$ as

\begin{equation}\label{eq95}
\delta^{1}_{\modl nd\modl}:~\Omg^{1}(\modl nd\modl)\mapto\modl
nd\modl
\end{equation}

\noindent we define

\begin{equation}\label{eq96}
{\mathcal{S}}_{\conn}:=\conn+\mathrm{ker}\delta^{1}_{\modl
nd\modl}\equiv\{ \conn
+u\in\sconn_{\struc}(\modl):~\delta^{1}_{\modl nd\modl}(u)=0\}
\end{equation}

\noindent for $u\in\Omg^{1}(\modl nd\modl)(X)$. Of course, for
$u=0\in\Omg^{1}(\modl nd\modl)(X)$, one sees that $\conn$ belongs
to ${\mathcal{S}}_{\conn}$, so that

\begin{quotation}
\noindent ${\mathcal{S}}_{\conn}$ {\em is a subspace of}
$\sconn_{\struc}(\modl)$ {\em through} $\conn$. In fact, one can
show \cite{mall2,mall4} that ${\mathcal{S}}_{\conn}$ {\em is an
affine $\cons$-linear subspace of} $\sconn_{\struc}(\modl)$ {\em
through the point} $\conn$, {\em modelled after}
$(\mathrm{ker}\delta^{1}_{\modl
nd\modl})(X)$.\footnote{$(\mathrm{ker}\delta^{1}_{\modl
nd\modl})(X)$ being in fact a sub-$\struc(X)$-module of
$\Omg^{1}(\modl nd\modl)(X)$.}
\end{quotation}

\noindent Moreover, and this is crucial for defining
$T({\mathcal{O}}_{\conn},\conn)$, one is able to prove
\cite{mall2,mall4} that

\begin{equation}\label{eq97}
\mathrm{im}\conn_{\modl
nd\modl}\oplus\mathrm{ker}\delta^{1}_{\modl
nd\modl}=\Omg^{1}(\modl
nd\modl)(X)=:T(\sconn_{\struc}(\modl),\conn)
\end{equation}

\noindent for any local gauge $U$ of $\modl$.

{\it In toto}, since both $\conn_{\modl nd\modl}$ and
$\delta^{1}_{\modl nd\modl}$ are restricted on the gauge group
$\mathrm{Aut}\modl$, and in view of (\ref{eq84}), one realizes
that

\begin{equation}\label{eq98}
T({\mathcal{O}}_{\conn},\conn)=\mathrm{im}(\conn_{\modl nd\modl
|\mathrm{Aut}\modl})=\mathrm{ker}(\delta^{1}_{\modl nd\modl
|\mathrm{Aut}\modl})^{\perp}
\end{equation}

\noindent where `$\perp$' designates `orthogonal subspace' with
respect to the $\struc$-metric $\rho$ on $\modl$. Thus,

\begin{quotation}
\noindent ${\mathcal{S}}_{\conn}$ {\em is the orthogonal
complement of the tangent space $T({\mathcal{O}}_{\conn},\conn)$
to the orbit ${\mathcal{O}}_{\conn}$ of $\conn$ at the point
$\conn$ of} $\sconn_{\struc}(\modl)$.
\end{quotation}

\noindent At the same time, for `infinitesimal variations'
$u\in\Omg^{1}(\modl nd\modl)(X)$ around
$\conn\in\sconn_{\struc}(\modl)$, one can show \cite{mall2,mall4}

\begin{equation}\label{eq99}
\begin{array}{c}
T({\mathcal{O}}_{\conn+u},\conn+u)=\mathrm{im}((\conn+u)_{\modl
nd\modl |\mathrm{Aut}\modl})=\cr\mathrm{im}((\conn_{\modl
nd\modl}+u)|_{\mathrm{Aut}\modl})=\{(\conn_{\modl
nd\modl}+u)g:~g\in\mathrm{Aut}\modl\}
\end{array}
\end{equation}

Concomitantly, in order to arrive at
$T(M(\modl),{\mathcal{O}}_{\conn})$, one realizes
\cite{mall2,mall4} that

\begin{quotation}
\noindent {\em the gauge group} $\mathrm{Aut}\modl$ {\em acts on}
$\sconn_{\struc}(\modl)$ {\em in a way that is compatible with its
affine structure}.
\end{quotation}

\noindent That is to say, one has

\begin{equation}\label{eq100}
g(\conn+u)=g\conn+gu,~\forall
g\in\mathrm{Aut}\modl\,\mathrm{and}\, u\in\Omg^{1}(\modl
nd\modl)(X)
\end{equation}

\noindent The bottom-line of these remarks is that

\begin{quotation}
\noindent $M(\modl):=\sconn_{\struc}(\modl)/\mathrm{Aut}\modl$
{\em can still be construed as an affine space modelled
after}\linebreak $\Omg^{1}(\modl
nd\modl)(X)/\mathrm{Aut}\modl\simeq(\mathrm{im}(\conn_{\modl
nd\modl |\mathrm{Aut}\modl}))^{\perp}\simeq{\mathcal{S}}_{\conn}$.
\end{quotation}

\noindent Hence one concludes that

\begin{equation}\label{eq101}
T(M(\modl),{\mathcal{O}}_{\conn})\simeq{\mathcal{S}}_{\conn}
\end{equation}

Now that we have $M(\modl)$ we are in a position to define
similarly moduli spaces of (self-dual) spin-Lorentzian
connections. Of course, our definition of `tangent spaces' on
${\mathcal{O}}_{\conn}$ and on $M(\modl)$ above carries through,
virtually unaltered, to the particular (self-dual) Lorentzian
case. As noted above, this will become relevant in section 5
where, in view of certain problems that both the canonical and the
covariant quantization approaches to quantum general relativity
(based on the Ashtekar variables) encounter, the need to develop
differential geometric ideas and techniques on the moduli space of
(self-dual) spin-Lorentzian connections has arisen in the last
decade or so.

\subsubsection{Moduli space of (self-dual) spin-Lorentzian
connections $\con^{(+)}$}

The last remark prompts us to comment briefly on the space of
gauge-equivalent (self-dual) spin-Lorentzian connections on the
(real) Lorentzian vector sheaf $\modll=(\modl ,\rho)$ of rank $4$
which is of special interest to us in the present paper. When the
latter is endowed with a (self-dual) Lorentzian metric connection
$\con^{(+)}$ which ({\it ie}, whose curvature scalar
$\ricci^{(+)}(\con^{(+)})$) is a solution of (the self-dual
version of) (\ref{eq53}),\footnote{Which in turn means that
$(\modll ,\con^{(+)})\equiv(\modl ,\rho ,\con^{(+)})$ defines a
(self-dual) E-L field.} it is reasonable to enquire about other
gauge-equivalent (self-dual) E-L fields $(\modll
,\breve{\con}^{(+)})$, with
$\con^{(+)}\stackrel{g}{\sim}\breve{\con}^{(+)}~(g\in\struct=\aut_{\struc}{\modll})$.

From what has been said above, one readily obtains the local gauge
group of $\modll$

\begin{equation}\label{eq102}
\begin{array}{c}
\aut_{\struc}\modll(U)\equiv\aut_{\rho}\modl(U)=\mathrm{Aut}_{\rho}(\modl|_{U})=:
\orthl(U)\simeq\cr L^{\uparrow}\subset
M_{4}(\struc)^{^{\bull}}(U)=
\mathcal{GL}(4,\struc)(U)=\mathrm{GL}(4,\struc(U))
\end{array}
\end{equation}

\noindent and, like in (\ref{eq92}), we obtain the localized
moduli space (`orbifold sheaf') of gauge-equivalent (self-dual)
spin-Lorentzian $\struc$-connections $\con^{(+)}$ (or their gauge
potential parts $\aconn^{(+)}$) on $\modll$

\begin{equation}\label{eq103}
{\mathcal{M}}^{(+)}(\modll)=\sconn^{(+)}_{\struc}(\modll)/\aut_{\struc}\modll\equiv\sconn^{(+)}_{\struc}(\modll)/\aut_{\rho}\modl
\end{equation}

\noindent Finally, in a possible covariant quantization scenario
for vacuum Einstein Lorentzian gravity that we are going to
discuss in section 5, $\mathcal{M}(\modll)$ may be regarded as the
(quantum) configuration space of the theory in a way analogous to
the scheme that has been proposed in the context of Ashtekar's new
variables for non-perturbative canonical quantum gravity
\cite{ash,ashish,ashlew1,ashlew2}. In connection with the latter,
we note that since the main dynamical variable is a {\em
self-dual} spin-Lorentzian connection $\con^{+}$\footnote{Or
again, locally, its gauge potential part $\aconn^{+}$.} (see end
of 3.3), the corresponding moduli space is denoted by

\begin{equation}\label{eq104}
\mathcal{M}^{+}(\modll)=\sconn^{+}_{\struc}(\modll)/\aut_{\struc}\modll\equiv\sconn^{+}_{\struc}(\modll)/\aut_{\rho}\modl
\end{equation}

\noindent where, as we have already mentioned earlier, the (local)
orthochronous Lorentz structure (gauge) symmetries $\struct$ of
$\modll$ can be written as
$\aut_{\struc}\modll(U)\equiv\aut_{\rho}\modl(U)=L^{\uparrow}:=SO(1,3)^{\uparrow}\stackrel{\mathrm{locally}}{\simeq}SL(2,\com)\subset
M_{2}(\com)$.\footnote{Always remembering of course that
$L^{\uparrow}=SO(1,3)^{\uparrow}$ and its double covering
spin-group $SL(2,\com)$ are only locally ({\it ie}, Lie
algebra-wise) isomorphic ({\it ie}, $sl(2,\com)\simeq
so(1,3)^{\uparrow}$). Also, for a general (real) Lorentzian vector
sheaf $(\modl ,\rho)$ of rank $n$, which locally reduces to
$\struc^{n}$ ({\it ie}, it is a locally free $\struc$-module), its
local (structure) group of Lorentz transformations is
$\aut_{\rho}\modl(U)=\mathcal{S}\mathcal{L}(n,\struc)(U)\equiv\mathrm{SL}(n,\struc(U))\subset\aut_{\struc}\modl(U)=\mathcal{G}\mathcal{L}(n,\struc)(U)\equiv\mathrm{GL}(n,\struc(U))\equiv
M_{n}(\struc)^{\bull}(U)=(\modl nd_{\struc}\modl)^{\bull}(U)$.}

\section{Kinematics for a Finitary, Causal and Quantal
Loren- tzian Gravity}

One of our main main aims in this paper is to show that the
general ADG-theoretic concepts and results presented in the last
two sections are readily applicable in the particular case of the
{\em curved finsheaves of qausets} perspective on (the kinematics
of) Lorentzian gravity that has been developed in the two past
papers \cite{malrap1,malrap2}. In the present section, we recall
in some detail from \cite{malrap1}, always under the prism of ADG,
the main kinematical structures used for a locally finite, causal
and quantal version of vacuum Einstein Lorentzian gravity, thus we
prepare the ground for the dynamical equations to be described
`finitarily' in the next. In the last subsection (4.3), and with
the reader in mind, we give a concise {\it r\'{e}sum\'{e}}---a
`causal finitarity' manual so to speak---of some (mostly new) key
kinematical concepts and constructions to be described {\it en
passant} below.

More analytically, we will go as far as to present a finitary
version of the (self-dual) moduli space
$\mathcal{M}^{(+)}(\modll)$ in (\ref{eq103}) and (\ref{eq104})
above---arguably, {\em the} appropriate (quantum) kinematical
configuration space for a possible (quantum) theoresis of the
(self-dual) spin-Lorentzian connections $\aconn_{i}^{(+)}$
inhabiting the aforesaid finsheaves of qausets. We will also
present, based on recent results about projective and inductive
limits in the category $\ctriad$ of Mallios' differential triads
\cite{pap1,pap2}, as well as on results about projective limits of
inverse systems of principal sheaves endowed with Mallios'
$\struc$-connections \cite{vas1,vas2,vas3}, the recovery, at the
{\em projective limit} of infinite refinement (or localization) of
an {\em inverse system of principal finsheaves of qausets and
reticular spin-Lorentzian connections on them}, of a structure
that, from the ADG-theoretic perspective, comes very close to, but
does not reproduce exactly, the kinematical structure of classical
gravity in its gauge-theoretic guise---the principal orthochronous
Lorentzian fiber bundle $\princ^{\uparrow}$ over a
$\smooth$-smooth spacetime manifold $M$ endowed with a non-trivial
(self-dual) smooth spin-Lorentzian connection $\con^{(+)}$ on it
(subsection 4.2).\footnote{The word `emulates' above pertains to
the fact that our projective limit triad (as well as the principal
sheaf and spin-Lorentzian connection relative to it) will be seen
not to correspond precisely to the classical differential triad
$(\struc_{X}\equiv{}^{\K}\smooth_{X},\partial ,\Omg^{1})$, but to
one that in the context of the present ADG based paper may be
regarded as a `{\em generalized smooth}' triad (write
\texttt{smooth} for short). This \texttt{smooth} triad's structure
sheaf will be symbolized by $\struc_{X}\equiv{}^{\K}\ssmooth_{X}$
in order to distinguish it from the ${}^{\K}\smooth_{X}$ employed
in the classical case. On the other hand, we will be using the
same symbols for the flat $0$-th order nilpotent derivation
$d^{0}\equiv\partial$ as well as the $\struc$-module of $1$st
order differential forms $\Omg^{1}$ in the
$\ssmooth$-\texttt{smooth} and the usual $\smooth$-smooth triads.}
In this way, we are going to be able to make brief comparisons,
even if just preliminarily at this early stage of the development
of our theory, between a similar differential geometric scheme on
the moduli space of gauge-equivalent spin-Lorentzian connections
that has been worked out in \cite{ashlew2}, like ADG, {\em through
entirely algebraic methods}.\footnote{For, to recall Grauert and
Remmert: ``{\em The methods of sheaf theory are algebraic.}''
\cite{grarem}. The purely algebraic character of ADG has been
repeatedly emphasized in
\cite{mall1,mall2,mall3,malros1,malros2,malrap1,malrap2,mall7,mall4}.}
However, and this must be stressed from the start,

\begin{quotation}
\noindent unlike \cite{ashlew2}, where projective limit techniques
are used in order to endow (a completion of) the moduli space of
gauge-equivalent connections with a differential manifold-like
structure, thus (be able to) induce to it classical differential
geometric notions such as differential forms, exterior
derivatives, vector fields, volume forms {\it etc}, {\em we, with
the help of ADG, already possess those at the finitistic and
quantal level of the curved finsheaves of qausets. Moreover, our
projective limit result---the \texttt{smooth} differential triad,
Lorentzian principal sheaf and non-trivial connection on it which,
as noted above, closely resembles the classical
$\smooth$-diferential triad as well as the principal orthochronous
Lorentz sheaf (bundle) and its associated curved locally
Minkowskian vector sheaf (bundle) over the $\smooth$-smooth
manifold $M$ of general relativity---only illustrates the ability
of our discrete algebraic (quantal) structures to yield at the
(correspondence) limit of infinite localization or refinement of
the qausets a structure that emulates well the kinematical
structure of classical Lorentzian gravity}
\cite{rapzap1,malrap1,rapzap2,malrap2}. At the same time, and
perhaps more importantly, this indicates, in contrast to
\cite{ashlew2} where projective limits are employed in order to
produce `{\em like from like}' ({\it ie}, induce a classical
differential geometric structure from inverse systems of
differential manifolds), what we have repeatedly stressed here,
namely that, {\em to do differential geometry---the differential
geometric machinery so to speak---is not inextricably tied to the
$\smooth$-smooth manifold, so that we do not depend on the latter
to provide us with the standard, and by no means unique, necessary
or `preferred', differential mechanism usually supplied by the
algebra $\smooth(M)$ of smooth functions on the differential
manifold $M$ as in the classical case. Our differential geometric
machinery, as we shall see in the sequel, comes straight from the
(incidence) algebras inhabiting the stalks of vector, differential
module and algebra sheaves like the generic locally free
$\struc$-modules $\modl$ of ADG above, over a finitary topological
base space(time) without mentioning at all any differential
structure that this base space should {\it a priori} be equipped
with, and certainly not the classical $\smooth$-manifold one. In
other words, our differential geometric machinery does not come
from assuming $\smooth_{M}$ as structure sheaf in our finitary,
ADG-based constructions.\footnote{A similar point was made in
footnotes 8 and 64, for example. We will return to discuss it in
more detail in the concluding section.}}
\end{quotation}

\noindent We would like to distill this to the following slogan
that time and again we will encounter in the sequel:

\vskip 0.1in

\begin{quotation}
\noindent {\bf Slogan 1.} Differentiability derives from (algebras
in) the stalk (in point of fact, from the structure sheaf $\struc$
of coefficients or generalized arithmetics), not from the base
space.\footnote{As we have said many times, the classical case
corresponding to taking for base space $X$ (a region of) the
smooth manifold $M$ and for $\struc_{X}$ its structure sheaf
$\smooth_{X}$---the sheaf of germs of sections of infinitely
differentiable functions on $X$.}
\end{quotation}

Then, the upshot of our approach to all the structures to be
involved in the sequel is that

\begin{quotation}
\noindent in the spirit of ADG
\cite{mall1,mall2,mall3,malrap1,malrap2,malros1,malros2} and what
has been presented so far here along those lines, {\em everything
to be constructed below, whether kinematical or dynamical, is
manifestly independent of a background $\smooth$-smooth spacetime
manifold $M$, its `structure group' $\mathrm{Diff}(M)$ and, as a
result, of the usual differential geometry ({\it ie}, calculus)
that such a base space supports. In a nutshell, our (differential)
geometric constructions are genuinely background
$\smooth$-manifold free}.
\end{quotation}

\vskip 0.2in

\noindent Interestingly enough, such a position recurs time and
again, as a {\em leit motiv} so to speak, in the Ashtekar quantum
gravity program \cite{ash1,ash2}. But let us now go on to more
details.

\subsection{Principal Finsheaves and their Associated Finsheaves of Qausets}

First, we give a short account of the evolution of our ideas
leading to \cite{malrap1} and \cite{malrap2} which the present
paper is supposed to continue as it takes a step further into the
dynamical realm of qausets.\footnote{For a more detailed and
thorough description of the conceptual history of our work, as
well as of its relation with category and topos theory, the reader
is referred to the recent work \cite{rap6}. A topos-theoretic
treatment of finitary, causal and quantal Lorentzian gravity is
currently under way \cite{rap7}.}

\subsubsection{A brief history of finitary spacetime and
gravity}

Our entire project of developing a finitary, causal and quantal
picture of spacetime and gravity started with Sorkin's work on
discrete approximations of continuous spacetime topology
\cite{sork0}. Briefly, Sorkin showed that when one substitutes the
point-events of a bounded region $X$ of a topological ({\it ie},
$\cont$) spacetime manifold $M$ by `coarse' regions ({\it ie},
open sets) $U$ about them belonging to a locally finite open cover
$\gauge_{i}$ of $X$, one can effectively replace the latter by
locally finite partially ordered sets (posets) $P_{i}$ which are
$T_{0}$-topological spaces in their own right and, effectively,
topologically equivalent to $X$. Then, these posets were seen to
constitute inverse systems $\inv=(P_{i},\succeq)$ of finitary
topological spaces, with the relation $P_{j}\succeq P_{i}$ being
interpreted as `the act of topological refinement or resolution of
$P_{i}$ to $P_{j}$'.\footnote{Meaning essentially that the open
covering $\gauge_{i}$ of $X$ from which $P_{i}$ derives is a
subcover of ({\it ie}, coarser than) $\gauge_{j}$. Roughly, the
latter contains more and `smaller' open sets about $X$'s points
than the former. In this sense, acts of `refinement',
`resolution', or `localization' are all synonymous notions. That
is, one refines the coarse open sets about $X$'s point-events and
in the process she localizes them ({\it ie}, she effectively
determines their locus) at higher resolution or `accuracy'. As
befits this picture, Sorkin explicitly assumes that ``{\em the
points of $X$ are the carriers of its topology}'' \cite{sork0}.}
Sorkin was also able to show, under reasonable assumptions about
$X$,\footnote{For instance, $X$ was assumed to be {\em relatively
compact} (open and bounded) and (at least) $T_{1}$.} that the
$P_{i}$s are indeed legitimate substitutes of it in that at the
inverse or projective limit of infinite refinement, resolution or
localization of the $\gauge_{i}$s and their associated $P_{i}$s,
one recovers the $\cont$-region $X$ (up to homeomorphism).
Formally one writes

\begin{equation}\label{eq105}
\underleftarrow{\lim}\inv\equiv\lim_{\infty\leftarrow
i}P_{i}\equiv P_{\infty}\stackrel{\mathrm{homeo}.}{\simeq}X
\end{equation}

Subsequently, by exploring ideas related to Gel'fand
duality,\footnote{We will comment further on Gel'fand duality in
the next section.} which had already been anticipated in
\cite{zap0}, Raptis and Zapatrin showed how to associate a finite
dimensional, associative and noncommutative {\em incidence Rota
$\K$-algebra} $\omg_{i}$ with every $P_{i}$ in $\inv$, and how
these algebras can be interpreted as discrete and quantum
topological spaces bearing a non-standard topology, called the
{\em Rota topology}, on their primitive spectra\footnote{That is,
the sets of the incidence algebras' primitive ideals which, in
turn, are kernels of irreducible representations of the
$\omg_{i}$s.} \cite{rapzap1}. They also showed, in a way
reminiscent of the Alexandrov-\v{C}ech construction of nerves
associated with locally finite open covers of manifolds, how {\em
the $P_{i}$s may be also viewed as simplicial
complexes}\footnote{See also \cite{zap1,zap3} about this.} as well
as, again by exploring a variant of Gel'fand duality, how there is
a contravariant functor between the category $\as$ of finitary
substitutes $P_{i}$ and poset morphisms\footnote{Monotone maps
continuous in the topology of the $P_{i}$s.} between them, and the
category $\rz$ of the incidence algebras $\omg_{i}$ associated
with the $P_{i}$s and injective algebra homomorphisms between
them. Below, we would like to highlight three issues from the
investigations in \cite{rapzap1}:

\begin{enumerate}

\item Since the $\omg_{i}$s are objects dual to the $P_{i}$s
which, in turn, are discrete homological objects ({\it ie},
finitary simplicial complexes) as mentioned above, they ({\it ie},
the incidence algebras) can be viewed as {\em discrete
differential manifolds} \cite{dimu1,dimu2,dimu4,zap3}. Indeed,
they were seen to be reticular spaces

\begin{equation}\label{eq106}
\omg_{i}=\bigoplus_{p\in\Z_{+}}\omg^{p}_{i}=
\stackrel{\alg_{i}}{\overbrace{\omg^{0}_{i}}}\oplus
\stackrel{\ring_{i}}{\overbrace{\omg^{1}_{i}\oplus\omg^{2}_{i}\oplus\ldots}}\equiv\alg_{i}\oplus\ring_{i}
\end{equation}

\noindent of $\Z_{+}$-graded $\alg_{i}$-bimodules $\ring_{i}$ of
(exterior) differential forms $\omg^{p}_{i}~(p\geq1)$\footnote{In
(\ref{eq106}), $\alg_{i}\equiv\omg^{0}_{i}$ is a commutative
subalgebra of $\omg_{i}$ called {\em the algebra of coordinate
functions in} $\omg_{i}$, while
$\ring_{i}\equiv\bigoplus_{i}^{p\geq1}\omg^{p}_{i}$ a linear
subspace of $\omg_{i}$ called {\em the module of differentials
over} $\alg_{i}$. The elements of each linear subspace
$\omg^{p}_{i}$ of $\omg_{i}$ in $\ring_{i}$ were seen to be
discrete analogues of (exterior) differential $p$-forms. We also
note that in the sequel we will use the same boldface symbol
`$\alg_{i}$' and `$\ring_{i}$' to denote the algebra of reticular
coordinates and the module of discrete exterior differentials over
it as well as the finsheaves thereof.} related within each
$\omg_{i}$ by nilpotent Cartan-K\"{a}hler-like (exterior)
differential operators
$d^{p}_{i}:~\omg^{p}_{i}\mapto\omg^{p+1}_{i}$.

\item Since now the $\omg_{i}$s are seen to be structures encoding not only topological, but also
differential geometric information, it was intuited that an
inverse---or more accurately, since the incidence algebras are
objects Gel'fand-dual to Sorkin's topological posets---a direct
system $\diromg=\{\omg_{i}\}$ of the $\omg_{i}$s should yield, now
at the {\em direct} or {\em inductive limit} of infinite
refinement of the $\gauge_{i}$s as in (\ref{eq105}), an algebra
$\omg_{\infty}$ whose commutative subalgebra part $\alg_{\infty}$
corresponds to ${}^{(\K)}\smooth(X)$---the algebra of ($\K=\R
,\com$-valued) smooth coordinates of the point-events of $X$,
while $\omg^{p}_{\infty}$ in $\ring_{\infty}$ to the
${}^{(\K)}\smooth(X)$-bimodules of smooth differential $p$-forms
cotangent at each and every point-event of $X$ which, in turn, can
now be regarded as being a smooth region of a $\smooth$-manifold
$M$.\footnote{In retrospect, and as we shall see in the sequel
from an ADG-theoretic perspective, that initial anticipation in
\cite{rapzap1,rapzap2}---that is, that at the inductive limit of
infinite localization of the $\omg_{i}$s one should recover the
classical smooth structure of a $\smooth$-manifold---was wrong, or
better, slightly misled by the classical $\smooth$-theory. In
fact, as noted earlier, based on ADG results about inverse and
direct limits of differential triads, we will argue subsequently
that at the continuum limit one recovers a \texttt{smooth} algebra
structure ${}^{\K}\ssmooth(X)$ and ${}^{\K}\ssmooth(X)$-bimodules
$\omg^{p}_{\infty}$ of \texttt{smooth} $p$-forms over it, and that
both of which may be regarded as `generalized', albeit close,
relatives of the corresponding classical $\smooth$-ones. Thus,
rather than directly anticipate that one should obtain the local
smooth structure of a $\smooth$-manifold at the inductive limit of
infinite refinement (of the incidence algebras), perhaps it is
more correct at this point just to emphasize that passing from the
poset to the incidence algebraic regime one catches a glimpse not
only of the topological, but also of the differential structure of
discretized spacetime. This essentially shows that {\em the
differential operator}---the heart and soul of differential
geometry---{\em comes straight from the algebraic structure}.
Equivalently, {\em incidence algebras provide us with a
(reticular) differential geometric mechanism}, something that the
`purely topological' finitary posets were unable to supply since
they are merely associative multiplication structures ({\it ie},
arrow semigroups, or monoids, or even poset categories) and not
linear structures ({\it ie}, one is not able to form differences
of elements in them). This remark will be of crucial importance
subsequently when we will apply ADG-theoretic ideas to these
discrete differential algebras.} We will return to discuss further
this limit in subsection 4.2.

\item The aforesaid continuum limit was physically interpreted
as Bohr's correspondence principle, in the following sense: the
local (differential) structure of classical $\smooth$-smooth
spacetime should emerge at the physically `ideal' (or
operationally `non-pragmatic') limit of infinite localization of
the alocal, discrete and quantal algebraic substrata
$\omg_{i}$.\footnote{For further remarks on this limiting
procedure and its physical interpretation, the reader is referred
to \cite{rapzap1,malrap1,rapzap2,malrap2,zap2}. We will return to
it in an ADG-theoretic context in the next subsection where, as
noted above, we will show that one does not actually get the
classical $\smooth$-smooth structure at the continuum limit, but a
$\ssmooth$-\texttt{smooth} one akin to it. We will also argue that
this ({\it ie}, that we do not get back the $\smooth$-smooth
spacetime manifold at the projective/inductive limit of our
finitary structures) is actually welcome when viewed from the ADG
perspective of the present paper.}

\end{enumerate}

In the sequel, following Sorkin's dramatic change of physical
interpretation of the locally finite posets $P_{i}$ in
\cite{sork0} from finitary topological spaces to {\em causal sets}
(causets) $\caus_{i}$ \cite{bomb87},\footnote{For a thorough
account of this semantic switch from posets as discrete topologies
to posets as locally finite causal spaces, the reader is referred
to \cite{sork1}.} the corresponding reticular and quantal
topological spaces $\omg_{i}$ where similarly interpreted as {\em
quantum causal sets} (qausets) $\qaus_{i}$
\cite{rap1}.\footnote{The reader should note that, in accordance
with our convention in \cite{rap1,malrap1,malrap2}, from now on
all our constructions referring to reticular {\em causal}
structures like the $\caus_{i}$s and their associated
$\qaus_{i}$s, will bear a right-pointing arrow over them just to
remind us of their causal interpretation. (Such causal arrows
should not be confused with the right-pointing arrows over
inductive systems.)} Qausets, like their causet counterparts, were
regarded as locally finite, causal and quantal substrata
underlying the classical Lorentzian spacetime manifold of
macroscopic gravity.\footnote{That causality, as a partial order,
determines not only the topology and differential structure of the
spacetime manifold as alluded to above, but also its conformal
Lorentzian metric structure of (absolute) signature $2$, has been
repeatedly emphasized in \cite{bomb87,sork4,sork2,sork3}.} On the
other hand, it was realized rather early, almost ever since their
inception in \cite{bomb87,sork2}, that causets are sound models of
the {\em kinematical structure} of (Lorentzian) spacetime in the
quantum deep, so that in order to address genuinely dynamical
issues {\it vis-\`a-vis} {\em quantum gravity}, causet theory
should also suggest a dynamics for causets. Thus,

\vskip 0.1in

\centerline{\em how can one vary a locally finite poset?}

\vskip 0.1in

\noindent has become the main question in the quest for a dynamics
for causets\footnote{Rafael Sorkin in private correspondence.}
\cite{rap6}.

It was roughly at that point, when the need to develop a dynamics
for causets arose, that ADG entered the picture. In a nutshell, we
intuited that a possible, rather general answer to the question
above, is

\vskip 0.1in

\centerline{\em by sheaf-theoretic means!}

\vskip 0.1in

\noindent in the sense that the fundamentally algebraic methods of
sheaf theory, as employed by ADG, could be somehow used to model a
realm of dynamically varying causets or, preferably, due to a
quantum theoresis of (local) causality and gravity that we had in
mind, of their qauset descendants.

However, in order to apply the concrete sheaf-theoretic ideas and
techniques of ADG to qausets, it was strongly felt that we should
somehow marry first Sorkin's original finitary posets in
\cite{sork0} with sheaves proper. Thus, {\em finitary spacetime
sheaves} (finsheaves) were defined as spaces $S_{i}$ of (algebras
of) continuous functions on Sorkin's $T_{0}$-posets $P_{i}$ that
were seen to be {\em locally homeomorphic} to each other
\cite{rap2}.\footnote{That is, one formally writes $P_{i}\ni U
\underset{\pi_{i}}{\buildrel{\sigma_{i}}\over{\rightleftarrows}}S_{i}(U)$,
where $\pi_{i}$ is the continuous projection map from the sheaf
space $S_{i}$ to the base topological poset $P_{i}$, $\sigma_{i}$
its inverse (continuous local section) map and $U$ an open subset
of $P_{i}$. In other words, for every open $U$ in $P_{i}$:
$\pi_{i}\circ\sigma_{i}(U)=U\Leftrightarrow[\forall U\in P_{i}:~
\sigma_{i}=\pi_{i}^{-1}]$ ({\it ie}, $\sigma_{i}$ is a local
homeomorphism having $\pi_{i}$ for inverse) \cite{rap2,mall1}.
Here we symbolize these finsheaves by $S_{i}\equiv S_{P_{i}}$.}
The definition of finsheaves can be captured by the following
commutative diagram which we borrow directly from \cite{rap2}

\begin{equation}\label{eq107}
\square[X`P_{i}`\cont_{X}`S_{P_{i}};f_{i}`\sigma\equiv
\pi^{-1}`\pi_{i}^{-1}\equiv\sigma_{i}`\hat{f}_{i}]
\end{equation}

\noindent where $\cont_{X}$ is the usual sheaf of germs of
continuous functions on $X$, while $f_{i}$ and $\hat{f}_{i}$ are
continuous surjections from the topological spaces $X$ and
$\cont_{X}$ to the finitary topological spaces $P_{i}$ and
$S_{i}$, respectively.

Now, the diagram (\ref{eq107}) above prompts us to mention that
the complete analogy between Sorkin's finitary topological posets
$P_{i}$ and finsheaves $S_{i}$ rests on the result that an inverse
system $\invs=(S_{i},\hat{\succeq)}$ of the latter was seen in
\cite{rap2} to possess a projective limit sheaf $S_{\infty}\equiv
S_{P_{\infty}}$\footnote{From (\ref{eq105}),
$P_{\infty}\stackrel{\mathrm{homeo}.}{\simeq}X$.} that is
homeomorphic to $\cont_{X}$---the sheaf of germs of sections of
continuous functions on the topological spacetime manifold $X$.
That is to say, similarly to (\ref{eq105}), one formally writes,

\begin{equation}\label{eq108}
\underleftarrow{\lim}\invs\equiv\lim_{\infty\leftarrow
i}S_{i}\equiv
S_{\infty}\stackrel{\mathrm{homeo}.}{\simeq}\cont_{X}
\end{equation}

\noindent One could cast the result above as a limit of
commutative diagrams like the one in (\ref{eq107}) which defines
finsheaves, as follows

\begin{equation}\label{eq109}
\begin{CD}
P_{i}@>\pi_{i}^{-1}>\sigma_{i}>S_{i}\\
@Vf_{ij}V\succeq_{ij}V        @V\hat{\succeq}_{ij}V\hat{f}_{ij}V\\
P_{j}@>\pi_{j}^{-1}>\sigma_{j}>S_{j}\\
\vdots & &\vdots\\
@Vf_{j\infty}\circ f_{ij}=:f_{i\infty}V\succeq_{i\infty}V@V
\hat{\succeq}_{i\infty}V\hat{f}_{i\infty}:=\hat{f}_{j\infty}\circ\hat{f}_{ij}V\\
\underset{\infty\leftarrow i}{\lim}P_{i}\equiv
P_{\infty}\stackrel{\mathrm{homeo}.}{\simeq} X @>\pi^{-1}
>\sigma>\cont_{X}\stackrel{\mathrm{homeo}.}{\simeq}S_{\infty}
\equiv\underset{\infty\leftarrow i}{\lim}S_{i}
\end{CD}
\end{equation}

\noindent with $f_{ij}$ and $\hat{f}_{ij}$ continuous
injections---the `refinement' or `localization arrows'---between
the $P_{i}$s in $\inv$ and the $S_{i}$s in $\invs$,
respectively.\footnote{These arrows capture precisely the partial
order (or net) refinement relations $\succeq$ and $\hat{\succeq}$
between the finitary posets in $\inv$ and their corresponding
finsheaves in $\invs$ respectively, as (\ref{eq109}) depicts ({\it
eg}, we formally write: $P_{i}\overset{f_{ij}}{\mapto}P_{j}\equiv
P_{j}\succeq_{ij}P_{i}$). Also from (\ref{eq109}), one notices
what we said earlier in connection with (\ref{eq105}) and
(\ref{eq108}), namely, that $X$ and $\cont_{X}$ are obtained at
the categorical limit of infinite (topological) refinement or
localization ($\succeq_{i\infty}$ and $\hat{\succeq}_{i\infty}$)
of the $P_{i}$s and the $S_{i}$s, respectively.}

Having finsheaves in hand, our next goal was to materialize
ADG-theoretically our general answer to Sorkin's question
mentioned above. The basic idea was the following:

\begin{quotation}
\noindent Since sheaves of (algebraic) objects of any kind may be
regarded as universes of variable objects \cite{mall1,macmo}, by
(sheaf-theoretically) localizing or `gauging' the incidence Rota
algebras modelling qausets over the finitary topological posets
$P_{i}$ or their locally finite causet descendants
$\caus_{i}$,\footnote{For instance, one could regard $\caus_{i}$
as a topological space proper by assigning a `causal topology' to
it, as for example, by basing such a topology on `open' sets of
the following kind: $I^{-}(x):=\{ y\in\caus_{i}:~y\rightarrow x\}$
($\forall x\in\caus_{i}$) (`lower' or `past-set topology'), or
dually on: $I^{+}(x):=\{ y\in\caus_{i}:~x\rightarrow y\}$ (`upper'
or `future-set topology'), or even on a combination of both---{\it
ie}, on `open' causal intervals of the following sort:
$A(x,y):=I^{+}(x)\cap I^{+}(y)$ (the so-called Alexandroff
topology). It is one of the basic assumptions about the causets of
Sorkin {\it et al.} that the cardinality of the Alexandroff sets
$A(x,y)$ is finite---the so-called {\em local finiteness} property
of causets \cite{bomb87}. As basic open sets generating the three
topologies above, one could take the so-called {\em covering
past}, {\em covering future} and {\em null Alexandroff} `open'
sets, respectively. These are $I^{-}_{c}(x):=\{
y\in\caus_{i}:~(y\rightarrow x)\wedge (\nexists
z\in\caus_{i}:~y\rightarrow z\rightarrow x)\}$, $I^{+}_{c}(x):=\{
y\in\caus_{i}:~(x\rightarrow y)\wedge (\nexists
z\in\caus_{i}:~x\rightarrow z\rightarrow y)\}$ and
$A_{0}(x,y)=\emptyset\Leftrightarrow (x\rightarrow
y)\wedge(\nexists z\in\caus_{i}:~x\rightarrow z\rightarrow y)$,
respectively. (Note: the {\em immediate arrows} in the Hasse
diagram of any poset $P$ appearing in the definition of
$I_{c}^{-}$, $I_{c}^{+}$ and $A_{n}(x,y)$ are called {\em covering
relations} or {\em links} and they correspond to the {\em
transitive reduction} of the partial order based at each vertex in
the directed and transitive graph of $P$. In turn, the three
topologies mentioned above can be obtained by taking the {\em
transitive closure} of these links \cite{rap1,malrap2}.)} the
resulting finsheaves would stand for worlds of variable
qausets---ones varying dynamically under the influence of a
locally finite, causal and quantal version of gravity {\it in
vacuo} which, in turn, could be concisely encoded in non-flat
connections on those finsheaves \cite{malrap1}. Moreover, and this
cannot be overemphasized here, by using the rather universal
sheaf-theoretic constructions of ADG, we could carry virtually all
the usual $\smooth$-differential geometric machinery on which the
mathematical formulation of general relativity rests, to the
locally finite setting of finsheaves of qausets
\cite{malrap2}---the principal differential geometric objects
being, of course, the aforesaid connections on the relevant
finsheaves, which implement the dynamics of qausets.
\end{quotation}

Thus, as a first step in this development, we set out to define
{\em (curved) principal finsheaves}
$\peel_{i}:=\qaut_{\Qalg_{i}}\Qaus_{\caus_{i}}\equiv\qaut_{i}\Qaus_{i}$
{\em of qausets, and their associated finsheaves}
$\Qaus_{\caus_{i}}\equiv\Qaus_{i}$, {\em over a causet}
$\caus_{i}$.\footnote{In what follows we will be often tempted to
use the same epithet, `{\em principal}', for both the $\peel_{i}$s
and their associated $\Qaus_{i}$s. We do hope that this slight
abuse of language will not confuse the reader. As we will see in
the sequel, this identification essentially rests on our assuming
a general Kleinian stance towards (physical) geometry whereby
`states' (of a physical system) and the `symmetry group of
transformations of those states' are regarded as being
equivalent.} By establishing finitary versions of the classical
general relativistic principles of equivalence and locality, we
realized that the (local) structure (gauge) symmetries of
$\Qaus_{i}$ are finitary correspondents of the orthochronous
Lorentz Lie group ({\it ie}, locally in $\caus_{i}$ one writes
formally: $\qaut_{\Qalg_{i}}\Qaus_{\caus_{i}}(U)=
SO(1,3)_{i}^{\uparrow}$),\footnote{Where $U$ is an open set in
$\caus_{i}$ regarded as a causal-topological space (see footnote
96 above).} and that they could thus be organized into the
aforesaid $\struct_{i}$-finsheaves $\peel_{i}$. Then, by
definition, the $\Qaus_{i}$s are the associated finsheaves of the
principal $\peel_{i}$s.

From the start we also realized that the localization or `gauging'
of qausets in the $\peel_{i}$s and their associated $\Qaus_{i}$s
meant that these finsheaves could be endowed with non-trivial
({\it ie}, non-flat) reticular spin-Lorentzian connections
$\conf_{i}$ {\it \`a la} ADG. Indeed, in complete analogy to the
general ADG case, after having defined reticular flat connections
as the following $\mathbf{K}$-linear and section-wise Leibniz
condition (\ref{eq2})-obeying finsheaf morphisms

\begin{equation}\label{eq110}
\vec{d}^{0}_{i}\equiv\vec{\partial}_{i}:~\Qaus^{0}_{i}\equiv\Qalg_{i}\mapto\Qaus^{1}_{i}
\end{equation}

\noindent as in (\ref{eq1}), as well as higher order extensions

\begin{equation}\label{eq111}
\vec{d}^{p}_{i}:~\Qaus^{p}_{i}\mapto\Qaus^{p+1}_{i},~(\N\ni
p\geq1)
\end{equation}

\noindent between the vector subsheaves $\Qaus^{p}_{i}$ of
$\Qaus_{i}$, we defined in \cite{malrap1} non-flat connections
$\conf_{i}$ on the finsheaves $\Qaus_{i}$ of finite dimensional
differential $\qalg_{i}$-bimodules $\qaus_{i}$\footnote{The reader
should have gathered by now that in the stalks of the structure
finsheaves $\Qalg_{i}$ dwell the (causal versions $\qalg_{i}$ of
the) abelian (sub)algebras $\alg_{i}$ (of $\omg_{i}$) in
(\ref{eq106}), while in the fibers of $\Qring_{i}$ the (causal
versions $\qring_{i}$ of the) $\alg_{i}$-modules $\ring_{i}$ in
(\ref{eq106}).} again as the following $\mathbf{K}$-linear and
section-wise Leibniz condition-obeying (\ref{eq4}) finsheaf
morphisms

\begin{equation}\label{eq112}
\conf_{i}:~\qmodl_{i}\equiv\Qaus_{i}^{*}\mapto\qmodl_{i}\otimes_{\Qalg_{i}}\Qaus_{i}\equiv\Qaus_{i}(\qmodl_{i})
\end{equation}

\noindent similarly to (\ref{eq3}).\footnote{The reader should
note in connection with (\ref{eq112}) that the `identification'
$\qmodl_{i}\equiv\Qaus_{i}^{*}$ tacitly assumes that there is a
(Lorentzian) metric $\qrho_{i}$ on the vector sheaves $\qmodl_{i}$
effecting canonical isomorphisms $\tilde{\qrho}_{i}$ between them
and their dual differential module (covector) finsheaves
$\Qaus_{i}$, as in (\ref{eq12}). We will give more details about
$\qrho_{i}$ and the implicit identification of the finitary
vectors in $\qmodl_{i}$ with their corresponding forms in
$\Qaus_{i}$ shortly. For the time being, we note that we would
like to call $\conf_{i}$ `{\em the (f)initary, (c)ausal and
(q)uantal (v)acuum dynamo}' (fcqv-dynamo) for a reason to be
explained in the next section.} Moreover, in complete analogy to
the local expression for the abstract $\conn$s in (\ref{eq8}), the
finitary $\conf_{i}$s were seen to split locally to

\begin{equation}\label{eq113}
\conf_{i}=\vec{\partial}_{i}+\qaconn_{i},
~(\qaconn_{i}\in\Qaus^{1}_{i}(U),~ U~\mathrm{open~in}~\caus_{i})
\end{equation}

\noindent and the reticular gauge potentials $\qaconn_{i}$ of the
$\conf_{i}$s above were readily seen to be $\qaut_{i}$-valued
local sections of $\Qaus^{1}_{i}$ ({\it ie}, `discrete'
$so(1,3)^{\uparrow}_{i}\simeq sl(2,\com)_{i}$-valued local
$1$-forms),\footnote{Of course, since the $\Qaus_{i}$s are curved,
they do not admit global sections \cite{mall1,malrap1}. In view of
the name `fcqv-dynamo' we have given to $\conf_{i}$ in the
previous footnote, its gauge potential part $\qaconn_{i}$ may be
fittingly coined a `{\em fcqv-potential}'. The fcqv-potential,
like its abstract analogue $\omega$ in (\ref{eq6})--(\ref{eq8}),
is an $n\times n$-matrix of sections of local reticular $1$-forms
({\it ie}, $\qaconn_{i}\equiv(\qaconn^{i}_{pq})\in
M^{i}_{n}(\Qaus^{1}_{i}(U)),~U~\mathrm{open~in}~\caus_{i}$). Also,
since the local structure of the gauge group $\struct_{i}$ of the
$\Qaus_{i}$s is the reticular orthochronous Lorentz Lie algebra
$so(1,3)^{\uparrow}_{i}$, we will denote the vector finsheaves
$\qmodl_{i}$ above as $\qmodll_{i}=(\qmodl_{i} ,\qrho_{i})$, in
accord with our notation earlier for the (real) orthochronous
Lorentzian vector sheaves $\modll=(\modl ,\rho)$ of rank $4$ in
the context of ADG. (However, to avoid uncontrollable
proliferation of symbols and eventual typographical congestion of
indices, superscripts {\it etc}, we will not denote the dual
spaces $\Qaus_{i}$s of the $\qmodll_{i}$s by
$\Qaus_{i}^{\uparrow}$.) Moreover, notice that, as it was
mentioned in \cite{malrap1}, the `finitarity index $i$' on
$so(1,3)^{\uparrow}_{i}$ indicates that the Lie group manifold
$SO(1,3)^{\uparrow}$ of (local) structure gauge symmetries of the
qausets is also subjected to discretization as well. It is
reasonable to assume that {\em finitary structures have finitary
symmetries} or equivalently and perhaps more popularly, {\em
discrete structures possess discrete symmetries}. This is in
accord with our abiding to a Kleinian conception of (physical)
geometry, as noted in footnote 97. On the other hand, we shall see
in the next section that the finitarity index indicates only that
our structures are `discrete' and {\em not} that they are
essentially dependent on the locally finite covering (gauge)
$\gauge_{i}$ of $X$. In fact, we will see that (from the dynamical
perspective) our constructions are {\em inherently gauge
$\gauge_{i}$-independent} and for this reason `{\em alocal}'
\cite{rapzap1,malrap1,rapzap2}. In other words, the (dynamical)
role played by the base localization causet $\caus_{i}$ and, {\it
in extenso}, by the region $X$ of the Lorentzian spacetime
manifold that the latter discretizes relative to $\gauge_{i}$, is
physically insignificant.} in analogy with both the classical and
the abstract (ADG) theory.

At this point, we must stress a couple of things about these
finitary spin-Lorentzian connections $\conf_{i}$ {\it vis-\`a-vis}
the general ADG theory presented in the previous two sections.

\begin{enumerate}

\item {\bf About the base space.} As it was mentioned in
\cite{rap2}, \cite{malrap1} and \cite{malrap2}, in our finitary
regime there are mild relaxations of the two basic conditions of
{\em paracompactness} and {\em Hausdorffness} ($T_{2}$-ness) that
ADG places on the base topological space $X$ on which the vector
sheaves $\modl$ bearing connections $\conn$ are soldered. As noted
in footnote 81, the starting region $X$ of the topological
spacetime manifold $M$ from which the $\caus_{i}$s (and their
associated $\qaus_{i}$s) come from was assumed in \cite{sork0} to
be {\em relatively compact} and (at least) $T_{1}$. If one relaxes
paracompactness to relative compactness, and $T_{2}$-ness to
$T_{1}$-ness (and we are indeed able to do so without any loss of
generality),\footnote{In fact, as noted in both \cite{rapzap1} and
\cite{rapzap2}, at the finitary poset level one must actually
insist on relaxing Hausdorffness, because a $T_{2}$-finitary
substitute in \cite{sork0} is automatically trivial as a
topological space---that is, it carries the discrete topology, or
equivalently, it is a completely disconnected set (no arrows
between its point vertices).} one is still able to carry out in
the locally finite regime the entire spectrum of the ADG-theoretic
constructions described in the last two sections.\footnote{In
fact, we could have directly started our finsheaf constructions
straight from a paracompact and Hausdorff $X$ without coming into
conflict with Sorkin's results. For instance, already in
\cite{malrap2} we applied the entire sheaf-cohomological panoply
of ADG to our finsheaves of qausets.}

\item {\bf About the stalk: Lorentzian metric and its orthochronous symmetries.}
The stalks of the $\Qaus_{i}$s are occupied by qausets
$\qaus_{i}$; in other words, they are the spaces where the (germs
of the) continuous local sections of the $\Qaus_{i}$s take values.
These qausets, as it has beeen argued in \cite{malrap1}, determine
a metric $\qrho_{i}$ of Lorentzian signature. Thus, as it was
emphasized in footnote 17 of 2.2, $\qrho_{i}$ is not carried by
the base space $\caus_{i}$, which is simply a topological space;
rather, it concerns directly the (objects living in the stalks of
the) relevant finsheaves {\it per se}. In fact, we may define this
metric to be the following finsheaf morphism:

\begin{equation}\label{eq114}
\qrho_{i}:~\qmodll_{i}\oplus\qmodll_{i}\mapto\Qalg_{i}
\end{equation}

\noindent which, like its abstract version $\rho$ in (\ref{eq11}),
is $\Qalg_{i}$-bilinear between the (differential)
$\Qalg_{i}$-modules $\Qaus_{i}$ concerned and (section-wise)
symmetric.\footnote{In connection with footnote 100, we note that
we tacitly assume that $\qmodll_{i}=(\qmodl_{i},\qrho_{i})$ in
(\ref{eq114}) is the dual to $\Qaus_{i}$ ({\it ie},
$\Qaus_{i}=\qmoddll_{i}=\Hom_{\Qalg_{i}}(\qmodll_{i},\Qalg_{i})$).
It is also implicitly assumed that $\qrho_{i}$ in (\ref{eq114})
induces a canonical isomorphism between $\qmodll_{i}$ and its dual
$\Qaus_{i}$ analogous to (\ref{eq12}). Thus, with a certain abuse
of language, but hopefully without causing any confusion, we will
assume that $\Qaus_{i}\equiv\qmodll_{i}$ ({\it ie}, we identify
via $\qrho_{i}$ finitary covectors and vectors) and use them
interchangeably in what follows.} It follows that the
$\Qalg_{i}$-metric preserving (local) automorphism group finsheaf
$\qaut_{\Qalg_{i}}\qmodll_{i}|_{U\in \caus_{i}}
\equiv\qaut_{\qrho_{i}}\qmodl_{i}|_{U\in \caus_{i}}$ is the
aforesaid principal $\qstruct$-finsheaf
$\peel_{i}(U)\equiv\qaut_{\qrho_{i}}\qmodl_{i}(U)\equiv\mathrm{
SO}(1,3;\Qalg_{i}(U))^{\uparrow}_{i}$ of reticular orthochronous
isometries of the (real) Lorentzian finsheaf
$\qmodll_{i}=(\qmodl_{i} ,\qrho_{i})$ of rank $4$.\footnote{Since,
as noted in footnote 15, specific dimensionality arguments do not
interest us here as long as the algebras involved in the stalks of
our finsheaves are (and they are indeed) finite dimensional, the
reader may feel free to choose an arbitrary, finite rank $n$ for
our finsheaves. Then, the reticular Lorentzian $\Qalg_{i}$-metric
$\qrho_{i}$ involved will be of absolute signature $n-2$ ({\it
ie},
$\qrho_{i}=\mathrm{diag}(-1,\stackrel{n-1}{\overbrace{+1,+1,\ldots
+1}})$) and its local invariance (structure) group
$\mathrm{SO}(1,n-1;\Qalg_{i}(U))^{\uparrow}$ ($U$ open in
$\caus_{i}$, as usual).}

Also, in accordance with Sorkin {\it et al.}'s remark in
\cite{bomb87} that a (locally finite) partial order determines not
only the topological and the metric structure of the Lorentzian
manifold of general relativity, but also its differential
structure, we witness here that the aforementioned nilpotent
Cartan-K\"{a}hler (exterior) differentials $\vec{d}_{i}^{p}$,
which as we saw in (\ref{eq111}) effect vector subsheaf morphisms
$\vec{d}_{i}^{p}:~\Qaus_{i}^{p}\mapto\Qaus_{i}^{p+1}$ ($\Z\ni
p\geq0$), derive directly from the algebraic structure of the
$\Qaus_{i}$s---that is to say, again straight from the stalk of
the finsheaves of qausets without any dependence on the base
causet $\caus_{i}$ which is simply a causal-topological space. We
cannot overemphasize this either:

\begin{quotation}
\noindent{\em Differentiability in our finitary scheme, and
according to ADG, does not depend on the base space (which is
assumed to be simply a topological space); the differential
mechanism comes staright from the stalk ({\it ie}, from the
algebraic objects dwelling in it) and, {\it a fortiori}, certainly
not from a classical, $\smooth$-smooth base spacetime manifold.}
\end{quotation}

\item {\bf About the physical interpretation.} We would like
to comment a bit on the physical interpretation of our principal
finsheaves of qausets and the reticular spin-Lorentzian
connections on them. First we must note that Sorkin {\it et al.},
after the significant change in physical interpretation of the
locally finite posets involved from topological in \cite{sork0} to
causal in \cite{bomb87,sork4,sork1,sork2,sork3} alluded to above,
insisted that, while the topological posets can be interpreted as
coarse approximations to the continuous spacetime manifold of
macroscopic physics, the causets should be regarded as being truly
fundamental structures in the sense that the macroscopic
Lorentzian manifold of general relativity is an approximation to
the deep locally finite causal order, not the other way around.

Our scheme strikes a certain balance between these two poles. For
instance, while we assume a base causet on which we solder our
incidence algebras modelling qausets, that causet is also assumed
to carry a certain topology---the `causal topology'\footnote{See
footnote 96.}---so that it can serve as the background topological
space on which to solder our algebraic structures, which in turn
enables us to apply ADG to them thus unveil potent differential
geometric traits of the qausets in the stalks, as described above.
This causal topology however, in contradistinction to Sorkin's
$T_{0}$-topological posets which model ``{\em thickened space-like
hypersurfaces}'' in continuous spacetime \cite{sork0}, is regarded
as a theory of `thickened' causal regions in space{\em time}
\cite{rap1,malrap1,rapzap2}.\footnote{For more on this, see
subsection 4.3 below.} Furthermore, as it has been emphasized in
\cite{malrap1}, while the non-flat reticular spin-Lorentzian
connections $\conf_{i}$ on the corresponding $\Qaus_{i}$s can be
interpreted as the fundamental operators encoding the {\em curving
of quantum causality} thus setting the kinematics for a
dynamically variable quantum causality, an inverse system
$\finv:=\{(\peel_{i},\conf_{i})\}$ was intuited to `converge' at
the operationally ideal ({\it ie}, non-pragmatic and `classical'
in Bohr's `correspondence principle' sense \cite{rapzap1}) limit
of infinite refinement or localization of both the base causets
and the associated qauset fibers over them to the classical
principal fiber bundle $(\princ^{\uparrow} ,\conn)$ of continuous
local orthochronous Lorentz symmetries $so(1,3)^{\uparrow}$ of the
$\smooth$-smooth spacetime manifold $M$ of general relativity and
the $sl(2,\com)$-valued spin-Lorentzian gravitational connection
$\conn$ on it.\footnote{For more technical details about the
projective limit of $\finv$, the reader must wait until the
following subsection. At this point it must be stressed up-front,
in connection with footnote 75, that what we actually get at the
projective limit of $\finv$ is a $\ssmooth$-\texttt{smooth}
principal bundle (and its spin-Lorentzian connection) over the
region $X$ of a `generalized differential manifold' ({\it ie},
$\ssmooth$-\texttt{smooth}) $M$.} Since $(\princ^{\uparrow}
,\conn)$ is the gauge-theoretic version of the kinematical
structure of general relativity---the dynamical theory of the
classical field of local causality $g_{\mu\nu}$,\footnote{For
recall that the spacetime metric $g_{\mu\nu}(x)$, for every $x\in
M$, delimits a Minkowski lightcone based at $x$ (by the
equivalence principle, the curved gravitational spacetime manifold
of general relativity is, locally, Minkowski space, {\it ie},
flat, and in this sense general relativity may be viewed as
special relativity being localized or `gauged'). Thus, the
Einstein equations of general relativity, which describe the
dynamics of $g_{\mu\nu}$ (which, in turn, can be interpreted as
the field of the ten gravitational potentials), effectively
describe the dynamical changes of (the field of) local causality.
All this was analyzed in detail in \cite{malrap1}.} each
individual member $(\peel_{i},\conf_{i})$ of the inverse system
$\finv$ was interpreted as the kinematics of a locally finite,
causal and quantal version of (vacuum) Einstein Lorentzian
gravity.\footnote{As we shall see in the next section, the actual
kinematical configuration space for the locally finite, causal and
quantal vacuum Einstein gravity is the moduli space $\fconn_{i}$
of finitary spin-Lorentzian connections $\conf_{i}$. As we shall
see, projective limit arguments also apply to an inverse system of
such reticular moduli spaces.} {\it In toto}, we have amalgamated
aspects from the interpretation of both the finitary substitutes
and the causets, as follows:\footnote{Further distillation and
elaboration on these ideas in subsection 4.3.}

\begin{quotation}
\noindent{\em `Coarse causal regions' are truly fundamental,
operationally sound and physically pragmatic, while the classical
pointed $\smooth$-smooth spacetime manifold ideal.\footnote{More
remarks on `coarse causal regions' will be made in subsection
4.3.} Curved finsheaves of qausets
$(\peel_{i}\equiv\qmodll_{i},\conf_{i})$ model the kinematics of
dynamical (local) quantum causality {\it in vacuo} as the latter
is encoded in the $fcqv$-dynamo $\conf_{i}$. A generalized ({\it
ie}, $\ssmooth$-\texttt{smooth}) version of the classical
kinematical structure of general relativity, $(\princ^{\uparrow}
,\conn)$, over the differential spacetime manifold $M$, arises at
the ideal and classical (Bohr's correspondence) limit of infinite
localization of the qausets---in point of fact, of
$\finv$.\footnote{This is a concise {\it r\'esum\'e} of a series
of papers \cite{rapzap1,rap1,malrap1,rapzap2,malrap2,rap6}. Of
course, `infinite localization' requires `infinite microscopic
power' ({\it ie}, energy of determination or `measurement' of
locution) which is certainly an ideal ({\it ie}, operationally
non-pragmatic and physically unattainable) requirement. This seems
to be in accord with the pragmatic cut-offs of quantum field
theory and the fundamental length $L_{P}$ (the Planck length) that
the `true' quantum gravity is expected to posit (and below which
it is expected to be valid!), for it is fairly accepted now that
one cannot determine the locus of a quantum particle with
uncertainty (error) less than $L_{P}\approx 10^{-35}m$ without
creating a black hole. This seems to be the {\it raison
d'\^{e}tre} of all the so-called `discrete' approaches to quantum
spacetime and gravity \cite{malrap1}.}}
\end{quotation}

\item {\bf About `reticular' differential geometry.} The basic
moral of our application of ADG to the finitary regime as
originally seen in \cite{malrap1} as well as here, but most
evidently in \cite{malrap2}, is that the fundamental differential
mechanism which is inherent in the differential geometry that we
all are familiar with\footnote{Albeit, just from the classical
({\it ie}, $\smooth$-smooth) perspective.} is independent of
$\smooth$-smoothness so that it can be applied in full to our
inherently reticular models, or equally surprisingly, to spaces
that appear to be ultra-singular and incurably pathological or
problematic when viewed from the differential manifold's viewpoint
\cite{malros1,malros2,ros}. In our case, what is startling indeed
is that none of the usual `discrete differential mathematics'
({\it eg}, difference calculus, finite elements or other related
Regge calculus-type of methods) is needed in order to address
issues of differentiability and to develop a full-fledged
differential geometry in a (locally) finite setting. For instance,
there appears to be no need for defining up-front `{\em discrete
differential manifolds}' and for developing {\it a priori} and,
admittedly, in a physically rather {\it ad hoc} manner a `{\em
discrete differential geometry}' on them\footnote{Like for example
the perspective adopted in \cite{dimu1,dimu2,dimu3,dimu4}.} in
order to investigate differential geometric properties of
`finitary' (ordered) spaces.\footnote{Like graphs (directed, like
our posets here, or undirected), or even finite structureless
sets.} For they too can be cast under the wider axiomatic,
algebraico-sheaf-theoretic prism of ADG as a particular
application of the general theory. All in all, it is quite
surprising indeed that the basic objects of the usual differential
geometry like `tangent' vectors (derivations), their dual forms,
exterior derivatives, Laplacians, volume forms {\it etc}, carry
through to the locally finite scene and none of their discrete
(difference calculus') analogues is needed, but this precisely
proves the point:
\end{enumerate}

\begin{quotation}
\noindent One feels, perhaps `instinctively' due to one's long
time familiarity with and the numerous `habitual' (but quite
successful!) applications of the usual smooth calculus where the
differential mechanism comes from the supporting space ({\it ie},
it is provided by the algebra $\smooth(M)$ of infinitely
differentiable functions on the differential manifold $M$), that
in the `discrete' case too some novel kind of `discrete
differential geometry' must come from a `discrete differential
manifold'-type of base space---as if {\em the differential
calculus follows from, or at least that it must be tailor-cut to
suit, space}. In other words, in our basic working philosophy we
have been misled by the habitual applications and the numerous
successes of the smooth continuum into thinking that
differentiability comes from, or that it is somehow vitally
dependent on, the supporting space. By the present application of
ADG to our reticular models we have witnessed how, quite on the
contrary, {\em differentiability comes from the stalk}---{\it ie},
from algebras dwelling in the fibers of the relevant
finsheaves---and it has nothing to do with the ambient space,
which only serves as an auxiliary, and in no way contributing to
the said differential mechanism, topological space for the
sheaf-theoretic localization of those algebraic objects. The usual
differential geometric concepts, objects and mechanism that
relates the latter still apply in our reticular environment and,
perhaps more importantly, in spite of it.
\end{quotation}

\subsection{Projective Limits of Inverse Systems of Principal Lorentzian Finsheaves}

Continuous limits of finitary simplicial complexes and their
associated incidence algebras, regarded as discrete and quantal
topological spaces \cite{rapzap1,rapzap2}, have been studied
recently in \cite{zap1,zap2}. In this subsection, always based on
ADG, we present the projective limit of the inverse system
$\finv=\{(\peel_{i},\conf_{i})\}$ of principal Lorentzian
finsheaves of qausets $\peel_{i}$ equipped with reticular
spin-Lorentzian connections $\conf_{i}$ which was supposed in
\cite{malrap1} to yield the classical kinematical structure of
general relativity in its gauge-theoretic guise---that is, the
principal orthochronous spin-Lorentzian bundle over the (region
$X$ of the) $\smooth$-smooth spacetime manifold $M$ of general
relativity locally supporting an $sl(2,\com)$-valued (self-dual)
smooth connection ({\it ie}, gauge potential) $1$-form
$\aconn^{(+)}$. We center our study on certain results from a
recent categorical account of projective and inductive limits in
the category $\ctriad$ of Mallios' differential triads in
\cite{pap1,pap2}, as well as on results from a treatment of
projective systems of principal sheaves (and their associated
vector sheaves) endowed with Mallios' $\struc$-connections in
\cite{vas1,vas2,vas3}. Then, we compare this inverse limit result,
at least at a conceptual level and in a way that emphasizes the
calculus-free methods and philosophy of ADG, with the projective
limit of a projective family $\ashlew$ of compact Hausdorff
differential manifolds employed in \cite{ashlew2} in order to
endow the moduli space $\aconn/\grouv$ of gauge-equivalent
non-abelian Y-M and gravitational connections with a differential
geometric structure. In fact, we will maintain that an inverse
system $\invmod$ of our finitary moduli spaces should yield at the
projective limit of infinite localization a generalized version
({\it ie}, a $\ssmooth$-\texttt{smooth} one) of the classical
moduli space $\infconn^{(+)}$ of gauge-equivalent (self-dual)
$\smooth$-connections on the region $X$ of the smooth spacetime
manifold $M$.

\medskip

The concept-pillar on which ADG stands is that of a {\em
differential triad} $\triad=(\struc ,\partial ,\Omg)$ associated
with a $\mathbf{K}=\mathbf{R},\cons$-algebraized space
$(X,\struc)$. In ADG, {\em differential triads specialize to
abstract differential spaces}, while the $\struc$s in them stand
for (structure sheaves of) {\em abstract differential algebras of
generalized smooth or differentiable coordinate functions}, and
they were originally born essentially out of realizing that

\begin{quotation}
\noindent {\em the classical differential geometry of a manifold
$X$ is deduced from its structure sheaf $\smooth_{X}$, the latter
being for the case at issue the result of the very topological
properties\footnote{Poincar\'{e} lemma \cite{malros1,malrap2}.} of
the underlying `smooth' manifold $X$.}
\end{quotation}

\noindent Thus, in effect, the first author originally, and
actually quite independently of any previous relevant work,
intuited, built and subsequently capitalized on the fact that the
algebra sheaf $\struc$ of generalized arithmetics (or abstract
coordinates) is precisely the structure that provides one with all
the basic differential operators and associated `intrinsic
differential mechanism' one needs to actually do differential
geometry---the classical, $\smooth$-smooth theory being obtained
precisely when one chooses $\smooth_{M}$ as one's structure sheaf
of coordinates.\footnote{Yet, we can still note herewith that the
first author arrived at the notion of a {\em differential triad}
as a particularization to the basic differentials of the classical
theory of the amply ascertained throughout the same theory
instrumental role played by the notion of an
$\struc$($\equiv\smooth_{X}$)-connection ({\it ie}, covariant
differentiation).} Thus, the objects dwelling in the stalks of
$\struc$ may be perceived as {\em algebras of generalized (or
abstract) `infinitely differentiable' (or `smooth') functions},
with the differential geometric character of the base localization
space $X$ left completely undetermined---in fact, it is regarded
as being totally irrelevant to ADG.\footnote{Of course, as also
noted earlier in footnote 64, in the classical case ({\it ie},
when one identifies $\struc_{M}\equiv\smooth_{M}$) there is a
confusion of the sort `{\em who came first the chick or the
egg?}', since one tends to identify the underlying space(time)
({\it ie}, the $\smooth$-smooth manifold $M$) with its structure
sheaf $\smooth_{M}$ of smooth functions and, more often than not,
one is (mis)led into thinking that {\em differentiability---the
intrinsic mechanism of differential geometry} so to speak---comes
(uniquely!) from the underlying smooth manifold. This is precisely
what ADG highlighted: {\em differentiability comes in fact from
the structure sheaf, so that if one chooses `suitable' or
`appropriate' (to the problem one chooses to address) algebras of
`generalized smooth' functions other than $\smooth(M)$, one is
still able to do differential geometry (albeit, of a generalized
or abstract sort) in spite of the classical, $\smooth$-smooth base
manifold}.}

In \cite{pap1}, the differential triads of ADG were seen to
constitute a category $\ctriad$---{\em the category of
differential triads}. Objects in $\ctriad$ are differential triads
and morphisms between them represent abstract differentiable maps.
In $\ctriad$ one is also able to form finite products and, unlike
the category of smooth manifolds where an arbitrary subset of a
(smooth) manifold is not a (smooth) manifold, one can show that
every object $\triad$ in $\ctriad$ has canonical subobjects
\cite{pap1}. More importantly however, in \cite{pap2} it was shown
that $\ctriad$ is complete with respect to taking projective and
inductive limits of projective and inductive systems of triads,
respectively.\footnote{In fact, Papatriantafillou showed that
projective/inductive systems of differential triads having either
a common, fixed base topological space $X$ (write
$\triad_{i}(X)$), or a projective/inductive system thereof indexed
by the same set of indices (write $\triad_{i}(X_{i})$), possess
projective/inductive limits. Below, we will see that our
projective/inductive system $\finv=\{(\peel_{i},\conf_{i})\}$ of
finitary posets (causets), (principal) finsheaves of incidence
algebras (qausets) over them and reticular spin-Lorentzian
connections on those finsheaves, are precisely of the second kind.
The reader should also note here that in the mathematics
literature, `projective', `inverse' and `categorical' limits are
synonymous terms; so are `inductive' and `direct' limits (also
known as `categorical colimits'). The result from \cite{pap2}
quoted above can be stated as follows: {\em the category}
$\ctriad$ {\em is complete and cocomplete}. This remark, that is
to say, that $\ctriad$ is (co)complete, will prove to be of great
importance in current research \cite{rap7} for showing that the
category of finsheaves of qausets---which is a subcategory of
$\ctriad$---is, in fact, an example of a structure known as a {\em
topos} \cite{macmo}---a topos with a non-Boolean (intuitionistic)
internal logic tailor-made to suit the finitary, causal and
quantal vacuum Einstein-Lorentzian gravity developed in the
present paper.} This is a characteristic difference between
$\ctriad$ and the category of manifolds where the projective limit
of an inverse system of manifolds is not, in general, a
manifold.\footnote{From a categorical point of view, this fact
alone suffices for regarding the abstract differential spaces (of
structure sheaves of generalized differential algebras of
functions and differential modules over them) that the
ADG-theoretic differential triads represent as being more powerful
and versatile differential geometric objects than
$\smooth$-manifolds. As also mentioned in \cite{pap2}, it was
precisely due to the aforesaid shortcomings of the category of
smooth manifolds that led many authors in the past to generalize
differential manifolds to {\em differential spaces} in which the
manifold structure is effectively redundant
\cite{sik1,sik2,most,hell}. In fact, the first author's
differential triads generalize both $\smooth$-manifolds and
differential spaces, and, perhaps more importantly for the
physical applications, they are general enough to include
non-smooth (`singular') spaces with the most general,
non-functional, structure sheaves \cite{malros1,malros2,ros}. On
the other hand, a little bit later we will allude to and, based on
ADG and its finitary application herein, comment on an example
from \cite{ashlew2} of an inverse system of differential manifolds
that yields a differential manifold at the projective limit.}
Moreover, Vassiliou, by applying ADG-theoretic ideas to principal
sheaves (whose associated sheaves are precisely the vector sheaves
of ADG) \cite{vas1,vas2,vas3}, has shown that when the flat
differentials $\dot{\partial}$ of the triads in the aforesaid
projective/inductive systems of Papatriantafillou are promoted
({\it ie}, `gauged' or `curved') to $\struc$-connections
$\dot{\conn}$ on principal sheaves, the corresponding
projective/inductive systems
$(\princ_{i},\dot{\conn}_{i})$\footnote{With
$({\mathcal{I}},\geq)$ a partially ordered, directed set (net) of
indices `$i$' labelling the elements of the inverse/direct system
$(\princ_{i},\dot{\conn}_{i})$. The systems
$(\princ_{i},\dot{\conn}_{i})$ are said to be (co)final with
respect to the index net $({\mathcal{I}},\geq)$. We remind the
reader that in our case `$i$' is the finitarity or localization
index ({\it ie}, locally finite open covers $\gauge_{i}$ of
$X\subset M$ form a net \cite{sork0,rap2,malrap1,malrap2}).} have
principal sheaves endowed with non-flat connections as
inverse/direct limits.

Thus, in our locally finite case, the triplet
$\vec{\triad}_{i}=(\Qalg_{\caus_{i}}\equiv\Qalg_{i}
,\Qring_{\caus_{i}}\equiv\Qring_{i},\vec{d}^{p}_{i})$ is an
ADG-theoretic differential triad of a (f)initary, (c)ausal and
(q)uantal kind. In other words, the category $\ctriad_{fcq}$
having for objects the differential triads $\vec{\triad}_{i}$ and
for arrows the finitary analogues of the triad-morphisms mentioned
above is a subcategory of $\ctriad$ called {\em the category of
fcq-differential triads}. So, we let
$\invtriad:=\{\vec{\triad}_{i}\}$ be the {\em mixed
projective-inductive} system of fcq-differential triads in
$\ctriad_{fcq}$.\footnote{The term `mixed projective-inductive'
(or equivalently, `{\em mixed inverse-direct'}) system pertains to
the fact that the family $\invtriad$ (implicitly) contains both
the projective system $\inv=\{ \caus_{i}\}$ of reticular base
causets, and the inductive system $\diromg$ of qausets
corresponding (by Gel'fand duality) to the aforesaid causets.
(Note that we refrain from putting right-pointing causal arrows
over $\inv$ and $\diromg$, in order to avoid notational
confusion.)} By straightforwardly applying Papatriantafillou's
results \cite{pap1,pap2} to the inverse-direct system $\invtriad$
we obtain a projective-inductive limit triad
$\sstriad=(\struc\equiv{}^{(\K)}\ssmooth_{X},\Omg^{p}_{\infty},d^{p}_{\infty})$
(write:
$\sstriad=\overrightarrow{\underleftarrow{\lim}}\invtriad\equiv
\overset{i\rightarrow\infty}{\underset{\infty\leftarrow
i}{\lim}}\{\vec{\triad}_{i}\}$), here called
`$\ssmooth$-\texttt{smooth} differential triad', consisting of the
structure sheaf $\ssmooth_{X}$ of generalized infinitely
differentiable ({\it ie}, $\ssmooth$-\texttt{smooth}) functions on
$X$, as well as of (sheaves $\Omg^{p}_{\infty}$ over $X$ of)
${}^{(\K)}\ssmooth(X)$-bimodules $\omg^{p}_{\infty}$ of
$\K$-valued differential forms related by exterior differentials
($\mathbf{K}$-linear sheaf morphisms) $d^{p}_{\infty}$.

We can then localize or gauge the Cartan-K\"ahler differentials of
the fcq-differential triads in $\ctriad_{fcq}$ as worked out in
\cite{malrap1}, thus obtain the inverse system $\finv=\{
(\peel_{i},\conf_{i})\}$ alluded to above.\footnote{We could have
chosen to present the collection $\{ (\peel_{i},\conf_{i})\}$ as
an inductive family of principal finsheaves and their finitary
connections, since the connections (of any order $p$)
$\conf^{p}_{i}$ in each of its terms are effectively obtained by
localizing or gauging the reticular differentials
$\vec{d}^{p}_{i}$ in each term of $\diromg$. However, that we
present $\finv$ dually, as an inverse system, is consistent with
our previous work \cite{malrap1,malrap2} and, as we shall see
shortly, it yields the same result at the continuum limit ({\it
ie}, the $\ssmooth$-principal bundle).} As mentioned earlier, the
limits of projective systems of principal sheaves equipped with
Mallios $\struc$-connections have been established in
\cite{vas1,vas2,vas3}. Hence, by straightforwardly carrying
Vassiliou's results to the finitary case, and as it was
anticipated in \cite{malrap1,malrap2}, we get that $\finv$ yields
at the projective limit a generalized classical principal
$\ssmooth$-\texttt{smooth} (spin-Lorentzian) fiber bundle (whose
associated bundle is the $\ssmooth$-\texttt{smooth} (co)tangent
vector bundle of ${}^{(\K)}\ssmooth(X)$-modules of $\K$-valued
differential forms) endowed with a \texttt{smooth}
$so(1,3)^{\uparrow}$-valued connection $1$-form $\aconn$ over a
(region $X$ of) the $\ssmooth$-smooth spacetime manifold
$M$\footnote{Write $({}^{(\K)}P_{\infty},^{(\K)}\con_{\infty})$
for the $\ssmooth$-\texttt{smooth} principal bundle and its
non-trivial spin-Lorentzian connection.} \cite{malrap1,malrap2}.
All in all, we formally write

\begin{equation}\label{add1}
\begin{array}{c}
\sstriad=(\struc_{X}\equiv{}^{\K}\ssmooth_{X},
\Omg^{p}_{\infty},d^{p}_{\infty})=\overrightarrow{\underleftarrow{\lim}}\invtriad\equiv
\overset{i\rightarrow\infty}{\underset{\infty\leftarrow
i}{\lim}}\{\vec{\triad}_{i}\}\equiv\overset{i\rightarrow\infty}{\underset{\infty\leftarrow
i}{\lim}}\{(\Qalg_{i},\Qring_{i},\vec{d}^{p}_{i})\}\cr
({}^{(\K)}\pee_{\infty}
,{}^{(\K)}\con_{\infty})=\underleftarrow{\lim}\finv\equiv\underset{\infty\leftarrow
i}{\lim}\{ (\peel_{i},\conf_{i})\}
\end{array}
\end{equation}

\noindent and diagrammatically one can depict these limiting
procedures as follows

\begin{equation}\label{add2}
\begin{CD}
\vec{\triad}_{i}@>\mathrm{gauging}>\vec{\partial}_{i}\mapto\vec{\con}_{i}=
\vec{\partial}_{i}+\qaconn_{i}>(\peel_{i},\conf_{i})\\
@V\mathrm{injective~tri-}V\mathrm{ad~morphism}V
@V\mathrm{injective~\struct_{i}-fin-}V\mathrm{sheaf~morphism}V\\
\vec{\triad}_{j}@>\mathrm{gauging}>\vec{\partial}_{j}\mapto\vec{\con}_{j}=
\vec{\partial}_{j}+\qaconn_{j}>(\peel_{j},\conf_{j})\\
\vdots & &\vdots\\ @V\mathrm{infinite}V\mathrm{refinement}V@V
\mathrm{infinite}V\mathrm{refinement}V\\
\sstriad=\overrightarrow{\underleftarrow{\lim}}\invtriad@>\mathrm{gauging}>\partial_{\infty}
\mapto\con_{\infty}=\partial_{\infty}+\aconn_{\infty}>
({}^{(\K)}\princ^{\uparrow}_{\infty}
,{}^{(\K)}\con_{\infty})=\underset{\infty\leftarrow i}{\lim}\finv
\end{CD}
\end{equation}

\noindent

\medskip

\subsubsection{A brief note on projective versus inductive limits}

We mentioned earlier the categorical duality between the category
$\as$ of finitary substitutes $P_{i}$ and poset morphisms between
them, and the category $\rz$ of the incidence algebras $\omg_{i}$
associated with the $P_{i}$s and injective algebra homomorphisms
between them, which duality is ultimately rooted in the general
notion of Gel'fand duality.\footnote{See our more analytical
comments on Gel'fand duality in the next section.} In a
topological context, the idea to substitute Sorkin's finitary
topological posets by incidence Rota algebras was originally aimed
at {\em `algebraizing space'} \cite{zap0}---that is to say, at
replacing `space' (of which, anyway, we have no physical
experience\footnote{Again, see more analytical comments on the
`unphysicality' of space(time) in the next section.}) by suitable
(algebraic) objects that may be perceived as living on that
`space' and, more importantly, from which objects this `space' may
be somehow derived by an appropriate procedure (Gel'fand
spatialization). In fact, as briefly described before, again in a
topological context and in the same spirit of Gel'fand duality,
the second author substituted Sorkin's $P_{i}$s by finsheaves
$S_{i}$ of (algebras of) continuous functions that, as we said,
are (locally) topologically equivalent ({\it ie}, locally
homeomorphic) spaces to the $P_{i}$s \cite{rap2}. Here too, the
basic idea was, in an operational spirit, to replace `space' by
suitable algebraic objects that live on `it', and it was observed
that the maximum localization (finest resolution) of the
point-events of the bounded region $X$ of the $\cont$-spacetime
manifold $M$ by coarse, open regions about them at the inverse
limit of a projective system of $P_{i}$s, corresponds (by Gel'fand
duality) to defining the stalks of $\cont_{X}$---the sheaf of
(germs of) continuous functions on the topological manifold
$X$---at the {\em direct limit} of (infinite localization of) {\em
an inductive system of the} $S_{i}$s.\footnote{And it should be
emphasized that the stalks of a sheaf are the `ultra-local' ({\it
ie}, maximally localized) point-like elements of the sheaf space
\cite{mall1,rap2}.} At the end of \cite{rap2} it was intuited that
if the stalks of the $S_{i}$s were assumed to be inhabited by
incidence algebras which are discrete differential manifolds as
explained above, at the inverse limit of infinite refinement or
localization of the projective system $\inv$ of Sorkin's
topological posets yielding the continuous base topological space
$X$, the corresponding (by Gel'fand duality) inverse-direct system
$\invtriad$ of finitary differential triads should yield the
classical structure sheaf $\struc_{X}\equiv{}^{(\com)}\smooth_{X}$
of germs of sections of (complex-valued)\footnote{In
\cite{rapzap1,rap1,rap2,rapzap2,malrap1} it was tacitly assumed
that we were considering incidence algebras over the field $\com$
of complex numbers.} smooth functions on $X$ and the sheaf
${}^{\com}\Omg_{X}$ of ${}^{(\com)}\smooth(X)$-bimodules of
(complex) differential forms, in accordance with Gel'fand duality.

There are two issues to be brought up here about this intuition at
the end of \cite{rap2}. First thing to mention is that, as alluded
to earlier, it is more accurate to say that, since the incidence
algebras are objects categorically or Gel'fand dual to Sorkin's
topological posets, and since the latter form an inverse or
projective system $\inv$, the former should be thought of as
constituting a direct or inductive system $\diromg$ of algebras
possessing ${}^{\K}\smooth(X)$ and ${}^{(\K)}\omg(X)$ over it as
an {\em inductive limit}.\footnote{Hence, precisely speaking, the
aforesaid fcq-differential triads constitute a mixed
inverse-direct system $\invtriad$ having the
$\ssmooth$-\texttt{smooth} differential triad $\sstriad$ as an
inductive limit \cite{pap1,pap2}.} In fact, as mentioned in the
previous paragraph, the stalks of ${}^{(\K)}\Omg_{X}$ (in fact, of
any sheaf! \cite{rap2}), which are inhabited by germs of sections
of $\smooth$-smooth ($\K=\R ,\com$-valued) differential forms, are
obtained precisely at that inductive limit. We may distill all
this to the following physical statement which foreshadows our
remarks on Gel'fand duality to be presented in the next section:

\begin{quotation}
\noindent{\em While `space(time)' is maximally (infinitely)
localized (to its points) by an inverse limit of a projective
system of Sorkin's finitary posets, the (algebraic) objects that
live on space(time) ({\it ie}, the various physical fields) are
maximally (infinitely) localized in the stalks of the finsheaves
that they constitute by a direct limit of an inductive system of
those finsheaves.} Equivalently stated, {\em `space(time)' is
categorically or Gel'fand dual to the physical fields that are
defined on `it'.}
\end{quotation}

The second thing that should be stressed here, and in connection
with footnote 75, is that we do not actually get the classical
differential geometric structure sheaf ${}^{\K}\smooth_{X}$ and
the corresponding sheaf ${}^{\K}\Omg_{X}$ of
${}^{\K}\smooth(X)$-modules of differential forms. {\it In toto},
we do not actually recover the classical $\smooth$-smooth
differential triad
$\striad:=(\struc_{X}\equiv{}^{\K}\smooth_{X},\partial
,\Omg^{1}_{X})$ at the limit of infinite localization of the
system $\invtriad$, but rather we get the {\em generalized smooth}
({\it ie}, what we call here $\ssmooth$-\texttt{smooth}) triad
$\sstriad=(\struc_{X}\equiv{}^{\K}\ssmooth_{X},
\Omg^{p}_{\infty},d^{p}_{\infty})$. Of course, by the general
theory ({\it ie}, ADG), we are guaranteed that the direct, cofinal
system $\invtriad$ of `generalized discrete differential
spaces'---that is, the fcq-triads
$\vec{\triad}_{i}=(\Qalg_{i},\Qring_{i},\vec{d}^{p}_{i})$---yields
a well defined differential structure at the categorical colimit
within $\ctriad$; moreover, according to ADG, it is quite
irrelevant whether the differential triad at the limit is the
classical smooth $\striad$ of the featureless $\smooth$-manifold
proper or one for example that is infested by singularities thus
most pathological and unmanageable when viewed from the classical
$\smooth$-manifold perspective
\cite{malros1,malros2,ros}.\footnote{See footnote 121.} The point
we wish to make here is simply that at the continuum limit we get
{\em a}, not {\em the} familiar $\smooth$-smooth, differential
structure on the continuous topological ($\cont$) spacetime
manifold $X$. This differential structure `for all practical
purposes' represents for us the classical, albeit `generalized',
differential manifold, and the direct limiting procedure that
recovers it a generalized version of Bohr's correspondence
principle advocated in \cite{rapzap1}. That this differential
structure obtained at the `classical limit' is indeed adequate for
accommodating the classical theory will become transparent in the
next section where we will see that based on $\sstriad$ we can
actually write the classical vacuum Einstein equations of general
relativity; albeit, in a generalized, ADG-theoretic way {\em
independently of the usual $\smooth$-manifold}. In fact, we will
see that these equations are obtained at the inverse limit of a
projective system $\inveinst$ of vacuum Einstein equations---one
for each member of $\invtriad$.

\subsubsection{Some comments on real versus complex spacetime and
the general use of the number fields $\mathbf{\R}$ and
$\mathbf{\com}$}

As it has been already anticipated in \cite{malrap1,rap6},
starting from principal finsheaves of {\em complex} $(\K=\com)$
incidence algebras carrying non-flat reticular spin-Lorentzian
$\Qalg_{i}$-connections $\conf_{i}$ as $\cons$-linear finsheaf
morphisms between the `discrete' differential
$\Qalg_{i}$-bimodules $\vec{\Omg}^{p}_{i}~(p\geq1)$ in
$\Qring_{i}$, {\em complex} (bundles of) \texttt{smooth}
coordinate algebras, modules of differential forms over
them\footnote{That is to say, the generalized `classical',
$\ssmooth$-\texttt{smooth} differential triad $\sstriad$ mentioned
above.} and \texttt{smooth} $so(1,3)^{\uparrow}_{\com}$-valued
connection $1$-forms $\aconn$ (over a \texttt{smooth} complex
manifold) are expected to emerge at the inductive/projective limit
of infinite refinement and localization of the qausets and the
principal finsheaves thereof.\footnote{Indeed, in the context of
non-perturbative (canonical) quantum gravity using Ashtekar's new
gravitational connection variables, we will see in the next
section how a holomorphic Lorentzian spacetime manifold and
smooth, complex (self-dual) connections on it are the basic
dynamical elements of the theory.} Thus it may be inferred that in
order to recover the real spacetime continuum of macroscopic
relativistic gravity (general relativity), some sort of {\em
reality conditions} must be imposed after the projective limit,
the technical details of which have not been fully investigated
yet \cite{zap1,zap2}. The nature of these conditions is a highly
non-trivial and subtle issue in current quantum gravity research
\cite{baez}.

On the other hand, starting from incidence algebras over $\R$
$(\K=\R)$, one should be able recover a {\em real}
$\ssmooth$-\texttt{smooth} manifold instead of a complex one at
the projective/inductive classical limit', but then one would not
be faithful to the conventional quantum theory with its continuous
coherent superpositions over $\com$.\footnote{And indeed, in
\cite{rap1,rapzap2,malrap2} the $\com$-linear combinations of
elements of the incidence algebras where physically interpreted as
{\em coherent quantum superpositions} of the causal-topological
arrow connections between the event-vertices in the corresponding
causets. In fact, it is precisely this $\com$-linear structure of
the qausets that qualifies them as sound {\em quantum} algebraic
analogues of causets, which are just associative multiplication
structures (arrow semigroups or monoids or even poset categories).
Also, in connection with footnote 87, we emphasize that it is the
linear structure of qausets (prominently absent from causets) that
gives them both their differential (geometric) and their quantum
character.} On the other hand, {\it prima facie} it appears to be
begging the question to maintain that we have an `innately' or
`intrinsically finitistic' model for the kinematical structure of
Lorentzian quantum spacetime and gravity (and, as we shall contend
in the following section, also for the dynamics) when its
(noncommutative) algebraic representation employs {\it ab initio}
the continuum of complex numbers as the field of (probability)
amplitudes.

For example, in the light of application of ideas from presheaves
and topos theory to quantum gravity, Butterfield and Isham
\cite{buttish}, and more recently Isham \cite{ish3}, have also
explicitly doubted and criticized the {\it a priori} assumption
and use of the continuum of either the reals or, {\it a fortiori},
of the complexes in quantum theory {\it vis-\`a-vis} the quest for
a genuinely quantum theoresis of spacetime structure and gravity.
In \cite{ish3} in particular, Isham maintains that the use of the
arithmetic continua of $\R$ (modelling probabilities and the
values of physical quantities) and $\com$ (probability amplitudes)
in standard quantum mechanics is intimately related (in fact,
ultimately due) to the {\it a priori} assumption of a classical
stance against the `nature' of space and time---{\it ie}, the
assumption of the classical spacetime continuum. In the sequel, in
order to make clear-cut remarks on this in relation to ADG, as
well as to avoid as much as we can `vague dark apostrophes', by
`spacetime continuum' we understand the locally Euclidean arena
({\it ie}, the manifold) that (macroscopic) physics uses up-front
to model spacetime. Our contention then is that {\em Isham
questions the use of $\R$ and $\com$ in quantum theory precisely
because he is motivated by the quest for a genuinely quantum
theoresis of spacetime and gravity}, for in quantum gravity
research it has long been maintained that the classical spacetime
continuum ({\it ie}, the manifold) must be abandoned in the
sub-Planckian regime where quantum gravitational effects are
expected to be significant.\footnote{For instance, see the two
opening quotations.} Thus, his basic feeling is that the
conventional quantum theory, with its continuous superpositions
over $\com$ and probabilities in $\R$, which it basically inherits
from the classical spacetime manifold, must be modified {\it
vis-\`a-vis} quantum gravity. {\it In toto}, {\em if the manifold
has to go in the quantum deep, so must the number fields $\R$ and
$\com$ of the usual quantum mechanics, with a concomitant
relatively drastic modification of the usual quantum formalism to
suit the non-continuum base space(time)}.\footnote{See below.}
Perhaps the use from the beginning of one of the finite number
fields ${\Z}_{p}$\footnote{With `$p$' a prime integer.} for
$c$-numbers would be a more suitable choice for our reticular
models, but then again, what kind of quantum theory can one make
out of them?\footnote{Chris Isham in private communication.} The
contents of this paragraph are captured nicely by the following
excerpt from \cite{ish3}:

\begin{quotation}
\noindent ``{\small\em ...These number systems {\rm [{\it ie},
$\R$ and $\com$]} have a variety of relevant mathematical
properties, but the one of particular interest here is that they
are continua, by which---in the present context---is meant not
only that $\R$ and $\com$ have the appropriate cardinality, but
also that they come equipped with the familiar topology and
differential structure that makes them manifolds of real dimension
one and two respectively. My concern is that the use of these
numbers may be problematic in the context of a quantum gravity
theory whose underlying notion of space and time is different from
that of a smooth manifold. The danger is that by imposing a
continuum structure in the quantum theory a priori, one may be
creating a theoretical system that is fundamentally unsuitable for
the incorporation of spatio-temporal concepts of a non-continuum
nature: this would be the theoretical-physics analogue of what a
philosopher might call a `category error'...}''
\end{quotation}

\noindent while, two years earlier \cite{buttish}, Butterfield and
Isham made even more sweeping remarks about the use of smooth
manifolds in physics in general, and their inappropriateness {\it
vis-\`a-vis} quantum gravity:

\begin{quotation}
\noindent ``{\small\em ...the first point to recognise is of
course that the whole edifice of physics, both classical and
quantum, depends upon applying calculus and its higher
developments (for example, functional analysis and differential
geometry) to the values of physical quantities...why should space
be modelled using $\R$? More specifically, we ask, in the light of
{\rm [our remarks above about the use of the continuum of the real
numbers as the values of physical quantities]:} Can any reason be
given apart from the (admittedly, immense) `instrumental utility'
of doing so, in the physical theories we have so far developed? In
short, our answer is No. In particular, we believe there is no
good {\it a priori} reason why space should be a continuum;
similarly, {\it mutatis mutandis} for time. But then the crucial
question arises of how this possibility of a non-continuum space
should be reflected in our basic theories, in particular in
quantum theory itself, which is one of the central ingredients of
quantum gravity...}''\footnote{Excerpt from ``{\itshape Whence the
Continuum?}'' in \cite{buttish}. These remarks clearly pronounce
our application here of ADG, which totally evades the usual
$\smooth$-calculus, to finitary Lorentzian quantum gravity (see
also remarks below).}
\end{quotation}

At this point it must be emphasized that in ADG, $\R$ and $\com$
enter the theory through the generalized arithmetics---the
structure sheaf $\struc_{X}$, which, as noted earlier, is supposed
to be a sheaf of commutative $\K=\R ,\com$-algebras ({\it ie},
$\mathbf{K}=\mathbf{R},\cons\hookrightarrow\struc$). In turn,
these arithmetics are invoked only when one wishes to represent
local measurements and do with them general calculations with the
vector sheaves $\modl$ employed by ADG.\footnote{See sections 2
and 3, and in particular the discussion in 4.3 next.} It is at
this point that the basic assumption of ADG that the $\modl$s
involved are locally free $\struc$-modules of finite rank
$n$--that is to say, locally isomorphic to $\struc^{n}$---comes in
handy, for all our local measurements and calculations involve
$\struc$, $\struc^{n}$ and, {\it in extenso}, the latter's natural
local transformation matrix group $\aut\modl(U)=\modl
nd\modl(U)^{\bullet}\equiv M_{n}(\struc(U))^{\bull}$. Thus, {\em
real and complex numbers enter our theory through `the back-door
of measurement and calculation'}, {\it in toto}, through
`geometry' as understood by ADG.\footnote{This is in accord with
our view of $\struc$ mentioned earlier as the structure carrying
information about the `geometry', about our own measurements of
`it all' (see footnotes 17, 41, the end of 2.3 and subsection 4.3
next). In agreement with Isham's remarks in \cite{ish3} briefly
mentioned above, {\em it is we, with our classical manifold
conception of space and time, who bring $\R$ and $\com$ into our
models of the quantum realm}. The quantum deep itself has no
`numbers' as such, and it is only our observations,
measurements---in effect, `geometrizations---of `it all' that
employs such $c$-numbers (Bohr's correspondence principle). {\em
Nature has no number or metric; we dress Her in such, admittedly
ingenious, artifacts} (see footnote 17 and also the following
one). Based on ADG and its finitary application to Lorentzian
quantum gravity here, shortly we will go a step further than Isham
and altogether question the very notion of `spacetime' in the
quantum realm. Thus, under the prism of ADG, the question whether
spacetime is `classical' or `quantum' should be put aside and the
doubts of using $\R$ and $\com$ in quantum theory should not be
dependent in any way on the answer to that question.}

On the other hand, and in connection with the last footnote, since
the constructions of ADG are genuinely independent of (the usual
calculus on) $\smooth$-manifolds,\footnote{In fact, of any
`background spacetime structure', whether `continuous' or
`discrete'.}, whether real (analytic) or complex (holomorphic),
\cite{mall1,mall2,malros1,malros2,malrap1,mall3,malrap2,mall7},
Isham's remarks that the appearance of the arithmentic continua in
quantum theory are due to the {\it a priori} assumption of a
classical spacetime continuum---a locally Euclidean manifold---do
not affect ADG. Of course, we would actually like to have at our
disposal the usual number fields in order to be able to carry out
numerical calculations (and arithmetize our abstract algebraic
sheaf theory) especially in the (quantum) {\em physical}
applications of ADG that we have in mind.\footnote{For recall
Feynman: ``{\em The whole purpose of physics is to find a number,
with decimal points {\it etc}. Otherwise, you haven't done
anything.}'' \cite{feyn1}---and arguably, numbers are obtained by
measurements, observations and the general
`instrumental/operational-geometrical activity' that physicists
exercise (in their local laboratories, `with clocks and rulers' so
to speak) on Nature. {\em Numbers are not Nature's own}. Thus,
both the arithmetics, as encoded in the abelian algebra structure
sheaf $\struc$, and the $\struc$-metric $\rho$ relative to it, lie
on the observer's ({\it ie}, the classical) side of the quantum
divide and are not `properties' of quantum systems---they are our
own `devices' (see footnote 17). This brings to mind Aeschylus'
remark in `Prometheus Bound': ``{\em Number, the most ingenious of
human inventions}'' \cite{aeschylus} (notwithstanding of course
the innumerable modern debates among the philosophers of
mathematics whether `number' is a creation of the human mind or
whether it exists, in a non-physical Platonic world of Ideas, `out
there').} We may distill all this to the following:

\begin{quotation}
\noindent In the general ADG theory, and in its particular
finitary application to quantum gravity here, the commutative
number fields, which happen to be locally Euclidean continua ({\it
ie}, the manifolds $\R\simeq\R^{1}$ and $\com\simeq\R^{2}$ being
equipped with the usual differential geometric---{\it ie},
$\smooth$-smooth---structure), do not appear in the theory from
assuming up-front a background spacetime manifold.\footnote{For,
as we have time and again emphasized in this paper, ADG evades
precisely this: doing the usual differential geometry (calculus)
on a classical $\smooth$-smooth background manifold
\cite{mall1,mall2,malrap1,malrap2}.} Rather, they are only built
into our generalized arithmetic algebra sheaf $\struc_{X}$, thus
they are of sole use in our local calculations and `physical
geometrization' ({\it ie}, `analysis of measurement operations')
of the abstract algebraic theory. As such, they are not actually
liable to Isham's criticism and doubts,\footnote{That is, again,
that the use of the fields of real (probabilities) and complex
(probability amplitudes) numbers in quantum theory is basically
due to the {\it a priori} assumption of a classical spacetime
manifold.} for ADG totally evades the base geometric spacetime
manifold.

For instance, from our ADG-theoretic perspective, this
independence of measurement from an `ambient' spacetime continuum
and its focus solely on the (physical) objects (fields) {\it per
se} that live on that background `space(time)'---and perhaps more
importantly, regardless of whether the latter is a `discrete' or a
continuous manifold base arena---may be seen as a
`post-anticipation' of Riemann's words in \cite{riemann} which we
quote {\it verbatim} from \cite{mall7}:

\begin{quotation}
\noindent ``{\small\sl Ma\ss~bestimungen erfordern eine
Unabh\"{a}ngichkeit der Gr\"{o}\ss en vom Ort, die in mehr als
einer Weise stattfinden kann.}'' : ``{\small\em Specifications
{\rm [:\;measurements]} of mass require an independence of
quantity from position, which can happen in more than one way.}''
\end{quotation}

\noindent Moreover, as we shall see subsequently, and in contrast
to Isham, we do not aim for a non-continuum theoresis of spacetime
(and gravity) in order to abolish the {\it a priori} use of $\R$
and $\com$ in the usual quantum theory\footnote{At least, as long
as we abide to the operational idea that our quantal operations,
which classically involve (ideal) clocks and measuring rods
\cite{einst3,sklar,grunbaum} which, in turn, are admittedly
modelled after $\R$ \cite{ish3}, are organized into
(noncommutative) algebras ({\it ie}, in line with Heisenberg's
conception of an algebraically implemented `quantum
operationality' \cite{malrap1}) as well as that upon measurement
they yield commutative numbers in the base field (Bohr).} with a
concomitant modification of the latter to suit the non-continuum
base spacetime, for there is no (background) `spacetime' (whether
`discrete' or `continuous') as such in the quantum deep and in ADG
the (structural) role played the base (topological) space is a
(physically) atrophic, inactive, dynamically non-participatory
one.
\end{quotation}

The last remark also prompts us to highlight from \cite{ish3}
another remark of Isham that is quite relevant to our present
work:\footnote{The excerpt below is taken from section 2.2 in
\cite{ish3} titled `{\itshape Space-time dependent quantum
theory}'.}

\begin{quotation}
\noindent ``{\em\small The main conclusion I wish to draw from the
discussion above is that a number of {\it a priori} assumptions
about the nature of space and time are present in the mathematical
formalism of standard quantum theory, and it may therefore be
necessary to seek a major restructuring of this formalism in
situations {\rm [like for example those motivated by quantum
gravity ideas\footnote{Our addition to tie the text with what
Isham was discussing prior to it.}]} where the underlying
spatio-temporal concepts (if there are any at all) are different
from the standard ones which are represented mathematically with
the aid of differential geometry.\footnote{And, of course, Isham
refers to the usual differential geometry of $\smooth$-manifolds.}

A good example would be to consider from scratch how to construct
a quantum theory when space-time is a finite causal set: either a
single such---which then forms a fixed, but non-standard,
spatio-temporal background---or else a collection of such sets in
the context of a type of quantum gravity theory. In the case of a
fixed background, this new quantum formalism should be adapted to
the precise structure of the background, and can be expected to
involve a substantial departure from the standard formalism: in
particular to the use of real numbers as the values of physical
quantities and probabilities.}''

\end{quotation}

In the next section we will see exactly how, with the help of ADG,
we can write the vacuum Einstein equations for Lorentzian gravity
over a causet and, in contradistinction to Isham's remarks above,
without having to radically modify quantum theory---in particular,
in its use of $\R$ and $\com$---in order to suit that discrete,
non-continuum background spacetime. As a matter of fact, we will
see that this base causet plays no physically significant role
apart from serving as a (fin)sheaf-theoretic localization
scaffolding in our theory; moreover, no quantum theory proper
(either the standard one, or a modified one intuited by Isham
above) will be employed to quantize the classical theory ({\it
ie}, Einstein's equations on the smooth manifold). All in all, as
we will witness in the sequel, in a strong sense our ADG-based
finitary vacuum Einstein gravity may be perceived as being
`inherently' or `already quantum', `fully covariant'---{\it ie},
as involving only the dynamical fields and not being dependent in
any way on an external, base spacetime, be it granular or a smooth
continuum, and certainly as not being the outcome of applying
quantum theory ({\it ie}, `formally quantizing') the classical
theory of gravity on a spacetime manifold ({\it ie}, general
relativity).

\subsubsection{A brief `critique' of the Ashtekar-Lewandowski
projective limit scheme}

In \cite{ashlew2}, a projective system $\ashlew$ of compact
Hausdorff manifolds labelled by graphs---which can be physically
interpreted as `floating lattices'---was employed in order to
endow, at the projective limit of that family of manifolds, the
moduli space $\infconn/\grouv$ of $\smooth$-smooth gauge
($\grouv$)-equivalent Y-M or (self-dual) gravitational connections
with a differential geometric structure including vector fields,
differential forms, exterior derivatives, metric volume forms,
Laplace operators and their measures, as well as the rest of the
familiar $\smooth$-smooth differential geometric entities. As we
shall see in the next section, there has been an ever growing need
in current approaches to non-perturbative canonical (Hamiltonian,
loop variables-based) or covariant (Lagrangian, action-based)
quantum gravity, to acquire a firm tangent bundle perspective on
$\infconn/\grouv$ ({\it ie}, have a mathematically well defined
$T(\infconn/\grouv)$ object), since $T(\infconn/\grouv)$ can serve
as the physical phase space of quantum Y-M theories and gravity in
its gauge-theoretic form in terms of Ashtekar's (self-dual)
connection variables \cite{ash} and one would like to do
differential geometry on that space. Thus, the basic idea is that
if such a mathematically rigorous differential geometric status is
first established on the moduli space, one could then hope to
tackle deep quantum gravity problems such as the Hilbert space
inner product (and measure) problem, the problem of time, the
non-trivial character of $\infconn/\grouv$ when regarded as a
$\grouv$-bundle, the problem of physical Wilson loop observables
{\it etc}\footnote{See 5.3 for a more analytical exposition and
discussion of some of these problems.} by the conventional
calculus-based ({\it ie}, the usual $\smooth$-differential
geometric) methods of the canonical or the covariant approaches to
quantum field theory.

Although, admittedly, algebraic methods were used in
\cite{ashlew2} towards endowing the moduli space of connections
with the conventional differential geometric apparatus, the very
nature ({\it ie}, the $\smooth$-smooth character) of each member
of $\ashlew$ shows the original intention of Ashtekar and
Lewandowski: in order to induce the usual $\smooth$-differential
geometric structure on $\infconn/\grouv$ at the projective limit,
one must secure that each member of the inverse system $\ashlew$
comes equipped with such a structure---that is to say, it is a
differential manifold itself. In other words, as it was already
mentioned in the beginning of this section, the essence of
\cite{ashlew2} is that {\em like differential structure yields}
({\it ie}, induces at the inverse limit) {\em like differential
structure}. Now, in view of the fact that some (if not all!) of
the aforementioned problems of $T(\infconn/\grouv)$ come precisely
from the $\smooth$-smoothness of the spacetime manifold and,
concomitantly, from the group $\mathrm{Diff}(M)$ of its `structure
symmetries'\footnote{See next section.}, it appears to us that
this endeavor is to some extent `begging the
question'.\footnote{The quest(ion) being for (about) a quantum
gravitational scheme that is finitistic, but more importantly,
{\em genuinely background $\smooth$-smooth spacetime
manifold-free} (see following section).} Of course, it is quite
understandable with `general relativity or $\smooth$-smooth
spacetime manifold-conservative' approaches to quantum gravity,
such as the canonical or the covariant (path-integral)
ones,\footnote{See category 1 in the prologue to this paper.} to
maintain that the differential geometric mechanism is intimately
tied to (or comes from) the differential manifold, for, after all,
{\em manifolds were created for the tangent bundle}.\footnote{In
the next section we will return to comment further on this in
connection with (\ref{add7}).}

However, this is precisely the point of ADG: the intrinsic,
`inherent' mechanism of differential geometry has nothing to do
with $\smooth$-smoothness, nothing to do with $\smooth$-smooth
manifolds, and the latter (in fact, its structure sheaf
$\smooth_{M}$) provide us with just a (the classical, and by no
means {\em the} `preferred', one) `mechanism for
differentiating'.\footnote{See the concluding section about `the
relativity of differentiability'.} For instance, as we saw in
sections 2 and 3, one can develop a full-fledged differential
geometry, entirely by algebraic ({\it ie}, sheaf-theoretic) means
and completely independently of $\smooth$-smoothness, on the
affine space of connections as well as on the moduli space of
gauge-equivalent connections.\footnote{For the full development of
differential geometry {\it \`a la} ADG on gauge-theoretic moduli
spaces, the reader is referred to \cite{mall4}.} In the finitary
case of interest here, and in striking contradistinction to
\cite{ashlew2}, we have seen above (and in the past
\cite{malrap1,malrap2}) how each principal finsheaf $\peel_{i}$ of
qausets in the projective system $\finv$ carries virtually all the
differential geometric panoply without being dependent at all on
the classical $\smooth$-manifold. In fact, in the next section we
will see how such a $\smooth$-smooth spacetime manifold-free
scenario will not prevent us at all from writing a locally finite
version of the usual Einstein equations for vacuum Lorentzian
gravity. Quite on the contrary, it will enable us to evade
altogether $\mathrm{Diff}(M)$ as well as some of the aforesaid
problems that the latter group creates in our search for a cogent
non-perturbative quantum gravity, whether canonical or covariant,
on the moduli space of gravitational connections. Moreover, we
will see how we can recover the $\ssmooth$-\texttt{smooth} vacuum
Einstein equations at the projective limit of an inverse system
$\inveinst$ of fcqv-ones. Already at a kinematical level, at the
end of the next subsection we will argue ADG-theoretically how the
`generalized classical' $\ssmooth$-\texttt{smooth} moduli space of
gauge-equivalent (self-dual) spin-Lorentzian connections can be
obtained at the inverse limit of an inverse system $\invmod$ of
fcqv-moduli spaces.

But before we do this, let us recapitulate and dwell a bit longer
on some central kinematical ideas that were mentioned {\it en
passant} above.

\subsection{Remarks on the `Operational' Conception of Finitary Quantum Causality: a Summary of
Key Kinematical Notions for Finitary, Causal and Quantal Vacuum
Einstein-Lorentzian Gravity}

Our main aim in this subsection is to highlight the principal new
kinematical notions, of a strong operational-algebraic flavour,
about `finitary causality' originally introduced in
\cite{malrap1}. In this way, we are going to emphasize even more
the characteristic contrast between our `{\em operational and
quantal}'---in fact, `{\em observer-dependent}'---conception of
locally finite causality via qausets, and Sorkin {\it et al.}'s
more `{\em realistic}' causet theory proper. As a main source for
drawing this comparison of our approach against causet theory we
are going to use \cite{sork1}. Also, by this review we hope to
make clearer to the reader the intimate connection between central
ADG-theoretic notions such as `open gauge', `structure sheaf of
generalized arithmetics/coordinates or measurements' {\it etc},
and some primitive notions of the finitary approach to spacetime
(topology) as initially presented by Sorkin in \cite{sork0}.

With \cite{malrap1} as our main reference and compass to orientate
us in this short review, we provide below a list of primitive
assumptions, already explicitly or implicitly made in
\cite{sork0}, that figure prominently in all our ADG-based trilogy
({\it ie}, in \cite{malrap1,malrap2} and here) on finitary
spacetime and Lorentzian quantum gravity:

\begin{enumerate}

\item The basic intuitive and heuristic
assumption is the following identification we made in
\cite{malrap1}:

\begin{equation}\label{ad1}
\mathrm{``(coarse)~localization"}\equiv
\mathrm{``(coarse)~measurement/observation"}
\end{equation}

\noindent For the moment, assuming with Sorkin that topology is a
`predicate' or property of the (quantum) physical system
`spacetime', in the sense that ``{\em the points of the manifold
are the carriers of its topology}'' \cite{sork0}, {\em we model
our coarse measurements of (the topological relations between)
spacetime point-events by `regions' or `open sets' about them}.
Conversely, the open sets of a covering separate or distinguish
the points of $X$. We thus have, for a bounded region $X$ of a
classical $\cont$-spacetime manifold $M$,\footnote{As explained in
\cite{rapzap1}, the assumption of a bounded spacetime region $X$
rests on the fact that actual or `realistic' experiments are
carried out in laboratories of finite size and are of finite
duration.} and a locally finite open cover $\gauge_{i}$ of
it,\footnote{Again, as explained in \cite{rapzap1}, the assumption
of a locally finite open covering $\gauge_{i}$ rests on the
experimental fact that we always record, coarsely, a finite number
of events.}

\begin{equation}\label{ad2}
\mathrm{``(coarse)~determination~of}~x\in X"\equiv
\mathrm{``open~set}~U\in\gauge_{i}~\mathrm{about~}x"
\end{equation}

\item Operationally speaking, it is widely recognized that {\em localization involves `microscopic energy', and
measurement a gauge}. We thus identify again (nomenclature-wise)

\begin{equation}\label{ad3}
\mathrm{``open~set}~U\in\gauge_{i}~\mathrm{about}~x"\equiv
\mathrm{``open~gauge}~U~\mathrm{of}~x"
\end{equation}

\noindent and note that this---{\it ie}, `{\em open gauge}'--- is
precisely the name ADG gives to the sets of the open coverings of
the base topological or localization space $X$ involved in a
differential triad \cite{mall1,mall2}.

\item Of course, the better ({\it ie}, more accurate or sharp) the localization,
the higher the microscopic energy of resolution (of $X$ into its
point-events). Thus, we suppose that the locally finite open
coverings of $X$ form an inverse system or net ({\it ie}, a
partially ordered set itself) with respect to the relation
`$\succeq$' of {\em fine graining}. Roughly, better (more accurate
or sharper) localization of $x$ involves smaller and more numerous
open sets about it, thus higher microscopic energy of resolution.

\item With these operational assumptions, Sorkin's two main results in \cite{sork0} can be interpreted
then as follows:

--- i) Sorkin's `algorithm'---{\it ie}, the extraction of a
$T_{0}$-topological poset $P_{i}$ from $X$ relative to a locally
finite open cover $\gauge_{i}$---involves separating and grouping
together into equivalence classes (of `observational
indistinguishability') the point-events of $X$ relative to the
open gauges $U$ in $\gauge_{i}$.\footnote{See
\cite{sork0,rap1,rap2,malrap1,malrap2} for more details about
Sorkin's algorithm.} {\em Point-events in the same equivalence
class (which is a vertex) in $P_{i}$ are interpreted as being
indistinguishable relative to our coarse measurements or
`observations' in $\gauge_{i}$}, and

--- ii) Sorkin's inverse limit of the projective system of
topological posets $\inv$ can now be interpreted as the recovering
of the locally Euclidean $\cont$-topology of $X$ at the finest
resolution or `ultra localization' of $X$ into its point-events.
{\em In this sense, the continuous manifold topology is,
operationally speaking, an ideal or `non-pragmatic'}
\cite{rapzap1} {\em situation involving infinite (microscopic)
energy of localization or measurement}.

\item Then came Sorkin's radical re-interpretation of the locally
finite partial orders involved from topological to causal
\cite{sork1}, which essentially planted the seed for causet
theory. We recall from \cite{sork1} a telling account of this
conceptual sea-change:

\begin{quotation}

\noindent ``{\small\em ...Still, the order inhering in the finite
topological space seemed to be very different from the so-called
causal order defining past and future. It had only a topological
meaning but not (directly anyway) a causal one. In fact the big
problem with the finite topological space was that it seemed to
lack the information which would allow it to give rise to the
continuum in all its aspects, not just in the topological aspect,
but with its metrical (and therefore its causal) properties as
well...The way out of the impasse involved a conceptual jump in
which the formal mathematical structure remained constant, but its
physical interpretation changed from a topological to a causal
one...The essential realization then was that, although order
interpreted as topology seemed to lack the metric information
needed to describe gravity, the very same order reinterpreted as a
causal relationship, did possess information in a quite
straightforward sense...In fact it took me several years to give
up the idea of order-as-topology and adopt the causal set
alternative as the one I had been searching for...}

\end{quotation}

\item Now, the basic idea in \cite{rap1}, but most explicitly
in \cite{malrap1} under the light of ADG, is that, in spite of
Sorkin's semantic switch above, and in order to retain our picture
of finitary posets as graded discrete differential manifolds (or
homological objects/simplicial complexes),\footnote{So that we
could apply the differential geometric ideas of ADG, in a
(fin)sheaf-theoretic context \cite{rap2}, at the reticular level
of causets \cite{malrap2}. Indeed, the fundamental reason that we
insist that the locally finite posets we are using are {\em
simplicial complexes} is that the construction of the incidence
algebras from such posets is manifestly {\em functorial}
\cite{rapzap1,rapzap2,zap1}, which in turn secures that the
(fin)sheaves over them {\em exist}. Had we, like Sorkin {\it et
al.} insisted on {\em arbitrary} (locally finite) posets (see
below), the correspondence `finitary posets'$\mapto$`incidence
algebras' would not be functorial, and the (fin)sheaves that we
would be talking about would not actually exist. Furthermore, the
bonus from working with (locally finite) posets that are {\it a
fortiori} simplicial complexes is that the (incidence algebras of
the) latter, again as shown in \cite{rapzap1,rapzap2,zap1}, have a
rich (discrete) graded differential structure, which has opened
the possibility of applying ADG-theoretic ideas to the
(fin)sheaves thereof.} we felt we had to give a more
operational-algebraic (thus more easily interpretable quantum
mechanically \cite{rap1}) definition of finitary causality than
causets. We read from \cite{malrap1} what this operational,
observation $\gauge_{i}$-dependent conception of (quantum)
causality involved:

\begin{quotation}
\noindent ``{\small...All in all, (quantum) causality is
operationally defined and interpreted as a `{\em power
relationship}' between spacetime events relative to our coarse
observations (or approximate operations of local determination or
`measurement') of them, namely, if events $x$ and $y$ are coarsely
determined by ${\mathcal{N}}(x)$ and ${\mathcal{N}}(y)$ with
respect to $\gauge_{i}$, and ${\mathcal{N}}(y)\subset
{\mathcal{N}}(x)$,\footnote{Where ${\mathcal{N}}(x)$ is
effectively the \v{C}ech-Alexandrov `nerve-cell' \cite{cech,alex}
of $x$ relative to $\gauge_{i}$, namely, the smallest open set
$\bigcap\{ U\in\gauge_{i}:~x\in U\}$ in the subtopology of $X$
(generated by countable unions of finite intersections of the open
gauges $U$ in $\gauge_{i}$) which includes $x$ (see also
\cite{malrap2}). By such cells one builds up (abstract) simplicial
complexes (nerves) which, as noted before, are isomorphic to
Sorkin's finitary $T_{0}$-topological posets in \cite{sork0}
essentially under two additional conditions on $\gauge_{i}$: that
it is {\em generic} ({\it ie}, all non-trivial intersections of
its open sets are different) and {\em minimal} ({\it ie}, if any
of its open sets is omitted, it ceases being a covering of $X$)
\cite{rapzap2,porter}. (This footnote is not included in
\cite{malrap1}).} then we say that `{\em $x$ causes $y$}'. The
attractive feature of such a definition and interpretation of
causality is that, by making it relative to $\gauge_{i}$, we
render it `{\em frame-}' or `{\em gauge-}' or even `{\em
observation-dependent}'...}''\footnote{Such a cellular
(simplicial), but more importantly to our physical interpretation
here, `coarse observation-dependent' (`perturbing
operations-sensitive'), decomposition of spacetime, apart from
Regge's celebrated paper \cite{regge}, has been worked out by Cole
\cite{cole} and very recently by \cite{porter}. (This footnote is
also not included in \cite{malrap1}).}
\end{quotation}

\noindent Of course, the open sets in $\gauge_{i}$ now stand for
`{\em coarse causal regions}' or rough operations of `observation'
or ` measurement' of the causal relations between events in the
curved space{\em time} manifold \cite{malrap1}, not just coarse
approximations of the topological relations proper between events.
Thus, in view of Sorkin's semantic switch quoted above from
\cite{sork1}, as well as his assumption in \cite{sork0} that the
points of $X$ are the carriers of its topology, we assume a more
operational and at the same time less `realistic' stance than
Sorkin \cite{sork1} by maintaining that {\em the point-events of
$X$ are the carriers of causality in relation to our coarse and
perturbing observations (open gauges) $U$ in $\gauge_{i}$}
\cite{malrap1}.

\item Having secured that our structures now enjoy both a causal
and an operational interpretation, it became evident to us that
our scheme differs fundamentally from Sorkin {\it et al.}'s causet
scenario at least in the following two ways:

--- i) unlike the case in causet theory, which posits up-front a
`locally finite poset democracy', in our theoretical scheme not
all locally finite posets and their incidence Rota algebras may
qualify as being `operationally sound qausets'. Only posets coming
from coarse causal gauges $\qgauge_{i}$\footnote{Again, the
right-pointing arrow over the covering $\gauge_{i}$ indicates the
causal semantics `{\em coarse causal regions}' given to the open
sets $U$ in it above.} and their incidence algebras are admissible
as qausets proper. As mentioned above, this secures that the
locally finite posets extracted by Sorkin's algorithm from the
$\qgauge_{i}$s (which are now causally interpreted) can be viewed
as (causal) simplicial complexes\footnote{It must be noted here
that it was Finkelstein who first insisted, in a reticular and
algebraic setting not very different from ours called `the causal
net', for {\em a causal version of (algebraic) topology and its
associated (co)homology theory} \cite{df2}.} and, {\em in
extenso}, the incidence algebras (qausets) associated with them
can be viewed as graded discrete differential algebras (manifolds)
\cite{rapzap1,rapzap2,zap3,zap1} thus allowing the entire
ADG-theoretic panoply to be applied on (the finsheaves of) them
\cite{malrap1,malrap2}, and

--- ii) as noted before, our operational scheme is in glaring contrast to Sorkin {\it et al.}'s more `realistic' conception
of dynamical (local) causality (gravity). For example, we recall
from \cite{sork1} that for Sorkin, in contradistinction to the
rather standard operationalist or `instrumentalist' interpretation
of general relativity according to which the gravitational
potentials, as represented by the ten components of the metric
tensor $g_{\mu\nu}$, provide ``{\em a summary of the behaviour of
idealized clocks and measuring rods}''
\cite{einst3,grunbaum,sklar,torreti}, the gravitational
field---the dynamical field of `locality' or `local causality'
\cite{malrap1,rapzap2}---``{\em is an independent substance, whose
interaction with our instruments gives rise to clock-readings,
etc}''. This alone justifies the realist or `Platonic'
(ontological) causet hypothesis according to which ``{\em
spacetime, at small scales, \underline{is} a locally finite
poset}'' \cite{bomb87}---a realm quite detached from and
independent of (the operationalist or `pragmatist' \cite{df1}
philosophy according to which {\em all that there is and matters
is}) ``{\em what we actually do to produce spacetime by our
measurements}'' \cite{sork1}---whose partial order is the discrete
analogue of the relation that distinguishes past and future events
in the (undoubtedly realistic or `Platonic') macroscopic,
geometrical spacetime continuum of general relativity.

\item We now come to the ADG-theoretic assumption of `arithmetizing' or `coordinatizing' our
coarse localizations or measurements. This is represented by
assuming that the base topological space $X$, which we have
charted by covering it by the open gauges $U$ in $\gauge_{i}$ (or
equivalently, in $\qgauge_{i}$), is $\mathbf{K}$-algebraized in
the sense that we localize sheaf-theoretically over it abelian
$\K=\R ,\com$-algebras which comprise the structure sheaf
$\struc_{X}$. The latter is supposed to be the commutative algebra
sheaf of `generalized arithmetics' in our theory---the realm in
which our coarse local measurements, represented by the local
sections of $\struc$ (in
$\Gamma(U,\struc)\equiv\struc(U),~U\in\gauge_{i}$), take
values---the readings on our abstract gauges so to speak. That we
choose the stalks of $\struc$ to be inhabited by {\em abelian}
algebras is in accord with Bohr's quantum theoretic imperative
according to which our measurements always yield commutative,
$c$-numbers.\footnote{See also footnote 41 and
\cite{malrap1,malrap2}.} Furthermore, as it was also emphasized in
the previous subsection, since the constant sheaf
$\mathbf{K}=\mathbf{R},\mathbf{C}$ of the reals or the complexes
is canonically injected into $\struc$, we realize again that {\em
the usual numerical continua $\R$ and $\com$ enter into our theory
via the process of abstract coordinatization and local
measurement, and not by assuming that the base topological
space(time) $X$ is a classical, locally Euclidean continuum ({\it
ie}, a manifold)}. Finally, we must also emphasize here, as it was
noted throughout the previous sections, that {\em in ADG all our
(local) calculations reduce to expressions involving (local
sections of) $\struc$---in particular, all our vector sheaves
$\modl$ of rank $n$ are (locally) of the form
$\struc^{n}$\footnote{And rather fittingly, the {\em local
(coordinate) gauge} $e^{U}\equiv\{ U;~(e_{i})_{0\leq i\leq n-1}\}$
($U\in\gauge_{i}$) of the vector sheaf $\modl$ of rank $n$ in
footnote 19, which consists of local sections of $\modl$ (in
$\modl(U)\equiv(\struc(U))^{n}\equiv\struc^{n}(U)$), can be
equivalently called {\em a local frame of $\modl$} \cite{mall1}.}
and, as a result, their (local) structure symmetries comprise the
matrix group $(\modl nd\modl(U))^{\bull}\equiv
M_{n}(\struc(U))^{\bull}$}.

\item Finally, anticipating our comments on an abstract, essentially
categorical, version of gauge invariance and covariance of the
gravitational dynamics of qausets in terms of finsheaf morphisms
to be given subsequently, we note here that, although our
kinematical, operational-algebraic conception of finitary quantum
causality above is apparently observation or gauge
$\gauge_{i}$-dependent \cite{malrap1}, the dynamics, which is
expressed in terms of the principal (fin)sheaf morphism---the
finitary gravitational spin-Lorentzian connection $\conf_{i}$ and
its scalar curvature $\fricci(\conf_{i})$, will be seen to be
manifestly $\gauge_{i}$-independent. Thus, while quantum causality
is kinematically expressed as a power relationship between events
relative to our own coarse observations (gauges) of them in
$\gauge_{i}$, its dynamical law of motion is characteristically
independent of the latter \cite{malrap1}. We will comment further
on this apparent paradox in 5.1.1.

\end{enumerate}

\subsubsection{Projective limits of fcqv-moduli spaces}

In closing the present section, we would like to make some final
kinematical remarks. These concern inverse limits of moduli spaces
$\qmod_{i}^{(+)}(\qmodll_{i})$ of (self-dual) fcqv-spin-Lorentzian
connections (dynamos) $\conf^{(+)}_{i}$ on the Lorentzian
finsheaves $\qmodll_{i}:=(\qmodl_{i},\qrho_{i})$. These spaces are
defined as follows

\begin{equation}\label{eqad1}
\qmod_{i}^{(+)}(\qmodll_{i}):=\fconn_{i}^{(+)}(\qmodll_{i})/\qaut_{i}\qmodll_{i}
\end{equation}

\noindent and they are the fcq-analogues of the ADG-theoretic
moduli spaces defined in (\ref{eq92}) in general, as well as in
(\ref{eq103}) and (\ref{eq104}) in the particular case of
self-dual connections.\footnote{In (\ref{eqad1}),
$\fconn_{i}^{(+)}(\qmodll_{i})$ is the fcq- (and self-dual)
version of the abstract affine space $\sconn_{\struc}(\modl)$ of
$\struc$-connections $\conn$ on a vector sheaf $\modl$ in
(\ref{eq54}).} $\qmod_{i}^{(+)}(\qmodll_{i})$, as we shall see in
the next section, plays the role of the quantum configuration
space for our theory which regards the (self-dual) fcqv-dynamos
$\conf^{(+)}_{i}$ as (the sole) fundamental (quantum) dynamical
variables.

Now, one such moduli space corresponds to ({\it ie}, is based on)
each and every member of the direct system
$\invtriad=\{\triad_{i}\}$ of fcq-differential triads and, {\it in
extenso}, to each member of the inverse system
$\finv=\{(\peel_{i},\conf^{(+)}_{i})\}$ of principal Lorentzian
finsheaves of qausets and their reticular (self-dual)
spin-Lorentzian connections.\footnote{In fact, as we shall present
in 5.5.2, a tower of numerous important inverse/direct systems of
structures can be based on $\invtriad$. This just shows the
importance of the notion of differential triad in ADG and its
finitary application here.} Thus, we can similarly define the
projective system $\invmod:=\{ \qmod_{i}^{(+)}(\qmodll_{i})\}$ of
(self-dual) fcqv-moduli spaces like the one in (\ref{eqad1}) and,
according to the general ADG theory \cite{pap1,pap2}, take its
categorical limit, which yields

\begin{equation}\label{eqad2}
{\mathcal{M}}_{\infty}^{(+)}(\modll_{\infty})=\underset{\infty\leftarrow
i}{\lim}\invmod\equiv\underset{\infty\leftarrow
i}{\lim}\{\qmod_{i}^{(+)}(\qmodll_{i})\}
\end{equation}

\noindent the $\ssmooth$-\texttt{smooth} moduli space of
$\ssmooth(X)$-automorphism equivalent \texttt{smooth} (self-dual)
spin-Lorentzian connections ${}^{(\K)}\con_{\infty}^{(+)}$ on the
Lorentzian vector bundle/sheaf $\modll_{\infty}$ associated to the
principal orthochronous Lorentzian bundle/sheaf
${}^{(\K)}\princ^{\uparrow}\equiv\peel_{\infty}$ over the region
$X$ of the $\ssmooth$-\texttt{smooth} $\K$-manifold $M$. As noted
before, ${\mathcal{M}}_{\infty}^{(+)}(\modll_{\infty})$
corresponds to a generalized version ({\it ie}, a
$\ssmooth$-\texttt{smooth} one) of the classical moduli space
$\infconn^{(+)}$ of gauge-equivalent (self-dual) $\smooth$-smooth
spin-Lorentzian connections on the region $X$ of the usual
differential ({\it ie}, $\smooth$-smooth) spacetime $\K$-manifold
$M$.

\section{Locally Finite, Causal and Quantal Vacuum Einstein Equations}

This is the neuralgic section of the present paper. Surprisingly,
it is also the simplest one as it is essentially a straightforward
transcription of the ADG constructions and results of sections 2
and 3 to the locally finite case of curved finsheaves of qausets
$\qmodll_{i}$ and their reticular spin-Lorentzian connections
$\conf_{i}$. So, without further ado, we are going to present a
locally finite, causal and quantal version of the vacuum Einstein
equations (\ref{eq53}) for Lorentzian gravity emphasizing in
particular their physical interpretation. We also derive these
equations from an action principle.

\subsection{Finitary, Causal and Quantal Vacuum
Einstein-Lorentzian Gravity}

First we note that the $\Qalg_{i}$-connection $\conf_{i}$ on
$\qmodll_{i}$ is assumed to be compatible with the finsheaf
morphism $\qrho_{i}$ in (\ref{eq114}), as follows

\begin{equation}\label{eq115}
\conf^{i}_{\Hom_{\Qalg_{i}}(\qmodll_{i},\qmoddll_{i})}(\qrho_{i})=0
\end{equation}

\noindent which is the finitary analogue of (\ref{eq17}) implying
that the connection $\conf_{i}$ is torsionless.\footnote{Note that
in (\ref{eq115}), to avoid subscript congestion on $\conf$, we
have raised the refinement or finitarity index `$i$' to a
superscript.} $\conf_{i}$ is a reticular Lorentzian {\em metric}
connection.

Then, analogously to the abstract expressions (\ref{eq36}) and
(\ref{eq37}), and for the corresponding first prolongation
$\conf_{i}^{1}$ of $\conf_{i}(\equiv\conf_{i}^{0})$ as in
(\ref{eq33}) ({\it ie},
$\conf_{i}^{1}:~\Qaus_{i}^{1}\mapto\Qaus_{i}^{2}$), we define the
non-zero curvature $\fcurv_{i}$ of the reticular connection
$\conf_{i}$ on $\qmodll_{i}$ as the following $\modl
nd\qmodll_{i}$-valued reticular $2$-form

\begin{equation}\label{eq116}
\begin{array}{c}
\fcurv_{i}(\conf_{i}):=\conf^{1}_{i}\circ\conf_{i}\not=0\cr
\fcurv_{i}\in\mathrm{Hom}_{\Qalg_{i}}(\qmodll_{i},\Qaus^{2})=\Hom_{\Qalg_{i}}(\qmodll
,\Qaus_{i}^{2})(\caus_{i})=\Qaus^{2}(\modl
nd\qmodll_{i})(\caus_{i})
\end{array}
\end{equation}

\noindent emphasizing also that it is an $\Qalg_{i}$-morphism.
Thus, we can also define the associated Ricci tensor
$\fric_{i}\in\modl nd\qmodll_{i}$ as in (\ref{eq51}) and the
traced Ricci tensor corresponding to the reticular
$\Qalg_{i}$-valued Ricci scalar curvature $\fricci_{i}$ as in
(\ref{eq52}).\footnote{Of course, we assume that, locally in the
finsheaves, $\fric_{i}$ is a $0$-cocycle of $n\times n$-matrices
having for entries local sections of $\Qaus^{2}_{i}$---that is to
say, local $2$-forms on $\caus_{i}$, similarly to (\ref{eq38}).}

So, we are now in a position to write, at least formally, the
locally finite, causal and quantal version of the vacuum Einstein
equations for Lorentzian gravity (\ref{eq53}), as follows

\begin{equation}\label{eq117}
\fricci_{i}(\qmodll_{i})=0
\end{equation}

\noindent coining the pair $(\qmodl_{i},\conf_{i})$ consisting of
a curved finsheaf of qausets $\qmodl_{i}$ and the non-trivial
fcqv-dynamo\footnote{See footnotes 100 and 101. We note here that
one can straightforwardly write (\ref{eq117}) in terms of a {\em
self-dual} finitary spin-Lorentzian connection $\conf_{i}^{+}$ and
its Ricci curvature scalar $\fricci_{i}^{+}$. We will return to
self-dual connections in 5.3 where we will discuss a possible
`fully covariant' quantization scheme for vacuum Einstein
Lorentzian gravity.} $\conf_{i}$ on it effecting that curvature, a
{\em (f)initary, (c)ausal and (q)uantal (v)acuum Einstein field}
(fcqv-E-field) and, {\it in extenso}, the triplet
$(\qmodl_{i},\qrho_{i},\conf_{i})\equiv(\qmodll,\conf_{i})$ an
{\em fcqv Einstein-Lorentz field} (fcqv-E-L-field). In turn, the
latter prompts us to call the corresponding pentad
$(\Qalg_{i},\vec{\partial}_{i}\equiv\vec{d}^{0}_{i},\Qaus_{i}^{1},\vec{\kd}_{i}\equiv\vec{d}^{1}_{i},\Qaus^{2}_{i})$
an {\em fcqv-E-L-curvature space}, which, in turn, makes the base
causet $\caus_{i}$ a {\em fcqv-E-space}.

\subsubsection{Various interpretational matters}

Now that we have formulated the vacuum Einstein equations for
Lorentzian gravity on $\qmodll_{i}\equiv(\qmodl_{i},\qrho_{i})$,
we wish to comment briefly on their physical meaning and other
related issues of interpretation.

\begin{enumerate}

\item {\bf Differentiability is independent of
$\smooth$-smoothness.\footnote{This is the concluding slogan 2 in
\cite{malrap2}. We will elaborate further on it in the last
section.}} First we note, in keeping with our comments about
`reticular differential geometry' in part 4 of 4.1, that
(\ref{eq117}) is not a `discrete differential' ({\it eg}, a
difference) equation. Rather, it is a genuine, albeit abstract,
differential equation. The discreteness of the base causet
$\caus_{i}$---the fcqv-E-space---does not prevent us from
formulating genuine differential equations over it. As noted
repeatedly earlier, $\caus_{i}$ is merely a localization base
(topological) for the qausets (living in the stalks of
$\qmodll_{i}$) playing no role at all in the differential
geometric structure of our theory. In other words, our
differentials ({\it viz.}, connections) do not derive from the
background  space(time). {\em Space(time) does not dictate to us
the character of the differential mechanism}, as we would be
(mis)led to believe if we based ourselves on the classical
differential geometry according to which differentiability comes
from the $\smooth$-smooth manifold $M$ or equivalently, from the
coordinate algebras $\smooth(M)$ thereof. That our base space is
`discrete' does not mean at all that the differential geometric
mechanism should also be so.

\item {\bf A categorical dynamics and an abstract (generalized) principle
of general covariance independent of $\mathrm{Diff}(M)$.} Related
to 1, and as it was anticipated in \cite{malrap1}, the dynamics of
local quantum causality, as depicted in (\ref{eq117}), is
expressed solely in terms of (fin){\em sheaf morphisms}---the main
finsheaf morphism being the $\cons$-linear fcqv-dynamo
$\conf_{i}$. In fact, the fcqv-E-equations involve the curvature
$\fric_{i}$ of the connection $\conf_{i}$, which moreover is an
$\Qalg_{i}$-sheaf morphism. In other words, and in view of the
physical interpretation that ADG gives to the commutative algebra
sheaf $\struc$ of generalized coefficients,\footnote{See
discussion around footnote 41.} {\em the law for the
fcqv-E-gravity is independent both of our (local) `measurements'
or `geometry'} (as encoded in the structure sheaf of coefficients
$\Qalg_{i}(V)$) {\em and of our (local) gauges} (represented by
the open sets $U$ in the open covering $\gauge_{i}$ that we employ
to coarsely localize the events of $X$ and `measure' them in
$\Qalg_{i}(V)$; $V$ open in $\caus_{i}$). This is reflected in the
(local) gauge invariance of (\ref{eq117}) under (local)
transformations in $\qaut_{i}\qmodll_{i}(V)\simeq
M_{n}^{i}(\Qalg_{i}(V))^{\bull})$---the reticular (local)
structure (gauge) group of $\qmodll_{i}(V)\simeq\qalg_{i}^{n}(V)$.
This invariance, in turn, is a consequence of the fact that both
$\fricci_{i}$ and its contraction $\fric_{i}$ are gauge-covariant
as they obey a reticular analogue of the homogeneous gauge
transformation law for the gauge field strengths (\ref{eq39}).
Thus, as it has been already highlighted in \cite{malrap1}, our
scheme supports the following abstract categorical version of the
principle of general covariance of general
relativity:\footnote{The epithet `categorical' pertaining
precisely to that both $\conf_{i}$ and $\fric_{i}(\conf_{i})$ are
morphisms ($\mathbf{K}$- and $\Qalg_{i}$-morphisms, respectively)
in the relevant category of finsheaves of incidence algebras
(qausets) over locally finite posets (causets).}

\begin{quotation}
\noindent {\em The fcqv-dynamics, as expressed in} (\ref{eq117}),
{\em is gauge $\gauge_{i}$-independent}. Accordingly, {\em the
underlying topological spacetime $X$ and its causal discretization
$\caus_{i}$ based on the locally finite open cover $\gauge_{i}$
play no role in the dynamics of local quantum causality}
\cite{malrap1}.
\end{quotation}

\noindent It is reasonable to expect this since the fcqv-dynamo
$\conf_{i}$, or equivalently its fcqv-potential $\qaconn_{i}$, can
be viewed as the `generator' of the fcqv-dynamics.\footnote{In the
sense that the curvature $\fric_{i}(\conf_{i})$---the dynamical
variable in (\ref{eq106})---may be regarded as the `measurable,
geometric effect' since it is an $\Qalg_{i}$-morphism ({\it ie},
it respects our measurements), while $\conf_{i}$, from which
$\fric_{i}$ derives and which is not an $\Qalg_{i}$-morphism ({\it
ie}, it eludes our measurements!), as its `original, algebraic
cause'. That is why we called $\conf_{i}$ the fcqv-dynamo in the
first place: it is the generator of the fcqv-dynamics
(\ref{eq106})---the operator in terms of which the
fcqv-E-equations are formulated. Subsequently, we will see how
$\conf_{i}$ can be regarded as the main quantum configuration
variable and $\fconn_{i}$, the affine space of all such
fcqv-dynamos, the corresponding kinematical space of quantum
configurations (of $\conf_{i}$) in our theory.} and, as we argued
in 1 above, differentiability is independent of the background
causal-topological space $\caus_{i}$\footnote{The connection
$\conf_{i}$ being in effect a generalized differential operator
(derivation) of an essentially algebraic character
\cite{mall1,mall2,malrap1,malrap2}.} Thus, {\it a fortiori}

\begin{quotation}
\noindent {\em the fcqv-dynamics, as expressed in} (\ref{eq117}),
{\em is gauge $\gauge_{i}$-independent}. Accordingly, {\em the
underlying topological spacetime $X$ and its causal discretization
$\caus_{i}$ based on the locally finite open cover
$\gauge_{i}$\footnote{See subsection 4.3 above.} play no role in
the dynamics of local quantum causality as encoded in the
fcqv-dynamo $\conf_{i}$ or in its potential $\qaconn_{i}$}
\cite{malrap1}.
\end{quotation}

\noindent Plainly then, the reticular invariance (gauge) group of
(the vacuum dynamics of qausets (\ref{eq117}) generated by
$\conf_{i}$ on) $\qmodll_{i}$---the structure group
$\qaut_{i}\qmodll_{i}$---has no relation whatsoever with the
invariance group $\mathrm{Diff}(M)$ of the classical differential
spacetime manifold $M$ of general relativity. For instance,
$\mathrm{Diff}(M)$, which implements the principle of general
covariance in Einstein's classical theory of gravity, is precisely
the group that preserves the differential ({\it ie},
$\smooth$-smooth) structure of the underlying spacetime manifold.
In contradistinction, $\qaut_{i}\qmodll_{i}$, which locally is
isomorphic to $M^{i}_{n}(\Qalg_{i}(U))^{\bull}$,\footnote{And
$M^{i}_{n=4}(\Qalg_{i}(U))^{\bull}\simeq sl(2,\com)_{i}\simeq
so(1,3)^{\uparrow}_{i}$ \cite{malrap1}.} is the group that
preserves the local incidence algebraic structure of qausets
stalk-wise in their finsheaf $\qmodll_{i}$ thus {\em it has
nothing to do with the underlying topological base causet}
$\caus_{i}$ {\it per se}.\footnote{In other words,
$\qaut_{i}\qmodll_{i}$ acts directly on the (local) objects that
live on `space(time)' ({\it ie}, on the local sections of
$\qmodll_{i}$---the qausets), not on `space(time)' itself.} Of
course, since, as we argued earlier, differentiability in ADG, and
in our finitary theory in particular, derives from the stalk ({\it
ie}, from the incidence algebras modelling qausets), the (local)
gauge group $\qaut_{i}\qmodll_{i}$ of incidence algebra
automorphisms, {\em like its classical analogue}
$\mathrm{Diff}(M)$, {\em respects the reticular differential
structure}, {\em but unlike} $\mathrm{Diff}(M)$, {\em it (and the
reticular differential structure that it respects) does not come
from the background causal-topological space} $\caus_{i}$. All in
all,

\begin{quotation}
\noindent{\em Dynamics in our ADG-based theory, as expressed in}
(\ref{eq117}), {\em is genuinely background spacetime-free},
whether the latter is a smooth continuum, or a locally finite
causal space like a causet, or pretty much whatever else.
\end{quotation}

\item {\bf Everything comes from dynamics: no a priori spacetime.}
The last remarks in section 2 and the ones above bring to mind
Einstein's philosophical remark

\vskip 0.1in

\centerline{``{\small\em Time and space are modes by which we
think, not conditions in which we live}'' \cite{einst2}.}

\vskip 0.1in

\noindent as well as Antonio Machado's insightful poetic verse

\vskip 0.1in

\centerline{``{\em Traveller there are no paths; paths are made by
walking}'' \cite{machado}.}

\vskip 0.1in

\noindent in the sense that our theory (and ADG in general)
indicates that spacetime is not something `physically real'---{\it
ie}, it is not an active substance that participates in the
dynamics of Nature. {\em The only physically significant entity in
our theory is the dynamical fcqv-E-field}
$(\qmodll_{i},\conf_{i})$,\footnote{In subsections 5.3 and 5.4
this remark will prove to be of great import since we will argue
that our theory is `fully covariant' and, in a subtle sense that
we will explain, `innately quantum'.} which does not depend at all
on a supporting space(time) (of any sort, `discrete' or
`continuous') for its dynamical subsistence and propagation. This
is in glaring contrast to the classical theory (general
relativity) where spacetime is fixed {\it a priori},\footnote{That
is to say, {\em there are paths!}} once and forever so to speak,
by the theorist\footnote{That is to say, ``time and space are
modes by which {\em we} think...''---our own theoretical
constructs or figments.} to a background $\smooth$-smooth arena
and it does not get involved into the dynamics\footnote{That is to
say, spacetime is not an active, dynamical, `living' so to speak,
condition.} ({\it ie}, in the Einstein equations).

However, Machado's insight seems to go a bit further, for it
intuits not only that space(time) is (physically) non-existent
(because it is dynamically non-participatory), but also that it is
actually the `result' of dynamics.\footnote{That is, ``{\em paths
are made by walking}''.} How can we understand this in the context
of ADG and what we have said so far? To give a preliminary answer
to this question, we may have to address it first from a
kinematical and then from a deeper dynamical perspective.

\begin{itemize}

\item {\bf i) Spacetime from `algebraic kinematics'.} The kinematical emergence of
`space' from incidence algebras modelling discrete quantum
topological spaces and of `space{\em time}' from the same
structures, but when the locally finite partial orders from which
they come from are interpreted {\it \`a la} Sorkin \cite{sork1} as
causal sets rather than as finitary topological spaces, has been
worked out in \cite{rapzap1,rapzap2}. Especially in the second
reference, the kinematics of a reticular, dynamically variable
quantum spacetime topology---a Wheelerian foam-like structure so
to speak---was worked out entirely algebraically based on a
variant of Gel'fand duality\footnote{The reader will have to wait
until the following subsection for more comments on Gel'fand
duality.} coined {\em Gel'fand spatialization}. The latter
pertains to an extraction of {\em points} and the concomitant
assignment of a suitable {\em topology} on them, by exploiting the
structure and representation theory of (finite dimensional)
non-abelian associative algebras like our incidence Rota algebras
$\qaus_{i}$ modelling qausets. Such a procedure, quite standard in
algebraic geometry \cite{shaf}, is essentially based on first
identifying points with {\em kernels of} {\em irreducible
representations of the} $\qaus_{i}$s which, in turn, are {\em
primitive ideals} in the $\qaus_{i}$s, and then endowing the
collection of these ideals---the so-called {\em primitive spectra
of the incidence algebras} $Spec(\qaus_{i})$---with a non-trivial
topology.\footnote{For the incidence algebras in focus such a
topology is the {\em Rota topology} \cite{rapzap1,rap1,rapzap2}.}
Subsequently in \cite{malrap1}, we heuristically argued that the
very definition of the principal finsheaves $\peel_{i}$ of qausets
over Sorkin {\it et al.}'s causets, which are interpreted as the
kinematical structures of a locally finite, causal and quantal
theoresis of Lorentzian spacetime and vacuum Einstein gravity, is
essentially {\em schematic}.\footnote{In (noncommutative)
algebraic geometry, schemes---a particular kind of `ringed
spaces'---are sheaves of (noncommutative) rings or algebras over
their prime spectra usually endowed with the standard Zariski
topology \cite{shaf}. Incidentally, in ADG, the pair $(X,\struc)$,
which has been coined `$\mathbf{K}$-algebraized space', maybe
thought of as such (commutatively) ringed space \cite{mall1}. The
schematic aspects of our theory and their affinity to similar
noncommutative, quantal topological spaces known as {\em
quantales}, as well as to sheaves over such quantales (and the
topoi thereof), have been explored in \cite{rap3} and recently
reviewed in \cite{rap4,rap6}.} The general lesson we have learned
from this work is that

\begin{quotation}
\noindent {\em `space(time)' and its geometry\footnote{We use the
term `geometry' in a general sense which includes for instance
`topology' and other qualities of `space'.} is secondary,
derivative from a deeper, purely algebraic theoresis of Physis},
{\em inherent} already {\em in the initial}, so to say thus far
{\em `geometrical' aspect}.\footnote{We tacitly abide to the broad
`definition' of geometry as `{\em the analysis of algebraic
structure}'. It must also be noted here that Finkelstein has long
maintained in a spirit akin to ours that {\em spacetime,
causality, gauge fields and gravity} are {\em emergent notions
from a more basic, purely algebraic (and finitistic!) theory}
\cite{df0,df2,sel1,sel2,sel3,df1,sel4}; hence, {\em innately
`quantal'}!.}
\end{quotation}

\item {\bf ii) Spacetime from `algebraic dynamics'.} The idea that spacetime
and gravity come from an algebraically modelled (quantum) dynamics
is a deeper one than {\bf i)}. Presumably, in Machado's verse
quoted above,

\begin{quotation}
\noindent {\em it is exactly the particles, fields and their
mutual interrelations ({\it ie}, interactions) that `do the
walking', and by their dynamics they `define' ({\it ie}, delimit)
`spacetime'}.\footnote{From this perspective, the standard
procedure of first laying down the kinematics of a theory ({\it
eg}, the space of kinematical histories or paths of the system)
and then the dynamics, appears to be upside down. Dynamics
(`cause') comes first, the kinematical space (`effect') second.
This already points to a significant departure of our scheme from
Sorkin {\it et al.}'s causet theory whose development followed
Taketani and Sakata's methodological paradigm for the construction
of a physical theory according to which {\em one must first
develop (and understand!) the kinematics of a physical theory and
then proceed to formulate the dynamics} \cite{sork1}. Perhaps this
is the way {\em we} have so far practiced and understood
physics---{\it ie}, by first delimiting what can possibly happen
(kinematics) and then describing what actually happens
(dynamics)---but Physis herself may not work that way after all
\cite{mall7}.}
\end{quotation}

\noindent It must be noted that, still at the kinematical level of
description, Euclidean geometry is an abstraction from the motions
of, as well as the congruence and incidence relations between,
rigid bodies. However, Einstein was the first to realize that
geometry should not be regarded as an entity fixed {\it ab initio}
by the theoretician, but it must be made part of the general
physical process thus be subjected to dynamical changes
\cite{einst6}, hence he arrived at general relativity the
dynamical theory of the spacetime metric $g_{\mu\nu}$
\cite{einst3}. On the other hand, very early on Einstein also
realized that even though general relativity relativized the
spacetime metric and successfully described it as a dynamical
variable, the smooth geometric spacetime continuum was still lying
at the background as an inert, non-dynamical, ether-like substance
{\it a priori} fixed by the theorist \cite{einst5,einst1};
consequently, and intrigued by the dramatic paradigm-shift in
physical theory that quantum mechanics brought about, he intuited
soon after the formulation of general relativity that

\begin{quotation}
\noindent ``{\small\em ...The problem seems to me how one can
formulate statements about a discontinuum without calling upon a
continuum space-time as an aid; the latter should be banned from
theory as a supplementary construction not justified by the
essence of the problem---a construction which corresponds to
nothing real. But we still lack the mathematical structure
unfortunately...}'' (1916)\footnote{This quotation of Einstein can
be found in \cite{stachel}.}
\end{quotation}

\noindent and a year before his death, that

\begin{quotation}
\noindent ``{\small\em ...An algebraic theory of physics is
affected with just the inverted advantages and weaknesses} {\small
[than a continuum theory]}\footnote{In square brackets and
non-emphasized are our own completions of the text in order to
enhance continuity and facilitate understanding.}...{\small\em It
would be especially difficult to derive something like a
spatio-temporal quasi-order from such a schema...But I hold it
entirely possible that the development will lead there}...{\small
[that is,] {\em against a continuum with its infinitely many
degrees of freedom.}}'' (1954)\footnote{This quotation of Einstein
can be found in \cite{stachel}.}
\end{quotation}

\noindent Also, again motivated by the quantum paradigm, he
intuited that

\begin{quotation}
\noindent ``{\small\em  ...Perhaps the success of the Heisenberg
method points to a purely algebraic method of description of
nature, that is to the elimination of continuous functions from
physics.}'' (1936) \cite{einst4}
\end{quotation}

\noindent and, in the concluding remarks in the last appendix of
{\itshape The Meaning of Relativity}, that

\begin{quotation}

\noindent ``{\small ...[Quantum phenomena do] {\em not seem to be
in accordance with a continuum theory, and must lead to an attempt
to find a purely algebraic theory for the description of
reality.}}'' (1956) \cite{einst3}
\end{quotation}

\noindent In our theory, which rests on the intrinsically
algebraic sheaf-theoretic axiomatics of ADG \cite{mall1,mall2},
spacetime as such, especially in its classical $\smooth$-smooth
guise, plays no operative role in the formulation of the
fcqv-E-dynamics (\ref{eq117}). All that is of mathematical import
and physical significance in our scheme is the fcqv-E-field
$(\Qaus_{i},\conf_{i})$ the connection part of which---the
fcqv-dynamo $\conf_{i}$---being of purely categorico-algebraic
character. All that is physically meaningful in our model is
$(\Qaus_{i}\equiv\qmodll_{i},\conf_{i})$ and the dynamics
(\ref{eq117}) which it obeys. Furthermore, the quanta of the
fcqv-E-field, which have been called {\em causons} in
\cite{malrap1,malrap2}, represent the dynamical `elementary
particles' of the (gauge) fcqv-potential field $\qaconn_{i}$ of
quantum causality,\footnote{The reader should wait until our
remarks on geometric (pre)quantization in subsection 5.4 where we
make more explicit this `fields$\longleftrightarrow$particles
(quanta)' correspondence.} and by their algebraico-categorical
dynamics they {\em define} the quantum gravitational vacuum
without being dependent in any sense on an ambient spacetime---a
background stage that just passively supports their
dynamics.\footnote{We argued earlier that the role the base
topological causet---the fcqv-E-space $\caus_{i}$---plays in our
theory is merely an auxiliary one: $\caus_{i}$ is a substrate or
`scaffolding' that avails itself only for the sheaf-theoretic
localizations of the dynamically variable qausets; nothing else.}
At the same time, one may think of $\qaut_{i}\qmodll_{i}$---the
structure group of $\Qaus_{i}$ where the reticular connection
$1$-form $\qaconn_{i}$ takes values---as the algebraic
self-transmutations of the causon defining some sort of `{\em
quantum causal foam}' \cite{rapzap2}.\footnote{In a Kleinian
sense, the geometry of the causon---the quantum of the algebraic
fcqv-dynamo $\conf_{i}$ representing dynamical changes of (local)
quantum causality in (the stalks of, {\it ie}, the sections of)
$\qmodll_{i}\equiv\Qaus_{i}$---is encoded in the (structure) group
$\qaut_{i}\Qaus_{i}$ of its incidence algebraic automorphisms.}
Thus, we seem to find ourselves in accord with the quotation of
Feynman in the previous section, since

\begin{quotation}
\noindent{\em we actually avoid defining up-front the physical
meaning of quantum geometry, fluctuating topology, space-time
foam, {\it etc.}, and instead we give geometrical meaning after
quantization (algebraization).\footnote{This remark hints at {\em
our maintaining that our theory is}, to a great extent, {\em
already or innately quantum} (so that the usual formal procedure
of quantization of a classical theory, like general relativity, in
order to arrive at a quantum theory of gravity---regarded as
`quantum general relativity---is `begging the question' when
viewed from the ADG-based perspective of our theory). After
subsections 5.3 and 5.4, this claim of ours will become more
transparent.} In broad terms, algebra precedes geometry, since the
(algebraic dynamics of the) quantum precedes (geometrical)
`space'.}
\end{quotation}

In a similar vain, we note that, in the context of ADG, the
fundamental difference noted at the end of subsection 2.3 between
the notion of connection $\conn$---a purely algebraic notion
since, for instance, $\aconn$ transforms affinely
(inhomogeneously) under the gauge group,\footnote{That is to say,
it does not respect our local measurements of ({\it ie}, the
geometry of) the causon in $\Qalg_{i}(U)$.} and its curvature
$\curv(\conn)$---a purely geometric notion since it transforms
tensorially under the automorphism group of the vector
sheaf,\footnote{That is to say, it respects our local measurements
of the causon in $\Qalg_{i}(U)$.} becomes very relevant here. For
example, in connection with (\ref{eq117}), we note that
$\conf_{i}$ may be viewed as the generalized algebraic
differential operator in terms of which one sets up the
fcqv-E-equations, while its curvature $\fric_{i}(\conf_{i})$ as
the geometry ({\it ie}, the solution `space') of those equations.
Loosely speaking,

\begin{quotation}
\noindent {\em $\conn$ stands to $\curv(\conn)$ as the `cause'
(algebra/dynamics) stands to the `effect'
(geometry/kinematics).\footnote{See footnote 175.}}
\end{quotation}

\noindent Indeed, in \cite{malrap2}, and based on the abstract
version of the Chern-Weil theorem and the associated Chern
isomorphism {\it \`a la} ADG, we similarly argued that the purely
algebraic notion of connection $\conn$ lies on the quantal side of
the quantum divide (Heisenberg Scnhitt), while its geometric,
`observable' ({\it ie}, measurable) consequence---the curvature
$\curv(\conn)$---on the classical side.\footnote{Revisit footnote
41.} Moreover, in \cite{malrap2}, based on general geometric
pre-quantization arguments \cite{mall1,mall2,mall5}, we saw how
the algebraic causon---the quantum of the connection
$\conf_{i}$---eludes our measurements, so that what we always
measure is its field strength $\fricci(\conf_{i})$, never the
connection itself. In a Bohrian sense, the classical, geometrical
(because $\struc$-respecting) field strength is the result of our
measuring the quantum (because $\struc$-eluding), algebraic
connection.

In closing {\bf ii)}, we would like to mention, also in connection
with {\bf i)} above, that even string theory, which purports to
derive the classical spacetime manifold and Einstein's equations
from a deeper quantum string dynamics, has recently focused on
defining (spacetime) points and on deriving a topology for them by
entirely algebraico-categorical means not very different, at least
in spirit, from ours \cite{asp}.

\item {\bf iii) No topology and no metric on `space': an apparent
paradox from categorical dynamics.} We would like to mention
briefly the following apparently paradoxical feature of our theory
which has already been mentioned and resolved in \cite{malrap1}.
While we started by covering the spacetime region $X$ by the
`coarse' open gauges $U$ in $\gauge_{i}$ thus we associated with
the latter the base causal-topological space $\caus_{i}$ and
interpreted them as coarse observations or `rough chartings' of
the causal relations between events in $X$ \cite{malrap1}, at the
end, that is to say, at the dynamical level, the dynamics of
qausets over $\caus_{i}$ is gauge $\gauge_{i}$-independent since
it is expressed categorically in terms of the finsheaf morphisms
$\conf_{i}$.\footnote{Equivalently, the curvature finsheaf
morphism $\fric_{i}(\conf_{i})$ in (\ref{eq117}) is gauge
$\gauge_{i}$-covariant.} Thus, in the end the background
space(time) seems to `disappear' from the physical processes in
the quantum deep as it plays no role in the gauge invariant
dynamics of qausets. That this is only apparently and not really
paradoxical has been explained in detail at the end of
\cite{malrap1}. Here, and in connection with footnote 101, we
would like to bring to the attention of the reader that the
finitarity index (the degree of localization of our qausets) `$i$'
in (\ref{eq117}) should not be mistaken as indicating that
$\conf_{i}$ or its curvature $\fric_{i}$ are intimately dependent
on the gauge $\gauge_{i}$, for, as we repeatedly argued before,
{\em they are not}.\footnote{Quite on the contrary, as we said,
since they are finsheaf morphisms, they show that they are
$\gauge_{i}$-independent entities!} The index merely indicates
that our structures are discrete and that (\ref{eq117}) is the
finitary analogue of the ADG-theoretic expression
(\ref{eq53}).\footnote{As it were, the finitarity index shows that
our theory is a concrete application of ADG to the locally finite
regime of qausets; it is of no other physical significance.} The
corresponding statement that {\em the localization index is
physically insignificant} is precisely what it was meant in
\cite{rapzap1,malrap1,rapzap2} when we said that {\em the
incidence algebras}, whether they are taken to model discrete
quantum topological spaces proper \cite{rapzap1,rapzap2} or their
causal analogues---qausets \cite{rap1,malrap1}, {\em are alocal
structures} ({\it ie}, they are not vitally dependent on any
pre-existent or {\it a priori} postulated and physically
significant space(time)).

Now that we have shown both that the causal topology of the base
causet $\caus_{i}$ plays no role in the dynamics of qausets
(\ref{eq117}) and that differentiability comes from the incidence
algebras in the stalks of the curved $\Qaus_{i}$s, we are also in
a position to return to footnote 17, the comparison between
$\conn$ and $\curv(\conn)$ in 2.4, as well as to our comments on
the metric $\qrho_{i}$ in `about the stalk' in 4.1, and note that
in our algebraic connection-based ({\it ie}, gauge-theoretic)
scenario

\begin{quotation}
\noindent {\em fcqv-E-L-gravity does not describe the dynamics of
a vacuum spacetime metric as such in the way the original theory
({\it ie}, general relativity) does. Like the generalized
differential $\conf_{i}$, the $\Qalg_{i}$ metric $\qrho_{i}$ is a
finsheaf morphism, thus it is about the local (stalk-wise)
algebraic structure of the gauged qausets, not about the
underlying causal-topological $\caus_{i}$ {\it per se}. Hence, on
the face of} (\ref{eq117}), {\em we agree with Feynman's hunch in}
3.1 {\em that ``the fact that a massless spin-$2$ field can be
interpreted as a metric was simply a coincidence that might be
understood as representing some kind of gauge invariance''}.
\end{quotation}

\noindent Of course, it is again plain that the finitarity index
on the reticular metric $\qrho_{i}$ is of no physical (dynamical)
significance since it, like the geometrical notion of curvature,
is an $\Qalg_{i}$-respecting finsheaf morphism. Thus, $\qrho_{i}$,
like $\fricci_{i}$, lies on the classical (geometrical) side of
the quantum divide.\footnote{As it should, since it is {\em
us}---the observers---that carry on local acts of measurement on
`it' ({\it ie}, the quantum system `spacetime') and obtain
$c$-numbers in the process all of which are effectively encoded in
$\rho$. Indeed, geometry (and measurement) without a metric sounds
as absurd as convergence (and continuity) without a topology.}

\end{itemize}

\end{enumerate}

\subsection{Derivation of fcqv-E-L Gravity from an Action Principle}

We wish to emulate the situation in the abstract theory and derive
(\ref{eq117}) from the variation of a reticular, causal and
quantal version $\qeh_{i}$ of the E-H action functional $\eh$. In
the same way that the latter is a functional on the affine space
$\sconn^{(+)}_{\struc}(\modll)$ of (self-dual) Lorentzian
$\struc$-connections $\con$ on $\modll$ taking values in the space
$\struc(X)$ of global sections of $\struc$ (\ref{eq65}),
$\qeh_{i}$ is a functional on the space
$\fconn^{(+)}_{i}(\qmodll_{i})$\footnote{We write
$\fconn^{(+)}_{i}$ for $\fconn^{(+)}_{\Qalg_{i}}$. We met earlier
$\fconn_{i}^{(+)}$ in connection with the definition of the
reticular moduli spaces $\qmod_{i}^{(+)}(\qmodll_{i})$ in
(\ref{eqad1}).} of the (self-dual) fcqv-E-L-dynamos
$\conf^{(+)}_{i}$ on $\Qaus_{i}$ taking values in
$\Qalg_{i}(\caus_{i})$, as follows

\begin{equation}\label{eq118}
\qeh_{i}:~\fconn^{(+)}_{i}(\Qaus_{i})\mapto\Qalg_{i}(\caus_{i})
\end{equation}

\noindent reading `point-wise' in $\fconn^{(+)}_{i}(\Qaus_{i})$

\begin{equation}\label{eq119}
\fconn^{(+)}_{i}(\Qaus_{i})\ni\conf^{(+)}_{i}\mapsto\qeh_{i}(\conf_{i})\,
:=\fricci^{(+)}_{i}(\conf^{(+)}_{i})\, :=
tr\fric^{(+)}_{i}(\conf^{(+)}_{i})
\end{equation}

\noindent where, plainly, $\fricci^{(+)}_{i}$ is a global section
of the structure finsheaf $\Qalg_{i}$ of reticular coefficients
over the base causet $\caus_{i}$ ({\it ie},
$\fricci^{(+)}_{i}\in\Qalg_{i}(\caus_{i})$).\footnote{In what
follows, we will forget for a while the epithet `self-dual' (and
the corresponding notation) for the gravitational connection and
its curvature. We will return to self-dual $\conf_{i}$s a bit
later.}

At this point we recall the basic argument from 3.3: to be able to
derive (\ref{eq117}) from the variation (extremization) of
$\qeh_{i}$ with respect to $\conf_{i}\in\fconn_{i}(\Qaus_{i})$,
all we have to secure is that the derivative
$\stackrel{\centerdot}{\overbrace{\qeh_{i}(\conf_{i}\gamma(t))}}|_{t=0}$,
for a path $\gamma(t)$ in the reticular spin-Lorentzian connection
space $\fconn_{i}(\qmodll_{i})$
($\gamma:~\R\mapto\fconn_{i}(\Qaus_{i})$), is well defined. The
latter means in turn that there should be a well defined notion of
convergence, limit and, of course, a suitable topology on the
structure sheaf $\Qalg_{i}$ relative to which these two notions
make sense.

We recall from \cite{rapzap1,rap1,rap2,rapzap2,malrap1} that the
abelian (structure) subalgebras $\Qalg_{i}$ of the incidence
algebras $\qaus_{i}$ modelling the qausets in the stalks of the
$\Qaus_{i}$s can then be construed as carrying a (natural)
topology---the so-called {\em Rota topology}---provided by the
$\qaus_{i}$s' structure (primitive ideal) spaces (Gel'fand
duality).\footnote{In the next subsection we will comment further
on the rich import that Gel'fand duality has in our theory.} With
respect to the (now quantum causally interpreted) Rota topology,
it has been shown that there is a well defined notion of
(discrete) convergence and, {\em in extenso}, of limits
\cite{sork0,zap0,brezap,rapzap1,rap1,rap2,rapzap2}. Thus,
$\stackrel{\centerdot}{\overbrace{\qeh_{i}(\conf_{i}\gamma(t))}}|_{t=0}$
is well defined.

\subsection{Towards a Possible Covariant Quantum Dynamics for the Finitary
Spin-Lorentzian Connections}

We have seen how general relativity can be cast as a Y-M-type of
gauge theory in finitary terms, that is to say, how it may be
expressed solely as the dynamics of a fcqv-spin-Lorentzian
connection variable---the dynamo $\conf_{i}$. These dynamos have
been already `{\em kinematically quantized}' \cite{malrap1} and
`{\em geometrically (pre)-quantized}' to {\em causons}
\cite{malrap2}\footnote{With a concomitant sheaf-cohomological
classification of the corresponding associated curved line sheaves
$\lsh$ inhabited by these causons. We will return to make more
comments on geometric (pre)quantization in subsection 5.4.2.}
along the lines of ADG \cite{mall1,mall2,mall5}.

In the present subsection we discuss the possibility of developing
a covariant path integral-type of {\em quantum dynamics} for the
finitary spin-Lorentzian dynamos $\conf_{i}$ on the respective
$\qmodll_{i}\equiv\Qaus_{i}$s. As a first step, we wish to emulate
formally the usual practice in the quantum gauge theories of
matter ({\it ie}, QED, QCD and higher dimensional Y-M theories of
a semi-simple and compact Lie structure group $\struct$) whereby a
covariant quantum dynamics is represented by a path integral over
the space of the relevant connections on the corresponding
principal fiber bundles over a $\smooth$-smooth spacetime manifold
$M$ (a $\struct=U(1)$-bundle for QED, a $\struct=SU(3)$-bundle for
QCD and $\struct=SU(N)$-bundles for general Y-M theories). Thus,
in our case too, we intuit that the main object of study should be
the following `heuristic device'

\begin{equation}\label{eq120}
\QPI:=\int_{\fconn_{i}(\qmodll_{i})}e^{i\qeh_{i}}d\qaconn_{i}
\end{equation}

\noindent where $\fconn_{i}(\qmodll_{i})$ is the affine space of
finitary spin-Lorentzian connections $\conf_{i}$ on the curved
orthochronous Lorentzian finsheaves $\qmodll_{i}\equiv\Qaus_{i}$
of qausets which is thus being regarded as the (quantum)
kinematical configuration space (of `fcqv-dynamo or causon quantum
histories') of our theory. More precisely, due to the local
reticular gauge invariance of our theory, the actual physical
configuration space is the fcqv-analogue
$\qmod_{i}(\qmodll_{i}):=\fconn_{i}(\qmodll_{i})/\qaut_{i}(\qmodll_{i})$
of the moduli space in (\ref{eq103}) that we defined earlier in
(\ref{eqad1}), and it consists of finitary gauge-equivalent
fcqv-connections $\conf_{i}$. We thus recast (\ref{eq120}) as
follows

\begin{equation}\label{eq121}
\QPI:=\int_{\qmod_{i}}e^{i\qeh_{i}}d([\qaconn_{i}]_{\qaut_{i}\qmodll_{i}})
\end{equation}

\noindent where $[\qaconn_{i}]_{\qaut_{i}(\qmodll_{i})}$ denotes
the gauge $\qaut_{i}\qmodll_{i}$-equivalence classes of
fcqv-gravitational connections $\conf_{i}$ on $\qmodll_{i}$---the
elements of $\qmod_{i}(\qmodll_{i})$.

In what follows we enumerate our anticipations and various remarks
about $\QPI$ in (\ref{eq121}) by gathering information from both
the canonical ({\it ie}, Hamiltonian) approach to quantum general
relativity and the covariant path integral ({\it ie}, Lagrangian
or action-based) approach to Lorentzian quantum gravity. In
particular, and in connection with the former approach, we discuss
issues arising from Ashtekar's self-dual connection variables
scenario for both classical and quantum gravity \cite{ash} as well
as from their $\smooth$-smooth loop holonomies---the so-called
loop formulation of (canonical) quantum gravity
\cite{rovsm}\footnote{For reviews of the loop approach to quantum
gravity and relevant references, the reader is referred to
\cite{loll,rovelli0}.}---especially viewed under the functional
analytic ($C^{*}$-algebraic) prism of Ashtekar and Isham
\cite{ashish,ashlew1}. We thus commence our exposition with a
brief review of both the Hamiltonian (canonical) and the
Lagrangian (covariant) approaches to Lorentzian quantum gravity.

\subsubsection{The canonical (Hamiltonian) approach: Ashtekar
variables}

More than fifteen years ago, Ashtekar \cite{ash} proposed a new
set of variables for both classical and quantum general relativity
essentially based on a complex spacetime manifold and a self-dual
connection version of the Palatini comoving $4$-frame ({\it
vierbein}) formulation of gravity. The main assumptions were the
following:

\begin{itemize}

\item A $4$-dimensional, complex, orientable, $\smooth$-smooth spacetime
manifold $M$ of Lorentzian signature.

\item The basic gravitational variable
$\aconn^{+}_{\infty}$,\footnote{The index `$\infty$' just
indicates that $\aconn$ {\em is a $\smooth$-smooth connection on}
$M$.} which is a $so(1,3)_{\com}$-valued self-dual connection
$1$-form.

\item The {\it vierbein} variable $e$, which defines a vector
space isomorphism between the tangent space of $M$ and a fixed
`internal space' $\mink$ equipped with the usual Minkowski metric
$\eta$ and the completely antisymmetric tensor $\epsilon$.
$\aconn^{+}_{\infty}$ is self-dual with respect to
$\epsilon$.\footnote{More analytically and in bundle-theoretic
terms (note: most of the items to be mentioned in this footnote
should be compared one-by-one with the corresponding ADG-theoretic
ones defined earlier and the reader must convince herself that,
ADG-theoretically, we possess all the classical smooth vector
bundle-theoretic notions and constructions {\em without any notion
of $\smooth$-smoothness being used}. This observation will prove
crucial in the sequel---see comparison between our ADG-based
finitary scheme and the usual $\smooth$-approaches to
non-perturbative canonical or covariant Lorentzian quantum gravity
that the present footnote will trigger after (\ref{add7})), one
lets $\mathcal{T}$---equipped with a pseudo-Riemannian metric
$\eta$ and fixed orientation $\EuScript{O}$---be an `internal
Minkowskian bundle space' isomorphic to the tangent bundle $TM$.
$\EuScript{O}$ and $\eta$ define a nowhere vanishing global
section $\epsilon$ of $\wedge^{4}{\mathcal{T}}^{*}$. The aforesaid
fiber bundle isomorphism is symbolized as
$e:~TM\mapto{\mathcal{T}}$, and its inverse $e^{-1}$ is the
comoving $4$-frame field ({\it vierbein}) mentioned above (by
pushing-forward $e$ one can also define a volume form $\vol$ on
$M$, while $TM$ inherits via $e^{-1}$ the metric $\eta$ from
$\mathcal{T}$). $\eta$ similarly defines an isomorphism between
$\mathcal{T}$ and its dual ${\mathcal{T}}^{*}$. Fortunately, in
$4$ dimensions, $\eta$ and $\epsilon$ determine a unipotent
Hodge-$\star$ operator:
$\star:~\wedge^{2}{\mathcal{T}}\mapto\wedge^{2}{\mathcal{T}}$. One
then regards as basic dynamical fields in Ashtekar's theory the
aforementioned spin-Lorentzian metric ({\it ie},
$\eta$-preserving) connection $1$-form $\aconn^{+}_{\infty}$
(whose curvature $\curv_{\infty}^{+}$ is a section of
$\wedge^{2}{\mathcal{T}}\otimes\wedge^{2}{\mathcal{T}}^{*}$ and
satisfies relative to $\star$ the self-duality relation:
$\star\curv_{\infty}^{+}=\curv_{\infty}^{+}$) and the frame field
$e$ (which is a ${\mathcal{T}}$-valued $1$-form:
$e\in\Omega^{1}({\mathcal{T}})$). (Of course, one can also
transfer via $e^{-1}$ the connection $\aconn_{\infty}^{+}$ from
$\mathcal{T}$ to $TM$.)}

\end{itemize}

In the new variables $\aconn^{+}_{\infty}$ and $e$, the
gravitational action functional assumes the following so-called
{\em first-order form}

\begin{equation}\label{eq122}
S_{ash}[\aconn^{+}_{\infty},e]=\frac{1}{2}\int_{M}\epsilon(e\wee
e\wee\curv_{\infty}^{+})
\end{equation}

\noindent which may be readily compared with the usual Palatini
action

\begin{equation}\label{eq123}
S_{pal}[\aconn_{\infty},e]=\frac{1}{2}\int_{M}\epsilon(e\wee
e\wee\curv_{\infty})
\end{equation}

\noindent and directly see that $S_{ash}$ is $S_{pal}$'s self-dual
version.\footnote{Plainly, $\curv_{\infty}^{(+)}$ in both
(\ref{eq122}) and (\ref{eq123}) is the curvature of the
(self-dual) connection $\aconn^{(+)}_{\infty}$.} We also note
that, upon variation of both $S_{ash}$ and $S_{pal}$ with $e$, one
obtains the vacuum Einstein equations ({\it ie}, that
$\aconn^{(+)}_{\infty}$ is Ricci-flat), while upon variation with
$\aconn^{(+)}_{\infty}$, one obtains the metric-compatibility
condition for $\aconn^{(+)}_{\infty}$ ({\it ie}, that it is the
gauge potential part of the Levi-Civita connection of the metric).

The attractive feature of Ashtekar's new variables is that in
terms of them one can simplify and write neatly the Hamiltonian
constraints for gravity, thus one obtains a clear picture of how
to proceed and canonically quantize the theory {\it \`a la} Dirac.
To revisit briefly the Hamiltonian approach, one assumes that $M$
factors into two submanifolds:
$M=\Sigma^{3}\times\R$,\footnote{Assuming also that the `spatial'
or `spacelike' $3$-submanifold $\Sigma^{3}$ is orientable and
compact.} thus securing the $3+1$ decomposition needed in order to
approach quantum gravity canonically. Then, one assumes as {\em
configuration space} of the theory the affine space
${}^{3}\infconn^{+}$ of complex, smooth, self-dual,
$so(3)_{\com}$-valued connections ${}^{3}\aconn^{+}_{\infty}$ on
$\Sigma^{3}$,\footnote{Thus, in this picture gravity may be
thought of as an $SO(3)_{\com}$-gauge theory---the dynamical
theory of ${}^{3}\aconn^{+}_{\infty}$ in the connection space
${}^{3}\infconn^{+}$. Shortly we will see that gravity is actually
a `larger' theory transformation-wise: {\em it is an
$SO(3)_{\com}$-gauge theory together with
$\mathrm{Diff}(M)$-constraints coming from assuming up-front that
there is an external background $\smooth$-smooth spacetime
manifold}.} and as {\em phase space} the cotangent bundle
$T^{*}({}^{3}\infconn^{+})$ coordinatized by canonically conjugate
pairs
$({}^{3}\aconn^{+}_{\infty},{}^{3}E_{\infty})$\footnote{Where
${}^{3}E_{\infty}$ is a smooth vector density representing a
generalized electric field on $\Sigma^{3}$.} obeying the following
Poisson bracket relations\footnote{In (\ref{eq124}), we present
indexless symplectic relations. The reader is referred to
\cite{loll} for the more elaborate indexed relations.}

\begin{equation}\label{eq124}
\{ {}^{3}\aconn^{+}_{\infty},
{}^{3}E_{\infty}\}=\delta^{3}(x-y);~(x,y\in\Sigma^{3})
\end{equation}

\noindent In terms of these variables, the Hamiltonian for gravity
can be shown to be \footnote{Again, all indices, including the
ones for $\aconn$ and $E$ above, are omitted in (\ref{eq125}).}

\begin{equation}\label{eq125}
H(\aconn ,E)=\int_{\Sigma^{3}}(\frac{1}{2}\lambda_{l}\epsilon\curv
E^{2}+i\lambda_{s}\curv E)d^{3}x
\end{equation}

\noindent with $\lambda_{l}$ and $\lambda_{s}$ being Lagrange
multipliers corresponding to the well known lapse and shift
functions in the canonical formulation of gravity.

On the other hand, since the theory has internal (gauge)
$SO(3)_{\com}$-symmetries and external (spacetime)
$\mathrm{Diff}(M)$-symmetries, not all points (classical states)
in the phase space $T^{*}({}^{3}\infconn^{+})$ can be regarded as
being physical. This is tantamount to the existence of the
following five first-class constraints for gravity\footnote{Again,
all indices are suppressed for symbolic economy and clarity.}

\begin{equation}\label{eq126}
\begin{array}{c}
\mathrm{one~Gauss~divergence~constraint~(internal):}~\con E=0\cr
\mathrm{three~spatial~diffeos~constraints~(external):}~\curv
E=0\cr
\mathrm{one~temporal~diffeo~constraint~(external):}~\epsilon\curv
E^{2}=0
\end{array}
\end{equation}

\noindent which must be satisfied by the (classical) physical
states.\footnote{In (\ref{eq126}) the temporal-diffeomorphisms
constraint is commonly known as the {\em Hamiltonian constraint}.}
At the same time, $\con E$, $\curv E$ and $\epsilon\curv E^{2}$
can be seen to generate local gauge transformations in the
internal gauge space, as well as $\Sigma^{3}$-spatial diffeos and
$\R$-temporal diffeos respectively in the external
$M=\Sigma^{3}\times\R$-spacetime manifold,\footnote{The
Hamiltonian constraint generates the smooth time-evolution of
$\Sigma^{3}$ in $M$.} thus they transform between physically
indistinguishable (equivalent) configurations. It is important to
note here that pure Y-M theory also has the internal Gauss gauge
constraint, but not the other four external `spacetime
diffeomorphism' $\mathrm{Diff}(M)$-constraints. Due to this fact,
Loll points out for example that ``{\em pure gravity may be
interpreted as a Yang-Mills theory with gauge group}
$\struct=SO(3)_{\com}$, {\em subject to four additional
constraints in each point of $\Sigma$''}\footnote{Which we call
$\Sigma^{3}$ here.} \cite{loll}. We will return to this remark
soon. One should also notice here that since the integrand of
$H(\aconn ,E)$ in (\ref{eq124}) is an expression involving
precisely these four external spacetime
$\mathrm{Diff}(M)$-constraints, the Hamiltonian vanishes on
physical states.\footnote{This is characteristic of gravity
regarded as a gauge theory on a $\smooth$-smooth spacetime
manifold $M$, namely, $\mathrm{Diff}(M)$, which implements the
principle of general covariance, is (part of) gravity's gauge
(structure) group $\grouv$.} Since, as noted in footnote 221, $H$
is the generator of the smooth time-evolution of $\Sigma^{3}$ in
the spacetime manifold $M$, one says (even at the classical level)
that, at least from the canonical viewpoint, gravity is
`inherently' a no-time (`time-less') theory.

A straightforward canonical quantization of gravity {\it \`a la}
Dirac would then proceed by the following standard formal
replacement of the Poisson bracket relations in (\ref{eq124}) by
commutators

\begin{equation}\label{eq127}
\{ \aconn , E\}=\delta^{3}(x-y)\mapto
[\widehat{\aconn},\widehat{E}]=i\delta^{3}(x-y);~(x,y\in\Sigma^{3})
\end{equation}

\noindent with the hatted symbols standing now for field operators
acting on the unphysical phase space
$T^{*}({}^{3}\infconn^{+})$\footnote{As can be read from
\cite{loll} for instance, there are (technical) reasons for using
$T^{*}({}^{3}\infconn^{+})$ instead of the physical cotangent
bundle $T^{*}({}^{3}\infconn^{+}/\grouv)$ on the $3$-connection
moduli space ${}^{3}\infconn^{+}/\grouv$. We will comment on some
of them subsequently when we will emphasize the need to develop a
differential geometry on the moduli space of gauge-equivalent
connections.} which is suitably `Hilbertized'. The latter pertains
essentially to the promotion of the space
$\EuScript{F}({}^{3}\infconn^{+})=\{\Psi(\aconn)\}$ of
$\com$-valued functions on ${}^{3}\infconn^{+}$ to a Hilbert space
$\hil$ of physical states. This is usually done in two steps:

\begin{itemize}

\item First, in order to take into account the gauge and diffeomerphism invariance of the theory,
one projects out of $\EuScript{F}({}^{3}\infconn^{+})$ all the
wave functions lying in the kernel of the corresponding operator
expressions of the gravitational constraints in (\ref{eq126}).
These are precisely the {\em physical} quantum states-to-be, as
they satisfy operator versions of the constraints ({\it ie}, they
are annihilated by them). They comprise the following subspace
$\EuScript{F}_{p}$ of (p)hysical wave functions in $\EuScript{F}$

\begin{equation}\label{eq128}
\EuScript{F}_{p}:=\{\Psi(\aconn):~\widehat{\con
E}\Psi(\aconn)=\widehat{\curv
E}\Psi(\aconn)=\widehat{\epsilon\curv E^{2}}\Psi(\aconn)=0\}
\end{equation}

\noindent where the hatted symbols denote operators.

\item Then, one promotes $\EuScript{F}_{p}$ to a Hilbert space $\hil_{p}$
by endowing it with the following hermitian inner product
structure

\begin{equation}\label{eq129}
<\Psi_{2}(\aconn)|\Psi_{1}(\aconn)>:=
\int_{{}^{3}\infconn^{+}/\grouv}\Psi_{2}^{*}(\aconn)\Psi_{1}(\aconn)[d\aconn]_{\grouv}
\end{equation}

\noindent thus essentially by insisting that the wave functions
$\Psi(\aconn)$ are square-integrable with respect to $<.\, |\,
.>$. However,

\begin{quotation}
\noindent {\em so far one has not been able to find a fully
$\grouv$-invariant ({\it ie}, $SO(3)_{\com}$-gauge and
$\mathrm{Diff}(M)$-invariant) integration measure
$[d\aconn]_{\grouv}$ on} ${}^{3}\infconn^{+}/\grouv$.\footnote{Of
course, the tough problem is finding a
$\mathrm{Diff}(M)$-invariant measure, not an $SO(3)_{\com}$ one.
Ingenious ideas, involving abstract or generalized integration
theory, have been used in order to actually construct such a
$\mathrm{Diff}(M)$-invariant measure \cite{baez2,baez3}. We will
return shortly to comment a bit more on abstract integration
theory and generalized measures. Also, motivated by this remark
about $\mathrm{Diff}(M)$-invariant measures, from now on we will
abuse notation and identify the gauge (structure) group $\grouv$
of gravity only with its external smooth spacetime manifold
symmetries ({\it ie}, $\grouv\equiv\mathrm{Diff}(M)$) and forget
about its internal, `purely gauge', $SO(3)_{\com}$-invariances. }
\end{quotation}

\noindent This is essentially the content of the so-called {\em
inner product problem} in the canonical approach to quantum
general relativity.
\end{itemize}

Before we move on to discuss briefly the covariant path integral
approach to quantum gravity, which, as we shall see, also
encounters a similar `diffeomorphism-invariant measure over the
moduli space of connections' problem, we wish to present some
elements from the Ashtekar-Isham analysis of the loop approach to
canonical quantum gravity \cite{rovsm,ashish,ashlew1}. Of
particular interest to us, without going into any technical
detail, are two general features of this analysis: (i) the
application of a version of Gel'fand duality on the space of Y-M
and (self-dual) gravitational connections in a spirit not so
different from how we use Gel'fand duality in our
algebraico-sheaf-theoretic approach to causal sets here, and, as a
result of this application, (ii) its pointing to a generalized
integration theory over the moduli space
${}^{3}\infconn^{+}/\grouv$ in order to deal with the
`$\mathrm{Diff}(M)$-invariant measure problem' mentioned in
connection with the Hilbert space inner product in (\ref{eq129}).

In \cite{rovsm}, Rovelli and Smolin used non-local,
gauge-invariant Wilson loops---the traces of holonomies of
connections around closed loops in $\Sigma^{3}$\footnote{One
defines a Wilson loop as follows:
$W_{\aconn^{(+)}}^{\rho}(\ell):=tr\exp_{po}(\oint_{\ell\in\Sigma^{3}}\aconn^{(+)})$,
where $\ell$ is a spatial loop (in $\Sigma^{3}$), $\rho$ is a
(finite dimensional, complex) matrix representation of the Lie
algebra $\mathbf{g}$ of the gauge group $\struct$ where the
(self-dual) connection $\aconn^{(+)}$ takes values (in our case,
$so(3)_{\com}$, and the index `$po$' to $\exp$ denotes `{\em path
ordered}' \cite{loll}. For the sake of completeness, we note that
Rovelli and Smolin, based on Ashtekar's new variables $(\aconn
,e)$, actually defined an `adjoint' set of Wilson loop variables
that reads:
$W_{e}^{\rho}(\ell):=tr[e(\ell)\exp_{po}(\oint_{\ell\in\Sigma^{3}}\aconn^{(+)})]$.
}---and found physical states for canonical quantum gravity, that
is to say, ones that are annihilated by the aforementioned
operator constraints. Remarkably enough, they found that such
states can be expressed in terms of knot and link-invariants
(which themselves are $\com$-valued functions on knots and links
that are invariant under spatial diffeos), thus they opened new
paths for exploring the apparently intimate relations that exist
between gauge theories, (quantum) gravity, knot theory and, {\em
in extenso}, the geometry of low-dimensional
manifolds.\footnote{Refer to \cite{baez} for a thorough exposition
of the close interplay and the fertile exchange of ideas between
knot theory, gauge theory and (quantum) gravity.} Such promising
new research possibilities aside, what we would like to highlight
here are certain features in the aforesaid work of Ashtekar and
Isham which put Rovelli and Smolin's loop variables on a firm and
rigorous mathematical footing, and, in particular, opened the way
towards finding $\grouv$-invariant measures (as well as
generalized integrals to go with them) that could help us resolve
problems like the one of the inner product mentioned above.

Our first remark concerns the general moduli space
$\infconn/\grouv$ of gauge theories and gravity. We have seen
above what a crucial role it plays both in the classical and the
quantum descriptions of these theories. For one thing, it is the
classical configuration space of the theories in their
connection-based formulation. As we have said, to get the
classical phase space, one deals with the cotangent bundle
$T^{*}(\infconn/\grouv)$.\footnote{The elements of
$T^{*}(\infconn/\grouv)$ are the classical physical observables of
the theories.} In their quantum versions, the moduli space
$\infconn/\grouv$ is supposed to give way to the Hilbert space
$L^{2}(\infconn/\grouv ,d\mu)$ of $\com$-valued, square-integrable
functions on $\infconn/\grouv$ with respect to some measure
$d\mu$, which is in turn expected to be $\grouv$-invariant.
However, due to $\infconn/\grouv$'s infinite dimensionality,
non-linear nature and rather `complicated' topology,\footnote{This
refers to the usual $\smooth$ (Schwartz) topology \cite{mall0}.}
there are significant (technical) obstacles in finding ({\it ie},
actually constructing!) such a $d\mu$. Moreover, in the canonical
approach, the loop variables of Rovelli and Smolin provide us with
a set of manifestly $\grouv$-invariant configuration observables,
but we lack analogous gauge-invariant momentum observables not
least because the differential geometry of the moduli space
$\infconn/\grouv$ (and {\it in extenso} of the cotangent bundle
$T^{*}(\infconn/\grouv)$) has not been well developed or
understood.\footnote{Principally motivated by this ellipsis, and
as we noted earlier, \cite{ashlew2} explores further the
possibility of developing classical ({\it ie}, $\smooth$-smooth)
differential geometry on $\infconn/\grouv$.} These are some of the
technical difficulties one encounters in trying to develop
classical $\smooth$-smooth differential geometric ideas on spaces
of gauge-equivalent connections and exactly because of them one
could `justify' the ADG-theoretic perspective we have adopted in
the present paper.\footnote{The reader is referred to \cite{mall4}
for a more elaborate ADG-theoretic treatment of moduli spaces of
connections {\it vis-\`{a}-vis} gauge theories and gravity.}

Now, what Ashtekar and Isham did in order to deal with some of the
problems mentioned in the previous paragraph is to `downplay' the
structure of the space $\infconn/\grouv$ {\it per se} and rather
{\em work directly with the functions that live on that
space}.\footnote{This is well in line with the general philosophy
of ADG which we have repeatedly emphasized throughout this paper
and according to which, in order to gather more information and
gain more insight about (the structure of) `space'---whatever that
may be---one should look for an `appropriate' algebra that encodes
that information in its very structure. Then, in order to recover
`space' and perform the ever-so-useful in physics calculations
({\it ie}, `geometrize' or `arithmetize' the abstract algebraic
theory so to speak), one should look for suitable representations
of this algebra.} Thus, they defined the so-called {\em holonomy
$C^{*}$-algebra} $\gel=\EuScript{F}(\infconn/\grouv)$ of
$\com$-valued functions on $\infconn/\grouv$ generated by Wilson
loops $W(\ell)$ like the ones mentioned in footnote
226.\footnote{It must be noted however that {\em real} connections
$\aconn$ were employed in \cite{ashish}. The reader should not be
concerned about this technical detail here.} $\gel$ was
straightforwardly seen to be abelian, thus by using the well known
Gel'fand-Naimark representation theorem they identified $\gel$
with the commutative $C^{*}$-algebra $\overline{\EuScript{F}}$ of
continuous $\com$-valued functions on a compact Hausdorff
topological space $\gelsp\equiv Spec(\gel)$---the so-called {\em
Gel'fand spectrum of} $\gel$.\footnote{The points of $Spec(\gel)$
are kernels of (irreducible) representations of $\gel$ to $\com$
({\it ie}, homomorphisms of $\gel$ to $\com$ commonly known as
`characters'), with the latter being the `standard' abelian
involutive algebra. In turn, these kernels are maximal ideals in
$\gel$, so that equivalently one writes $\mathrm{Max}(\gel)$ for
$\gelsp\equiv Spec(\gel)$ (in the sequel, we will use
$Spec(\gel)$, $\gelsp$ and $\mathrm{Max}(\gel)$ interchangeably).
$\mathrm{Max}(\gel)$ carries the standard Gel'fand topology and
the elements of $\overline{\EuScript{F}}$ are continuous with
respect to it. (Memo: the Gel'fand topology on $\gelsp$ is the
weakest (coarsest) topology with respect to which the functions in
$\overline{\EuScript{F}}$ are continuous \cite{mall0}.)} In turn,
every (continuous and cyclic) representation
$\overline{\EuScript{F}}$ of $\gel\equiv \EuScript{F}$ has
$L^{2}(\mathrm{Max}(\gel))$ as carrier Hilbert space with respect
to some regular measure $d\mu$ on $\gelsp$ and, plainly, the
representatives of the $\com$-valued Wilson loop operators in
$\gel$ act on the elements $\Psi$ of $L^{2}(\gelsp)$ by
multiplication.

Thus, while $\infconn/\grouv$ is the classical configuration
space, quantum states $\Psi$ naturally live on
$\mathrm{Max}(\gel)$ and can be thought of as `generalized'
gauge-equivalent connections. In fact, Rovelli and Smolin
conceived of a deep correspondence between the spaces of
(functions on) gauge-equivalent connections and (functions on)
loops, which could be mathematically implemented by the following
heuristic integral `device'

\begin{equation}\label{eq130}
\gelt[\Psi(\ell)]:=\int_{\infconn/\grouv}tr(\exp_{po}\oint_{\ell}\aconn)\Psi([\aconn]_{\grouv})
d\mu([\aconn]_{\grouv})
\end{equation}

\noindent called the (non-linear and in general
non-invertible\footnote{The loop transform is supposed to carry
one from the connection to the loop picture, and back via
$\gelt^{-1}$. However, for $\gelt^{-1}$ to exist, a set of
(algebraic) constraints---the so-called Mandelstam
constraints---must be satisfied by Wilson loops
\cite{ashish,loll}.}) {\em loop transform}---a variant of the
usual functional-analytic Gel'fand transform.\footnote{The
Gel'fand transform may be viewed as a generalized Fourier
transform \cite{mall0}. The reader is encouraged to read from
\cite{ashish} a suggestive comparison made between the loop and
the Fourier transform. For an ADG-theoretic use of the Gel'fand
transform, in case $\struc$ is a topological algebra sheaf (the
`canonical' example of a unital, commutative topological algebra
being, of course, $\smooth(M)$---see remarks on Gel'fand duality
in 5.5.1), the reader is referred to \cite{mall1,mall2}.} Again,
in $\gelt(\Psi)$ we witness the need to find measures on
$\infconn/\grouv$.\footnote{In (\ref{eq130}), $[\aconn]_{\grouv}$
represents a class of $\grouv$-equivalent connections in
$\infconn$---an element of the moduli space $\infconn/\grouv$.}

This last remark brings us to the main point we wish to make about
the importance of the (abelian) $C^{*}$-algebraic point of view
(and the application of the Gel'fand spectral theory that goes
with it) on the moduli space of connections adopted by Ashtekar
and Isham based on the Rovelli-Smolin loop representation of
Ashtekar's new variables in the context of canonical quantum
general relativity:

\begin{quotation}
\noindent the holonomy $C^{*}$-algebraic perspective on
$\infconn/\grouv$ makes it clear that {\em one must adopt a
`generalized integration theory'}\footnote{The reader should refer
to \cite{baez2,baez3} for a relatively recent treatment of
generalized $\mathrm{Diff}(M)$-invariant measures on moduli spaces
of non-abelian Y-M and gravitational connections.} in order to
cope with integrals such as (\ref{eq129}) and (\ref{eq130}) and
with the measures involved in them.
\end{quotation}

\noindent The idea to use `generalized' or `abstract measures'
becomes `natural' in Ashtekar and Isham's work as follows: as we
noted above, the holonomy $C^{*}$-algebra
$\gel=\EuScript{F}(\infconn/\grouv)$ is first transcribed by the
Gel'fand-Naimark representation to the $C^{*}$-algebra
$\overline{\EuScript{F}}$ of bounded, continuous, $\com$-valued
functions on $\gel$'s spectrum $\mathrm{Max}(\gel)$ having for
carrier Hilbert space $L^{2}(\mathrm{Max}(\gel),d\mu)$. {\em How
can we realize the measure} $d\mu$ {\em and the integral with
respect to it}?

\begin{quotation}
\noindent The aforesaid idea of `generalized measures' can be
materialized in the $C^{*}$-algebraic context by identifying
$\int[\cdot]d\mu$ with {\em a state $\sigma$ on}
$\overline{\EuScript{F}}$---a (normalized, positive) linear form
on $\overline{\EuScript{F}}$, which is a member of
$\overline{\EuScript{F}}^{*}$. Then one thinks of $\sigma(f)$ as
an abstract expression of $\int f\, d\mu$
($f\in\overline{\EuScript{F}}$). In turn, having this integral in
hand, the inner product on $L^{2}(\mathrm{Max}(\gel))$ can be
realized as
$<\Psi_{2}|\Psi_{1}>=\int\Psi_{2}^{*}\Psi_{1}d\mu=s(\Psi_{2}^{*}\Psi_{1})$.\footnote{With
$\Psi_{2}^{*}$ the complex conjugate of $\Psi_{2}$ (note: the
reader should not confuse this $*$-star with the linear dual
$*$-star in $\overline{\EuScript{F}}^{*}$.}
\end{quotation}

\medskip

We now move on to discuss briefly the covariant path integral
(Lagrangian) approach to quantum gravity, so that afterwards we
can comment `cumulatively' from an ADG-theoretic viewpoint on the
heuristic integral $\QPI$ appearing in (\ref{eq121}) in comparison
with what we have said about both the canonical and the covariant
quantization schemes for gravity.\footnote{Since both of these
schemes are essentially based on the classical differential
geometry of the $\smooth$-smooth spacetime manifold $M$ ({\it ie},
they belong to category 1 in the prologue---in other words, {\em
they are `$\smooth$-smoothness conservative'}) which ADG evades,
such a comparison is relevant here and well worth the effort.}

\subsubsection{The covariant (Lagrangian) approach: the
$\mathrm{Diff}(M)$-invariant path integral measure problem}

One of the main disadvantages of any approach to the quantization
of gravity based on the canonical formalism is the latter's
breaking of full covariance by the unphysical $3+1$ space-time
split that it mandates. In the Ashtekar approach for instance, one
must choose a time-slicing by arbitrarily foliating spacetime into
spacelike hypersurfaces on which the self-dual connection
variables $\aconn_{\infty}^{+}$---the main dynamical variables of
the theory---are defined and canonical Poisson bracket (classical)
(\ref{eq124}) or commutator (quantum) (\ref{eq127}) relations are
imposed.\footnote{Also, by such a $3+1$ decomposition one secures
a well defined Cauchy problem for the dynamical equations (global
hyperbolicity).} The basic idea of a path integral quantization of
gravity is not to force any such physically {\it ad hoc} $3+1$
split, thus retain full covariance of the theory.

In a Lagrangian, (self-dual) connection-based formulation of
gravity in a $\smooth$-smooth spacetime manifold (like Ashtekar's
in (\ref{eq122}), but in all four spacetime dimensions), the path
integral would be the following heuristic object

\begin{equation}\label{eq131}
\PI_{\infty}=\int_{{}^{4}\infconn^{(+)}}e^{i[{}^{4}S^{(+)}_{ash}]}d\aconn
\end{equation}

\noindent where the integral is taken now over all the (self-dual)
$\smooth$-connections ${}^{4}\aconn_{\infty}^{(+)}$ over the whole
$4$-dimensional spacetime manifold $M$, and $S^{(+)}_{ash}$ is the
$4$-dimensional version of the Ashtekar action (\ref{eq122}) of
the (self-dual) smooth connection variable
${}^{4}\aconn_{\infty}^{(+)}$.\footnote{However, it must be
emphasized here that a $3+1$ space-time split is in a sense also
implicit here. $\PI_{\infty}$ in (\ref{eq131}) is normally
regarded as a {\em transition amplitude} and the dynamical
transition that it pertains to is between `boundary spatial
configuration $3$-geometries'---say,
$\Phi_{1}[{}^{3}\aconn_{1}^{(+)}]_{\Sigma^{3}_{1}}$ and
$\Phi_{2}[{}^{3}\aconn_{2}^{(+)}]_{\Sigma^{3}_{2}}$---with the
bulk $4$-spacetime geometry interpolating between them. One
usually writes $\PI_{\infty}|_{\Phi_{1}}^{\Phi_{2}}\equiv
<\Phi_{2}|\Phi_{1}>=\int_{\Phi_{1}}^{\Phi_{2}}e^{i[{}^{4}S^{(+)}_{ash}]}d\aconn$.}
Of course, again due to the
$\grouv\equiv\mathrm{Diff}(M)$-invariance of the theory, one would
expect the `physical' path integral to be

\begin{equation}\label{eq132}
\PI_{\infty}=\int_{{}^{4}\infconn^{(+)}/\grouv}e^{i[{}^{4}S^{(+)}_{ash}]}d([\aconn]_{\grouv})
\end{equation}

\noindent which, however, in order to make sense (even if only
`heuristically'!) care must be taken to make sure that one
integrates over a single member ${}^{4}\aconn_{\infty}^{(+)}$ from
each gauge equivalence class
$[{}^{4}\aconn_{\infty}^{(+)}]_{\grouv}$ in
${}^{4}\infconn^{(+)}/\grouv$. Among the aforementioned problems
of developing differential (and now integral) calculus on the
moduli space of (non-abelian) gauge (Y-M) theories and gravity, is
the fact that $\pi:~\infconn\mapto\infconn/\grouv$, regarded as a
principal $\grouv$-bundle, is non-trivial, that is to say, it has
no continuous global sections, which in turn means that there is
no unique gauge choice, no unique fixing or selecting a single
${}^{4}\aconn_{\infty}^{(+)}$ from each
$[{}^{4}\aconn_{\infty}^{(+)}]_{\grouv}$ in
${}^{4}\infconn^{(+)}/\grouv$. This is essentially the content of
the well known Gribov ambiguity in the usual $\smooth$-fiber
bundle-theoretic treatment of gauge theories.\footnote{The reader
should refer to \cite{mall2} for a more elaborate, albeit formal,
treatment, from an ADG-theoretic perspective, of the Gribov
ambiguity {\it \`a la} Singer \cite{singer}. What must be briefly
mentioned here is that the ADG-theoretic treatment of the Gribov
ambiguity in \cite{mall2} marks the commencement of the
development of a full-fledged differential geometry---again of a
non-classical, non-$\smooth$-smooth type---on the moduli space of
gauge-equivalent connections. For instance, one could take as
starting point for this development the following motivating
question: {\em what is the structure of the `tangent space'
$T({\mathcal{O}}_{\conn},\conn)$ to an orbit
${\mathcal{O}}_{\conn}$ of a connection $\conn$ in the affine
space $\sconn_{\struc}(\modl)$ of $\struc$-connections on a vector
sheaf $\modl$?} For example, in 3.4 we saw that,
ADG-theoretically, $T({\mathcal{O}}_{\conn},\conn)$ can be
identified with ${\mathcal{S}}_{\conn}^{\perp}$ (\ref{eq98}) and,
as a result, $T(M(\modl),{\mathcal{O}}_{\conn})$ with
$T({\mathcal{O}}_{\conn},\conn)$'s orthogonal complement ({\it
ie}, ${\mathcal{S}}_{\conn}$!) (\ref{eq101}). However, for the
latest results from the most analytical ADG-theoretic treatment of
moduli spaces of connections, the reader should refer to
\cite{mall4}.}

\medskip

All in all, however, again it all boils down to finding a measure
$d([\aconn]_{\grouv})$---in fact, a $\mathrm{Diff}(M)$-invariant
one, since (\ref{eq132}) involves smooth connections on a
$\smooth$-spacetime manifold $M$---on the moduli space
${}^{4}\infconn^{(+)}/\grouv$. Thus, we see how both the
non-perturbative canonical and the covariant approaches to quantum
gravity, whose formulation vitally depends on the classical
differential geometric apparatus provided by the $\smooth$-smooth
manifold (in fact, by the structure coordinate ring $\smooth(M)$)
and its structure group $\mathrm{Diff}(M)$, encounter the problem
of finding a $\mathrm{Diff}(M)$-invariant measure on their
respective moduli spaces. Below we argue how the ADG-theoretic
basis, on which our finitary, causal and quantal vacuum Einstein
gravity (\ref{eq117}) and its possible covariant path integral
quantization (\ref{eq121}) rest, bypasses completely significant
obstacles that these `conventional'
approaches\footnote{`Conventional' here means `classical', in the
sense that all these approaches are based on the usual
differential geometry of $\smooth$-manifolds. As we time and again
said before, these are approaches that belong to the category 1 of
`general relativity and manifold conservative' scenarios mentioned
in the prologue.} to quantum general relativity encounter.
Altogether, we emphasize that our approach is genuinely background
$\smooth$-smooth spacetime-free, fully covariant and that, based
on the fact that arguably all diseases ({\it ie}, singularities,
unrenormalizable infinities and other classical differential
geometric anomalies) come from assuming up-front $M$, it is
doubtful whether any `$\smooth$-conservative' attempt to quantize
general relativity (by essentially retaining $M$) will be able to
succeed.\footnote{Even more iconoclastically, in the following
subsection we will maintain that {\em our scheme is already
quantum}, so that {\em the quest for a quantization of gravity is
in effect `begging the question'}.}

In connection with the last remarks, cogent arguments coming from
\cite{df2,df5,jacob} further support the position that the attempt
to quantize gravity by directly quantizing general relativity
({\it ie}, by trying to quantize Einstein's equations in order to
arrive at the quantum of the gravitational force field---the
graviton) is futile, if one considers the following telling
analogy:

\begin{quotation}

\noindent {\em it is as if one tries to arrive at the fine
structure of the water molecule by quantizing the Navier-Stokes
equations of hydrodynamics}.

\end{quotation}

\noindent We definitely agree with this position; however, as we
saw before and we will crystallize in the next subsection, we
would not go as far as to maintain that in order to arrive at a
genuinely quantum theoresis of gravity one should first arrive at
a quantum description of (the background) spacetime structure
itself, for {\em spacetime does not exist} ({\it ie}, it has no
physical meaning). Rather, going quite against the grain of
theories that advocate either a `continuous' (classical) or a
`discrete' (quantum) spacetime, we will hold that a genuinely
covariant approach to quantum gravity should involve solely the
dynamical fields (and their quanta) without any dependence on an
external `spacetime substrate', whether the latter is assumed to
be `discrete' or `continuous'. This is what we mean by a `{\em
fully covariant}' (and `{\em already quantum}') picture of
gravity: {\em only the dynamical gravitational field (and its
quanta), and no ambient (external/background) spacetime which
forces one to consider its ({\it ie}, the spacetime's)
quantization, exists}.

\subsection{Cutting the Gordian Knot: No $\smooth$-Smooth Base Spacetime Manifold $M$, no
$\mathrm{Diff}(M)$, No Inner Product Problem, no Problem of Time,
a `Fully Covariant', `Purely Gauge-Theoretic' Lorentzian Quantum
Gravity}

In the present section we show how our finitary, ADG-based scheme
for `discrete' Lorentzian quantum gravity totally avoids three
huge problems that the differential manifold $M$,\footnote{Or
ADG-theoretically, the assumption of $\smooth_{M}$ for structure
sheaf $\struc$.}, or more precisely, its `structure group'
$\grouv\equiv\mathrm{Diff}(M)$\footnote{Here the term `structure
group' is not used exactly in the usual principal bundle and
gauge-theoretic sense. Rather fittingly, it pertains to the
`symmetries' of the structure sheaf $\struc$, which in the
classical case is identified with $\smooth_{M}$.} presents to both
the canonical and the covariant $\smooth$-manifold based
approaches to quantum gravity.

First, we would like to state up-front the main lesson we have
learned from ADG, which lesson, continuing the trend started in
\cite{malrap2}, we wish to promote to the following
slogan:\footnote{This is the second slogan in the present paper.
Recall the first one from the beginning of section 4.}

\begin{quotation}
\noindent {\bf Slogan 2.} {\em One can do differential geometry
without using any notion of calculus}; or what amounts to the
same, {\em without using at all (background) differential ({\it
ie}, $\smooth$-smooth) manifolds}
\cite{mall1,mall2,malros1,malros2,mall3,mall5,malrap1,mall6,malrap2,mall7,mall4}.
\end{quotation}

\noindent Thus, in the present paper, where ADG was applied to the
finitary-algebraic regime to formulate a causal and quantal
version of vacuum Einstein Lorentzian gravity, {\em no classical
differential geometric concept, construction, or result, and, of
course, no background (or base) $\smooth$-smooth spacetime
manifold, was used}. Precisely in this sense, our formulation of
(\ref{eq117}) and its covariant quantum version (\ref{eq121}) {\em
is genuinely background manifold-free or $\smooth$-smoothness
independent}.

Another basic moral of ADG which is invaluable for its direct
application to (quantum) gravity and (quantum) Y-M theories, and
which nicely shows its manifest evasion of the classical
differential geometry of $\smooth$-manifolds, can be expressed
diagrammatically as follows

\begin{equation}\label{add7}
{\fontsize{0.14in}{0.15in}
\begin{CD}
\boxed{\mathrm{CDG}\equiv\smooth\mathrm{-Manifolds}}@>(a)>\struc\equiv\smooth_{X}>\boxed{\mathrm{Tangent~
Bundles}}@>(b)>
\struc\equiv\smooth_{X}>\boxed{\smooth\mathrm{-Vector~Fields}}\\
@A(c^{'})A\struc\equiv\smooth_{X}~(X\subset M)A
@A(d^{'})A\struc\equiv\smooth_{X}~(X\subset M)A
@V(c)V\stackrel{(\smooth-\mathrm{connections})}{\smooth\mathrm{-derivations}}V\\
\boxed{\mathrm{ADG}\equiv\mathrm{arbitrary~base}~X}@>\mathrm{arbitrary}~\struc_{X}
>(a^{'})>\boxed{\mathrm{Vector~Sheaves}}@>\stackrel{(\struc-\mathrm{connections})}{\mathrm{\footnotesize
sheaf~morphisms}}>(b^{'})>\boxed{\mathrm{Differential~Equations}}
\end{CD}}
\end{equation}

\noindent which we can put into words again in the form of a
slogan:

\begin{quotation}
\noindent {\bf Slogan 3.} Unlike the Classical Differential
Geometry (CDG), whose (conceptual) development followed the path

\[
\begin{array}{c}
\mathrm{CDG\equiv Smooth~
Manifolds}\stackrel{(a)}{\mapto}\mathrm{Tangent~
bundles}\stackrel{(b)}{\mapto}\cr \mathrm{Smooth~Vector~
Fields}\stackrel{(c)}{\mapto}\mathrm{Differential~
Equations}(\equiv\mathrm{Physical~Laws})
\end{array}
\]

\noindent schematically described in (\ref{add7}), and which can
be read as follows: {\em the smooth manifold was made for the
tangent bundle, which in turn was made for the vector fields,
which were finally made for the differential equations (modelling
the local laws of classical physics;\footnote{In the concluding
section we will return to comment further on the fact that the
assumption of a differential manifold ensures precisely that the
dynamical laws of physics obey the classical principle of
locality.})} in contradistinction, the development of ADG followed
the path

\[
\mathrm{ADG}\stackrel{(a^{'})}{\mapto}\mathrm{Vector~Sheaves}
\stackrel{(b^{'})}{\mapto}\mathrm{Differential~Equations}
\]

\noindent which can be read as follows: {\em ADG refers in an
algebraico-categorical way directly to the dynamical
fields---represented by pairs such as `$(\modl ,\conn)$'---without
the intervention (neither conceptually nor technically) of any
notion of (background geometric manifold) space(time), or
equivalently, independently of any intervening coordinates}. In
other words, {\em ADG deals directly with the differential
equations (the laws of physics), which now are `categorical
equations' between sheaf morphisms---the $\struc$-connections
$\conn$ acting on the (local) sections of vector sheaves $\modl$
under consideration.} Of course, one can recover CDG from ADG by
identifying one's structure sheaf $\struc$ with
$\smooth_{M}$\footnote{$(a)$ in (\ref{add7}).} thus, in effect,
`descend' from abstract, algebraic in nature, vector sheaves to
the usual smooth vector or frame (tangent) bundles over (to) the
geometrical base spacetime $\smooth$-manifold $M$ ($c^{'},d^{'}$).
\end{quotation}

\subsubsection{Avoiding the problems of $\mathrm{Diff}(M)$ by
avoiding $M$}

Below, we mention three problems that our finitary-algebraic,
ADG-based perspective on quantum gravity manages to evade
completely. We choose to pronounce these problems via a comparison
between the canonical and the covariant $\smooth$-manifold based
approaches to quantum general relativity described above, and our
ADG-theoretic locally finite, causal and quantal Lorentzian vacuum
Einstein gravity. In particular, we initiate this comparison by
basing our arguments on the contents of footnote 212 which makes
it clear what the essential assumptions about the
$\smooth$-approaches to quantum gravity are, and it also
highlights their characteristic absence from our ADG-founded
theory. In this way, the value of the slogans 1--3 above can be
appreciated even more.

\begin{enumerate}

\item The fundamental assumption of all the non-perturbative $\smooth$-conservative
approaches to quantum gravity, whether Hamiltonian or Lagrangian,
is that {\em there is a background geometrical spacetime which is
modelled after a $\smooth$-smooth base manifold} $M$. Thus, the
point-events of $M$ are coordinatized by $\smooth$-smooth
functions whose germs generate the classical structure sheaf
$\struc\equiv\smooth_{M}$; hence, the natural `structure group' of
all those $M$-based scenarios is $\grouv\equiv\mathrm{Diff}(M)$.

\item The next assumption (of great import especially to the
canonical approach via the Ashtekar variables) we can read
directly from footnote 212: there is a (frame) bundle isomorphism
$e$ between $TM$ and an `internal' Minkowskian bundle
$\mathcal{T}$,\footnote{We may coin $e$ {\em the (local)
`external' Lorentzian $\smooth$-manifold $M$ soldering form}. It
may be thought of as the `umbilical cord' that ties (and feeds!)
all the differential geometric constructions used in
non-perturbative canonical or covariant quantum general relativity
with (from) the background smooth manifold $M$.} whose inverse
$e^{-1}$ defines a local {\it vierbein} ($4$-frame) field variable
on $M$\footnote{By abusing notation, we also denote the {\it
vierbein} by $e$.} and secures the faithful transference of the
classical $\smooth$-differential geometric structures, such as the
smooth (self-dual) connections $\aconn^{(+)}_{\infty}$, the smooth
Lorentzian metric $\eta$, the volume form $\vol$, the smooth
vector fields (derivations) and covectors (differential forms)
{\it etc}, from $\mathcal{T}$ to $TM$.\footnote{Hence our calling
$e$ above a (local) `external' Lorentzian $\smooth$-manifold $M$
soldering form. (Recall also from footnote 212 that $\eta$, which
is pulled back by $e^{-1}$ from $\mathcal{T}$ to $TM$, effects the
canonical isomorphism between $TM$---inhabited by
vectors/derivations tangent to $M$, and its dual
$TM^{*}$---inhabited by covectors/forms cotangent to $M$.)} In a
nutshell, $e^{-1}$ ensures that $TM$ comes fully equipped with the
classical (tangent bundle) differential geometric apparatus.

\item When it comes to (especially the canonical) dynamics, one can easily see
how this $\smooth$-spacetime bound language gives independent
physical existence and `reality' to the background ({\it ie},
`external' to the dynamical fields themselves) geometrical smooth
spacetime continuum itself, by statements such as,

\begin{quotation}
\noindent ``{\small\em In this approach\footnote{That is, the
canonical approach to quantum general relativity {\it \`{a} la}
Ashtekar.} the action of diffeomorphism group gives rise to two
constraints on initial data: the diffeomorphism constraint, which
generates diffeomorphisms preserving the spacelike hypersurface,
and the Hamiltonian constraint, which generates diffeomorphisms
that move the surface in a timelike direction.}''\footnote{Taken
from the preface of the book ``{\sl Knots and Quantum Gravity}''
where \cite{ashlew1} and \cite{loll} belong. The constraints
mentioned in this excerpt are precisely the four `external'
$\smooth$-smooth spacetime manifold $\mathrm{Diff}(M)$-constraints
in (\ref{eq126}).}
\end{quotation}

\noindent In the canonical Ashtekar approach, this is concisely
encoded in the assumption that the smooth $4$-frame field $e$ is
an independent (local) dynamical variable along with the
(self-dual) smooth spin-Lorentzian connection $1$-form
$\aconn^{(+)}_{\infty}$.\footnote{And recall from (\ref{eq122})
and (\ref{eq123}) that the vacuum Einstein equations are obtained
from deriving the Palatini-Ashtekar action functionals with
respect to $e$.}
\end{enumerate}

By striking contrast, our finitary, causal and quantal ADG-based
approach to Lorentzian vacuum Einstein gravity assumes neither
$M$\footnote{Thus it gives the smooth spacetime manifold no
independent physical (dynamical) reality `external' to the
dynamical gravitational gauge field itself (represented by the
connection).} (and, as a result, no $\mathrm{Diff}(M)$ either),
but perhaps more importantly, nor $e$. ADG in a sense cuts the
`umbilical cord' ($e$) that ties (and sustains differential
geometrically) the $\smooth$-conservative approaches to (by) the
background spacetime manifold $M$, and it concentrates directly on
(the physical laws for) the dynamical objects---in our case, the
(self-dual) fcqv-dynamos $\conf^{(+)}_{i}$---that live and
propagate on `it'. All in all we must emphasize that

\begin{quotation}
\noindent {\em the sole dynamical variable in our scheme is the
reticular (self-dual) spin-Lorentzian connection variable}
$\conf^{(+)}_{i}$ (in fact, {\em the fcqv-E-L-field}
$(\qmodll_{i},\conf_{i})$) {\em and ADG enables us to formulate
directly the dynamical equations for it without having to account
for ({\it ie}, without the mediation and support of) a background
geometrical smooth spacetime manifold $M$. In this sense, our
ADG-theoretic, connection-based approach is more algebraic and
more `pure gauge-theoretic' ({\it ie}, `fully covariant'---see
below) than the approaches to gravity which are based on the
classical $\smooth$-differential geometry of the smooth spacetime
manifold ({\it eg}, Ashtekar's). At the same time, since there is
no `external' spacetime manifold, there is no need either to
perform the necessary for the canonical quantization procedure
$3+1$ space-time split which, as we contended earlier, breaks
manifest covariance. Furthermore, $\mathrm{Diff}(M)$ is now
replaced, in a Kleinian sense} \cite{mall7}, {\em by the structure
group $\qaut_{i}$ of $\Qalg_{i}$-automorphisms of $\modll_{i}$
({\it ie}, the group of the reticular transformations of the
causon field itself---its dynamical self-transmutations so to
speak\footnote{It must be stressed that, according to the
geometric (pre)quantization axiomatics
\cite{mall2,mall5,mall6,mall7,mall4} that we subjected our causon
field $\conf_{i}$, or better, its associated fcqv-dynamo E-L field
$(\qmodll_{i},\conf_{i})$ in \cite{malrap2}, we can identify the
latter with its quanta (`particles')---the causons ({\it eg},
states of `bare' or free causons, when regarded as bosons---the
`carriers' of the dynamical field of quantum causality, are
represented by sections of line bundles $\vec{{\mathcal{L}}_{i}}$
associated with the $\peel_{i}$s \cite{malrap2}). Thus, one can
also think of $\qaut_{i}$ as acting directly on the dynamical
quanta of quantum causality---the causons. Shortly, we will
revisit some basic geometric (pre)quantization arguments from
\cite{mall6} in order to further support these remarks.}). All in
all, our approach is fully (gauge) covariant.\footnote{We are
tempted to call our scheme, after Einstein, {\em `unitary' field
theory}, since all that there is in it are the dynamical fields
(plus their associated quanta and their automorphisms) and no
ambient, external spacetime present. Because we have formulated
gravity purely gauge-theoretically ({\it ie}, as the dynamics
solely of the connection), we may alternatively coin our scheme
{\em `pure gauge' field theory}.}}
\end{quotation}

Now that we have stated, and analyzed in glaring contrast to the
$\smooth$-conservative canonical and covariant approaches to
quantum general relativity, the three slogans underlying our
fcqv-approach to Lorentzian vacuum Einstein gravity, we are in a
position to show how our theory simply evades the following three
caustic issues for non-perturbative quantum gravity:

\begin{enumerate}
\item {\sl The inner product problem}: in the canonical approach, this refers to the problem
of fixing the inner product in the Hilbert space of physical
states by requiring that it is invariant under $\mathrm{Diff}(M)$.
As noted earlier, in effect it is the problem of finding a
$\mathrm{Diff}(M)$-invariant measure. The same technical problem
({\it ie}, the problem of finding a $\mathrm{Diff}(M)$-invariant
measure) essentially persists in the fully covariant path integral
quantization approach to quantum general relativity (\ref{eq131})
and (\ref{eq132}). Since our theory is genuinely $\smooth$-smooth
manifold $M$-free, thus also manifestly
$\mathrm{Diff}(M)$-independent, it simply avoids the inner product
problem. We thus write

\begin{equation}\label{add8}
\Ovalbox{No~smooth~manifold~{\it
M}}\Rightarrow\Ovalbox{No~Diff({\it
M})}\Rightarrow\Ovalbox{No~inner~product~problem}
\end{equation}

\noindent However, it must be said that if one employs finite
dimensional (Hilbert) space representations for the incidence
algebras modelling qausets as in \cite{zap0,rapzap2}\footnote{But
note that in these works the incidence algebras are of a
topological, not a directly causal, nature.} and one regards the
latter spaces as inhabiting the stalks of associated finsheaves to
the $\peel_{i}$s, or even if one just works with the aforesaid
associated line sheaves $\vec{{\mathcal{L}}_{i}}$ of states of
`bare' or free causons, the issue of finding well defined
integration measures on them still persists. Generalized
integration theory \cite{bourbaki} and {\em Radon-type of
measures} on vector sheaves similar to the aforesaid `cylindrical'
ones employed by Ashtekar and Lewandowski (using Gel'fand's
spectral theory) in the context of the holonomy $C^{*}$-algebraic
approach to canonical quantum general relativity
\cite{ashish,ashlew1,ashlew2}, are currently under intense
development by ADG-theoretic means \cite{mall4}. Such measures are
expected to figure prominently in (and make mathematical sense of)
heuristic (path) integrals like (\ref{eq129})--(\ref{eq132}) and,
in the finitary case, like (\ref{eq120}) and
(\ref{eq121}).\footnote{Indeed, of special interest to ADG is to
develop a general and mathematically sound integral calculus on
the moduli spaces of gauge-equivalent connections on vector
sheaves (those in particular that appear in the ADG-theoretic
treatment of Y-M theories and gravity \cite{mall1,mall2,mall3})
again, {\em independently of the classical, differential
manifold-based, theory} \cite{mall4}. Such an abstract or
generalized integration theory could be regarded as the
ADG-theoretic analogue of the generalized integration and measure
theory that has been developed (albeit, still in the
$\smooth$-context!) in \cite{baez2,baez3}.}

\item {\sl The problem of time}: again in the context of canonical quantum general relativity,
this refers to the problem of requiring that the dynamics is
encoded in the action of $\mathrm{Diff}(M)$ on the (Hilbert) space
of physical states. Here too, our evasion of this problem is
rather immediate:

\begin{equation}\label{add9}
\Ovalbox{No~smooth~space{\em time}~manifold~{\it
M}}\Rightarrow\Ovalbox{No~Diff({\it
M})}\Rightarrow\Ovalbox{No~problem~of~time}
\end{equation}

\noindent For, as we have repeatedly argued above, our theory
deals directly with the dynamical physical objects
$(\conf_{i},\qmodll_{i})$ themselves and their
(self-)transformations (`structure symmetries') $\qaut_{i}$, and
does not posit the existence of an external (background) spacetime
continuum, let alone regard the latter as being physically
significant in any way.\footnote{The reader should refer to the
concluding section where further criticism is made of the base
spacetime manifold $M$ and its differentiable automorphisms
$\mathrm{Diff}(M)$, as both are regarded as the last relics of an
absolute, ambient, inert (non-dynamical), ether-like substance.}
In our scheme, $\qaut_{i}$ acts directly, via its representations
alluded to in 1 above, on the associated (line) sheaves of bare
causon states \cite{rapzap2}.\footnote{See further remarks on
geometric (pre)quantization that follow shortly.}

\item {\sl The problem of `full covariance'}: as in 2 above, this problem essentially
comes from assuming that the external, background spacetime
manifold is a physical entity---and not paying attention just to
the dynamical objects (fields and their particles) that live on
that `spacetime' which, anyway, are the only `physically real'
(`observable') entities. One is tempted to say here that the
reason for this (problem) was in effect the lack of having thus
far an appropriate framework to develop differential geometry---at
least to the extent that ADG for instance has
developed---different from that of the classical theory. In this
respect, we may still recall here Einstein's `confession' in
\cite{einst2}:

\begin{quotation}
\noindent {\small\em ``...Adhering to the continuum originates
with me not in a prejudice, but arises out of the fact that I have
been unable to think up anything organic to take its place...''}
\end{quotation}

\noindent which we will mention again in 6.1 in connection with
the singularities that assail the classical theory. In other
words, the desirable scenario here is

\begin{quotation}
\noindent {\em the formulation of the (quantum) gravitational
dynamics solely in terms of the connection} $\con$, or more
completely, {\em in terms of the `full', `unitary' or `pure' E-L
field $(\modll ,\con)$, and nothing else---in particular, without
referring to an external (background) spacetime (whether the
latter is assumed to be discrete or a continuum)}.
\end{quotation}

As we saw earlier, in the canonical (Hamiltonian) approach to
quantum general relativity there is a manifest breaking of
covariance by the necessary $3+1$ dissection of the (external)
spacetime continuum into space and time. Also, in a supposedly
covariant path-integral-type of quantization scenario for
Lorentzian gravity like (\ref{eq131}) or (\ref{eq132}), although
there is no such an explicit external space-time split, there
still persists however (built into the very CDG-formalism
employed) the assumption of an external (background) geometrical
$M$ experiencing, for instance, problems like 1.\footnote{Let
alone that in the actual implementation and interpretation of the
path integral as a dynamical transition amplitude in the
kinematical (moduli) space of gravitation $4$-connections,
`boundary $3$-geometries', which break full covariance, are
implicitly fixed at the end-points of the otherwise indefinite
integral (see footnote 242).}

\end{enumerate}

\subsubsection{A brief note on geometric (pre)quantization}

Now that we have argued about how our theory can evade completely
the inner product (Hilbert space) problem and the problem of time
essentially by avoiding altogether the background $M$ and its
`structure group' $\mathrm{Diff}(M)$, as well as how it may used
to formulate a `fully covariant' (quantum) dynamics for finitary
and causal vacuum Einstein Lorentzian gravity, we would like to
say a few words about another concrete application of ADG which
further supports those arguments. This application concerns the
subject of the so-called {\em Geometric (Pre)quantization} (GPQ)
\cite{mall2,mall5,mall6}.\footnote{In what follows, we do not
intend to present any technical details from
\cite{mall2,mall5,mall6}; rather, we would like to give a brief
outline of certain syllogisms and results of this application that
further vindicate the aforesaid evasion by our ADG-based theory of
the three problems of the background spacetime manifold based
quantum general relativity theories whether they are Hamiltonian
(canonical) or Lagrangian (path integral). As noted in footnote
258, we gather results mainly from \cite{mall6}.}

We read from \cite{mall6} that the main aim of GPQ is to arrive at
a quantum model of a relativistic particle---which is assumed to
be in the spectrum ({\it ie}, a so-called quantum particle
excitation) of a corresponding quantum field---{\em without having
to first quantize the corresponding classical mechanical system}
\cite{simms}. In other words, GPQ aspires to a quantum description
of elementary particles by referring directly to their (`second
quantized') fields ({\it ie}, without the mediation of the
procedure of first quantization of the classical mechanical or
field theory and of the conventional Hilbert space formalism that
accompanies it). On the other hand, it is well known that GPQ
heavily rests on the usual differential calculus of
$\smooth$-smooth (symplectic) manifolds\footnote{See remarks of
Isham from \cite{ish3} in the concluding paragraph of this section
on GPQ.}; hence, it is no surprise that ADG could be used to
generalize the foundations of GPQ, thus gain more insight into the
theory.

For instance, as we witnessed above, ADG completely circumvents
the underlying $\smooth$-smooth spacetime manifold and deals
directly with the (algebraic) objects that live on `it'. These
objects are the dynamical fields themselves (without recourse to
an external base spacetime manifold) or equivalently, in a purely
second quantized sense, the elementary particles (quanta) of these
fields. In fact, the main objective of applying ADG-theoretic
ideas to GPQ, basically motivated by certain fiber bundle
axiomatics originally laid down by Selesnick in \cite{sel}, is to
show that

\begin{quotation}
\noindent {\em elementary particles---the quanta of the dynamical
fields---can be classified according to their spin in terms of
appropriate vector sheaves $\modl$.}
\end{quotation}

\noindent In this respect, the main result of ADG applied to GPQ
is that

\begin{quotation}
\noindent {\em states of bare (free) bosons can be identified with
local sections of line sheaves $\lsh$,\footnote{That is to say,
vector sheaves of rank 1.} while states of bare (free) fermions
with local sections of vector sheaves $\modl$ of rank greater than
$1$} \cite{mall2,mall5,mall6}.
\end{quotation}

In order to arrive at that result, the first author had to posit
the following identifications, or better, make the following
bijective correspondences (`equivalences'), which we readily read
from \cite{mall6}:

\begin{enumerate}

\item States of elementary particles can be associated with
(local) sections of appropriate vector sheaves $\modl$, the latter
being provided in the classical theory by the sheaves of sections
of vector bundles over the spacetime manifold $M$ {\it \`a la}
Selesnick \cite{sel}.\footnote{By Selesnick's work \cite{sel},
these bundles correspond to finitely generated projective modules
over the topological algebra $\smooth(M)$ of the smooth spacetime
manifold $M$. ADG's primitive assumption of a general structure
sheaf $\struc$ other than $\smooth_{M}$ generalizes Selesnick's
bundles to vector sheaves $\modl$ that are locally free
$\struc$-modules of finite rank, as we saw before.}

\item An elementary particle---the irreducible constituent of matter---corresponds
uniquely, in a second quantized sense, to the quantum of a
particle field\footnote{The notion of `field' being regarded here
as an irreducible ({\it ur}) element of the theory, in the same
way that Einstein thought of it as ``{\em an independent, not
further reducible, fundamental concept}'' \cite{einst3}.}; one
writes:

\begin{equation}\label{add11}
\ovalbox{physical particle}\longleftrightarrow\ovalbox{particle
field}
\end{equation}

\item {\em A field, hence its quanta (elementary particles), is completely determined by its states}. The latter,
within the axiomatic framework of ADG, correspond to local
sections of suitably defined vector sheaves $\modl$. All in all,
one writes

\begin{equation}\label{add12}
\ovalbox{particle}\longleftrightarrow\ovalbox{field}\longleftrightarrow\ovalbox{states}\longleftrightarrow\ovalbox{local
sections}\longleftrightarrow\ovalbox{vector sheaf}
\end{equation}

\noindent with the latter identification ($\mathrm{local~sections}
\longleftrightarrow\mathrm{vector~sheaf}$) being, as a matter of
fact, a well known theorem in sheaf theory.\footnote{That is to
say, any (vector) sheaf is completely determined by its (local)
sections \cite{mall1,mall2}. In fact, in \cite{mall1} this has
been promoted to the following important slogan: ``{\em a sheaf is
its sections}''. So, there is a very close physico-mathematical
analogy lurking in (\ref{add12}): {\em in the same way that a
sheaf is completely determined by its sections, an elementary
particle}---{\it ie}, the quantum of a field---{\em is completely
determined by its states}.}

\item In fact, as we saw earlier, by `field' ADG
understands the pair $(\modl ,\conn)$.\footnote{This vector
sheaf-theoretic conception of a field by ADG comes as an
abstraction and vector sheaf-theoretic generalization of Manin's
fiber bundle-theoretic definition of the Maxwell's field of
electrodynamics as the pair $(\lsh_{Max},\conn_{Max})$ consisting
of a ($U(1)$) connection $\conn_{Max}$ on a line bundle
$\lsh_{Max}$ of `photon states' \cite{manin}. It is also important
to remark here that, semantically, ADG regards the connection
$\conn$ as `{\em the dynamical field proper}', while $\modl$ as
`{\em the carrier (state) space of (the particles or quanta of)
the field}'. In fact, both $\conn$ and $\modl$ are needed for
formulating the laws of nature (`differential equations') as
$\modl$ provides us with the sections (states of the
particle---the `Being' of the particle so to speak) on which
$\conn$ acts ({\it ie}, dynamically transforms the particle---the
`Becoming' of the particle so to speak). It is conceptually lame,
perhaps even `wrong', from the ADG-theoretic perspective to think
of $\modl$ (`state') apart from $\conn$ (`transformation of
state') and vice versa. The concept of field in ADG, as the pair
$(\modl,\conn)$, is a `holistic', `unitary' or `coherent' one, not
separable or `dissectible' into its two constituents.}

\item Finally, and very briefly, starting from work by
Selesnick in \cite{sel}, the first author was led to realize that
one can model the collection of quantum states of free elementary
particles by {\em finitely generated projective
$A$-modules}\footnote{Finiteness pertaining to the finite
dimensionality of the representations of the particles' compact
structure (symmetry) gauge group.} and then, depending on their
spin, classify them to {\em free bosons} whose states comprise
projective $A$-modules of rank 1, and {\em free fermions} having
for states elements of projective $A$-modules of rank greater than
or equal to 2.\footnote{In particular, by taking $A$ to be
$\smooth(M)$ \cite{mall6}.} Then, the transition to locally finite
$\struc$-modules $\modl$ of finite rank ({\it ie}, the vector
sheaves of ADG) was accomplished by using the Serre-Swan theorem
(suitably extended from the Banach algebra $A=\cont(M)$ on a
compact Hausdorff manifold $M$ to general topological non-normed
(non-Banachable) algebras such as $\smooth(M)$) in order to go
from the aforesaid finitely generated $\smooth(M)$-modules to
smooth vector bundles on $M$. Then the latter can provide us with
the (local) sections we need to build our $\modl$s.

\item All in all, the general result of applying ADG to GPQ is the
following `categorical' statement \cite{mall2,mall5,malrap2}

\begin{quotation}
\noindent {\em every (free) elementary particle is
(pre)quantizable ({\it ie}, it admits a (pre)quantizing line
sheaf)}.
\end{quotation}

\end{enumerate}

It must be noted here that the sheaf-cohomological classification
of our fcqv-E-L fields $(\conf_{i},\qmodll_{i})$ and their quanta
(causons) in \cite{malrap2} is essentially an application of the
results of the ADG-theoretic perspective on GPQ above to the
finitary, causal and quantal regime. {\it In toto}, and this is
the main reason we briefly alluded to ADG {\it vis-\`a-vis} GPQ
here,

\begin{quotation}
\noindent{\em being able, by circumventing ADG-theoretically the
classical external $\smooth$-spacetime manifold $M$, to refer
directly to the dynamical objects (fields), we can show not only
that (the dynamics of) these objects are `fully covariant', but
also that they are `intrinsically' of a quantum
nature,\footnote{That is, dealing directly and exclusively with
the propagating field is equivalent to dealing directly and solely
with its dynamical quantum (particle).} so that the quest for a
`blindfolded', head-on quantization of spacetime and general
relativity\footnote{That is, of the dynamics of the smooth
gravitational field (whether this is represented by the metric or
the connection-{\it cum}-frame field) propagating on a
$\smooth$-spacetime manifold.} appears to be begging the question.
Indeed, since our scheme is `fully covariant', `inherently
quantum'\footnote{In fact, we are tempted to regard these two
characterizations of our theory ({\it ie}, `fully covariant' and
`intrinsically quantum') as being equivalent, for ADG refers
directly to the dynamical fields and their quanta. Some strong
conceptual resonances with Einstein's vision of a unitary field
theory (which can `explain' quantum phenomena) are pretty obvious
here.} and it certainly does not arise from `quantizing somehow
the classical theory', we strongly doubt whether actually
quantizing a classical theory is physically meaningful at
all.\footnote{For instance, since first quantization is totally
bypassed by GPQ, there is {\it prima facie} no need for reasoning
`conventionally' ({\it ie}, by using Hilbert spaces, `observables'
and the rest of the conventional jargon, methods and technical
baggage of quantum mechanics) about causons and their dynamics. In
fact, the correspondence principle advocated initially in
\cite{rapzap1,rapzap2} about the incidence algebras modelling
discrete and quantum topological spaces should by no means be
regarded as a `consistency' or `physicality check' of our theory
({\it ie}, as if our theory {\em should} yield classical gravity
as a `low energy or weak gravitational field limit' in the same
way that the other discrete spacetime or continuum-based
approaches to quantum gravity are expected to). From the purely
ADG-theoretic point of view, immediate contact with the classical
theory is established simply by setting
$\struc\equiv\smooth_{M}$.}} Thus, with respect to the ADG-based
theory for fcqv-E-L gravity propounded here, to this last question
whether quantizing a classical theory (in our case, general
relativity) is physically meaningful at all, one might respond by
remarking that {\em this always depends on the type of theory that
one employs in order to describe the physical laws through the
corresponding (differential) equations.}
\end{quotation}

\noindent The last remarks would strike one, who is used to the
idea that one should be able to arrive at a quantum theory of
gravity by quantizing somehow general relativity ({\it ie}, by
employing a formal quantization procedure involving the usual
quantum mechanical concepts and mathematical structures such as
`observables', Hilbert spaces {\it etc} while still retaining the
classical calculus-based framework for both an external spacetime
and the dynamical laws for the now quantized fields on it), as
being at best odd, if we also quote the following passage from a
celebrated textbook that has nurtured generations and generations
of theoretical physicists \cite{lanlif}:

\begin{quotation}
\noindent ``{\small Quantum mechanics occupies a very unusual
place among physical theories: it contains classical mechanics as
a limiting case, yet at the same time {\em requires}\footnote{Our
emphasis.} this limiting case for its own formulation}.''
\end{quotation}

\noindent the emphasized`{\em requires}' being here the `operative
word'---precisely the one we have challenged and doubted in the
present paper.\footnote{In our case, one should substitute the
word `mechanics' by `gravity' or even by `general relativity' in
the quotation above in order to get a better feeling of the point
we wish to make. (Of course, this is an imaginary, `wishful
thinking' situation in which we are talking about quantum gravity
as if it has already been formulated!)} For, as it was noted at
the end of 5.3.2, we already have strong indications that trying
to quantize head-on general relativity is perhaps not the right
way to a quantum theory of gravity \cite{df2,df5,jacob}. In a
nutshell then, {\em we doubt that quantum gravity is,} or better,
{\em will prove to be \underline{quantized} gravity}.

We would like to close this discussion of the ADG-theoretic
perspective on GPQ with some very pertinent remarks of Isham in
his latest paper \cite{ish3}\footnote{The excerpts below are taken
from 2.1.1 in \cite{ish3}.} which emphasize precisely how the
(geometric) quantization of a classical theory is fundamentally
(and quite {\it a priori}ly, {\it ad hoc}, thus
inappropriately---especially for quantum gravity research) based
on the classical differential geometry of smooth manifolds
(essentially because the conventional quantum theory itself, which
we apply when we wish to quantize a classical theory, is based on
the manifold model for spacetime\footnote{See again related
comments in our discussion of the use of $\R$ and $\com$ in our
theory in 5.1.}).

\begin{quotation}
\noindent ``{\em\small ...In general, {\rm [when we start from a
classical theory and then `quantise' it]}, the configuration space
(if there is one) $Q$ for a classical system is modelled
mathematically by a differentiable manifold and the classical
state space is the co-tangent bundle $T^{*}Q$. The physical
motivation for using a manifold to represent $Q$ again reduces to
the fact that we represent physical space with a manifold...

Thus, in assuming that the state space of a classical system of
the form $T^{*}Q$ we are importing into the classical theory a
powerful a priori picture of physical space: namely, that it is a
differentiable manifold.\footnote{``There may be cases [like those
arising in the context of geometric quantization theory] where
$\mathcal S$ is a symplectic manifold that is not a cotangent
bundle; for example, ${\mathcal{S}}=S^{2}$. However, I would argue
that the reason $\mathcal S$ is assumed to be a {\em manifold} is
still ultimately grounded in an {\it a priori} assumption about
the nature of physical space (and time).'' (Our addition in square
brackets.)} This then carries across to the corresponding quantum
theory. For example, if `quantization' is construed to mean
defining the quantum states to be cross-sections of some flat
vector bundle over $Q$, then the domain of these state functions
is the continuum space $Q$...''}
\end{quotation}

\noindent This is more or less how (second) `quantization' was
originally construed fiber bundle-theoretically in \cite{sel} and
then was treated ADG-theoretically to suit GPQ ideas---albeit, in
the characteristic absence of a $\smooth$-smooth base spacetime
continuum (domain)---in \cite{mall2,mall5,mall6} and, in the
finitary spacetime and gravity case, in \cite{malrap2}. From this
point of view, this is another indication that our finitistic
theory for vacuum Einstein-Lorentzian gravity here may be regarded
as being `already quantized' (better, `inherently
quantum')---albeit, not at all `conventionally' in Isham's sense
of the word (which means that one applies the usual quantum
theory, with its classical manifold conception of space and time,
to an already existing classical theory).

\subsubsection{Remarks on Einstein's `new ether' and unitary field theory vis-\`{a}-vis `full covariance'}

Here we would like to bring together certain ideas that were
expressed above---in particular, in connection with the full
covariance of our theory, the identifications (\ref{add11}) and
{\ref{add12}) in the context of geometric (pre)quantization, as
well as with some allusions made earlier to our hunch that our
scheme is `already quantum', as it were, not in need of quantizing
({\it ie}, applying quantum theory to) the classical theory of
gravity (general relativity)---and some of Einstein's searching
thoughts about a new conception of `ether' in the light of his
continuous unitary field theory, singularities and the quantum
paradigm.\footnote{By unitary field theory we do not refer so much
to the more well known, life-long endeavor of Einstein to unify
gravity with electromagnetism and regard material particles as
being special states of condensed energy of ({\it ie},
`singularities' or `discontinuities' in) the (continuous) unified
field \cite{berg1}, as to his general intuition---which is of
course closely related to his well known unitary field theory
project---that all physical actions (including quantum matter)
must be described in terms of (continuous) fields. However, below
we are also going to comment on unified field theory in the more
popular sense of the term.} We will see how Einstein (i) tried to
respect as much as he could general relativity which posits an
ether-like spacetime background in the form of the differential
manifold and the smooth metric field imposed on this spacetime
continuum, (ii) always kept in mind the earlier abolition of the
material ether by special relativity so that he was careful not to
attribute mechanical properties to the ambient geometrical
spacetime continuum,\footnote{In a sense, field theory is not
mechanistic.} and (iii) was deeply impressed by the discontinuous
actions of (matter) quanta, and he intuited---at times in an
`oxymoronic' way which reflects precisely the opposite tension in
his mind between the continuous/geometrical actions of (special
and) general relativity and the discrete/algebraic ones of quantum
theory---a new kind of `ether' intimately related to the spacetime
continuum which may be cumulatively referred to as `{\em the
continuous unitary field}'. Then, we will discuss the affinities
and the fundamental differences between the latter, continuum
spacetime metric field-based (geometrical) and our ADG-theoretic,
connection-based `fully covariant' and `inherently quantum'
(reticular-algebraic) vacuum Einstein-Lorentz gravity. Along with
the Einstein references at the back, in the sequel we borrow some
of Einstein's quotations and various ideas about this rebirth of
the notion of ether from \cite{kostro}.

We commence with a quotation of Einstein, as early as 1924, in
which, in spite of the abolition of the `material' and
`mechanical' luminipherous ether by the special theory of
relativity already almost two decades earlier, he insists that in
the context of a continuous field theory on a spacetime continuum
the notion of ether (even if a generalized, non-mechanistic or
non-material one) is physically quite indespensible. For example,
he concludes the article `{\itshape \"{U}ber den \"{A}ther}'
\cite{einst1} as follows:

\begin{quotation}
\noindent ``{\em\small ...But even if these possibilities should
mature into genuine theories, we will not be able to do without
the ether in theoretical physics, {\it ie}, a continuum which is
equipped with physical properties; for the general theory of
relativity, whose basic points of view surely will always
maintain, excludes direct distant action. But every contiguous
action theory presumes continuous fields, and therefore also the
existence of an `ether'.}''\footnote{While, already four years
earlier \cite{einst5}, he had stressed the `ether imperative' in
physics as follows: ``{\em ...The ether hypothesis must always
play a part in the thinking of physicists, even if only a latent
part.}''.}
\end{quotation}

\noindent Therefore, for Einstein, the spacetime continuum,
supporting continuous fields, provides a {\em new ether} paradigm.
At the same time, he readily and repeatedly denied the independent
physical existence of space(time) apart from the continuous field
and the (in his own words) `physical continuum' ({\it ie}, the
ether) that supports or `carries' it, much as follows:

\begin{quotation}
\noindent ``{\em\small ...According to general relativity, the
concept of space detached from any physical content does not
exist. The physical reality of space is represented by a field
whose components are continuous functions of four independent
variables---the coordinates of space and time. It is just this
particular kind of dependence that expresses the spatial character
of physical reality.}'' \cite{kostro}\footnote{Page 175 and
reference therein.}
\end{quotation}

\noindent and

\begin{quotation}
\noindent ``{\em\small ...If the laws of this field are in general
covariant, that is, are not dependent on a particular choice of
coordinate system, then the introduction of an independent
(absolute) space is no longer necessary. That which constitutes
the spatial character of reality is simply the four-dimensionality
of the field. There is no `empty' space, that is, there is no
space without a field.}'' \cite{kostro}\footnote{Again, page 175
and reference therein.}
\end{quotation}

\noindent and, in a sense that was emphasized throughout the
present paper, he essentially maintained that {\em (the)
space(time) continuum and, concomitantly, the (new) ether is
inherent in the (gravitational) field}\footnote{Which, unlike in
our algebraic, connection-based theory however, he identified with
(the components of) the metric tensor $g_{\mu\nu}$.}:

\begin{quotation}
\noindent ``{\em\small ...No space and no portion of space can be
conceived of without gravitational potentials; for these give it
its metrical properties without which it is not thinkable at all.
The existence of the gravitational field is directly bound up with
the existence of space...}'' \cite{einst5}
\end{quotation}

\noindent also

\begin{quotation}
\noindent ``{\em\small ...according to the general theory of
relativity even empty space has physical qualities, which are
characterized mathematically by the components of the
gravitational potential.}'' \cite{kostro}\footnote{Again, page 111
and reference therein.}
\end{quotation}

\noindent and

\begin{quotation}
\noindent ``{\em\small Thus, once again `empty' space appears as
endowed with physical properties, {\it i.e.}, no longer as
physically empty, as seemed to be the case according to special
relativity. One can thus say that the ether is resurrected in the
general theory of relativity, though in a more sublimated form.}''
\cite{kostro}\footnote{Page 111 and reference therein.}
\end{quotation}

\noindent furthermore

\begin{quotation}
\noindent ``{\em\small ...There is no such thing as empty space,
{\it i.e.}, a space without field. Space-time does not claim
existence on its own, but only as a structural quality of the
field...}'' \cite{einst8}
\end{quotation}

\noindent and

\begin{quotation}
\noindent ``{\em\small ...space has lost its independent physical
existence, becoming only a property of the field...}''
\cite{einst3}\footnote{This brings to mind the remarks, albeit in
the context of the flat spacetime (quantum) field theory of
matter, of Denisov and Logunov: ``{\em\small ...Minkowski was the
first to discover that the space-time, in which all physical
processes occur, is unified and has a pseudo-Euclidean geometry.
Subsequent study of strong, electromagnetic, and weak interactions
has demonstrated that the pseudo-Euclidean geometry is inherent in
the fields associated with these interactions...Pseudo-Euclidean
space-time is not a priori, i.e., given from the start, or having
an independent existence. It is an integral part of the existence
of matter,...it is [always] the geometry by which matter is
transformed...}'' \cite{denisov}. Indeed, back in 5.1.1, and
shortly in our comments on Gel'fand duality (5.5.1), we argue how
the geometrical structure of what one might call `spacetime'
(including its topology and differential structure) is inherent in
the algebraic-dynamical field of quantum causality in the same way
that the geometrical notion of curvature is already inherent
(ultimately, derives from) the dynamical connection field, which
is the sole physically meaningful entity in our theory.}
\end{quotation}

\noindent while, for the sake of operationality or instrumentality
({\it ie}, for the existence of measuring rods and
clocks)\footnote{And this shows just how important for the
physical interpretation of the theory Einstein thought the
operational foundations of general relativity are.}

\begin{quotation}
\noindent ``{\em\small ...According to the general theory of
relativity, space without ether is unthinkable; for in such space,
not only would there be no propagation of light, but also no
possibility of existence for standards of space and time
(measuring rods and clocks), nor therefore any space-time
intervals in the physical sense...}'' \cite{einst5}
\end{quotation}

\noindent Thus, eventually, he was led to make the following
(telling for us) conceptual identification:

\begin{quotation}
\noindent ``{\em\small ...Physical space and the ether are only
different expressions for one and the same thing...}''
\cite{kostro}\footnote{Page 174 and reference therein.}
\end{quotation}

\noindent Moreover, keeping the identification above in mind, we
note that Kostro, in \cite{kostro}\footnote{Bottom of page 105 and
top of page 106.}, expresses concisely how this new ether may
culminate in the formulation and serve as the basic underlying
concept of a unified field theory (in the more popular sense), as
follows:

\begin{quotation}
\noindent ``{\small\em The last step in the development of the
relativistic concept of the ether would be the creation of a
unified field theory in which a unification of gravitational and
electromagnetic interactions is achieved and in which matter
consisting of particles would constitute special states of
physical space. Thus far, the attempts to develop such a theory
have been unsuccessful, the reason lying not in physical reality,
but in the deficiencies of our theories. It would be ideal to
develop such a unified field theory in which all the objects of
physics would come under the concept of the ether. Einstein
pointed out this problem at the very beginning of his
article:\footnote{Einstein's article Kostro is referring to is
`{\itshape \"Uber den \"Ather}' \cite{einst1}.}}

\vskip 0.1in

\noindent {\small\em ...one can defend the view that this notion}
[{\it ie}, the ether] {\small\em includes all objects of physics,
since according to a consistent field theory, ponderable matter
and the elementary particles from which it is built also have to
be regarded as `fields' of a particular kind or as particular
`states' of space.}''
\end{quotation}

\noindent This prompts us to cast, in complete analogy to the
ADG-theoretic identifications in the context of geometric
(pre)quantization in (\ref{add11}) and (\ref{add12}), Einstein's
conceptual identifications above as a {\it r\'esum\'e} of his
unitary field theory program, as follows:

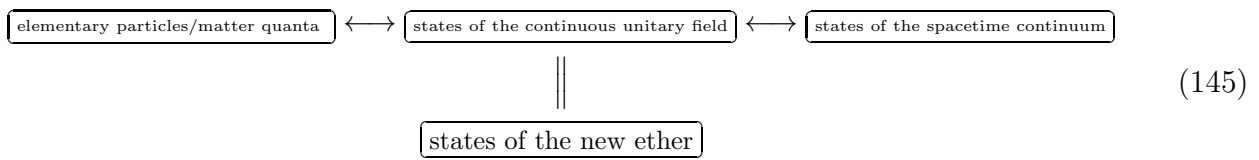
\begin{equation}\label{add13}
{\fontsize{0.15in}{0.15in}
\begin{CD}
\ovalbox{\tiny elementary particles/matter quanta
}\longleftrightarrow\ovalbox{\tiny states of the continuous
unitary field}\longleftrightarrow\ovalbox{\tiny states of the
spacetime
continuum}\\
@| \\
\ovalbox{\footnotesize states of the new ether}
\end{CD}}
\end{equation}

\noindent In comparison with our identifications in (\ref{add12}),
we note that since our ADG-theoretic perspective on finitary,
causal and quantal vacuum Einstein-Lorentzian gravity completely
evades the smooth background spacetime continuum and is based
solely on the fcqv-E-L field $\conf_{i}$, {\em our (arguably more
quantal, because reticular-algebraic) version of Einstein's new
ether above could be taken to be the `carrier' of this causon
field, namely, the vector sheaf $\qmodll_{i}$ itself}. {\em The
latter}, in close analogy to the inextricable relationship between
the ether, the (continuous) space(time) and the (gravitational)
field that Einstein intuited, but with the prominent absence of an
external, background $\smooth$-spacetime and our undermining of
the physical role played by the smooth gravitational metric field
$g_{\mu\nu}$ supported by it, {\em cannot be thought of
independently of the fcqv-gravitational connection that it carries
and vice versa}.\footnote{See again footnote 271 about this
`holistic' or, quite fittingly, `unitary' ADG-theoretic conception
of the gravitational connection and the vector sheaf (of states of
causons in our finitary theory) that carries it---our version of
Einstein's `new ether'.}

Now, since Einstein was well aware of the problem of singularities
that plague his geometric spacetime continuum based theory of
gravity\footnote{See quotations of Einstein subsequently and our
discussion in the epilogue.}, and at the same time he was `in awe'
of the (successes of the) quantum revolution, he on the one hand
asked:

\begin{quotation}
\noindent ``{\em\small Is it conceivable that a field theory
permits one to understand the atomistic and quantum structure of
reality?}'' \cite{einst3}
\end{quotation}

\noindent and on the other, quite paradoxically if we consider the
conceptual importance that he placed on the continuous field and
the spacetime continuum ({\it ie}, the new ether) supporting it,
he repeatedly doubted in the algebraic light of the quantum the
very geometrical ether ({\it ie}, the $\smooth$-smooth spacetime
continuum and the smooth metric field $g_{\mu\nu}$ that it
supports) that he so feverously propounded in the quotes
above.\footnote{See quotations in 5.1 and more extended ones in
\cite{malrap1,malrap2}.} For instance, until the very end of his
life he doubted the harmonious coexistence of the (continuous)
field together with its particles (quanta) in the light of the
singularities that assail the spacetime continuum, much as
follows:

\begin{quotation}
\noindent ``{\em\small ...Your objections regarding the existence
of singularity-free solutions which could represent the field
together with the particles I find most justified. I also share
this doubt. If it should finally turn out to be the case, then I
doubt in general the existence of a rational and physically useful
continuous field theory. But what then? Heine's classical line
comes to mind: `And a fool waits for the answer'...}" (1954)
\cite{stachel}
\end{quotation}

How can we explain and understand this apparently `paradoxical'
and `self-contradictory' stance of his against the spacetime
continuum {\it vis-\`a-vis} singularities and the
quantum?\footnote{That is, on the one hand to argue for the
geometrical spacetime continuum, in the guise of the new ether,
which is inherent in the continuous unitary field representing the
field together with its quanta---the particles that may in turn be
conceived as `singularities in the field', and at the same time on
the other, exactly due to those singularities ({\it eg}, the
infinities of fields right at their point-particle `sources') of
the manifold and the discontinuous, algebraically represented
actions of quanta, to urge us to abandon the geometrical
continuous field theory and look for ``{\em a purely algebraic
theory for the description of reality}''
\cite{einst3,malrap1}---one whose ``{\em statements are about a
discontinuum without calling upon a continuum space-time as an
aid}'' and according to which ``{\em the continuum space-time
construction corresponds to nothing real}'' (1916)
\cite{stachel,malrap2}.} Perhaps we can understand his apparently
`circular' and `ambiguous' attitude if we expressed the whole
`oxymoron' in a positive way, as follows: {\em we believe that
Einstein would have readily abandoned the continuous field theory
and the geometric spacetime continuum of general relativity in
view of the `granular' actions of quantum theory if he had an
`organic'\footnote{See quotation from \cite{einst2} and in 6.1.}
finitistic-algebraic theory to take its place}. Alas, again in his
own words just a year after he concluded the general theory of
relativity and at the very end of his life:

\begin{quotation}
\noindent ``{\small\em ...But we still lack the mathematical
structure unfortunately.}'' (1916)\footnote{For the whole
quotation, see \cite{malrap2}.}
\end{quotation}

\noindent and

\begin{quotation}
\noindent ``{\small\em ...But nobody knows how to obtain the basis
of such a {\rm [finitistic-algebraic]} theory.}''
(1955)\footnote{This is the last sentence, in the last section of
the last appendix of `{\itshape The Meaning of Relativity}'
\cite{einst3} appended in 1954. The whole quotation can be found
directly at the end of \cite{malrap1}.}
\end{quotation}

\noindent We would like to think that ADG, especially in its
particular finitistic-algebraic application here to Lorentzian
vacuum Einstein gravity, goes some way towards qualifying as a
candidate for the (mathematical) theory that Einstein was
searching for. Since we are talking about Einstein's unitary field
theory and the mathematics that he was searching for in order to
implement it, we give below a very fitting passage which concludes
Ernst Straus' reminiscences of Einstein in \cite{straus}:

\begin{quotation}
\noindent ``{\small\em ...Einstein's quest for the central problem
for the ultimate correct field theory is generally considered to
have failed. I think that this did not really surprise Einstein,
because he often entertained the idea that vastly new mathematical
models would be needed, that possibly the field-theoretical
approach through the kind of mathematics that he knew and in which
he could do research would not, could not, lead to the ultimate
answer,\footnote{See remarks by Bergmann and Kostro that follow
shortly; especially Kostro's words in footnote 311 about the
mathematics that Einstein knew and used in order to model his
unitary field theory.} that the ultimate answer would require a
kind of mathematics that probably does not yet exist and may not
exist for a long time. However, he did not have the slightest
doubt that an ultimate theory does exist and can be discovered.}''
\end{quotation}

We sum-up this discussion of Einstein's new ether by commenting on
and counterpointing some remarks of Peter Bergmann and Ludwik
Kostro in \cite{kostro}\footnote{Pages 164 and 165.} which
apparently maintain that what Einstein had in mind when he talked
about this new ether in the context of unitary field theory was
not the $\smooth$-smooth spacetime manifold {\it per se}, but the
extra structures (such as the metric, for example) that are
imposed on it.

First, Kostro asked:\footnote{In a talk titled `{\itshape Outline
of the history of Einstein's relativistic ether conception}'
delivered at the International Conference on the History of
General Relativity, Luminy, France (1988) \cite{kostro}.}

\begin{quotation}
\noindent ``{\small\em ...Which mathematical structure of
contemporary theoretical physics represents the entity Einstein
called `the new ether'?}''
\end{quotation}

\noindent to which Bergmann replied:

\begin{quotation}
\noindent ``{\small\em ...In the last decades of his life Einstein
was concerned with unitary field theories of which he created a
large number of models. So I think he was very conscious of the
distinction between the differential manifold (though he did not
use that term)\footnote{Einstein most of the time used the term
`{\em (space-time) continuum}' (our footnote).} and the structure
you have to impose on the differential manifold (metric, affine or
otherwise) and that he conceived of this structure, or set of
structures, as potential carriers of physical distinctiveness and
including the dynamics of physics.

Now, whether it is fortunate or unfortunate to use for the latter
the term like ether? I think simply from the point of view of
Einstein and his ideas that in the distinction between the
differential manifold as such and the geometrical structures
imposed on it we could, if we want, use the term ether for the
latter.}''
\end{quotation}

\noindent and to which, in turn, Kostro further added:

\begin{quotation}
\noindent ``{\small\em I am certain that Bergmann was right when
he claimed that the differential manifold as such, which is used
to model space-time without imposing upon it such structures as
metrics, {\it etc.} cannot be treated as a mathematical structure
representing Einstein's relativistic ether.

Bergmann was right, because the four-dimensional differential
manifold as such is a mathematical structure of too general a
nature, and it cannot physically define distinctive features of
the space-time continuum without imposing metrics and other
structures upon it. It is too general, because it can serve as an
arena or background for any macroscopic physical theory (and even
perhaps a microscopic one, because the debate over the status of
the differential manifold in microphysics is ongoing). By the act
of imposing metrics (i.e., the recipe for measuring space and time
intervals) and other structures upon it, the structure enriched in
such a way turns into something that represents distinctive
physical features of the real space-time continuum...}''
\end{quotation}

\noindent We partially agree with Bergmann and Kostro insofar as
their comments above entail that the background differential
spacetime manifold itself is devoid of physical significance and
that what is of physical importance is the `geometrical' objects
that live on this base arena which, in Bergmann's words, ``{\em
include the dynamics of physics}''. On the other hand, from the
novel perspective of ADG, and we would like to believe that both
Bergmann and Kostro would agree with us had they been familiar
with the basic tenets of ADG, we maintain that:

\begin{enumerate}

\item In general relativity, the smooth spacetime manifold serves
as the {\em carrier} of the structures imposed on it---after all,
this is how the structures like metric, affine (Levi-Civita)
connection {\it etc} acquire the epithet `{\em smooth}' in front
and become {\em smooth metric}, {\em smooth connection} {\it
etc}.\footnote{With the important clarification, however, that it
is a rather common mistake (made nowadays especially by
theoretical physicists) to think that the metric was assigned
(originally by Gauss and Riemann) on the manifold itself. Rather,
it was imposed on (what we now call) the (fibers of the) tangent
bundle (tangent to whatever `space' they used as base space)!
(revisit footnote 17). Thus, the commonly used term `{\em
spacetime metric}' can be quite misleading.} As such, {\em it can
still be perceived as a passive, a priori fixed by the theorist,
absolute, ether-like substance which sets the classically
unequivocal `condition or criterion of differentiability' for the
dynamical variations of these `physical' structures imposed on
it}.\footnote{See our comments on the relativity of
differentiability in the epilogue.} For, surely, if Einstein did
not have the background $\smooth$-spacetime at his disposal, the
(classical) differentials that the latter provides one with and
the rules of the mathematical theory known as (classical)
differential geometry (calculus) of manifolds that these
differentials obey, how could he write the dynamical laws for the
aforesaid extra physical structures? And, arguably, in a
Wheelerian sense, {\em no theory is a physical theory unless it is
a dynamical theory}. Thus, the usual differential calculus
provided Einstein with the basic mathematical tools which enabled
him to write the dynamical equations for his continuous,
`ethereal' fields.

\item As noted above, one should not forget that Einstein's dissatisfaction with
the geometrical spacetime continuum (manifold) came basically from
two sources: the singularities that assail general relativity and,
of course, the discontinuous and algebraic character of quantum
mechanical actions. In fact, at the very end of his life, and in
the context of his unitary field theory, he came to intuit that
these two `problematic', when viewed from the spacetime continuum
perspective, sources may be in fact intimately
related:\footnote{The following quotation can be found again in
the last appendix of \cite{einst3}. It is the extended version of
the one given a few paragraphs above.}

\begin{quotation}
\noindent ``{\small\em ...Is it conceivable that a field theory
permits one to understand the atomistic and quantum structure of
reality? Almost everybody will answer this question with `no'. But
I believe that at the present time nobody knows anything reliable
about it. This is so because we cannot judge in what manner and
how strongly the exclusion of singularities reduces the manifold
of solutions. We do not possess any method at all to derive
systematically solutions that are free of singularities...}''
\end{quotation}

\noindent ADG, as applied here (and in \cite{malrap1,malrap2}) to
a locally finite, causal and quantal vacuum Einstein gravity,
`kills both birds above with one stone': on the one hand, it
evades the $\smooth$-manifold and `engulfs' or `absorbs'
singularities into whichever structure sheaf of generalized
arithmetics (or coordinates) one chooses to employ in order to
tackle the physical problem one wishes to study
\cite{malros2,mall7,malrap3},\footnote{Again, for more comments on
singularities, the reader should go to the epilogue of the present
paper.} and on the other, it offers us an entirely algebraic and
finitistic way of doing (the entire spectrum of the usual)
differential geometry
\cite{mall1,mall2,malros1,malrap1,malrap2,mall4}. All in all, it
is our contention that Einstein (implicitly) questioned the very
(pseudo-)Riemannian differential geometry, which, in turn,
fundamentally relies on the differential spacetime manifold.

\item From the ADG-based perspective of the present paper, {\em there
is nothing physical about either an external background spacetime
(be it discrete or continuous) or about the metric structure that
we impose on it}. On these grounds alone, Bergmann and Kostro's
contention above that these concepts may be regarded as
representing Einstein's new ether, appears to be unacceptable. On
the other hand, we believe that our entirely algebraic conception
of the (gravitational) connection can be seen as the sole
dynamical variable in a quantal theory of Lorentzian gravity.
Fittingly then, the (associated) vector sheaf (of states of
causons), which are not soldered ({\it ie}, localized) on any
$\smooth$-smooth spacetime manifold whatsoever, may be taken to be
as the ADG-theoretic analogue of Einstein's `new ether': it is the
carrier of the fcqv-E-L field.

\item Finally, in view of the words of Feynman and Isham in the
beginning of the present work, as well as what has been shown,
partially motivated by these (or rather, `post-anticipatorily'),
in the present paper, {\em we simply have to disagree with
Kostro's contention that there is still a possibility that the
smooth manifold can serve as a (spacetime) background for a
microphysical theory---in particular, in the (feverously sought
after) quantum theory of gravity}. Although, admittedly, Einstein
did not know and use the differential geometry of smooth manifolds
the way we do today ({\it eg}, fiber bundle theory),\footnote{And
at this point we agree with Kostro when he says that ``{\em ...In
the physical space-time continuum model in his Special Theory of
Relativity and General Theory of Relativity, and in his attempts
to formulate a unitary relativistic field theory, Einstein could
not apply the tools and methods of the contemporary theory of
differential manifolds and the structures we use with them,
because he simply did not know them in the form in which they are
taught and applied today...}'' (\cite{kostro}, p. 164).} he still
had the tremendous physical insight to anticipate and foreshadow
subsequent thinkers and workers in quantum gravity, like Feynman
and Isham for example, who have been led by their own quests to
conclude that {\em the $\smooth$-smooth model of spacetime fares
poorly}, to put it mildly,\footnote{Not to say `fails miserably'.}
{\em in the quantum (gravity) regime}.

\end{enumerate}

\subsubsection{Brief remarks on `the matter of the fact'}

Since we have just commented on Einstein's unitary field theory,
since in causet theory there has been a strong indication lately
that one can derive matter fields directly from causets
\cite{ridesork}, and also since our scheme so far has focused
solely on pure vacuum gravity ({\it ie}, without the inclusion of
matter actions and other gauge force fields), we would like to
conclude this subsection by making a very short comment on the
possibility of including matter and other gauge field actions in
our locally finite, causal and quantal theory. Our brief addendum
is simply that, {\it prima facie}, the inclusion of fermionic
matter fields ({\it eg}, electrons), their connections ({\it eg},
Dirac-like operators), as well as their relevant gauge potentials
({\it eg}, electromagnetic field), can be straightforwardly
implemented ADG-theoretically as follows:

\begin{enumerate}

\item In line with our comments earlier on GPQ, the (states of) other gauge (boson) and matter (fermion)
fields can be modelled by (local) sections of the relevant line
($\mathrm{rank}=1$) and vector ($\mathrm{rank}>1$) (fin)sheaves
(here, over a causet), respectively.

\item Their corresponding (gauge) connections will be modelled by
their relevant finsheaf morphisms, and their (free) dynamics by
equations involving (the field strengths of) these morphisms
(which dynamics, in turn, by the very categorical definition of
those finsheaf morphisms and the covariance of their corresponding
field strengths, will be manifestly gauge $\gauge_{i}$-invariant).

\item Interactions between the matter and their gauge fields will
be algebraic expressions involving the relevant finsheaf morphisms
coupled to ({\it ie}, acting on) the aforesaid (local) sections.

\item {\it In toto}, in the finitary case of interest here, 1--3 will be finitistic, causal, explicitly
independent of an external, underlying ({\it ie}, background)
$\smooth$-smooth spacetime continuum ({\it ie}, `fully
covariant'), `purely gauge-theoretic' and `inherently quantum', as
it was the case for the vacuum gravitational field elaborated in
the present paper.

\end{enumerate}

\noindent However, for more information about the general
ADG-theoretic treatment of (non-gravitational) gauge ({\it ie},
electromagnetic and non-abelian Yang-Mills) theories and their
quantum matter sources, the reader should refer to \cite{mall4}.

\subsection{Comments on Gel'fand Duality and the Power of Differential Triads}
We close the present section by commenting briefly on the notion
of Gel'fand duality---an idea that we repeatedly alluded to and
found of great conceptual import in the foregoing. We also
illustrate how powerful the basic ADG-theoretic notion of
differential triads is for establishing continuum (`classical')
limits for a host of (physically) important mathematical
structures that we encountered earlier during the {\it aufbau} of
our locally finite, causal and quantal vacuum Einstein gravity.

\subsubsection{Gel'fand duality: from algebras to geometric spaces
and back}

By Gel'fand duality we understand the general `functional
philosophy' according to which, informally speaking,

\begin{quotation}
\noindent {\em the variable (argument) becomes function and the
function  variable (argument)}.
\end{quotation}

\noindent One could symbolically represent this as follows

\begin{equation}\label{add3}
f(x)\mapto \hat{x}(f)
\end{equation}

\noindent For example, in the previous section we noted that our
work with (finsheaves of) incidence algebras associated with
(over) the finitary topological posets of Sorkin is essentially
based on Gel'fand duality so that, in discussing inverse and
inductive limits of those posets and (the finsheaves of) their
incidence algebras respectively, we ended up concluding that {\em
`space(time)' is categorically or Gel'fand dual to the physical
fields that are defined on `it'}. This is precisely the semantic
content of (\ref{add3}), but let us explicate further this by
starting from the classical and well understood theory.

From the classical manifold perspective, Gel'fand duality has an
immediate and widely known application: the (topological)
reconstruction of a $\smooth$-smooth manifold $M$ as the spectrum
$\gelsp$ of its algebra $\smooth(M)$ of smooth functions
\cite{mall0}. To describe briefly this, let $M$ be a differential
manifold and $x$ one of its points. Consider then the following
collection of smooth $\R$-valued functions on $M$

\begin{equation}\label{add4}
I_{x}=\{\phi:~M\mapto\R |~\phi(x)=0\}\subset{}^{\R}\smooth(M)
\end{equation}

\noindent It is straightforward to verify that $I_{x}$ is a
maximal ideal of ${}^{\R}\smooth(M)$ and that the quotient of the
latter by the former yields the reals:
${}^{\R}\smooth(M)/I_{x}=\R$. In fact, in complete analogy to the
space $\mathrm{Max}(\gel)$ that we encountered earlier in
connection with Ashtekar and Isham's commutative $C^{*}$-algebraic
approach to the loop formulation of canonical quantum gravity
which employs the Gel'fand-Naimark representation
theorem\footnote{It must be noted however that ${}^{\R}\smooth(M)$
is an abelian {\em topological algebra}, not a Banach, let alone a
$C^{*}$-, algebra. In point of fact, it is well known that
${}^{\R}\smooth(M)$ is not `normable' or `Banachable' (\v{S}ilov)
\cite{mall0}. On the other hand, ${}^{\com}\cont(M)$, for a
compact manifold $M$, is the `archetypal' commutative
$C^{*}$-algebra---the very one Ashtekar and Isham used in
\cite{ashish} to represent $\gel$.}---it too a straightforward
application of Gel'fand duality,\footnote{For example, the
Gel'fand transform in (\ref{eq132}) is a precise mathematical
expression of a Gel'fand duality between the space of connections
and the space of loops involved in that theory
\cite{ashish,ashlew2}. Furthermore, to `justify' the notation in
(\ref{add3}), we note how in \cite{mall2} the Gel'fand transform
is defined (in the case of a topological algebra $A$): let $A$ be
a (unital, commutative, locally $m$-convex) topological algebra,
whose spectrum ({\it ie}, the set of non-zero, continuous,
multiplicative linear functionals on $A$) is $\gelsp(A)$. The
latter is equipped with the usual Gel'fand topology relative to
which the maps $\hat{x}:~\gelsp(A)\mapto A$, with
$\hat{x}(f):=f(x)$, are continuous. Then, the Gel'fand transform
algebra of $A$, is defined as: $A\,\hat{}:=\{ \hat{x}:~x\in A\}$.}
the set $Spec[{}^{\R}\smooth(M)]\equiv\gelsp[{}^{\R}\smooth(M)]$
of all maximal ideals $I_{x}~(x\in M)$ of ${}^{\R}\smooth(M)$ such
that

\begin{equation}\label{add5}
\R\hookrightarrow{}^{\R}\smooth(M)\mapto{}^{\R}\smooth(M)/I_{x}
\end{equation}

\noindent (within an isomorphism of the first term), is called
{\em the real (Gel'fand) spectrum of} ${}^{\R}\smooth(M)$.
Furthermore, if ${}^{\R}\smooth(M)$---regarded algebraic
geometrically as a commutative ring---is endowed with the
so-called Zariski topology \cite{harts}, or equivalently, with the
usual Gel'fand topology,\footnote{The coincidence between the
Gel'fand and the Zariski topology on $\gelsp[{}^{\R}\smooth(M)]$
is due to the fact that ${}^{\R}\smooth(M)$ is a regular
topological algebra \cite{mall0}.} then the `point-wise' map

\begin{equation}\label{add6}
M\ni x\mapsto I_{x}\in\gelsp[{}^{\R}\smooth(M)]
\end{equation}

\noindent can be shown to be a homeomorphism between the
$\cont$-topology of $M$ ({\it ie}, $M$ being regarded simply as a
topological manifold and the Gel'fand (Zariski) topology of
$\gelsp[{}^{\R}\smooth(M)]$. {\it In toto}, the essential idea of
Gel'fand duality here is to substitute the (topology of the)
underlying space(time) continuum by the (algebras of) objects
(functions/fields) that live on it, and then recover it by a
suitable technique, which we may coin {\em Gel'fand
spatialization}.

As noted before, in the finitary context too, incidence Rota
algebras'---ones taken to model finitary topological spaces, not
qausets---Gel'fand duality and, in particular, the aforesaid
method of Gel'fand spatialization was first applied in \cite{zap0}
and then further explored in \cite{rapzap1,rapzap2}. The basic
idea there was to substitute the continuous spacetime
poset-discretizations $P_{i}$ of Sorkin in \cite{sork0} by
functional-like algebraic structures $\omg_{i}$, assign a topology
to the latter, and then show how the original finitary poset
topology may be identified with the latter. Thus, in complete
analogy to the classical continuum case above, we used Gel'fand
spatialization and:

\begin{enumerate}

\item Defined `points' in the $\omg_{i}$s as (kernels of finite dimensional)
irreducible (Hilbert space) representations of them---that is, as
elements of their primitive (maximal) spectra
$\mathrm{Max}\omg_{i}$.

\item Assigned a suitable topology on those primitive ideals.\footnote{This is the
aforementioned `non-standard' Rota topology, since it was argued
that the Gel'fand (or the Zariski) topology on
$\mathrm{Max}\omg_{i}$ is trivial ({\it ie}, the
discrete---Hausdorff or $T_{2}$---topology)
\cite{zap0,rapzap1,rapzap2}.}

\item Identified the Rota topology on the primitive spectra of the $\omg_{i}$s with the
Sorkin topology of the $P_{i}$s.

\end{enumerate}

\noindent That the $\omg_{i}$s are Gel'fand dual to the $P_{i}$s
is concisely encoded in the result quoted in section 4 that there
is a (contravariant) functorial correspondence between the
respective categories $\rz$ and $\as$.\footnote{As also mentioned
in footnote 160 in 4.3, the correspondence (construction)
`finitary posets'$\mapto$`incidence algebras' is functorial
precisely because the $P_{i}$s are simplicial complexes
\cite{rapzap1,rapzap2,zap1}.} In effect, this is precisely the
correspondence that enables one to go from categorical (inverse,
projective) limits in $\as$ to categorical co- (direct, inductive)
limits in (finsheaves of incidence algebras in) $\rz$ mentioned
above.\footnote{As also noted in footnote 160, precisely due to
the functoriality of the correspondence (construction)
`finitary~posets'$\mapto$`incidence~algebras', finsheaves in the
sense of \cite{rap2} exist.} Furthermore, it was evident by the
very structure of the $\omg_{i}$s (as $\Z$-graded discrete
differential manifolds) that, in the $P_{i}$-dual picture of
incidence algebras, differential properties of the underlying
space(time) could be studied, not just topological. In other
words, in the finitary setting, Gel'fand duality revealed a
differential structure that is encoded in the $\omg_{i}$s which
was `masked' by the purely topological posets of Sorkin. With
respect to the classical continuum paradigm of Gel'fand duality
mentioned above, the analogy is clear:

\begin{quotation}

\noindent The $P_{i}$s are the reticular analogues of $M$ regarded
solely as a $\cont$-manifold, while the $\omg_{i}$s as the
reticular analogues of $M$ regarded as a differential manifold
\cite{rapzap1,rapzap2}.

\end{quotation}

\noindent In fact, precisely due to this suggestive analogy it was
intuited in \cite{rapzap1,rapzap2} that at the limit of infinite
refinement of the locally finite coverings of the bounded region
of $X$ not only the topological, but also the differential
structure of the continuum could be recovered. Heuristically
speaking, the $\omg_{i}$s' revealing of differential geometric
attributes suggested to us that also `change',\footnote{For any
differential operator `$d$' models change!} not only `static'
topological or `spatial' relations, could be modelled
algebraically and by finitary means.

Thus, as it was described in the previous section, in the sequel,
Gel'fand duality associating incidence algebras (qausets) to
locally finite posets modelling causets was first exploited in
\cite{rap1} by using Sorkin's fundamental insight in \cite{sork1}
that it is more physical to think of a partial order as causality
({\it ie}, as a `temporal' structure) than as topology ({\it ie},
as a `spatial' structure). Furthermore, Sorkin's demand for a
dynamical scenario for causets almost mandated to us the use of
sheaf theory---that is, to organize the incidence algebras
modelling qausets to sheaves of an appropriate, finitary kind
\cite{rap2}. Thus, curved finsheaves of incidence algebras were
born as kinematical spaces for the dynamical variations of qausets
out of blending this causal version of Gel'fand duality with the
ideas, working philosophy and technical panoply of ADG
\cite{malrap1,malrap2}.

The bottom line of all this is that the semantic essence of
Gel'fand duality---{\it ie}, to substitute the topology of the
background `space(time)' by the functions that live on
`it'---found its natural home in ADG, which, as we emphasized
repeatedly above, similarly directs one to pay more attention on
the objects (fields) that live on space(time) rather than on
spacetime {\it per se}, independently of whether the latter is
taken to be a reticular base topological space or a continuum. In
fact, we may further hold that

\begin{quotation}
\noindent at a differential geometric, not just at a topological,
level, ADG in some sense `breaks' Gel'fand
duality,\footnote{Gel'fand duality understood here as a
`topological symmetry' between the underlying space(time) and the
objects (functions) that dwell on it.} since it tells us that {\em
the differential geometric structure (mechanism) comes directly
from the (algebraic) objects that live (in the stalks of the
algebra sheaves on) space(time), not from the base space(time)
itself.\footnote{Thus, when one is interested solely in the
topological structure of the continuum $M$, the aforedescribed
classical `reconstruction result' of the manifold $M$ from the
algebra ${}^{\R}\smooth(M)$ shows precisely that the
$\cont$-topology of $M$ can be recovered from ${}^{\R}\smooth(M)$
by Gel'fand spatialization, while the differential structure
inherent in ${}^{\R}\smooth(M)$ is not essentially involved.
Similarly, at the finitary level, we saw above how the $\omg_{i}$
revealed a rich differential geometric structure that the purely
topological finitary posets of Sorkin in \cite{sork0} simply
lacked. Of course, it must be noted here that since the spectrum
of ${}^{\R}\smooth(M)$ can be identified (by Gel'fand duality)
with $M$ set-theoretically ({\it ie}, by a bijective map, which
moreover is a homeomorphism) one can also automatically transfer
from $M$ to $\gelsp[{}^{\R}\smooth(M)]$ the classical differential
({\it ie}, $\smooth$-smooth) structure. But this is another issue.
Notwithstanding (first author's hunch), there might be lurking
here an appropriate `{\em representation theorem}' that would
close the circle!}}
\end{quotation}

\noindent All in all, and from a causal perspective, Gel'fand
duality, coupled to ADG, allowed us to `differential geometrize'
and, as a result, (dynamically) vary Sorkin {\it et al.}'s causets
thus bring causet theory, which is a `{\em bottom-up}' approach to
quantum gravity, closer to other `{\em top-down}' approaches, such
as Ashtekar {\it et al.}'s.\footnote{This `bottom-up' and
`top-down' distinction of the approaches to quantum gravity is
borrowed from \cite{fay}. In relation to the three categories of
approaches mentioned in the prologue, category 1 may be thought of
as consisting of top-down approaches, while both categories 2 and
3 as consisting of bottom-up approaches.}

\subsubsection{Projective limits of fcqv-Einstein equations: the
power of differential triads}

We conclude this subsection by presenting an inverse system
$\inveinst$ of fcqv-Einstein equations like (\ref{eq117}) which
produces the generalized classical ({\it ie},
$\ssmooth$-\texttt{smooth}) vacuum Einstein equations for
Lorentzian gravity at the categorical (projective) limit of
infinite refinement or localization of the qausets. The discussion
below shows just how powerful the basic ADG-theoretic notion of a
differential triad is, since there is a hierarchy or `tower' of
projective/inductive systems of finitary structures which has at
its basis $\invtriad:=\{\vec{\triad}_{i}\}$---the inverse system
of fcq-differential triads (or its direct version $\invtriad$).

Anticipating some comments on singularities in the next section,
we also discuss the intriguing result that the
$\ssmooth$-\texttt{smooth} Einstein equations at the projective
limit hold over a `spacetime' that may be infested by
singularities---in other words, {\em the gravitational law does
not `break down' at the latter} since, anyway, an fcqv-version of
it appears to hold for every member of the system $\inveinst$ and
the latter are structures reticular, `singular' and quite remote
from the featureless smooth continuum. On the contrary,
singularities may be incorporated into (or absorbed by) the
structure sheaf of the $\ssmooth$-\texttt{smooth} differential
algebras so that the generalized differential geometric mechanism
continues to hold over them and the theory still enables one to
perform calculations in their presence\footnote{In the same way
that ADG enabled us earlier to `see through' the fundamental
discreteness of the base causets and write a perfectly legitimate
{\em differential} (Einstein) equation over them, in spite of
them.} \cite{malros1,malros2,mall7}.

But let us present straight away the aforesaid hierarchy of
projective/inductive families of finitary structures, commenting
in particular on the projective system $\inveinst$ mentioned
above. The diagram below as well as the discussion that follows it
will also help us recapitulate and summarize certain facts about
the plethora of inverse and direct systems we have encountered
throughout the present paper.

\medskip


\centerline{\doublebox{\bf The $\mathbf{11}$-storeys' tower of
finitary inverse/direct systems}}

\medskip

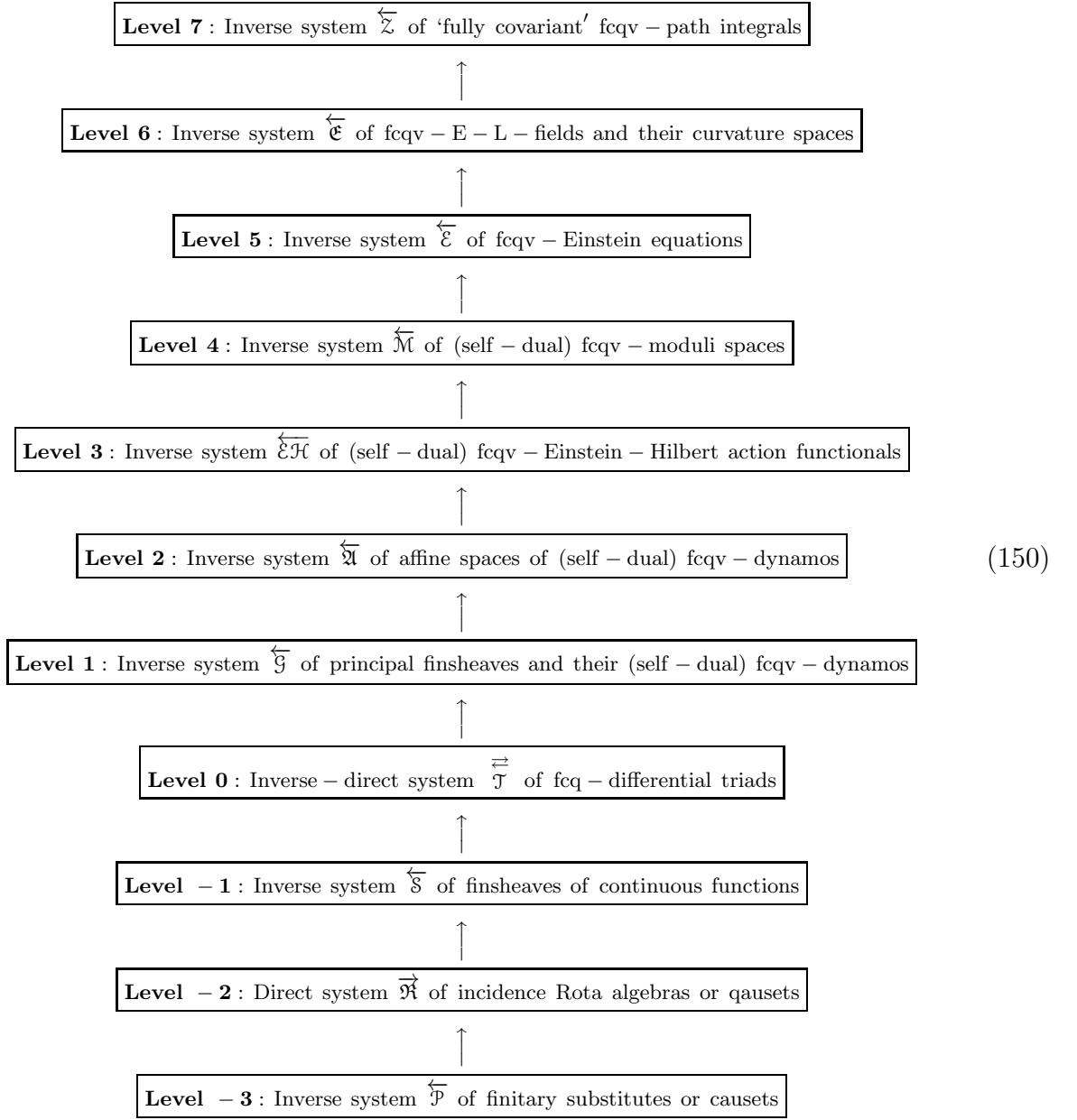
\begin{equation}\label{eqad7}
{\fontsize{0.13in}{0.13in}
\begin{CD}
\boxed{\mathbf{Level~7:}~\mathrm{Inverse~system~\invcq~of~`fully~covariant'~fcqv-path~integrals}}\\
@AAA\\
\boxed{\mathbf{Level~6:}~\mathrm{Inverse~system~\invel~of~fcqv-E-L-fields~and~their~curvature~spaces}}\\
@AAA\\
\boxed{\mathbf{Level~5:}~\mathrm{Inverse~system~\inveinst~of~fcqv-Einstein~equations}}\\
@AAA\\
\boxed{\mathbf{Level~4:}~\mathrm{Inverse~system~\invmod~of~(self-dual)~fcqv-moduli~spaces}}\\
@AAA\\
\boxed{\mathbf{Level~3:}~\mathrm{Inverse~system~\inveh~of~(self-dual)~fcqv-Einstein-Hilbert~action~functionals}}\\
@AAA\\
\boxed{\mathbf{Level~2:}~\mathrm{Inverse~system~\invsconn~of~affine~spaces~of~(self-dual)~fcqv-dynamos}}\\
@AAA\\
\boxed{\mathbf{Level~1:}~\mathrm{Inverse~system~\finv~of~principal~finsheaves~and~their~(self-dual)~fcqv-dynamos}}\\
@AAA\\
\boxed{\mathbf{Level~0:}~\mathrm{Inverse-direct~system~\invtriad~of~fcq-differential~triads}}\\
@AAA\\
\boxed{\mathbf{Level~-1:}~\mathrm{Inverse~system~\invs~of~finsheaves~of~continuous~functions}}\\
@AAA\\
\boxed{\mathbf{Level~-2:}~\mathrm{Direct~system~\diromg~of~incidence~Rota~algebras~or~qausets}}\\
@AAA\\
\boxed{\mathbf{Level~-3:}~\mathrm{Inverse~system~\inv~of~finitary~substitutes~or~causets}}
\end{CD}}
\end{equation}

\vskip 0.3in

\centerline{\doublebox{\bf Short stories about the eleven
storeys}}

\begin{itemize}

\item \fbox{\bf Levels $\mathbf{-3}$ to $\mathbf{-1}$:} The first three `underground
levels' can be thought of as assembling the fundamental one at
level zero. Indeed, as explained in section 4, each member
$\caus_{i}$ of $\inv$ (now causally interpreted as a causet)
comprises the base causal-topological space of each
fcq-differential triad $\vec{\triad}_{i}$ in $\invtriad$ bearing
the same finitarity index (level $-3$). Correspondingly (by
Gel'fand duality), each member (qauset) $\qaus_{i}$ of $\diromg$
comprises the reticular coordinate algebras, the bimodules of
differentials over them and the differential operators linking
spaces of discrete differential forms of consecutive grade (level
$-2$) that, when organized as finsheaves (level $-1$) over the
base causets of level $-3$, yield the inverse-direct system
$\invtriad$ of fcq-differential triads of level $0$.

\item \fbox{\bf Level $\mathbf{0}$:} This is the fundamental, `ground level' of the theory
in the sense that all the inverse systems at levels $\geq1$ have
at their basis $\invtriad$, as follows:

\item \fbox{\bf Level $\mathbf{1}$:} The inverse system $\finv$ of principal Lorentzian
finsheaves $\peel_{i}$ of (reticular orthochronous spin-Lorentzian
or causal symmetries of) qausets and their non-trivial ({\it ie},
non-flat, as well as self-dual) fcqv-dynamos $\conf_{i}^{(+)}$ can
be obtained directly from $\invtriad$ by (sheaf-theoretically)
localizing or `gauging' qausets in the stalks of the finsheaves in
the corresponding ({\it ie}, of the same finitarity index)
fcq-triads $\vec{\triad}_{i}\in\invtriad$ \cite{malrap1}.

\item \fbox{\bf Level $\mathbf{2}$:} The projective system
$\invsconn$ of affine spaces $\fconn_{i}^{(+)}$ of (self-dual)
fcqv-dynamos $\conf_{i}^{(+)}$ on the $\peel_{i}$s (or better, on
the $\qmodll_{i}$s associated with the $\peel_{i}$s) can be
obtained straightforwardly from $\finv$.

\item \fbox{\bf Level $\mathbf{3}$:} The inverse system $\inveh$ of
(self-dual) fcqv-Einstein-Hilbert action functionals can be easily
obtained from $\invsconn$ if we recall from
(\ref{eq118}--\ref{eq119}) the finitary version $\qeh_{i}^{(+)}$
of the ADG-theoretic definition of the E-H action functional $\eh$
in  (\ref{eq65}--\ref{eq66}).

\item \fbox{\bf Level $\mathbf{4}$:} Similarly to $\inveh$, the inverse system $\invmod$ of (self-dual)
fcqv-moduli spaces $\qmod_{i}^{(+)}$ in (\ref{eqad1}) can be
obtained from the inverse system $\invsconn$ member-wise, that is
to say, by quotienting each $\fconn_{i}^{(+)}$ in $\invsconn$ by
the automorphism group $\qaut_{i}\qmodll_{i}$ of the causon.

\item \fbox{\bf Level $\mathbf{5}$:} The projective system $\inveinst$ of fcqv-E-equations
as in (\ref{eq117}) is the main one we wish to discuss here. It
can be readily obtained, again member-wise from $\inveh$, by
varying each $\qeh_{i}^{(+)}$ in the latter collection with
respect to the (self-dual) fcqv-dynamo $\conf_{i}^{(+)}$ in each
member of $\finv$, as in (the finitary version of)
(\ref{eq67}--\ref{eq70}). The important thing to mention here is
that the inverse, continuum, `correspondence limit'
\cite{rapzap1,malrap1,rapzap2,malrap2} of these fcqv-E-equations
yields the `generalized classical' vacuum Einstein equations for
Lorentzian gravity on the $\ssmooth$-\texttt{smooth} spacetime
manifold $M$ which ({\it ie}, whose coordinate structure sheaf
$\ssmooth_{M}$), {\it prima facie}, may have singularities, other
general pathologies and anomalies of all sorts. We thus infer
that, by ADG-theoretic means,

\begin{quotation}
\noindent{\em we are able to write the law of gravitation over a
spacetime that may be teeming with singularities.}
\end{quotation}

\noindent In other words, and in characteristic contradistinction
to the classical $\smooth$-manifold based general relativity, {\em
the Einstein equations do not `break down' near singularities, and
the gravitational field does not stumble or `blow-up' at them.
Rather, it evades them, it `engulfs' or `incorporates' them
them,\footnote{This is so because the observable the gravitational
field strength is an $\struc$-morphism ({\it ie}, it respects the
generalized arithmetics in $\struc$), and the generalized
coordinate algebras in the structure sheaf may include arbitrarily
potent singularities.} it holds over them and, as a result, we are
able to calculate over them} \cite{mall7}.\footnote{We are going
to comment further on this in the next section.} Indeed, it has
been shown \cite{mall3} that with the help of ADG one can write
the gravitational vacuum Einstein equations over the most
pathological, especially when viewed from the
$\smooth$-perspective, space(time) $M$---one whose structure sheaf
$\struc_{M}$ consists of Rosinger's differential algebras of
generalized functions which, as noted earlier, have singularities
on arbitrary closed nowhere dense subsets of $M$ or even, more
generally, on {\em arbitrary sets with dense complements}
\cite{malros1,malros2,ros}.

\item \fbox{\bf Levels $\mathbf{6}$ and $\mathbf{7}$:} We will briefly comment on the last
two remaining projective systems, $\invel$ and $\invcq$. The first
is supposed to consist of (self-dual) fcqv-E-L fields
$(\qmodll_{i},\conf_{i}^{(+)})$ and their corresponding curvature
space pentads
$(\Qalg_{i},\vec{\partial}_{i}\equiv\vec{d}^{0}_{i},\Qaus_{i}^{1},\vec{\kd}_{i}
\equiv\vec{d}^{1}_{i},\Qaus^{2}_{i})$.\footnote{Which in turn, as
noted in 5.1, makes the base causet $\caus_{i}$ an {\em
fcqv-E-space}.} In line with footnote 58, we suppose that these
fcqv-curvature spaces and the fcqv-E-spaces $\caus_{i}$ supporting
them are the `solution spaces' of the corresponding equations in
$\inveinst$. At the same time, it must be noted that this
`gedanken supposition'---that is, that curvature spaces refer
directly to solutions of the fcqv-E-equations---is made to further
emphasize the point made at the previous level, namely, that: in
case one obtains an fcqv-E-L-field $(\qmodll_{i},\conf_{i}^{(+)})$
(and therefore its curvature $\fricci_{i}^{(+)}(\conf_{i}^{(+)})$)
that is a solution of (\ref{eq117}), then the projective,
$\ssmooth$-continuum limit of these solutions may be infested by
singularities ,{\em but still be a legitimate solution of ({\it
ie}, satisfy) the \texttt{smooth} vacuum Einstein equations and
the singularities did not in any way `inhibit' the physical law or
our calculations with it}.\footnote{These solutions being, in
fact, the results of our calculations in the presence of the
singularities incorporated in our own arithmetics $\struc$!} We
can summarize all this with the following statement quoted almost
{\it verbatim} from \cite{mall7}:

\begin{quotation}
\noindent {\em A physical law cannot be dependent on, let alone be
restricted by, singularities}.\footnote{Equivalently, {\em Nature
has no singularities} (see next section).}
\end{quotation}

\noindent This may be perceived as further support to Einstein's
doubts in \cite{einst3}:

\begin{quotation}
\noindent ``{\small\em It does not seem reasonable to me to
introduce into a continuum theory points (or lines {\it etc.}) for
which the field equations do not hold.}''\footnote{And Einstein's
doubts are remarkable indeed if one considers that they are
expressed in the context of classical field theory on a
$\smooth$-smooth spacetime manifold $M$ with the unavoidable
singularities that infest its coordinate structure sheaf
$\smooth_{M}$.}
\end{quotation}

\noindent As for the inverse system $\invcq$ whose members are
heuristic covariant fcqv-path integrals $\QPI$ {\it \`a la}
(\ref{eq121}), our comments for its projective continuum limit
must wait for results from the ADG-theoretic treatment of
functional integration in gauge theories currently under
development in \cite{mall4}. Our hunch is that, if the
fcqv-E-H-action $\qeh_{i}$ involved in the integrand of $\QPI$ is
taken to be a functional of the self-dual fcqv-dynamo
$\conf_{i}^{+}$ (write: $\qeh_{i}^{+}$ and, {\it in extenso},
$\QPI^{+}$), the continuum limit should yield the generalized
$\ssmooth$-version of the $\smooth$-path integral involving the
exponential of the \texttt{smooth} analogue of the smooth Asthekar
action $S_{ash}$ in (\ref{eq122}).
\end{itemize}

\section{Epilogue: the Wider Physical Significance of ADG}

In this concluding section we would like to discuss the wider
physical implications of our work here and of ADG in general. We
concentrate on two aspects: on the one hand, how ADG may
potentially help us evade the notorious $\smooth$-singularities,
thus we prepare the ground for a paper that is currently in
preparation \cite{malrap3}, and on the other, how ADG points to a
`relativized' notion of differentiability.

\subsection{Towards Evading $\smooth$-Smooth Singularities}

We would like to commence our brief comments on smooth
singularities, anticipating a more elaborate treatment in
\cite{malrap3}, with the following two quotations of Isham:

\begin{quotation}
\noindent {\small\em ``...A major conceptual problem of quantum
gravity is...the extent to which classical geometrical concepts
can, or should, be maintained in the quantum theory...''}
\cite{ish2}
\end{quotation}

\noindent [principally because]\footnote{Our addition in order to
link the two together.}

\begin{quotation}
\noindent {\small\em ``...The classical theory of general
relativity is notorious for the existence of unavoidable spacetime
singularities...''} \cite{ish1}
\end{quotation}

\noindent which are completely analogous to the two quotations in
the beginning of the paper. For instance, one could combine
Feynman's and Isham's words in the following way:

\begin{quotation}
\noindent {\em one cannot apply classical differential geometry in
quantum gravity, because one gets infinities and other
difficulties}.
\end{quotation}

Indeed, it is generally accepted that if one wishes to approach
the problem of quantum gravity by assuming up-front that spacetime
is (modelled after) a $\smooth$-smooth manifold,\footnote{Such an
approach would belong to the `calculus conservative' category 1
mentioned in the prologue.} one's theory would be plagued by
singularities well before quantization proper becomes an
issue---that is, long before one had to address the problem of
actually quantizing the classical theory. In other words, the
problem of singularities is already a problem of the classical
theory of gravity that appears to halt the program of quantizing
general relativity already at stage zero. Even if one turned a
blind eye to the singularities of the classical theory and
proceeded to tackle quantum general relativity as another quantum
field theory based again on the classical spacetime continuum, one
would soon encounter gravitational infinities that, although
milder and less robust than the singularities of the classical
theory, they are strikingly {\em
non-renormalizable}\footnote{Essentially due to the
dimensionfulness of Newton's gravitational constant.} in
contradistinction to the infinities of the quantum field theories
of gauge matter which are perturbatively finite. Altogether, it is
the $\smooth$-manifold $M$ (with its structure ring $\smooth(M)$
of infinitely differentiable functions) employed by the usual
differential geometry supporting both the classical and the
quantum general relativity which is responsible for the latter's
``{\em unavoidable spacetime singularities}'' and unremovable
infinities, and which makes classical (differential) geometric
concepts and constructions appear to be {\it prima facie}
inapplicable in the quantum deep.

On the other hand, the word `unavoidable' in Isham's quotation
\cite{ish1} above calls for further discussion, because it goes
against the grain of the very basic didactics of ADG {\it
vis-\`{a}-vis} singularities \cite{malros2,mall3,mall7}. It now
appears clear to us that the singularities of general relativity
come from assuming up-front $\smooth_{M}$ as the structure sheaf
$\struc$ of `coefficients' over which one applies the classical
differential geometric constructions to classical gravity. Since
the differential pathologies are due to $\smooth(M)$, the whole
enterprize of applying (differential) geometric concepts to
classical and, {\it in extenso}, to quantum gravity, seems to be
doomed from the start. On the other hand, ADG has taught us
precisely that {\em singularities are indeed avoidable if one uses
a different and more `suitable' to the physical problem at issue
structure arithmetics $\struc$ than $\smooth_{M}$}
\cite{mall2,malros2,mall3,mall7}. Moreover, ADG has time and again
shown that the `intrinsic mechanism' of the classical differential
geometry ($\struc_{X}\equiv\smooth_{X}$) can be carried over,
intact, to a generalized differential geometric setting afforded
by a general structure sheaf $\struc$ very different from
$\smooth_{M}$ \cite{malros1,malros2,mall3,malrap2}. Since $\struc$
can be taken to include arbitrary singularities, even of the most
extreme and classically unmanageable sort
\cite{malros1,malros2,ros}, it follows that {\em the said
differential mechanism is genuinely independent of singularities}.
That is to say,

\begin{quotation}
\noindent {\em not only we can avoid singularities
ADG-theoretically, but we can actually absorb or `engulf' them
into $\struc$} (provided of course these algebras are
`appropriate' or `suitable' for serving as the structure
arithmetics of the abstract differential geometry that has been
developed\footnote{That is to say, they can provide us with the
basis for defining differentials, connections, vectors, forms and
higher order $\otimes_{\struc}$-tensors, as well as the rest of
the `differential geometric apparatus' in much the same way that
$\smooth_{M}$ does, supported by the smooth manifold $M$, in the
classical theory.}) {\em and, as a result, calculate or perform
our (differential geometric) constructions over them, in spite of
their presence which thus becomes unproblematic}
\cite{mall7}.\footnote{In a straightforward way, ADG shows that
singularities can be integrated into the structure algebra sheaf
$\struc$ of our own `generalized measurements', `arithmetics' or
`coefficients', thus they should never be regarded as problems of
Physis. In other words, {\em Nature has no singularities, rather,
it is our own models of Her that are of limited applicability and
validity} ({\it eg}, in the classical case this pertains to the
$\smooth$-smooth manifold model $M$ for spacetime, the structure
sheaf $\struc\equiv\smooth_{M}$ that it supports, and the
$\smooth$-singularities that the latter hosts).}
\end{quotation}

\medskip

These remarks bring to mind Einstein's `apologetic
confession':\footnote{Which we encountered earlier in 5.4.1. We
too apologize for displaying this quotation twice, but we find it
very suggestive and relevant to one of the main points that we
would like to make in the present paper, namely, that {\em if
Einstein had a way} ({\it ie}, a theory and a working method) {\em
of doing field theory---and differential geometry in
general---independently of the pathological and unphysical
spacetime continuum, and, moreover, by finitistic-algebraic means
(in view of the quantum paradigm), he would readily abandon the
$\smooth$-smooth manifold} (see more remarks shortly). We claim
that ADG, especially in its finitary guise here, is such a
theory.}

\begin{quotation}
\noindent {\small\em ``...Adhering to the continuum originates
with me not in a prejudice, but arises out of the fact that I have
been unable to think up anything organic to take its place...''}
\cite{einst2},
\end{quotation}

\noindent in the sense that Einstein's commitment to the continuum
and, in effect, to the classical differential geometry supporting
his theory of gravitation, would not have been as strong or as
faithful\footnote{Quite remarkably though, considering that
general relativity enjoyed numerous successes and was
experimentally confirmed during Einstein's life.} had there been
an alternative (mathematical) scheme---perhaps one of a strong
algebraic character if one considers his life-long quest (in view
of quantum theory and the pathologies of the continuum) {\em for
an entirely algebraic description of reality}\footnote{See the
three quotations in 5.1.1.}---that worked as well as the
$\smooth$-differential geometry, yet, unlike the latter, was more
algebraic, not dependent on a dynamically inactive spacetime
continuum and, perhaps more importantly, it was not assailed by
singularities, infinities and other `differential geometric
diseases' coming from the {\it a priori} assumption of the smooth
background manifold.\footnote{In these terms we may understand the
epithet `{\em organic}' above.} We contend that ADG is a candidate
for the algebraic theory that Einstein had envisioned, for, as we
saw here and in a series of papers
\cite{malros1,malros2,malrap1,malrap2}, one can carry out all the
differential geometric constructions that are of use in the usual
differential geometry supporting general relativity with the help
of suitable vector and algebra sheaves over arbitrary base
spaces---even over ones that are extremely singular and reticular
when viewed from the perspective of the smooth continuum. Thus, in
effect,

\begin{quotation}
\noindent {\em according to ADG, (the intrinsic or inherent
mechanism of) differential geometry has nothing to do with the
background space so that, in particular, it is not affected by the
singularities of the manifold} \cite{mall7}.
\end{quotation}

For the sake of completeness, we would like to bring to the
attention of the reader two examples from the physics literature,
one old the other new, of theories that evade singularities in a
way that accords with the general spirit of ADG described above.

\begin{itemize}

\item {\bf Evading the exterior Schwarzschild singularity (old).}
The paradigm that illustrates best how a change in the coordinate
structure functions or generalized arithmetics $\struc$ may
effectively resolve a singularity is Finkelstein's early work on
the gravitational field of a point particle \cite{df}. It was well
known back then that the Schwarzschild solution of the Einstein
equations for the gravitational field of a point mass $m$ had two
singularities: an exterior one, at distance (radius) $r=2m$ from
$m$, and an interior right at the point mass ($r=0$). What
Finkelstein was able to show is that by an appropriate change of
coordinates\footnote{However, always in the context of a smooth
spacetime manifold $M$ ({\it ie}, still with the new coordinate
functions being members of $\struc\equiv\smooth_{M}$).}---the
so-called Eddington-Finkelstein frame, the exterior singularity is
`transformed away' revealing that the Schwarzschild spacetime acts
as a unidirectional, `semi-permeable', time-asymmetric membrane
allowing the outward propagation of particles and forbidding the
inward flux of antiparticles. For this, the $r=2m$ singularity was
coined `{\em coordinate singularity}' and was regarded as being
only a {\em `virtual' anomaly}---merely an indication that we had
laid down inappropriate coordinates to chart the gravitational
spacetime manifold.

On the other hand, it was also realized that the interior
singularity could not be gotten rid of by a similar coordinate
change,\footnote{Again though, still by remaining within the
$\smooth$-smooth manifold model.} thus it was held as being a {\em
`real'} or {\em `true' singularity}---an alarming indication that
general relativity is out of its depth when trying to calculate
the gravitational field right on its point source. Thus, ever
since Finkelstein's result, it has been hoped that only a genuine
quantum theory of gravity will be able to deal with the
gravitational field right at its source much in the same way that
the quantum theoresis of electrodynamics (QED) managed, even with
just the theoretically rather {\it ad hoc} method of `subtracting
infinities' (renormalization),\footnote{It is well known, for
instance, that Dirac expressed many times his dissatisfaction
about the renormalization program with its mathematically not well
founded and aesthetically unpleasing recipes: ``{\em Sensible
mathematics involves neglecting a quantity when it turns out to be
small---not neglecting it just because it is infinitely great and
you do not want it}'' \cite{dirac}.} to do meaningful physics
about the photon radiation field at its source---the electron.

According to this rationale however, notwithstanding the
perturbative non-renormaliza- bility of gravity due to the
dimensionality of Newton's constant, it has become obvious that
physicists have devotedly committed themselves so far to viewing
the spacetime point manifold as something physically `real' in the
sense that any of its points is regarded as potentially being the
host of a non-circumventable by $\smooth$-means singularity for a
physically important smooth field. That is, instead of reading
Finkelstein's result in a positive way, as for instance in the
following manner {\it \`a la} ADG,

\begin{quotation}
\noindent {\em when encountering any singularity, in order to
`resolve' it and be able to cope with ({\it ie}, calculate over)
it, one must look for an `appropriate' structure algebra of
coordinates that incorporates or `engulfs' it} \cite{mall7} {\em
and then one has to give a cogent physical interpretation of the
new picture},\footnote{The word `appropriate' meaning here in the
manner of ADG: a (differential) algebra of coordinates that
integrates the singularity (as a generalized coefficient) yet it
is still able to provide us with the basic differential mechanism
we need in order to set up the relevant dynamical equations over
it and calculate with them.}
\end{quotation}

\noindent physicists try instead to retain as much as they can
(admittedly, by ingenious methods at times) the differential
spacetime manifold $M$, its structure coordinates $\struc\equiv
\smooth_{M}$ and its structure symmetries
$\struct\equiv\mathrm{Diff}(M)$ as if they were physically real,
and at the same time quite falsely infer that the mechanism of
(classical) differential geometry does not apply over
singularities and, {\it in extenso}, in the quantum
deep.\footnote{Such an attitude was coined in \cite{malrap2} `{\em
$\smooth$-smooth manifold conservative}' and it is the spirit
underlying category 1 of approaches to quantum gravity mentioned
in the prologue. For instance, physicists try to isolate and
surgically cut-out of the spacetime manifold the offensive
singular points, thus continue the usual $\smooth$-differential
geometric practices in the remaining `effective manifold'. (In a
sense, they `artificially' remove, by hand and force as it were,
the ``{\em points, lines {\it etc} for which the field equations
do not hold}'', as we read in Einstein's quotation at the end of
the last section.) Current physics regards singularities as an
incurable disease of differential geometry. In contradistinction,
ADG maintains that they are unmanageable indeed by
$\smooth$-means, but also, more importantly, that the (algebraic
in nature) differential mechanism is not affected by them, so that
one should be able to continue `calculating' in their presence.}
All in all, it is as if:

\begin{enumerate}

\item The smooth spacetime manifold is a physically real substance to be
retained by all means.

\item The $\smooth$-singularities are also physically real as they are
Nature's ({\it ie}, the spacetime manifold's) own diseases---they
are real physical problems, `intrinsic' pathologies of Nature
(spacetime).

\item The (classical) differential calculus and the dynamical laws ({\it eg}, the Einstein
equations) supported by it break down at a singularity.

\item In order to retain the spacetime manifold so that one can continue doing calculus
({\it ie}, apply the usual differential geometric ideas and
techniques to physical situations---as it were, `continue the
validity of physical laws' and, in fact, {\em calcul}ate!),
singularities must be isolated and then somehow removed or
`surgically excised' from the manifold, leaving back an effective
spacetime manifold free of pathologies.

\end{enumerate}

At the same time, a natural follow-up of this line of thought is
the following basic hunch shared nowadays by almost all the
workers in the field of quantum gravity (string theorists aside)
looking for alternatives to the spacetime continuum of macroscopic
physics,\footnote{To name a few alternative schemes to the
spacetime continuum and to the classical theory of gravity that it
supports: simplicial (Regge) gravity, spin-networks, causet theory
{\it etc}.}

\begin{quotation}
\noindent{\em at strong gravitational fields near singularities,
or at Planck distances, the conventional image of spacetime as a
smooth continuum breaks down and should somehow give way to
something `discrete', `reticular', `inherently cut-off', and this
should be accompanied by a radical modification of the classical
differential geometry used to describe classical, `low energy'
Einstein gravity on $M$. At the core of this philosophy hibernates
the idea that the notion of spacetime---be it discrete or
continuous---must be retained at any cost, and that our methods of
calculation must be modified accordingly, as if all our
constructions must be tailor-cut to suit (or better, derive from)
a pre-existent background geometrical space(time).\footnote{In
spite of Einstein's serious doubts about the physical
meaningfulness of the concepts of space and time mentioned
earlier. Even more remarkably, in 4.2.2 we mentioned how Isham has
contemplated changing drastically the standard quantum theory
itself in order to suit non-continua spacetime backgrounds, such
as causal sets for example.}}
\end{quotation}

\item {\bf Passing through the initial singularity by ekpyrosis (new).} Together
with the interior Schwarzschild singularity, there is another one,
perhaps even more famous, which is a direct consequence of
Einstein's general theory of relativity, namely, the {\em initial
Big Bang singularity} marking the beginning of an expanding
Universe in the most successful of modern cosmological models. The
initial singularity, like the aforementioned interior
Schwarzschild one, is regarded as a fundamental, `true' spacetime
singularity and physics during the Planck epoch
($0$---$10^{-42}s$) is anticipated to be described consistently by
the ever elusive quantum theory of gravity. However, recently, in
the context of the string, membrane and M-theory approach to
quantum gravity, Khoury {\it et al.} have proposed a scenario
according to which one can actually evade the initial
singularity---as it were, do meaningful pre-Big Bang era physics
\cite{turok1,turok2,turok3}. Without going into any technical
details, we just note that their proposal basically involves a
(coordinate) field transformation,\footnote{Still assuming however
$\smooth$-smoothness for the various fields involved ({\it ie},
$\struc\equiv\smooth_{X}$ in our language; where $X$ is a
higher-dimensional differential manifold, {\it eg}, a Riemann
hypersurface).} completely analogous to Finkelstein's frame change
in \cite{df},\footnote{Neil Turok in private communication
\cite{turok}.} which enables one to go through the initial
singularity as if it was a diaphanous membrane. Thus, even the
most robust and least doubted singularity of all, the Big Bang,
has been shown (again, simply by the use of $\smooth$-means!) to
be no problem, no pathology of Nature at all, and that a rich
physics is to be discovered even for the period `before time
began'.
\end{itemize}

\subsection{The Relativity of Differentiability}

In connection with our brief remarks on $\smooth$-singularities
above, we wish to close the present paper with further remarks on
the opening two quotations of Feynman and Isham. In particular, in
line with the discussion of `gravity as a gauge theory' in section
3, we would like to emphasize that,

\begin{enumerate}

\item while we share Feynman's scepticism about the metric-formulation
of general relativity\footnote{After all, {\em the metric, as well
as the space hosting it, are our own ascriptions to Physis; they
are not Nature's own} (recall Einstein's quotation \cite{einst2}
in 5.1.1). ADG emphasizes that the $\struc$-metric $\rho$, as the
term suggests, is crucially dependent on our own measurements or
`generalized arithmetics' in $\struc$, so that, like the
singularities of the previous subsection, it is not Nature's own
property: {\em we ascribe it to Her!} (See footnote 17.) This is
in line with quantum theory's basic algebraico-operationalist
philosophy (and goes against the Platonic realist ideal of
classical physics) according to which, {\em quantum systems do not
possess physical properties of their own, that is, independently
of our acts of observing them. These acts, in turn, can be
suitably organized into algebras of physical operations,
generalized `measurements' so to speak, on the quantum system}.}
and his hunch that there is a fundamental gauge invariance lurking
there,

\item we do not share his apparently `negative' stance towards differential
geometry. Of course, his position is understandable to the extent
that he is referring to (and he is actually referring to!) the
usual calculus on $\smooth$-manifolds, but this is precisely the
point of ADG: {\em one should not question the `differential
mechanism' {\it per se} when encountering singularities,
infinities and other pathologies in classical differential
geometry}. For, loosely speaking, `{\em the mechanism is fine}',
{\em as it works, that is, as one can actually do differential
geometry in principle over any space, no matter how singular}.
Rather, {\em one should question the $\smooth$-smooth manifold $M$
itself whose only operative role in the said `differential
mechanism' is to provide us with the algebras (by no means unique
or `preferred' in any sense\footnote{See the {\em principle of
relativity of differentiability} to follow shortly.}) $\smooth(M)$
of infinitely differentiable functions (and the classical
differential geometric mechanism supported by them)} which, in
turn, {\em are the very hosts of the aforementioned singularities
and the other `classical differential geometric diseases'}.

Since Feynman's stance appears to accord with
Isham's,\footnote{See the two quotations opening the paper.} our
reply to the latter is similar; expressed somewhat differently,

\item we seem to be misled by the classical theory---the $\smooth$-differential
geometry---into thinking that the various `differential geometric
pathologies' are faults and shortcomings of the differential
mechanism, thus also infer that {\em differential geometry does
not apply in the quantum deep}. As noted earlier, it is perhaps
habit or long-time familiarity with smooth manifolds and their
numerous successful applications to physics, including general
relativity and the quantum field theories of matter, that makes us
think so,\footnote{See quotation of Einstein concluding the paper
below. At this point, to give an indication of this
attitude---{\it ie}, of the persistent, almost `religious'
adherence of some physicists to the spacetime manifold---we may
recall Hawking's opening words in \cite{hawk} where he discusses
singularities in general relativity {\it vis-\`a-vis} quantum
gravity: ``{\small\em ...Although there have been suggestions that
spacetime may have a discrete structure, I see no reason to
abandon the continuum theories that have been so successful.
General relativity is a beautiful theory that agrees with every
observation that has been made. It may require modifications on
the Planck scale, but I don't think that will affect many of the
predictions that can be obtained from it...}'' This appears to be
the {\em manifold-conservative} stance against singularities and
quantum gravity {\it par excellence}.} for ADG has shown us that
{\em the differential mechanism still applies effectively over any
space---even over ones that are much more singular (in a very
straightforward, but technical, sense)}
\cite{malros1,malros2,ros}, {\em or even over ones that are
manifestly discontinuous and more quantal} \cite{malrap1,malrap2},
{\em than the `featureless' differential manifold}. On the other
hand, ADG has also shown us that the `differential diseases' are
exactly due to our assuming up-front a differential manifold
background space to support our differential geometric
constructions, thus agreeing in that sense with Feynman and Isham.
However, in contradistinction to them,

\begin{quotation}
\noindent{\em in view of ADG, one does not need the differential
manifold in order to differentiate.}
\end{quotation}

\noindent All in all, ADG suggests that

\medskip

\begin{quotation}
\noindent {\em to heal the differential pathologies, one must
first kick the $\smooth$-smooth manifold habit.}
\end{quotation}

\medskip

\noindent Thus, continuing the `sloganeering' with which we
concluded \cite{malrap2}\footnote{Especially, see slogan 2 there.}
and expressed slogans 1--3 in the present paper, we may distill
the remarks above to the following `{\em relativity of
differentiability'} principle:

\item {\em The differential spacetime manifold by no means sets a
preferred ({\it ie}, unique) frame ({\it ie}, model) for
differentiating physical quantities. Differential equations,
modelling physical laws that obey the generalized principle of
locality,\footnote{Which maintains that physical laws should be
modelled after differential equations that depict the
cause-and-effect nexus between `infinitesimally' or `smoothly
separated' (`$\smooth$-contiguous') events---arguably what one
understands by `differential locality' ({\it ie}, local causality
in the $\smooth$-smooth spacetime manifold)
\cite{malrap1,rapzap2}.} can be also set up independently of the
$\smooth$-smooth manifold---in fact, as we saw in this paper,
regardless of any background (base) space(time)}. Since we have
repeatedly argued and witnessed in this paper that {\em
differentiability derives from the stalk} ({\it ie}, from the
algebraic objects dwelling in the relevant sheaves) and {\em not
from the underlying space(time)}, we may say that the `absolute'
and fixed differentiability of the smooth spacetime manifold,
which for Einstein represented the last relic of an inert,
`dynamically indifferent' ether-like substance
\cite{einst5,einst1} that ``{\em acts, but is not acted upon}''
\cite{einst3},\footnote{More precisely, Einstein's doubts about
the physical reality of the absolute, dynamically passive
spacetime continuum of the (special) theory of relativity were
expressed in \cite{einst3} (p. 55) as follows: ``{\em ...In this
latter statement {\rm [{\it ie}, that from the standpoint of
special relativity {\sl continuum spatii et temporis est
absolutum}]} {\sl absolutum} means not only `physically real', but
also `independent of its physical properties, having a physical
effect, but not itself influenced by physical conditions'...}''
Indeed so, in the special theory of relativity the metrical
properties of the spacetime continuum were not relativized, so
that the metric was not regarded as a dynamical variable. The
general theory of relativity viewed the metric---`the field of
locality' (local causality or local chronology)---as a dynamical
variable and effectively evaded the aforesaid `{\sl temporis est
absolutum}', but it must again be emphasized here that general
relativity in a sense came short of fully relativizing ({\it ie},
regarding as dynamical variables) the whole panoply of structures
(or `properties' in Einstein's words above) that the spacetime
continuum comes equipped with. For instance, the continuum's
structures which are arguably `deeper' than the metrical, such as
the topological and the differential, are simply left absolute,
non-relativized (non-dynamical), `fixed by the theorist once and
forever as the differential manifold background'. As noted
repeatedly earlier and in previous works
\cite{rapzap1,rapzap2,malrap1,malrap2}, in a genuinely (fully)
quantum theoresis of spacetime structure and dynamics even the
topological and the differential structures are expected to be
subjected to relativization and dynamical variability---thus {\em
become `observables', `in principle measurable' dynamical
entities}. For it has been extensively argued that {\em the common
denominator of both relativity (`relativization') and the quantum
(`quantization') is dynamics (`dynamical variation')} \cite{df1}.
So that ``{\em all is quantum}'' (see footnote 3) means
essentially that ``{\em all is dynamical}''. But then, if
everything is in constant flux in the quantum deep, {\em whence
space?}, and, {\it mutatis mutandis}, {\em whence time?} Totally,
{\em is there any spacetime at all?}, and even more doubtfully,
{\em whence the spacetime manifold?}.} `{\em relativized}' with
respect to the algebraic objects that live on whatever
`spacetime'\footnote{The inverted commas over `spacetime' remind
one of the physically dubious (especially at Planck scale)
significance of this concept.} we have used as a base space
`scaffolding' to localize sheaf-theoretically those physically
significant algebraic objects. We may figuratively refer to the
abstract algebraico-sheaf-theoretic differentiability properties
(of the system `quantum spacetime'---or better, of the very
dynamical quanta in which that `spacetime' is inherent) as `{\em
differentiables}', in analogy to the standard algebraically
represented `observables' or even the `beables' of the usual
(material) quantum physical systems. Thus, to wrap things up,

\begin{quotation}
\noindent {\em `Differentiables' are properties of ({\it ie},
derive from) the algebraic structure of the objects (sections of
algebra sheaves) that live on `space(time)', not from
`space(time)' itself which, especially in its classical
$\smooth$-smooth manifold guise, is doubtful whether it has any
physical significance at all}
\cite{ish2,ish1,rapzap1,malrap1,buttish,rapzap2,malrap2,mall7,ish3}.
\end{quotation}

\noindent However, since we have repeatedly quoted above
Einstein's doubts about the smooth geometric spacetime continuum
{\it vis-\`a-vis} singularities and the quantum, we would like to
end the paper with another telling quotation of his which
sensitizes us to the fact that successful, therefore {\it a
priori} assumed and habitually or uncritically applied,
theoretical concepts and mathematical structures,\footnote{Like,
in our case, the $\smooth$-smooth spacetime manifold and its
anomalies. As we also mentioned in 4.2.2, in \cite{buttish,ish3}
too the dangers of assuming {\it a priori} the spacetime continuum
in our excursions into the quantum deep are explicitly
pronounced.} can exercize so much power on us that they often mask
their true origin and pragmatic usefulness---{\it ie}, that {\em
they simply are our own theoretical constructs of limited
applicability and validity}---and mislead us into thinking that
they are `unavoidable necessities' and, what's worse, Nature's own
traits:

\begin{quotation}
\noindent ``{\small\em ...Concepts which have proved useful for
ordering things easily assume so great an authority over us, that
we forget their terrestrial origin and accept them as unalterable
facts. They then become labelled as `conceptual necessities', `a
priori situations', etc. The road of scientific progress is
frequently blocked for long periods by such errors. It is
therefore not just an idle game to exercise our ability to analyse
familiar concepts, and to demonstrate the conditions on which
their justification and usefulness depend, and the way in which
these developed, little by little...}'' (1916) \cite{einst7}
\end{quotation}

\end{enumerate}

\section*{Acknowledgments}

The second author (IR) is indebted to Chris Isham for numerous
discussions on quantum gravity, but more importantly, for his
unceasing moral encouragement and support. He would also like to
acknowledge helpful exchanges on finitary posets, simplicial
complexes and causal sets with Raquel Garcia and Yousef
Ghazi-Tabatabai, and to express his thanks to the European Union
for funding this work via a Marie Curie postdoctoral research
fellowship held at Imperial College.

\end{document}